Dissertation

Davood Momeni

# Optimization of epitaxial Graphene Growth for Quantum Metrology

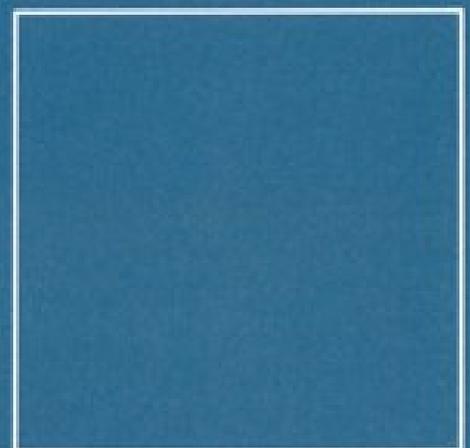

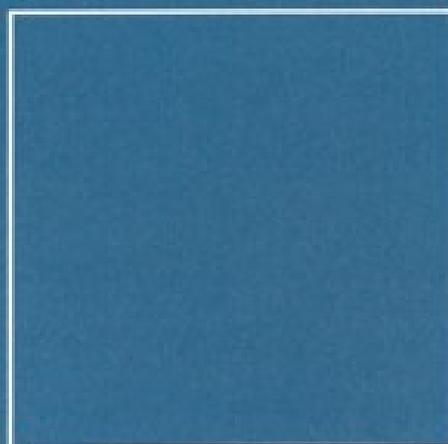

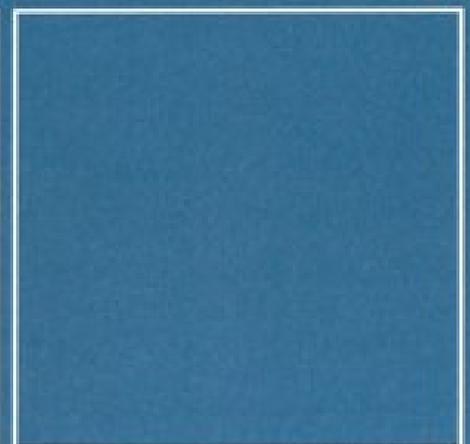

# Optimization of Epitaxial Graphene Growth for Quantum Metrology

Der QUEST-Leibniz-Forschungsschule

der Gottfried Wilhelm Leibniz Universität Hannover

zur Erlangung des akademischen Grades

Doktor der Naturwissenschaften

Dr. rer. nat.

genehmigte Dissertation von

## M.Sc. Davood Momeni Pakdehi

2020

# Physikalisch-Technische Bundesanstalt

Elektrizität
PTB-E-117

Davood Momeni

# Optimization of Epitaxial Graphene Growth for Quantum Metrology

# Optimization of Epitaxial Graphene Growth for Quantum Metrology

Referent: PD Dr. rer. nat. Hans W. Schumacher
*Physikalisch-Technische Bundesanstalt, Braunschweig*

Korreferent: Prof. Dr. rer. nat. Rolf Haug
*Gottfried Wilhelm Leibniz Universität Hannover*

Korreferent: Prof. Dr. Christoph Tegenkamp
*Technische Universität Chemnitz*

# Abstract

The electrical quantum standards have played a decisive role in modern metrology, particularly since the introduction of the revised International System of Units (SI) in May 2019. By adapting the basic units to exactly defined natural constants, the quantized Hall resistance (QHR) standards are also given precisely. The Von Klitzing constant $R_K = h/e^2$ ($h$ Planck's constant and $e$ elementary charge) can be measured precisely using the quantum Hall effect (QHE) and is thus the primary representation of the ohm. Currently, the QHR standard based on GaAs/AlGaAs heterostructure has succeeded in yielding robust resistance measurements with high accuracy $<10^{-9}$.

In recent years, graphene has been vastly investigated due to its potential in QHR metrology. This single-layer hexagonal carbon crystal forms a two-dimensional electron gas system and exhibits the QHE, due to its properties, even at higher temperatures. Thereby, in the future the QHR standards could be realized in more simplified experimental conditions that can be used at higher temperatures and currents as well as smaller magnetic fields than is feasible in conventional GaAs/AlGaAs QHR.

The quality of the graphene is of significant importance to the QHR standards application. The epitaxial graphene growth on silicon carbide (SiC) offers decisive advantages among the known fabrication methods. It enables the production of large-area graphene layers that are already electron-doped and do not have to be transferred to another substrate. However, there are fundamental challenges in epitaxial graphene growth. During the high-temperature growth process, the steps on the SiC surface bunch together and form terraces with high steps. This so-called step-bunching gives rise to the graphene thickness inhomogeneity (e.g., the bilayer formation) and extrinsic resistance anisotropy, which both deteriorate the performance of electronic devices made from it.

In this thesis, the process conditions of the epitaxial graphene growth through a so-called polymer-assisted sublimation growth method are minutely investigated. Atomic force microscopy (AFM) is used to show that the previously neglected flow-rate of the argon process gas has a significant influence on the morphology of the SiC substrate and atop carbon layers. The results can be well explained using a simple model for the thermodynamic conditions at the layer adjacent to the surface. The resulting control option of step-bunching on the sub-nanometer scales is used to produce the ultra-flat, monolayer graphene layers without the bilayer inclusions that exhibit the vanishing of the resistance anisotropy. The comparison of four-point and scanning tunneling potentiometry measurements shows that the remaining small anisotropy represents the ultimate limit, which is given solely by the remaining resistances at the SiC terrace steps.

Thanks to the advanced growth control, also large-area homogenous quasi-freestanding monolayer and bilayer graphene sheets are fabricated. The Raman spectroscopy and scanning tunneling microscopy reveal very low defect densities of the layers. In addition, the excellent quality of the produced freestanding layers is further evidenced by the four-point measurement showing low extrinsic resistance anisotropy in both micro- and millimeter-scales.

The precise control of step-bunching using the Ar flow also enables the preparation of periodic non-identical SiC surfaces under the graphene layer. Based on the work function measurements by Kelvin-Probe force microscopy and X-ray photoemission electron microscopy, it is shown for the first time that there is a doping variation in graphene, induced by a proximity effect of the different near-surface SiC stacks. The comparison of the AFM and low-energy electron microscopy measurements have enabled the exact assignment of the SiC stacks, and the examinations have led to an improved understanding of the surface restructuring in the framework of a step-flow model.

The knowledge gained can be further utilized to improve the performance of epitaxial graphene quantum resistance standard, and overall, the graphene-based electronic devices. Finally, the QHR measurements have been shown on the optimized graphene monolayers. In order to operate the graphene-based QHR at desirably low magnetic field ranges ($B < 5\,\text{T}$), two known charge tuning techniques are applied, and the results are discussed with a view to their further implementation in the QHR metrology.

# Kurzdarstellung

Elektrische Quantennormale spielen eine wichtige Rolle in der modernen Metrologie, besonders seit der Einführung des revidierten Einheitensystems (SI) im Mai 2019. Durch die Zurückführung der Basiseinheiten auf exakt definierte Naturkonstanten sind auch die quantisierten Werte von Widerstandsnormalen (QHR) exakt gegeben. Die Von-Klitzing-Konstante $R_K = h/e^2$ ($h$ Planck-Konstante und $e$ Elementarladung) lässt sich mittels des Quanten-Hall-Effekts (QHE) präzise messen und ist somit die primäre Darstellung des Ohm. Die Quanten-Widerstandsnormale bestehen aktuell aus robusten GaAs/AlGaAs-Heterostrukturen, die eine Genauigkeit $<10^{-9}$ für die Widerstands-Messung erlauben.

In den letzten Jahren wird verstärkt Graphen auf sein Potenzial für die Widerstandmetrologie untersucht. Der einlagige hexagonale Kohlenstoffkristall bildet ebenfalls ein zweidimensionales Elektrongas aus, das den Quanten-Hall-Effekt zeigt – und dies auf Grund seiner Eigenschaften schon bei höheren Temperaturen. Damit könnten in Zukunft Widerstandsnormale für vereinfachte experimentelle Bedingungen realisiert werden, die bei höheren Temperaturen und Strömen oder kleineren Magnetfeldern eingesetzt werden können, als es mit konventionellen GaAs/AlGaAs-QHR möglich ist.

Für den Einsatz als Widerstandsnormal ist die Qualität des Graphens von entscheidender Bedeutung. Unter den bekannten Herstellungsmethoden bietet das epitaktische Wachstum von Graphen auf Siliciumcarbid (SiC) entscheidende Vorteile. Es lassen sich damit großflächige Graphenschichten herstellen, die nicht auf ein anderes Substrat übertragen werden müssen. Allerdings gibt es grundlegende Herausforderungen beim epitaktischen Wachstum. So tritt bei hohen Prozesstemperaturen eine Bündelung der Kristallstufen auf der SiC-Substratoberfläche auf (Step-bunching), was zu einer bekannten extrinsischen Widerstandsanisotropie führt und darüber hinaus die Bildung von Bilagen-Graphen begünstigt. Beides verschlechtert die Eigenschaften der daraus hergestellten Widerstandsnormale.

In dieser Dissertation werden zunächst die Prozessbedingungen des mittels der sogenannten Polymer-Assisted-Sublimations-Growth-Methode hergestellten epitaktischen Graphens auf SiC genauer untersucht. Mithilfe der Rasterkraft-Mikroskopie (Atomic-Force-Microscopy, AFM) wird gezeigt, dass es einen erheblichen Einfluss der bisher wenig beachteten Flussrate des Prozessgases Argon auf die Morphologie des SiC-Substrates und der oberen Kohlenstoffschichten gibt. Anhand eines einfachen Modells für die thermodynamischen Verhältnisse in einer oberflächennahen Schicht lassen sich die Ergebnisse hervorragend erklären. Die sich daraus ergebende Kontrollmöglichkeit des Step-bunching auf Sub-Nanometer-Skalen wird genutzt, um ultraflache, monolagige Graphenschichten ohne Bilageneinschlüsse herzustellen, die eine verschwindende Widerstandsanisotropie aufweisen. Der Vergleich von Vierpunkt-Messungen und Scanning-Tunneling-Potentiometery-Messungen zeigt, dass die verbleibende geringe Anisotropie das ultimative Limit darstellt, die allein durch die verbleibenden Widerstände an den SiC-Terrassenstufen gegeben ist.

Dank der fortschrittlichen Wachstumskontrolle werden auch großflächige, homogene quasi-freistehende Monolage- und Bilage-Graphenschichten hergestellt. Die Raman-Spektroskopie und die Rastertunnel-Mikroskopie zeigen sehr geringe Defektdichten der Schichten. Darüber hinaus wird die hervorragende Qualität der hergestellten quasi-freistehenden Schichten durch die Vierpunkt-Messung unter Beweis gestellt, die eine geringe extrinsische Widerstandsanisotropie zeigt.

Die präzise Kontrolle des Step-bunching mittels Ar-Fluss ermöglicht auch die gezielte Präparation von periodischen, nicht-identischen SiC-Oberflächen unter der Graphenlage. Anhand von Messungen der Austrittsarbeit mit Kelvin-Probe-Force-Microscopy und X-ray Photoemission-Electron-Microscopy konnte erstmals gezeigt werden, dass es eine Variation der Graphendotierung, induziert durch einen Proximity Effekt der unterschiedlichen oberflächennahen SiC-Stapel, gibt. Der Vergleich von AFM und Low-Energy-Electron-Microscopy-Messungen ermöglicht die genaue Zuordnung der SiC-Stapel und die Untersuchungen führen insgesamt zu einem verbesserten Verständnis der Oberflächen-Umstrukturierung im Rahmen eines adäquaten Step-Flow-Modells.

Die gesammelten Erkenntnisse können zur Verbesserung der Eigenschaften von Graphen-Quantennormalen und auch allgemein von graphenbasierten Bauteilen genutzt werden. Abschließend werden QH-Widerstandsmessungen an optimierten Graphen-Monolagen gezeigt. Um den Magnetfeldbereich ($B < 5\,\mathrm{T}$) einzuschränken, werden zwei bekannte extrinsische Dotiertechniken verwendet und die Ergebnisse werden im Hinblick auf den weiteren Einsatz in der QH-Metrologie diskutiert.

**Schlüsselwörter:** Wachstum des epitaktischen Graphens, Siliciumcarbid, Argon-Flussrate, Widerstandsanisotropie, homogenes quasi-freistehendes Graphen

# Contents

# 1

## 1. Introduction

### Abstract

*T*his chapter covers a brief overview, including the short historical remarks, primary goals/challenges, and the structure of this thesis.

## 1.1. Historical background

The graphene discovery may go back to 1859 when Benjamin Brodie discovered a new form of carbon called "graphon" as the material we know as graphene oxide nowadays. [1] "Die Dünnste Kohlenstoff-Folien" literally meaning "the thinnest carbon layers," was the title of the paper published in 1962 by Boehm et al., who indeed looked at the single carbon layers. [2,3] For a single carbon layer of the graphitic structure, Boehm et al. in 1986 coined the term "graphene" derived from the first part of the graphite and the ending "ene" that refers to polycyclic aromatic hydrocarbons. [4]

On the other hand, in 1896, the first graphitization of silicon carbide was patented by E. G. Acheson. [5] This corresponds to the period that X-ray was discovered, which boosted delving into the details of crystal and surfaces in various materials. Importantly, the X-ray was used to characterize the graphite obtained by the thermal decomposition of SiC much later in 1965 by D. V. Badami. This happened almost a quarter-century after that P. R. Wallace [6] introduced the graphite's band structure. In his study, D. V. Badami identified epitaxial graphene on SiC and reported the relationship between the crystal orientation of graphite and underneath 6H-SiC. [7]

Ten years later, in 1975, A. J. Van Bommel et al. investigated the graphite on SiC (obtained by Si sublimation of SiC (0001) in high-vacuum) using the low-energy electron diffraction (LEED), which revealed a $(6\sqrt{3} \times 6\sqrt{3})R30°$ surface reconstruction, the so-called buffer layer. [8] The first intercalation (i.e., penetration of small species in, e.g., graphite layers) and separation (i.e., exfoliation) of graphite flakes were carried out by Schafhaeutl back in 1840. [9] The unique electronic properties of such flakes, the so-called freestanding layers, were not discovered yet at that period. The high-resolution transmission electron microscopy (HRTEM) could directly visualize a thin graphene layer on the Si-face of SiC in contrast to the SiC C-face. [10] A breakthrough happened in 2004 in a study carried out by C. Berger et al., in which two-dimensional electron gas properties of ultrathin epitaxial graphite were reported and led to an enormous expansion in this area. [11] The groundbreaking isolation of the graphene flakes, using mechanical exfoliation of highly ordered pyrolytic graphite (HOPG), was introduced by A. Geim and K. Novaselov in 2004. [12,13]

Even though the exfoliation of graphene from graphite was an easy-to-use technique and led to numerous studies demonstrating unique features of graphene [14,15], the main drawbacks have been complicated thickness control, reproducibility, and small flake sizes. These all have been the objective of further research considering the epitaxial graphene growth on SiC as a method with the potential to attain the large-area graphene. In about 2008, improved graphene layers were obtained by ex-situ (i.e., in argon ambient) growth instead of in-situ, which latter is known to result in rough graphene formation. [16,17] A year later, the so-called quasi-free-standing monolayer graphene (QFMLG) was fabricated, in which the buffer layer $(6\sqrt{3} \times 6\sqrt{3})R30°$ turned to monolayer graphene by annealing in a hydrogen atmosphere. [18]

A peculiar application of epitaxial graphene, the quantum resistance standard, was introduced by A. Tzalenchuk et al. in 2010. [19] This work inspired the metrology working groups to use graphene, e.g., as a replacement for the GaAs heterostructure for the QHR metrology purposes.

In the same year, in several studies, the graphene growth mechanism on SiC using the HRTEM and the buffer layer's atomic structure using scanning tunneling microscopy (STM) was revealed. [20,21] Using the angle-resolved photoemission electron microscopy (ARPES) electronic band structure of single and bilayer graphene was reported. [22] H. Hibino et al., by using the low-energy electron microscopy (LEEM) method, successfully counted the number of carbon layers on SiC. [23] The 100 GHz graphene-based transistors and later operation of wafer-scale graphene integrated circuits were reported by Y-M. Lin et. al. [24,25].

In 2011, graphene was produced on SiC using the CVD method by W. Strupinski et al. [26], following the earlier efforts of CVD graphene synthesis on other substrates. [27–29] The origin of doping, $n$-type in epitaxial graphene, and $p$-type after hydrogen intercalation (QFMLG) was explained by J. Ristein et al. [30,31]. Also, there were several attempts to transfer epitaxially grown graphene on other substrates. [32,33] The epitaxial graphene nanoribbons were reported as ballistic conductors by J. Baringhaus et al. [34]. A significant improvement in adlayer-free and sizeable graphene fabrication has been achieved through both the CVD [35] and epitaxial graphene growth [36–38] methods, and the latter has resulted in almost vanishing of a so-called resistance anisotropy induced by SiC substrate. [38,39] In 2014, epitaxial graphene was used as the basis for the growth of transition metal dichalcogenides (TMDs). [40]

In pursuit of growing other low-dimensional materials, epitaxial graphene provides an excellent platform, as shown by several recent studies. [41–45] This makes epigraphene a great material of choice for various theoretical and practical applications, as for metrological purposes, which is the primary focus of this thesis.

## 1.2. Thesis objective and structure

For the realization of robust and reliable performance of the graphene-based quantum Hall resistance (QHR) metrology, high-quality graphene samples are demanding. This thesis aims to investigate and understand the epitaxial graphene growth on silicon carbide, in order to improve the quality for a sizeable, reproducible, and pure monolayer graphene synthesis. This study further explores the fabrication of different graphene types and employs various characterization analysis to scrutinize both locally and macro-scales the samples' quality.

After this introductory in Chapter **1**, a brief overview of the graphene's fundamental properties, the mechanism of epitaxial graphene growth on SiC, the crystal structure of hexagonal SiC polytype, as well as the basics of the quantum Hall effect in graphene will be given in Chapter **2**. It also includes an introduction to a so-called buffer layer and intercalation technique for the fabrication of the so-called freestanding graphene layers.

Chapter **3** provides a short description of the characterization techniques employed in this thesis.

Chapter **4** presents the sample preparation, cleaning, different growth recipes, and microfabrication processes.

In Chapter **5**, the fabrication results of diverse graphene types including buffer layer, monolayer graphene, bilayer graphene, as well as freestanding-monolayer and bilayer graphene are discussed in detail. The so far neglected argon flow-rate and its impact on growth are described within a quasi-thermal equilibrium model supported by several characterization methods. Also, in a multi-perspective study, the influence of the miscut angle, as well as the SiC polytype are investigated.

Chapter **6** covers a systematic study of a so-called extrinsic resistance anisotropy of epitaxial monolayer, QFMLG, and QFBLG samples. An almost negligible resistance anisotropy of about 3% underlines the high-quality of the produced samples. This study importantly shows the minimum resistance anisotropy achievable in epigraphene on SiC. Moreover, not merely the step regions but also the terrace areas are investigated regarding the local resistance characteristics.

Chapter **7** reveals a special interaction between graphene and SiC terminations. The presented advanced graphene growth allows creating the self-patterned graphene layers on either identical or non-identical SiC terminations. This provides a novel playground to study the interplay of the SiC terminations and top carbon layers as the central part of the Chapter **7**. Furthermore, the SiC stacking terminations are identified and assigned within the framework of a so-called joint-cubic-hexagonal (JCH) step-flow model that describes the step-retraction SiC surfaces.

Chapter **8** presents the magneto-transport results in epigraphene with respect to the metrological purpose of the graphene-based quantum Hall resistance standard. The environmental influences, doping, encapsulation/ isolation, and charge tuning using different techniques are discussed and evaluated.

Finally, a brief overview of the results, conclusions, and outlooks are given in Chapter **9**.

# 2. Theory and concepts

**Abstract**

*T*his chapter presents a literature review and provides the supporting information for the discussions in the following chapters. The crystal structure of silicon carbide and graphene, growth methods, and mechanism, as well as the quantum Hall effect measurements, are briefly discussed.

## 2.1. Crystal and electronic properties

Graphene is a material that integrates several excellent electronic, optical, and thermal properties. This is so far unique to graphene because, for instance, one can find materials that are as conducting or transparent as graphene, but one can scarcely find any materials that are simultaneously electrically conducting, optically transparent (with a wide range of the optical spectrum from infrared to ultraviolet) and mechanically flexible. Graphene is an allotrope of carbon, which is the 4th most abundant element in nature. Moreover, it is low-cost, eco-friendly, sustainable, and importantly its 2-dimensional structure makes it compatible with the traditional semiconductor processing planar structure and can be practically integrated into the existing mainstream technology. [15]

The unit cell of graphene composed of two carbon atoms with $sp^2$-hybridization. This results in a honeycomb structure with three $sp^2$-orbitals located at an angle of 120° relating to each other. The carbon atoms in graphene have four electrons in their valance band (electron configurations: [He] $2s^2 2p^2$). Three of these electrons– one electron from s-orbital and two electrons from p-orbital ($p_x$ and $p_y$)–form three $sp^2$ hybridized planar orbitals and connect to their adjacent carbon atoms via σ-bonds. The fourth electron from unaffected p-orbital ($p_z$, which is perpendicular to the planar structure) covalently binds to the neighboring carbon atom with an out-of-plane π-binding, see **Figure 2.1a**. The strong σ-bonds makes graphene one of the most mechanically stable materials. Due to the Pauli principle, these bands have a filled shell and thus form a deep valence band. [46] In the case of π-band, since each p-orbital has one extra electron, the π band is half-filled. Half-filled bands are important as they lead to several interesting electronic properties, e.g., in transition elements. The π-states are also responsible for the electrical conductivity in graphene and interlayer coupling in graphite. [47] Importantly, delocalized electrons in its π-system are responsible for the excellent electronic properties. [15,48,49] **Figure 2.1b** depicts the hexagonal structure of graphene with a basis of two atoms per unit cell (marked as A and B). The unit cell vectors ($\vec{a}_1$, $\vec{a}_2$) can be written as:

$$\vec{a}_1 = \frac{a_{C-C}}{2}\left(3, \sqrt{3}\right), \quad \vec{a}_2 = \frac{a_{C-C}}{2}\left(3, -\sqrt{3}\right) \tag{2-1}$$

where $a_{C-C} \approx 1.42$ Å is the carbon-carbon distance, which results in a lattice constant of $a = 2.46$ Å. [48,49]

**Figure 2.1c** demonstrates the reciprocal lattice of graphene with reciprocal lattice vectors, which are given by:

$$\vec{b}_1 = \frac{2\pi}{3a_{C-C}}\left(1, \sqrt{3}\right), \quad \vec{b}_2 = \frac{2\pi}{3a_{C-C}}\left(1, -\sqrt{3}\right) \tag{2-2}$$

**Figure 2.1. Crystal and band structure of graphene.**

(a) σ-bond and π-bond formed by sp² hybridization.

(b) The crystal lattice of graphene, with two equivalent sublattices A (red) and B (blue) of the two carbon atoms in the unit cell (gray rhombus). Lattice vectors are $\vec{a}_1$ and $\vec{a}_2$.

(c) Schematic of reciprocal lattice of graphene with the reciprocal lattice vectors $\vec{b}_1$ and $\vec{b}_2$, the Brillouin zone (marked cyan area), and the high symmetry points $\Gamma$, $K$, and $K'$, and $M$.

(d) E-k relation showing linear dispersion in graphene at Dirac cones. The conductance band (green-shaded) touches the valence (red-shaded) band at the $K$ and $K'$ points.

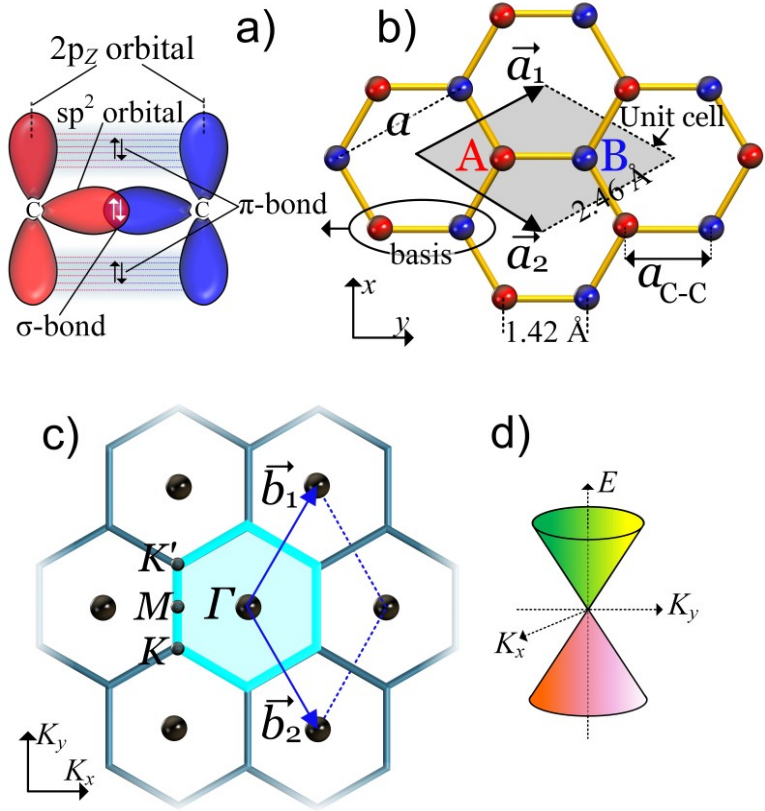

In 1947 P. R. Wallace first studied the unusual semi-metallic behavior of graphene and introduced an analytical expression for the dispersion relation $E(\vec{k})$ for graphene. [6] $K$ and $K'$ points (so-called Dirac points), located at the corners of the Brillouin zone of graphene, shown in **Figure 2.1c**. These two points are characteristics for the physics of graphene and have values of:

$$K = \frac{2\pi}{3a_{C-C}} \left(1, \frac{1}{\sqrt{3}}\right), \;\; K' = \frac{2\pi}{3a_{C-C}} \left(1, \frac{1}{\sqrt{3}}\right) \tag{2-3}$$

At the Dirac points, the graphene's band spectrum is closely similar to the Dirac spectrum for massless fermions. **Figure 2.1c** shows the graphene's conduction and valance bands touch each other at the $K$ and $K'$ points. Thus, graphene can be considered as a gapless semiconductor. Accordingly, the energy band dispersion relation in graphene shows a linear feature, see **Figure 2.1d**, where the energy dispersion $E \propto k$, is in strong contrast to $E \propto k^2$ as in the conventional semiconductors or metals. Considering the tight-binding Hamiltonian model for electrons in graphene, they can hop to both the nearest- and next-nearest-neighbor atoms. [50]

All that so far were briefly discussed are associated with the properties of single-layer graphene. When two or more graphene layers stack on top of each other, they form so-called bilayer (BLG) or few-layer (FLG) graphene, respectively. The interaction between the layers drastically changes the properties of carbon layers in the stack compared to that in a single layer. For instance, in BLG, a band splitting occurs, and the linear dispersion is lost. [51] Another example is the unconventional quantum Hall effect (QHE) of BLG compared to the QHE in the monolayer graphene. [52,53] In general, the electronic band dispersion near the Fermi level, and consequently, the nature of the charge carriers, is highly sensitive to the number of layers and the stacking geometry. This will be explored in multiple aspects from electronic transport (e.g., resistance anisotropy) to magneto-transport (e.g., VdP and QHE) within this work.

## 2.2. Magneto-transport

There are two distinct known kinds of integer quantum Hall effect: (i) the conventional quantum Hall effect which occurs in 2D semiconductors (e.g., GaAs/AlGaAs heterostructure, for example) [54], and (ii) a counterpart QHE which is observed in monolayer graphene resulting in shifted positions of the Hall plateaus. The latter is related to the massless relativistic characteristic of charge carriers in graphene, which mimics Dirac fermions. [14,55,56] The third type of QHE was reported for bilayer graphene. [52] Perhaps, a fourth kind could also be possibly addressed as recently was shown in graphite with a thickness of hundreds of atomic layers. Although QHE in 3-dimensions (3D) is forbidden (destruction of the quantization due to spreads of Landau levels into overlapping bands in 3D), the measured QHE was attributed to a dimensional reduction of electron dynamics in high magnetic fields. [57] In the following, the abovementioned QHE types (i) and (ii) are briefly described.

**Figure 2.2. The Hall effect.**

In the presence of a magnetic field ($B$) perpendicular to the applied current ($I$), electrons move in circle (cyclotron motion, as illustrated in the upper-left sketch) from one side and accumulate at the other side (indicated by curved arrow) and built up the Hall voltage $V_{Hall}$. The situation is similar for holes, with the exception that the sign of the Hall voltage changes.

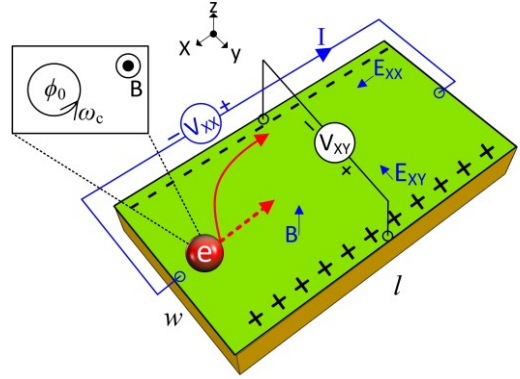

### 2.2.1. Integer quantum Hall effect

The classical Hall effect is observed when current flows through an electrical conductor in the presence of a perpendicular magnetic field. [58] The Hall effect arises from the fact that a magnetic field ($B$) causes charged particles with an effective mass of $m^*$ to move in a circle (cyclotron) with a fixed frequency equal to $\omega_c = eB/m^*$. A schematic of the Hall-measurement is depicted in **Figure 2.2**. Due to the Lorentz force, electrons are deviated and propagate to the edge of the sample. This leads to a transversal voltage $V_{xy} = V_{Hall}$, the so-called Hall-voltage. The Hall voltage depends on externally applied parameters (i.e., magnetic field $B$, and current $I$) and equals for one type of charge carrier in the implemented semiconductor (e.g., electrons) as follows:

$$V_{XY} = \frac{IB}{nte} \qquad (2\text{-}4)$$

where $n$ is three-dimensional charge carrier density, t is thickness, and $e$ is the electron charge. The Hall measurement allows calculating the charge carrier density of the material under investigation. From the Hall coefficient ($A_{Hall}$) in **(2-5)** the type of charge carrier can be identified; for the electron (holes), the sign of $A_{Hall}$ becomes negative (positive).

$$A_{Hall} = \frac{1}{ne} \qquad (2\text{-}5)$$

$$R_{XY} = \frac{V_{XY}}{I} = \frac{B}{n_S e} \qquad (2\text{-}6)$$

In the case of a two-dimensional electron gas; wherein the motion of the electrons is limited to the XY plane, the product $nt$ can be defined as a two-dimensional sheet carrier density $n_S$. This leads to the Hall resistance $R_H$ as written in equation **(2-6)**.

From the classical point of view, while longitudinal resistivity ($\rho_{XX}$) is constant, the transverse resistivity ($\rho_{XY}$) varies linearly with the external magnetic field $B$. This does not come true in the case of the quantum Hall effect, which is observed in 2DEG systems in a strong perpendicular magnetic field at low temperatures. Then oscillations (Shubnikov-De Haas (SdH) oscillations) in the longitudinal resistivity ($\rho_{XX}$) and so-called Hall plateaus in the transversal resistivity ($\rho_{XY}$) are observed. In 1980 Klaus von Klitzing found that the Hall resistivity plateaus show exact quantized values of fractions of $h/e^2$. [59] The longitudinal resistivity becomes vanishingly small ($\rho_{XX} \sim 0$) in that $B$ range at the plateaus, which means transport without dissipation (zero resistance). The quantized Hall resistance is calculated from the given equation as follows:

$$R_{XY} = \frac{V_{XY}}{I} = \frac{h}{\nu e^2} \tag{2-7}$$

where $I$ is the channel current, $V_{XY}$ is the Hall voltage, $h$ is the Planck's constant, $e$ is the elementary charge, and $\nu$ is known as "filling factor," which is an integer (1, 2, 3, …) (in integer quantum Hall effect–IQHE) that define the different plateaus. The center of each plateau occurs when the magnetic field takes the values given by **(2-8)**.

$$B = \frac{h n_s}{\nu e} = \frac{n_s}{\nu} \phi_0 \tag{2-8}$$

$$\phi_0 = \frac{h}{e} \tag{2-9}$$

$\phi_0$ is known as the flux quantum. The Hall resistivity should take the value **(2-7)** when $\nu$ Landau levels (LLs) are filled. This will be explained further in the following. As seen, $R_{XY}$ depends only on the fundamental constants proportional to $h/e^2$. [59,60] For this fundamental resistance, a new constant was defined and is known as the von Klitzing constant $R_K = h/e^2 = 25812.8074555\ \Omega$. [61]

The IQHE is explained in terms of single-particle orbitals of an electron in a magnetic field and is related to the Landau quantization. [54,62] In the presence of the magnetic field, the energy states are reconfigured to certain discrete energy levels known as Landau levels (LLs), which are the allowed energies for cyclotron orbits under quantization conditions (**Figure 2.3a, and b**). This means that when $\nu$ Landau levels are filled, there is a gap in the energy spectrum, so to occupy the next state it costs energy that is proportional to $\hbar\omega_c$ ($\omega_c$ is cyclotron frequency). LLs are formally calculated by solving the Schrödinger equation for free electrons in the presence of electric and magnetic fields. Accordingly, the problem reduces to that of a harmonic oscillator shifted by the magnetic length $l_B = \sqrt{\hbar/eB}$ ($l_B \sim$ 5-10 nm is the size of the cyclotron orbit and is independent of material parameters) with eigenvalues given by:

$$E_n = \hbar\omega_c(N + \frac{1}{2}) \qquad (2\text{-}10)$$

with the reduced Planck constant $\hbar = h/2\pi$, and $N$ an integer (zero included). [63,64]

When the Fermi level lies in a gap, the electronic scattering rate vanishes. Without scattering, the electrons cannot move along a direction perpendicular to both the electric and magnetic fields. Hence, the longitudinal resistance ($R_{XX}$) is zero, and Hall resistance is given by **(2-7)**. Theoretically, each LL is a heavily degenerate δ-function in the density of states (DOS) filled with localized states undergoing cyclotron motion. Therefore, from the equation **(2-7)**, it is expected that the quantization of resistance happens exactly when ν LLs are filled. Experimentally, nevertheless, even when the Fermi energy lies between two LLs, the resistance quantization (plateaus in the QHE regime) is still present. This is due to the broadening of LLs as a result of disorders existing in the system. [63,64] Disorders result in two different kinds of electronic states: localized and extended states at each LL. This situation is shown in **Figure 2.3a-b**.

The localized states do not carry current, while the extended states do. Although the existence of an impurity reduces the number of states which can carry the current, however, in an oversimplified model, it can be assumed that the electrons passing by an impurity will speed up to increase the current to compensate exactly the deficit of current due to the presence of the localized state. [63]

In practice, the realization of QHE can be achieved either by (i) flowing a constant current (fixed Fermi level, i.e., fixed carrier density) while varying the magnetic field or (ii) by varying the carrier density via a gate in a fixed magnetic field. In case (i) by increasing the magnetic field, the energy spacing between LLs enhances, and because the $E_F$ is fixed, higher LLs are emptied, and as each one is emptied, the $R_{XY}$ increases.

In case (ii) by increasing the electron density, the different electronics states are gradually filled up, which is equivalent to shifting the Fermi energy $E_F$ through the DOS. When $E_F$ moves in a mobility gap (localized states), the occupation of the extended states does not change, and since only these states carry the current, the Hall resistance will not change either, leading to an extension of Hall plateau. When $E_F$ reaches the next Landau level, dissipation occurs in the system, and the Hall resistance goes to the next plateau. Hence, the QHE can be regarded as a successive localization–delocalization transitions when the Fermi energy $E_F$ moves through the DOS. Peaks in $R_{XX}$ are observed each time the Fermi level crosses the center of an LL, and the plateaus in $R_{XY}$ and vanishing $R_{XX}$ are observed whenever the Fermi level lies in between the LLs. [65]

The steps in $R_{XY}$ in the QHE occurs when the LLs filled with the localized carriers undergo the cyclotron motion. As carriers undergo cyclotron motion, they include quantum of magnetic flux, thus the number of localized carriers per unit area ($n_{LL}$) can be calculated as written in **(2-11)**. This is indeed the density of electrons required to fill $\nu$ Landau levels. Thus, by knowing the $n_{LL}$, the number of completely filled LLs, i.e., the filling factor $\nu$ can be found by dividing the total density of electrons in the system ($n_S$) by the number of localized carriers, ($n_{LL}$), as written in **(2-12)**. [64]

$$n_{\mathrm{LL}} = \frac{B}{\phi_0} = \frac{eB}{h} \tag{2-11}$$

$$\nu = \frac{n_{\mathrm{s}}}{n_{\mathrm{LL}}} \tag{2-12}$$

When a LL is full, the Fermi level lies in a gap between occupied levels, the filling factor $\nu$ has an integer value, and vanishing of the longitudinal resistivity occurs.

## Figure 2.3. The quantum Hall effect.

(a) Formation of Landau levels (LLs) in the density of states (DOS) in conventional 2DEG as a result of the quantization of cyclotron motion. Left (orange-highlighted) shows a typical continuum of DOS at zero magnetic fields. When the magnetic field is applied, the states should theoretically become quantized (delta-function) in LLs for an ideal system without disorders and impurities (marked as B > 0, no disorder). In practice, systems include disorders (marked as B > 0, disorder). Disorders lead to the broadening of LLs and formation of localized (black/green hatched regions)- and extended states (cyan-colored). The localized states do not carry current, whereas extended states do. The energy spacing between LLs in conventional 2DEG is proportional to the magnetic field ($\Delta E_{LL} \propto B$), leading to an equidistant spacing sequence of the Landau levels.

(b) Schematic of the density of states of graphene under QHE conditions.

(c) The spectrum of LLs for graphene indicating a unique LLs sequence depending on ($\Delta E_{LL} \propto \sqrt{B}$). Also, there is zero-energy LL that is equally shared by electrons and holes. Regions of localized and extended states are illustrated. There is a large energy spacing between the 1st and 0th LL in graphene. The energy spacing in the LL spectrum of graphene is not equidistant in contrast to the conventional 2DEG shown in (a).

(d) Schematic comparison between the QHE in conventional 2DEG (blue) and graphene (red). The quantized values in graphene for the filling factors of $\nu = 2$, 6 are shown. [63,64]

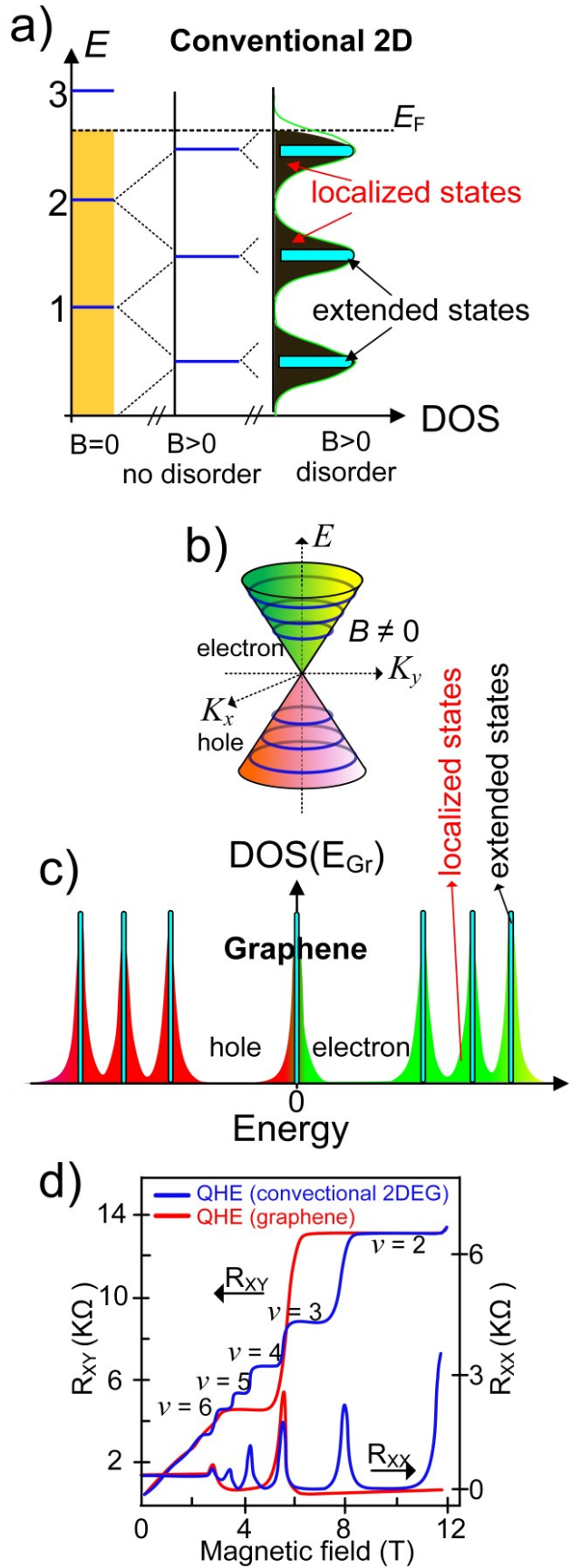

## 2.2.2. Quantum Hall effect in monolayer graphene

The linear energy dispersion of monolayer graphene leads to a modification of the magneto-transport in graphene. Accordingly, the energy gap between neighboring Landau levels is not equidistant, as with conventional semiconductors (e.g., GaAs/AlGaAs). In graphene, the energy dispersion of carriers around $K$ and $K'$ is similar to that of ultra-relativistic particles with zero mass $m_0$. Thus, the Dirac equation is used to describe carriers' behaviors, which mimic massless Dirac Fermions around the Dirac points. [15,64] Accordingly, the cyclotron frequency from the conventional 2D-systems is modified for Dirac Fermions, as shown in **(2-13)**. The LL spectrum is then given by **(2-14)**.

$$\omega_C = \sqrt{2}\frac{v_F}{l_B} = v_F\sqrt{\frac{2eB}{h}} \qquad (2\text{-}13)$$

$$E_{LL-Gr} = \pm\hbar\omega_c\sqrt{N} = v_F\sqrt{2\hbar eBN} \qquad (2\text{-}14)$$

$N$ is an integer number (including zero), and $v_F$ ($\sim 10^6 \ ms^{-1}$) is the Fermi velocity. [60]

There exist three main differences with conventional 2D systems. (i) The energy spacing of LL in graphene is not equidistant and is associated with ($\Delta E_{LL} \propto \sqrt{B}$) instead of $\Delta E_{LL} \propto B$ and evenly spaced LLs in conventional 2D systems. (ii) The degeneracy of each LL is duplicated. This is due to spin-up/spin-down (as in conventional 2D systems), and valley degeneracy ($K$ and $K'$), thereby each LL in graphene can take double as many electrons as LL do in the conventional 2D systems. (iii) There is a particular LL in graphene at zero energy ($E = 0$), which is shared equally by electrons and holes (only spin degenerate). As a result, it holds half as many states as other LLs.

These all drastically affect the characteristic of $R_{XY}$ in the quantum Hall regime resulting in monolayer graphene. For a conventional 2D-systems from **(2-11)**, the number of electrons (per unit area) needed to fill a single LL was shown to be $B/\phi_0$. In graphene, due to spin degeneracy and double valleys ($K$ and $K'$) degeneracy each filled LL contributes $4B/\phi_0$ electrons. However, for the abovementioned reason in (iii) for zero energy LL only contributes $2B/\phi_0$. Accordingly, the electron density $n_S$ corresponding to $N$ filled LL in graphene is as follows:

$$n_S = \boxed{N\frac{2eB}{h}}_K + \boxed{N\frac{2eB}{h}}_{K'} + \boxed{N\frac{2eB}{h}}_{E=0} = \frac{4eB}{h}\left(N+\frac{1}{2}\right) \qquad (2\text{-}15)$$

Thus, by substituting $n_S$ value of **(2-15)** into the equation for the conventional Hall effect in equation **(2-6)**, the $R_{XY}$ is given by:

$$R_{XY} = \frac{B}{en_S} = \frac{h}{4e^2(N + \frac{1}{2})} \tag{2-16}$$

Accordingly, the quantized values for graphene are multiples of $h/4e^2$ with filling factors of $\nu = 2, 6, 10, 14, \ldots$ (for $N \geq 0$).

From the experimental perspectives, the unique LL energy spacing of graphene ($\Delta E_{LL} \propto \sqrt{B}$), which is deduced from the expression given in **(2-14)** is significant. This technically means reaching to $\nu = 2$ ($R_{XY} = h/2e^2$) at a lower magnetic field compared to the conventional 2DEG ($\Delta E_{LL} \propto B$). The energy spacing between the lowest LL, $N=0$, and the first excited one, $N = \pm 1$, is very high and allows surviving the QHE up to room temperatures. [60]

The epitaxial graphene on SiC is highly $n$-doped ($\sim 10^{-13}\text{cm}^{-2}$) arising from the SiC substrate and buffer layer. The buffer layer with donor-like behavior has a strong density of states; when using a gate for charge tuning, it significantly degrades the gating efficiency, i.e., by remarkably diminishing the gate capacitance. [66] Therefore, gate tuning of SiC/G is not suitable for metrological applications. Besides, the charge transfer from the bottom SiC and buffer layer to the graphene results in a strong magnetic field-dependent carrier concentration leading to a giant quantized $\nu = 2$ Hall plateau. This is advantageous and indicates a robust Hall plateau for SiC/G at high magnetic fields (up to 50 T). [67]

In Chapters **4** and **8**, the charge carrier tuning of SiC/G close to the charge neutrality point will be discussed. Moreover, the electronic properties of an epitaxial SiC/G layer are strongly influenced by its surface morphology, in which there is a direct relationship between the resistivity of epitaxial graphene on SiC and its step-heights. [68–70] Therefore, producing graphene with smaller step-height distribution was intended in this study.

## 2.3. Epitaxial graphene on SiC (0001)

There exist several methods and principles to synthesize graphene which can be generally summarized into two categories: the top-down approach (e.g., mechanical exfoliation [13], liquid-phase exfoliation [71], and graphene oxide reduction [72]) and the bottom-up approach such as epitaxial graphene growth via silicon sublimation growth on SiC or chemical vapor deposition (CVD) method. [8,11,26,73,74] The epitaxial graphene through the CVD method can be grown on SiC but also on various other substrates such as Cu, Ni, Ir, and Ru. [75–77] The SiC is special since it is indeed the only substrate containing a vast amount of carbon, which allows epitaxial graphene production without the need for any reactive gas, i.e., by simple high-temperature graphitization under Ar or vacuum. A large-scale already doped graphene fabrication directly on insulator without the need for transfer makes epitaxial graphene growth on SiC special, among other techniques.

### 2.3.1. Growth mechanism

Hexagonal polytypes of SiC have two possible surface terminations, i.e., Si- or C-face, see **Figure 2.4a**. The termination substantially affects the graphene growth process. In this study, it is referred to as "silicon face growth" concerning epitaxial graphene grown on SiC (also as epigraphene or SiC/G). Epitaxial graphene on SiC is one of the best candidates for the wafer-scale production of graphene-based devices. [8,11,73]

Silicon carbide has a tendency towards graphitization when it is annealed under elevated temperature in vacuum or atmospheric pressure. The growth process is driven by the sublimation of silicon atoms from the SiC surface and the subsequent rearrangement of the remaining carbon atoms into a densely packed honeycomb structure. [75] The number of graphene layers, stacking layers, and coupling to the substrate varies with crystal face orientation and growth conditions. [20,21]

The graphitization of SiC (0001) can be understood as two successive steps, in which first, a so-called buffer layer is formed. Further decomposition provides carbon atoms under the buffer layer, and they form a new buffer layer. The original buffer layer loses its bonding with the SiC substrate and then turns into graphene. This is schematically shown in **Figure 2.4b and c**. During the material growth, the atoms are organized by themselves to reduce their total free energy. The atoms rearrange into their energetically most favorable sites in the given environment, provided sufficient thermal energy and time is allowed. [75]

**Figure 2.4. Schematic of different types of epitaxial graphene on SiC (0001).**

(a) Side-view of the hexagonal SiC sample, which indicates both Si- and C-face. (b) Structure of the epitaxial buffer layer. (c) Structure of the epitaxial graphene on SiC. Formation of the (d) QFMLG and (e) QFBLG using hydrogen intercalation technique on the buffer layer and epitaxial graphene, respectively. More info in ref. [38].

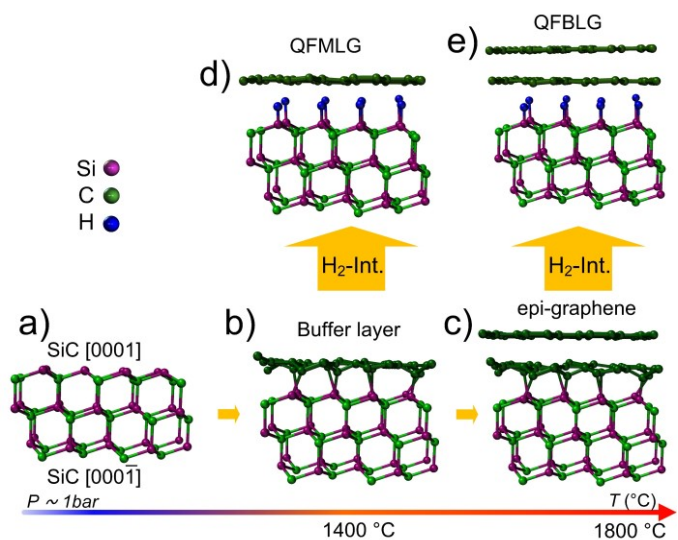

Precise control of growth conditions as well as careful selection of crystal type, face side, and miscut angle are required for the growing single layer of graphene. The temperature of the substrate plays an important role. The time of annealing is another parameter for controlling the growth rate. Also, annealing in the argon environment and the argon flow-rate significantly influence the graphitization process. [16,37,38,78]

During the graphitization process of a SiC(0001) surface, the Si-rich surface transforms into a C-rich surface, and the surface atoms arrangement typically from $(3 \times 3)$ to $(\sqrt{3} \times \sqrt{3})R30°$ and $(6\sqrt{3} \times 6\sqrt{3})R30°$ as the temperature increases. **Figure 2.5** illustrates the $(6\sqrt{3} \times 6\sqrt{3})R30°$ surface, often called buffer layer (or "zeroth layer"), consists of a hexagonally connected C layer. [79,80]

Since the buffer layer is partially covalently bound to the underlying Si (**Figure 2.4b**), it does not exhibit graphene-like electronic properties. [16,73,81,82] However, the buffer layer has a graphene-like lattice, with undistorted σ states but a distorted π band. The distortions are caused by covalent bonds forming between a part of the carbon atoms (about 1/3) in the buffer layer and the underlying silicon atoms. [83,84] Quasiperiodic $(6 \times 6)$ domain pattern emerges out of a larger commensurate $(6\sqrt{3} \times 6\sqrt{3})R30°$ periodic interfacial reconstruction. [79]

In a growth model by Norimatsu et al. [85], it was argued that because the area densities of the C atoms in ideal graphene and a single SiC bilayer have values of 32.9 nm⁻² and 10.5 nm⁻², respectively, the formation of one graphene layer requires about three SiC bilayers. This suggests that for the fabrication of graphene, including its buffer layer, at least about six Si-C bilayers will be consumed. Moreover, when the buffer layer has formed, proceeding the growth requires much higher energy.

### Figure 2.5. Epitaxial buffer layer on SiC (0001).

(a) Structural model of the $(6\sqrt{3} \times 6\sqrt{3})R30°$ reconstruction in top view showing the Si-face (1×1)-SiC substrate and the graphene-like lattice of the initial carbon layer. The unit cell of the buffer layer (blue), including the quasi- (6×6) corrugation (green) are shown. These are also depicted on the atomically resolved STM image of $(6\sqrt{3} \times 6\sqrt{3})R30°$ reconstruction of epitaxial graphene on 4H-SiC (0001). Figures are edited from ref. [86].

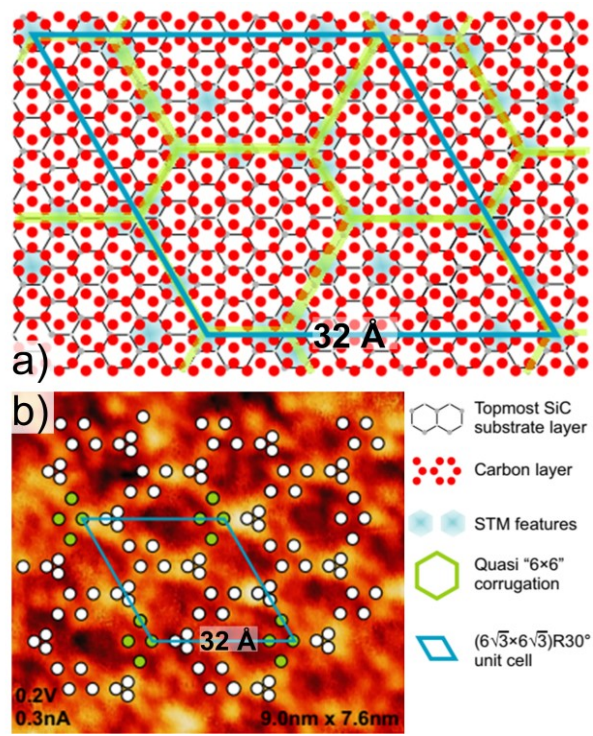

Also, after the formation of graphene, the growth rate drastically decreases. This happens because the overgrown carbon layers act as a Si-diffusion-barrier, hindering Si sublimation from the surface, and therefore proved a self-limiting feature to graphene growth on SiC(0001). [87] In contrast, the growth on the C-face of SiC is hardly controllable and leads to multilayer graphene formation. [88] This is important in growth kinetics when a high thickness control for fabrication pure graphene layers (e.g., BFL, MLG, or BLG) is desired. Furthermore, The Bernal stacked bilayer graphene that grows on SiC has a semi-conducting nature; therefore, depending on the charge carrier density can show either metallic or insulating behavior. [22,89–91] In the context of this thesis, such inhomogeneities (e.g., BLG inclusion) in graphene synthesis, electronic transport, as well as Hall bar fabrications, are investigated.

### 2.3.2. Intercalation

Since its discovery over a century ago, intercalation has enabled the creation of different materials for various applications. [1,92–94] Importantly, intercalation can be applied to the epitaxial buffer layer/SiC to physically decouple it from the substrate. This has been performed through a wide range of intercalating species, e.g., H, Ge, Ga, O, Li, Au. [18,41,43,95–101] Beyond that, epigraphene is an excellent platform to create sub-dimensional elemental and compound layers using the intercalation method. [42,102–106] The influence of the covalent bonding in the buffer layer is one of the primary suspects for the heavily reduced mobility in epigraphene compared to exfoliated graphene. The limited mobility was attributed to a combination of Coulomb and short-range scattering from charge traps in epitaxial graphene and strong electron-phonon scattering. [107–109] For the same reasons, the carrier mobility in graphene is firmly temperature dependence. [110]

This effect can be ameliorated by the intercalation technique, which leads to producing "quasi-freestanding" epitaxial graphene. **Figure 2.4b-e** depicts hydrogen intercalation applied on the epitaxial buffer layer and monolayer graphene resulting in quasi-freestanding monolayer graphene (QFMLG) and quasi-freestanding bilayer graphene (QFBLG), respectively. While epitaxial graphene on SiC is originally $n$-type ($\sim 10^{13}$ cm$^{-2}$) [16], depending on the served intercalants $p$-type, $n$-type, and charge-neutral graphene can be obtained. [94,95,100,111] Interestingly, for Au and Ge intercalants, depending on the thicknesses of the elemental layers below carbon layers, either $p$- and $n$-doped graphene can be produced. [101,112] Hydrogen intercalation, which yields a $p$-doped freestanding graphene layer, has vastly been studied by several groups. [18,38,108,113–121] However, the precise atomic intercalation mechanism in epitaxial graphene is still not completely understood.

Although the principle of intercalation is simple, in practice, a scalable freestanding layer is not readily achievable. The main challenge is the low quality of epitaxial buffer- or graphene layers, i.e., their thickness inhomogeneity. Moreover, the intercalation circumstance could induce incomplete decoupling or even etched areas. In this study, the high quality of large-area epitaxial buffer- and monolayer graphene layers are further tested by hydrogen intercalation. A hydrogen (5%) gas dissolved in argon (95%) was used, thereby also the influence of gas purity in the intercalation process is intensively investigated. These all will be discussed in Chapters **4** and **5**.

## 2.4. Silicon carbide

### 2.4.1. Crystal structure

The compound named ''Kohlensilicium'' (silicon carbide), with an exact stoichiometry of Si:C = 1:1, was discovered in 1824. [122] The SiC possesses extraordinary mechanical, nuclear, and electrical properties made it a great candidate to be used in a vast array of applications. Importantly, due to its broad bandgap semiconductor nature, the SiC is particularly useful for high power and high-frequency devices and thereby the linchpin to "green energy" that would replace less energy-efficient silicon-based technology switches (for high currents and voltages). [123–126] Reaching these widespread applications had been challenging for many years because of fundamental problems in growing high-quality single-crystal boules free of micropipe defects and micrometer-scale pinholes formed by dislocations, however nowadays have been substantially improved. [125]

From the crystallographic aspects, SiC is a special versatile substance in which more than 200 crystal modifications, so-called polytypes are so far known. [127] These structures differ both in the order (in which cubic and hexagonal layers are arranged) and in the number of layers in the unit cell. [128] Polytypes are often characterized by Ramsdell designations (mainly due to its formal shortness) [129], constituted by a natural number, equal to the number of layers in the period in the direction perpendicular to the basal plane, and a letter symbol characterizing the crystal system of the Bravais lattice: C, cubic; H, hexagonal; R, rhombohedral. [123,130] There is only one cubic polytype, 3C-SiC, sometimes referred to as β-SiC, while all the others are called α-SiC. [75] The 3C-SiC can be prepared at much lower temperatures (1473 to 2273 K) than hexagonal polytypes (2473 to 2773 K) and was shown to be beneficial for graphene sublimation growth. Nonetheless, it is not commercially available since it is complicated to grow high-quality material without double positioning boundaries, also called twin boundaries. [31]–[33]

In the binary compound SiC, each carbon atom is surrounded by four silicon atoms (and vice versa), which form tetrahedron crystal structure by strong $sp^3$ orbitals, as either $SiC_4$ or $CSi_4$, see **Figure 2.6a, b**. As seen in **Figure 2.6a, b** there exist two different kinds of subsequent Si-C-tetrahedra modifications: one resulting in a hexagonal "*h*" site and one resulting in a cubic "*k*" site, both just differing in a 60° twist of the upper tetrahedra. [130] This can be followed through the cyan-color lines in **Figure 2.6b,** which depicts the continued or linear stacking (cubic), which always have the same orientation (staggered orientation). The cyan-color line in **Figure 2.6a** indicates the 60° rotation in the hexagonal tetrahedron, i.e., the stacking from one to the next double layer is switched (also called switched stacking or eclipsed orientation). [134]

The $SiC_4$ or $CSi_4$ tetrahedra create stacks of Si–C bilayers consecutively in a discrete stacking order with three distinct sites (named A, B, C) to occupy, see **Figure 2.6c**. Depending on this stacking order, different polytypes with different crystal modifications will result. The Si–C bonding energy is 289 kJ/mol (~2.995 eV) with a distance $d = 189$ pm. The distance between two similar atoms is $a = 308$ pm. The spacing between two layers of identical atoms, which is the height of the tetrahedron, is 252 pm. **Figure 2.6d and e** sketch the unit cell of 4H-SiC and 6H–SiC, which are among the most common polytypes and are used in this thesis for graphene growth. In the following, the most important SiC substrate parameters for the growth are discussed.

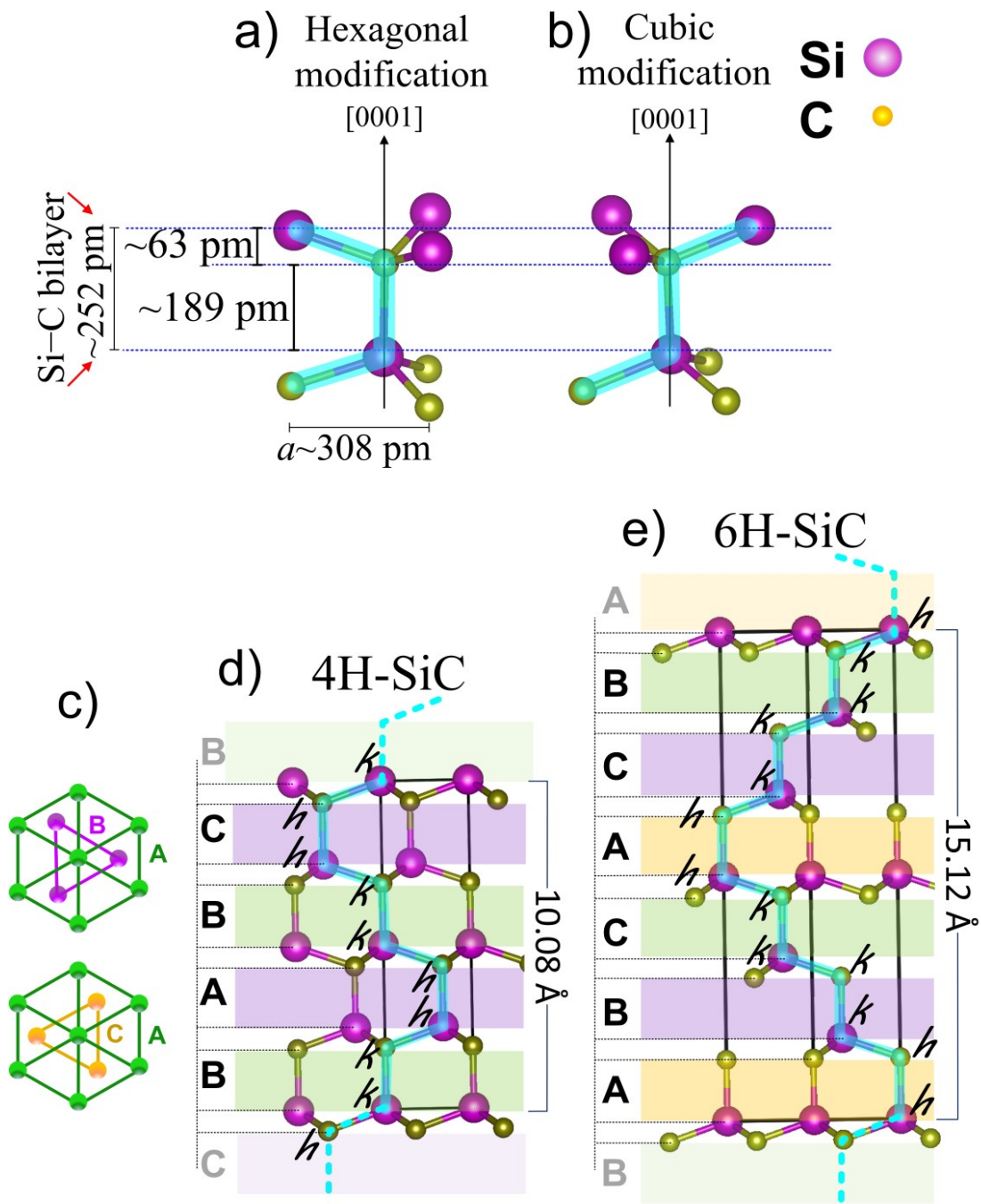

**Figure 2.6. The crystal structures of 4H- and 6H-SiC.**

The SiC crystal structure comprises two interconnected $Si_4C/C_4Si$ tetrahedra with either hexagonal or cubic configurations. In the stacks, the orientation of tetrahedra in the next layer can be rotated by 60° (hexagonal) (a) or be the same (cubic) (b). As depicted by the cyan-color line, it is evident that hexagonal sites are always located at the edges of the zigzag pattern, whereas cubic sites are always within straightforwardly directed rows. Jagodzinski's "$h$–$k$" notation [135] is also used to define the polytypes along the c-axis) (c) Three possible positions (top view) where the tetrahedra, as mentioned earlier, can occupy. This leads to different stacking scenarios and polytypism in SiC, as displayed for the 4H-SiC (d) and 6H-SiC (e) unit cells. [130,136]

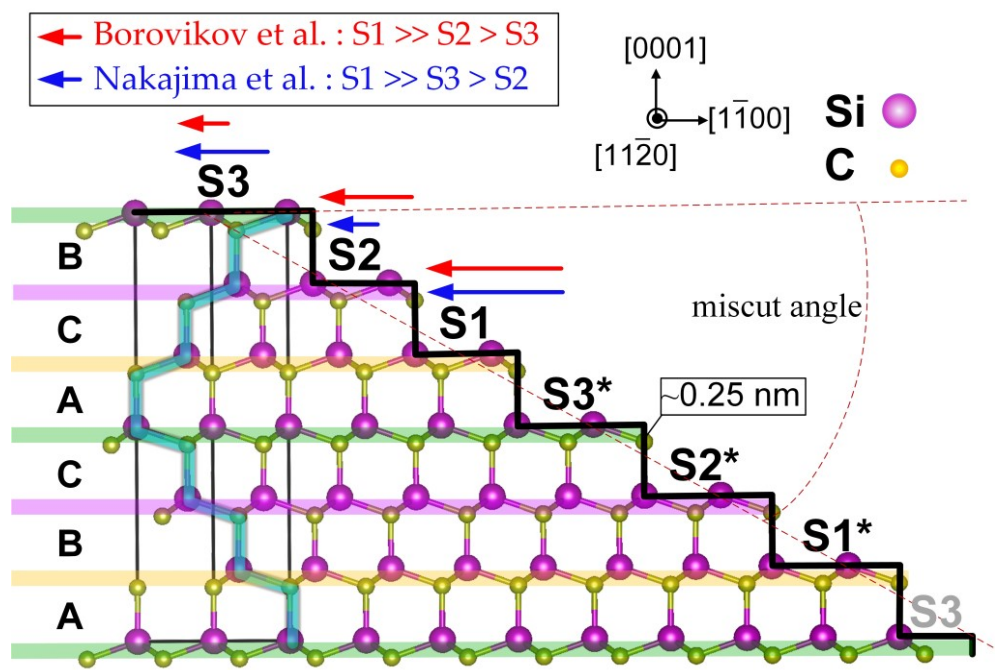

**Figure 2.7. Step flow etching on 6H-SiC (0001).**
Schematic of the as-received/polished 6H-SiC samples with a small miscut angle towards
[1$\bar{1}$00]. Since the unit cell of 6H is composed of six sequential Si−C bilayers stacks, the
truncation results in the appearance of 6 terrace types (with single Si-C bilayer step height),
denoted as (S1, S2, S3, S1*, S2*, and S3*). The Sn and Sn* are energetically similar but are
60° rotated related to each other. Accordingly, three terrace types with inequivalent
surface energies exist. Different length arrows indicate the different etch velocities (during
growth) of the steps. The blue arrows were proposed by Nakajima et al. [137] and red ones
by Borovikov et al. [138]. See the text for details.

### 2.4.2. Step flow

Both surfaces of 4H- and 6H-SiC polytypes are promising materials for epitaxial
graphene growth. For our purpose, the growth of single-layer graphene, Si-face,
gives higher growth thickness control while on the C-face leads to multilayer
graphene formation. [139] The 4H- and 6H-SiC wafers with polished and
protected (by a so-called epi-ready layer) surfaces are commercially available.
The wafers are cut from large SiC boules with a small miscut angle towards
definite crystal planes.

The cross-sectional structure and associated bilayer stacking sequences of 6H-
SiC (0001), is shown in **Figure 2.7**. After cutting the 6H-SiC polytype with a
misorientation angle, six possible terminations shape at the SiC surface (S1, S2,

S3, S1*, S2*, S3*). The Sn/Sn* nomenclature denotes different types of surface configurations, and "n" indicates the total number of identically orientated Si–C bilayers (n= 1, 2, 3). Equivalent layers, but rotated by 60° are labeled by (n*). [140] Thus, this in total gives three configurations at the very top of the surface.

During etching and growth on SiC, the surface undergoes a restructuring process in which the steps move and group together (so-called step-bunching process). The final shape of the SiC crystal surface is often determined by a minimization of the total surface free energy. Since the terraces have inequivalent surface energies (due to the physical relationship with the underlying bilayers), the etching/growth is driven by a "step flow."

The step-bunching mechanism, e.g., during graphitization or hydrogen etching of the SiC surfaces, has been vastly studied [137,138]. F.R Chein and S. R. Nutt [141] suggested a step bunching model during the growth of 6H–SiC. On account of this growth model, Nakajima et al. [137] and Borovikov et al. [138] proposed step flow models that seem to be not entirely confirming each other and therefore are discussed here.

F. R Chein and S. R. Nutt [141–143] calculated the extra energies ($\Delta U$) for depositing one new bilayer on the terraces (in meV per Si–C pair atoms) as follows:

S3 (6H1) terrace: $\Delta U_{growth}^{S3(6H1)} = E_0 + 1.33,$ (2-17)

S1 (6H2) terrace: $\Delta U_{growth}^{S1(6H2)} = E_0 - 6.56,$ (2-18)

S2 (6H3) terrace: $\Delta U_{growth}^{S2(6H3)} = E_0 - 2.34.$ (2-19)

The $E_0$ is the energy of a Si–C pair without interaction between neighboring layers. From these energy values, it can be expected that the S1 termination is the fastest grown surface (since it requires the lowest energy to be grown) while the S2 and S3 are the second and third fast grown surfaces, respectively. In their step flow model, Borovikov et al. [138] proposed that the terrace, which is growing faster, would then be slower removed during the etching process. This accounts for the energy minimization and stability reason, which means fast-growing steps are the slowest in etching and vice versa. Accordingly, the etching rate would be S1≫ S2 >S3. This situation is shown with the red-color arrows in **Figure 2.7**.

The step flow model from Nakajima et al. [137] is different and based on the abovementioned calculations, the variation energies with desorbing (etching) one bilayer are given as follows:

S3 (6H1) terrace: $\Delta U_{etching}^{S3(6H1)} = -\Delta U_{growth}^{S2(6H3)} = -E_0 + 2.34,$ ( 2-20)

S1 (6H2) terrace: $\Delta U_{etching}^{S1(6H2)} = -\Delta U_{growth}^{S3(6H1)} = -E_0 - 1.33,$ ( 2-21)

S2 (6H3) terrace: $\Delta U_{etching}^{S2(6H3)} = -\Delta U_{growth}^{S1(6H2)} = -E_0 + 5.56.$ ( 2-22)

In this model, the given energy to etch a surface is equal to the negative energy value of the surface below. These results indicate that the S2 (6H3) surface is the most difficult bilayer for desorbing, i.e., the most stable one. Correspondingly, the etch velocity is as follows: S1≫ S3 >S2, as shown with the blue-color arrows in **Figure 2.7**. Although in the literature, these two models are assumed identical, they indeed are not. Both models agree that the S1/S1* has the highest decomposition velocity. However, discord is seen for the second and third decomposition surface velocities, i.e., S2/S2* and S3/S3* layers in 6H-SiC. The high-controlled synthesis presented in this study allows us to experimentally evaluate these step-flow models. In Chapter **7**, the step-bunching is discussed within a framework of a step-flow model that considers a joint hexagonality and cubicity of bonds in off-bond and on-bond directions. It will be experimentally shown that the terrace S1 will first catch terrace S2 and form two Si–C bilayers (~0.5 nm). If the growth proceeds, the terrace S3 will advance and merge with the two bilayer steps and creates a three Si–C bilayers height (~0.75 nm). Also, for the 4H-SiC polytype, the step flow model can be conveniently applied since there are only two types of terminations. This will be studied in Chapters **6**, **7**, and appendix **A6**.

### 2.4.3. Miscut and crystal planes

Both miscut angle (also called tilt-angle, misorientation angle, or off-axis angle) and crystal orientation substantially influence the growth as well as electronic properties of overgrown carbon layers. Surface orientations are given by the miscut angle between the surface plane concerning a crystallographic plane. If truncating of SiC happens perpendicular to the c-axis with no inclination (absolute 0°-off), then it should turn out an entirely flat surface. The misorientation of the surface is the reason that flat terraces are formed interrupted by step edges of a certain height corresponding to the miscut angle. The misorientation angle plays a decisive role in graphene morphological and transport properties. [87,144] Additionally, unintentional variations due to technical reasons can occur, e.g., warp and bow of the surface. Local deviations from the given value can appear close to dislocations or crystal defects.

Moreover, the crystal plane is relevant to the growth process. For example, it was both theoretically and experimentally shown that quasi-freestanding graphene could be directly grown on nonpolar ($11\bar{2}0$) and ($1\bar{1}00$) planes. [145] On SiC ($1\bar{1}00$), the growth resembles the one on SiC($000\bar{1}$) exhibiting considerable rotational disorder. In the case of SiC ($11\bar{2}0$) graphene grows without the rotational disorder, even though no buffer layer is present. The size of the miscut angle [37,146–148] and the crystal direction [137,149–152] are important for the process of, e.g., hydrogen etching. The latter was shown on the surfaces tilted toward [$1\bar{1}00$], leading to a higher etching rate along with the steps than perpendicular to the steps. As a result, straight steps could be formed on these surfaces. In contrast, the etching rate was reported to be higher perpendicular to the steps on the surfaces tilted toward [$11\bar{2}0$] thus, therefore, zigzag like steps could be formed. The SiC crystal plane thus decisively influences both the etch and growth processes. [153–155]

### 2.4.4. Polarization doping

As shown in **Figure 2.6**, the structures of SiC polytypes can be viewed as stacking of cubic($k$) or hexagonal($h$) Si-C bilayers on top of another. The Si–C bond has predominantly a covalent character (88%) with small (12%) ionic contribution (Si positively, C negatively charged). [75] Except for the 3C-SiC (zero hexagonality), other SiC polytypes, i.e., hexagonal polytypes, with lower symmetry possess inequivalent Si-C bonds property. In the 3C-SiC, all four sp$^3$ orbitals have equivalent properties, whereas in hexagonal polytypes, there is a preferential c-axis, and the sp$^3$ orbital is extended along this axis, which differs from the other three. [156] For example, in 6H-SiC along the c-axis, the distance of $k$-$k$ double layer (i.e., Si-C bilayer with both C and Si owning cubic positions) was measured to be $2.5163 \pm 0.0008$ Å and for $h$-$k$ bilayers equal to $2.5212 \pm 0.0004$ Å. Therefore, the Si-C bonds parallel to the c-axis (on-bond) are longer than those inclined to it (off-bond). [157–159] Correspondingly, there is charge transfer from the weaker (longer) longitudinal bonds to the stronger (shorter) transverse bonds. These bond-to-bond charge transfer and structural (ionic) relaxation cause an intrinsic so-called spontaneous polarization (SP). [159] Therefore, the SP, which generally occurs in dielectric crystals, can be interpreted as a superposition of localized dipoles due to each of the stacking boundaries, and it is mainly due to the charge density redistribution. [159,160]

In bulk, the net electric field across the unit cell vanishes imposed by phase periodic boundary conditions. Nevertheless, at the surfaces, the translation symmetry is broken, and the dipoles may add up forming an uncompensated polarization field. As a result, the electrostatic (Hartree) potential will increase or decrease, leading to an electrostatically unstable surface unless a source of hole or electron trapping is attached to it, e.g., by attracting charges from the air. Indeed, this makes the SP of crystals very difficult to be determined both in practice and theory. This effect becomes vital when it comes to epitaxial graphene.

The SiC substrate influences the electronic properties of epitaxial graphene layers in multiple ways. QFMLG on hexagonal SiC obtained by H-intercalation is highly $p$-doped. This $p$-doping was attributed to the SP since: (i) QFMLG on 3C-SiC (SP is missing) is not $p$-type (but slightly $n$-type due to defects), (ii) the Si–H bonds are electronically inactive (no doping contribution), (iii) the passivated surface is supposed to be in principle defect-free (hence no doping contribution), and (iv) the depletion of the bulk doping of SiC (especially for semi-insulated SiC) is insufficient for the substantial doping (corresponding to areal $n_s < 5 \times 10^{10}$ cm$^{-2}$ in the depletion layer of SiC substrate), therefore the strong hole-doping is directly associated with the SP effect. Herein, the spontaneous polarization creates a pseudo-acceptor layer at the surface that is fully equivalent to real acceptors.

Furthermore, since the SP along the (0001) direction remains negative for all hexagonal SiC polytypes, only $p$-doping should be induced at the Si terminated quasi-freestanding graphene (QFG) systems. [30,160,161]

The SP also has a vital impact on the doping of epitaxial graphene grown on SiC (0001), which includes a buffer layer. The pseudo-acceptor charges from SiC are overcompensated by the broad donor-like states from the buffer layer superimposed with donor-states from the carbon and silicon dangling bonds on the surface.

Moreover, there is a linear relationship between the bandgap and degree of hexagonality in all SiC polytype: it varies from ~2.39 eV in the 3C structure (0% hexagonality) to ~3.30 eV in the 2H form (100% hexagonality). [157] For the polytypes of interest in this work, the 4H (50% hexagonal) and 6H (33% hexagonal), the bandgap is about 3.23 eV and 3.0 eV, respectively. Accordingly, higher doping (with a factor of ~1.5) of epitaxial graphene on 4H is typically expected than on 6H. [31]

Interestingly, for the most common 4H- and 6H-SiC polytypes, an extensive calculation study by Sławińska et al. [161] estimates a certain dependence of the doping on the precise location of the stacking defects (hexagonal bilayer) closest to the surface. Accordingly, for a given thickness, the doping decreases by $\sim 2 \times 10^{12}$ cm$^{-2}$, the deeper it (stacking defect) is buried, which is due to the depolarization effect of the crystalline layers at the surface, see **Figure 2.8**. This interesting calculation shows that an ample range of doping can be achieved almost continuously by controlling the number of stacking defects, their concentration, and their proximity to the surface top graphene layers. A similar effect was expected from other somewhat elder density functional theory (DFT) calculations. [157,159,162–164]

The experimental results presented in this thesis give evidence for the theoretical prediction by Sławińska et al. [161] demonstrating that the graphene doping depends not only on the bulk polarization but also on a SiC termination dependent polarization doping effect. This will be discussed in detail in Chapters **5** and **7**.

Please note that the SP is independent of strain (not like piezoelectric) and can be observed in other wurtzite crystals (e.g., III-nitrides). The SP makes it possible to directly manipulate the electronic properties (e.g., carrier concentration) in a semiconductor like conventional doping without using impurity dopants. [165,166] Moreover, the polarization-induced internal electric field leads to the so-called quantum-confined Stark effect in the quantum nanostructures as recently was shown in defects (e.g., vacancy) in SiC below epigraphene [166,167], indicating still much room for further studies.

**Figure 2.8. Spontaneous polarization of quasi-freestanding graphene on SiC.**

Doping of the graphene layer for all G/SiC$_n$ as a function of n (number of Si–C bilayers), polytype, and the stacking defect location closest to the surface. The left axis gives the surface charge density and the right axis (quadratic scale) the Dirac point shift $\Delta DP = DP - \mu$, ($\mu$ chemical potential). Horizontal lines at the right give the bulk calculated SP associated with each polytype. Figure edited from ref. [161].

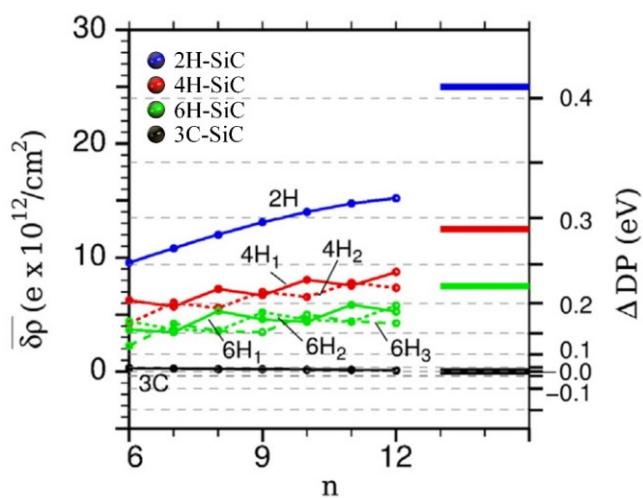

# 3. Characterization techniques

**Abstract**

*I*n the framework of this thesis, several versatile characterization techniques were employed to investigate more in detail the quality of different types of graphene samples considering their morphological, electronic transport, or magneto-transport properties. This chapter is devoted to showing the working principle of the implemented experimental techniques, which support the discussions in the next chapters.

## 3.1. Atomic force microscopy

Atomic force microscopy (AFM), as a valuable and versatile surface analysis technique, is often used to display the real space morphology of samples on μm-scales. In this thesis, the AFM as a standard technique was regularly used to characterize the samples' surface properties. The main component of AFM is a cantilever with a sharp tip. During measurement, the tip scans over the surface. Following the atomic force interaction between the tip and substrate, the AFM can generate informative images of the substrate. The NANOStation AFM (produced by S.I.S) used in this study provides a pronounced resolution that precisely resolves structures with height equals to ~0.25 nm. The SSS-NCLR or PPP-NCLR are the AFM silicon tips fabricated by Nanosensors™ that were used in this work. **Figure 3.1** indicates the principle of AFM used in this work. The device combines optical microscopy and scanning probe microscopy (SPM) in a Zeiss Axiotech microscope. The local deflection between the probe and substrate is measured by a fiber-optic interferometer. It is based on the principle of optical retardation between a reference wave and a detected wave. The diode laser provides laser light via a fiber-optic coupler to the fiber-optic cable mounted to the measuring head (cantilever). Some of the incident light is reflected at the fiber end, where it defines the reference wave, and at the top-side of the cantilever, where it defines the detected wave. The two waves interfere with each other and are transmitted back to the photodiode via the fiber-optical coupler. The resulting intensity of the created signal is detected in the photodiode to be further processed. As the AFM scan across the surface, the tip's motions (up, down, and side to side) are monitored via the laser beam reflected from the cantilever.

Technically, the contact (DC) mode and non-contact mode (AC) (applied in this study) are two modes that are applied for AFM measurement. In DC contact mode, the tip is permanently in contact with the substrate during scanning, which may damage the tip and cantilever as a result of a shear force. In the non-contact (AC) mode, the tip is adjusted appropriately in minimum contact with the substrate while the cantilever oscillates. This mode records information from the substrate's surface in which the tip is not in contact (or in an intermediate contact) with the substrate.

Important information about the substrates examined with the AFM (in AC mode) can be extracted from their phase images. The AFM generates the phase image by monitoring the phase lag between the signal, which drives the cantilever to oscillate (in non-contact mode), and the actual cantilever oscillation signal. From the phase contrasts in the AFM phase image, one can understand variation in surface properties (e.g., elasticity, stiffness, friction, and adhesion) and the existence of different materials, which all may cause the phase contrast. The AFM phase imaging is based on the detection of variation in the energy dissipated in a local region of the sample surface. [168] The AFM phase inspection may sometimes be considered as a non-trivial technique for studying the surfaces. However, it is extensively used in this work, for instance, to investigate the interaction between the graphene and bottom SiC surface terminations, as discussed in detail in Chapter **7**.

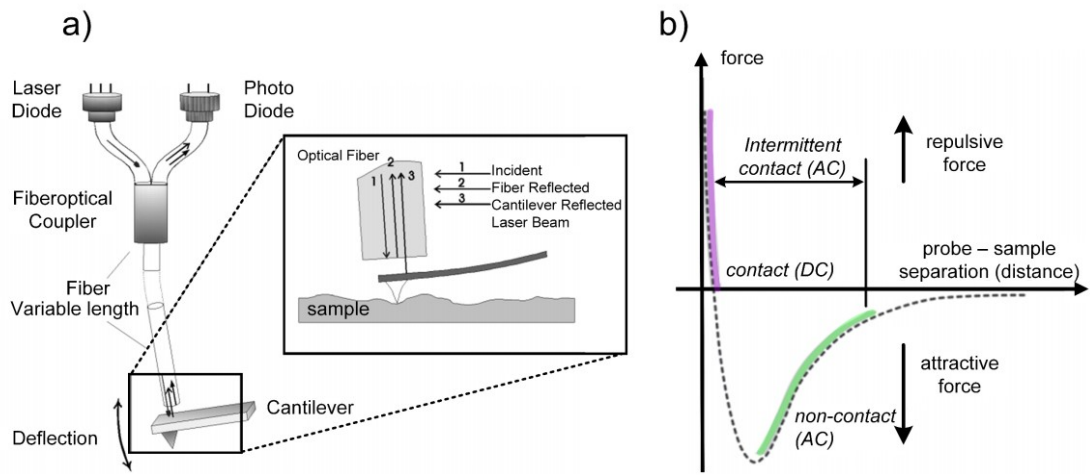

**Figure 3.1. Working principle of atomic force microscopy.**
a) Sketch of the AFM working principle. The AFM tip interacts with the substrate. The raster scanning motion of the AFM tip as it scans across the surface is monitored using a laser beam reflected off from the cantilever. The sensitive photodetector can track the reflected laser beam and collect the vertical and lateral motion of the probe. b) General operating regimes for different AFM imaging modes. Adapted from refs. [37,168]

## 3.2. Kelvin probe force microscopy

Kelvin probe force microscopy (KPFM) is primarily based on the instrumentation of an AFM. [169] The KPFM is implemented to investigate the graphene quality, considering the thickness and, more importantly, its interaction with the SiC substrate. The principle of the KPFM technique is shown in **Figure 3.2**. When two different conductors, here sample and tip, are brought into electrical contact, electrons will flow from the one with lower work function ($\phi$) to the one with higher work function, equalizing the Fermi energies. They could be assumed as a capacitor (parallel plates) if they are not in contact but proximity plate, thus equal, but opposite charges will be induced on the surfaces. The potential generated between these two surfaces is called the contact potential difference (CPD), or surface potential, which is equal to the work function difference of the two materials. This CPD can easily be measured by applying an external bias until the surface charges disappear, wherein the external bias equals the CPD ($V_{DC} = V_{CPD}$).

The KPFM is a tapping mode technique and is performed in this work in double-pass mode, in which at first trace the topography of the sample is recorded and then in a certain lift distance from the substrate the contact potential difference (V_CPD) is captured. The V_CPD directly depends on the work function of the sample and tip, thereby gives straightforward information about the Fermi levels of the sample and tip. Similar to the AFM which can detect atomic forces by amplitude or frequency modulations (AM or FM), KPFM can measure the electrostatic force ($F_ω$) either by AM or FM modes, the latter (FM mode) known to have much higher spatial resolution and accuracy. [169,170] The tip and sample form a capacitor with an electrostatic force (F_es) in between, which is:

$$F_{es}(z) = -\frac{1}{2}\frac{\partial C(z)}{\partial z}(\Delta V)^2 \qquad (3\text{-}1)$$

where $z$ is the direction normal to the sample surface, $\Delta V$ is the potential difference between $V_{CPD}$ and the voltage applied to the AFM tip ($\Delta V = V_{tip} \pm V_{CPD}$), and $\partial C/\partial z$ is the gradient of the capacitance between the tip and sample surface. [169]

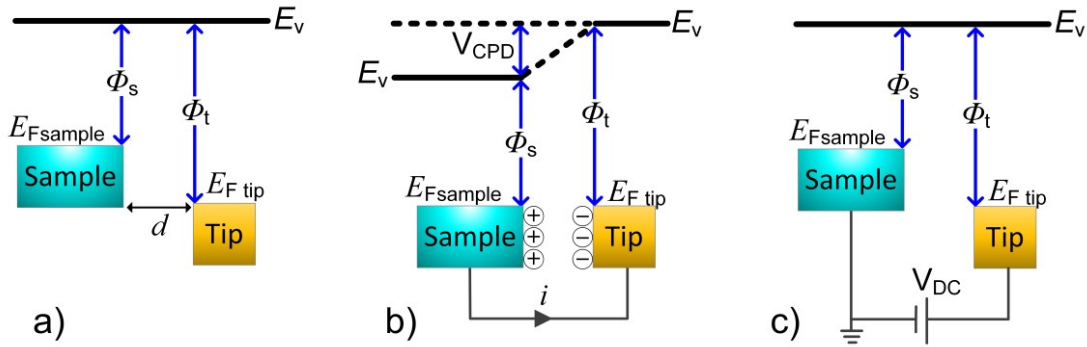

**Figure 3.2. Working principle of Kelvin probe microscopy.**
The illustration of energy and charge diagram of the Kelvin probe microscopy technique principle, where $E_v$ is the vacuum energy level, and $E_{Fsample}$ as well as $E_{Ftip}$ are Fermi energy levels of the sample and tip, respectively. a) Separated tip and sample with a distance and no electronic contact. b) Tip and sample are in electrical contact and the V_CPD forms in between. (c) External bias ($V_{DC}$) is applied between the tip and sample to nullify the $V_{CPD}$.

In the double pass model, the mechanical excitation of the cantilever is deactivated, and a voltage ($V_{AC}$ sin($\omega$t) + $V_{DC}$) is applied, therefore the $\Delta V = V_{tip} \pm V_{CPD} = (V_{DC} \pm V_{CPD}) + V_{AC}$ sin($\omega$t). By substituting $\Delta V$ in the equation **(3-1)**, the $F_{es}$ will be as follow, including three parts (where $\omega = 2\pi f_0$):

$$F_{DC} = -\frac{\partial C(z)}{\partial z}\left[\frac{1}{2}(V_{DC} \pm V_{CPD})^2\right] \qquad (3\text{-}2)$$

$$F_{\omega} = -\frac{\partial C(z)}{\partial z}(V_{DC} \pm V_{CPD})V_{AC}\,sin(\omega\tau) \qquad (3\text{-}3)$$

$$F_{2\omega} = \frac{\partial C(z)}{\partial z}\frac{1}{4}V_{AC}^2[\cos(2\omega\tau) - 1] \qquad (3\text{-}4)$$

The AC bias leads to oscillation force ($F\omega$ and $F2\omega$) at two harmonics of $\omega$ and $2\omega$ when the ($V_{DC} \pm V_{CPD}$) is not zero. AM-KPFM measures $F\omega$ directly from the amplitude of the cantilever oscillation at $\omega$ that is induced by $V_{CPD}$ and $V_{AC}$. The feedback loop in AM-KPFM tries to maintain the $F\omega$ by using the oscillation amplitude ($V_{AC}$) until the $V_{AC}$ drops to zero when $V_{DC}$ equals the CPD. With knowing the work function of the tip ($\phi_{tip}$), the work function of the sample can be determined as $\phi_{sample} \approx \phi_{tip} - e\Delta V_{CPD}$.

**Figure 3.3. Schematic depiction AM-KPFM operation in tapping mode.**

The cantilever measures surface topography on the first (main) scan (trace and retrace) (marked as 1). The cantilever ascends to lift scan height (marked as 2), then follows the stored surface topography at the lift height above the sample while responding to electric influences on the second scan (marked as 3).

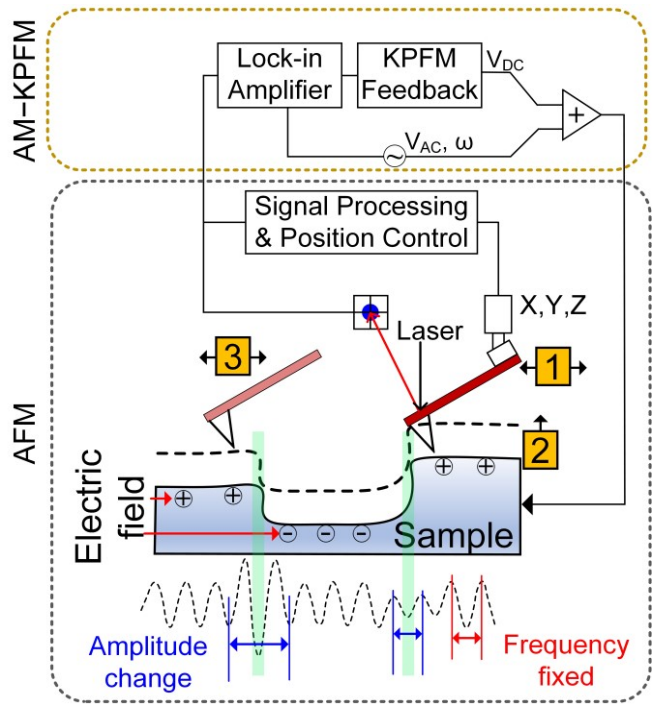

The frequency of the AC bias is typically selected to be the resonant frequency ($f_0$) of the AFM cantilever for enhanced sensitivity afforded by the cantilever's quality factor (Q). **Figure 3.3** displays a block diagram of our KPFM (AM) setup. A KPFM made by JPK Instruments AG was used to study the samples in this works. The KPFM was equipped with a conductive coated (Cr/Pt, 5/25 nm) micromachined monolithic silicon tips (Tap300E-G, r=25nm) from ''budget sensors''. Although the KPFM operation is relatively simple, however, the low spatial resolution, distance tip dependency, the apex of tip, geometry of conductive probe [171], measurement environment influences (e.g., air, vacuum), parasitic capacitance, tip quality all in particular for AM mode strongly limit a quantitative investigation of the sample. Therefore, generally, the technique suffers from a weak lateral resolution ~50–70 nm. [169,170,172]

## 3.3. Raman spectroscopy

As a non-destructive tool, Raman spectroscopy enables valuable information about the crystal's phononic properties. In this work, the Raman spectroscopy as a powerful method was implemented mainly for characterizing the thickness [173], doping [174], strain [175], and defects [176–178] of graphene samples.

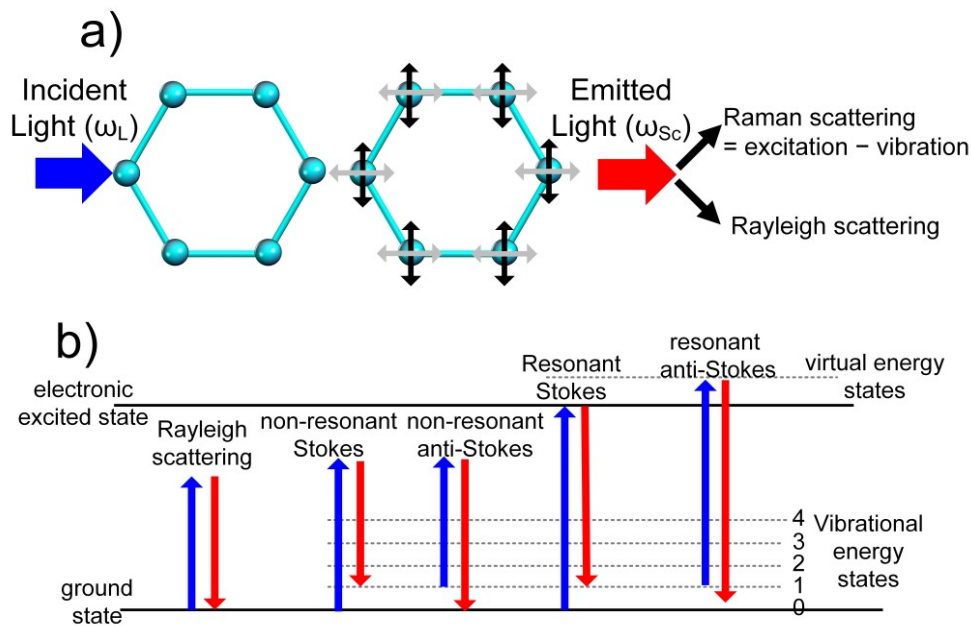

**Figure 3.4. Schematic representation of the Raman scattering mechanism.**
a) The incident light ($\omega_L$) impinges on the sample and excites an electron-hole pair. The pair decays into a phonon $\Omega$ and another electron-hole pair. The latter recombines and emits a photon ($\omega_{Sc}$). b) The resulting Rayleigh and Raman scattering in resonant and non-resonant conditions. Figures are partly taken from refs. [75,179].

The Raman spectroscopy technique is based on an inelastic light (photons) scattering (Raman scattering) by phonons. When the photons hit on a sample, a time-dependent perturbation occurs in the kinetic and potential (Hamiltonian) energies in the system. Because of the photon rapidly-changing electric field, only electrons respond to this perturbation. Herein, the measurement process involves the excitation of electrons in the crystal from a ground state ($E_{GS}$) through the monochromatic light ($\hbar\omega_L$) lead to an energy increase to $E_{GS} + \hbar\omega_L$ (virtual state).

Rayleigh scattering happens when the excited electron-hole pair returns to the ground state, and the frequency of the emitted and incident photons remains the same, but only the propagation direction of the photon may change (elastic scattering).

With a much lower probability compared to Rayleigh scattering, the Raman scattering happens when the incoming photon loses part of its energy while interacting with a phonon, therefore, leaving the sample with a lower energy of $\hbar\omega_{Sc}$. This is known as the Stokes process. The energy is equal to $\hbar\omega_L - \hbar\omega_{Sc} = \hbar\Omega$ (An incoming photon $\omega_L$ excites an electron-hole pair. The pair decays into a phonon $\Omega$ and another electron-hole pair).

If the incoming photon finds the sample in an excited vibrational state, after the interaction process, the system comes back to its ground state, and the photon can exit the sample with an increased energy of $\hbar\omega_{Sc} = \hbar\omega_L + \hbar\Omega$. This corresponds to the anti-Stokes process and has been reported to be less probable than the Stokes process. Therefore, the Raman scattering is displayed as the Stokes measurement intensity of the scattered light due to the difference between the incident and scattered photon energy known as the "Raman shift." The unit of the Raman shift is usually written as $cm^{-1}$, which can be converted in meV as $1 meV = 8.0655447$ $cm^{-1}$ ($1$ $cm^{-1} \approx 0.124$ meV). [75,179,180] The non-resonant Raman scattering is when $E_{GS} + \hbar\omega_L$ does not correspond to a stationary state, as is indeed the case for most materials. [179]

The principles of Raman spectroscopy in graphene can be better understood considering the origin of the different vibrational modes, as shown in the phonon band dispersion in **Figure 3.5**. Since graphene has two atoms in its unit cell, six phonon branches exist, including three acoustic (A) and three optical (O) phonons. Four of these phonons (two O and two A) are in-plane (i), and the other two phonons (one A and one O) are out-plane (o). If the vibration direction is perpendicular to or parallels with the carbon-carbon bonds, then the modes are classified as transversal (T) or longitudinal (L), respectively. The Raman bands of graphene mainly are due to the iLO and iTO vibration modes. [178,181]

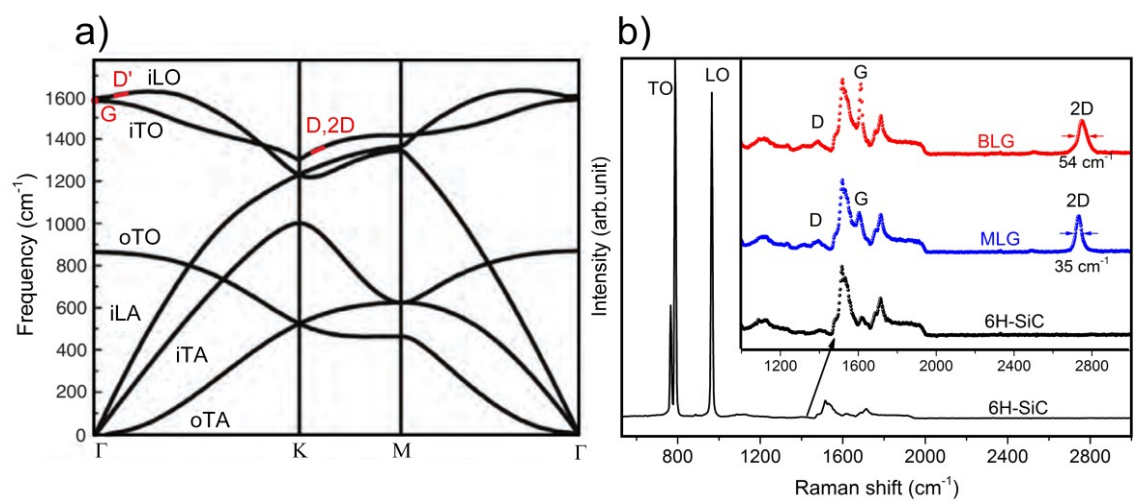

**Figure 3.5. Raman spectroscopy on epitaxial graphene.**
a) The phonon dispersion relation of graphene indicating the oTA, iTA, iLA, oTO, iTO iLO modes (adapted from ref. [181]). (b)Typical Raman spectra and characteristic features (D, G, 2D peaks) of SiC substrate (black), monolayer graphene (blue), and bilayer graphene (red).

Considering the Raman spectra of epitaxial graphene, three prominent features are essential: the so-called D-peak, G-peak, and 2D-peak, as demonstrated in **Figure 3.5b**. The G-Peak is a first-order non-dispersive Raman scattering band, originated from the in-plane vibrations of $sp^2$ bonded carbon atoms in graphene. As shown in **Figure 3.6a-c**, by impinging the incident photon (blue arrow), an electron-hole pair is excited at the Γ-point of the first Brillouin zone. The electron-hole pair is scattered by the iTO or iLO phonons and recombines by emitting a photon. Since the G-peak is directly related to graphene's Fermi energy, it is experimentally significant for investigating the carrier density in graphene. [174]

The D-peak, the second-order dispersive Raman scattering is due to the breathing-like modes of six-atom rings, see **Figure 3.6d, e**. The D-peak appears in the presence of point defects or grain boundaries. The intensity ratio of D/G reflects the number of defects and grain size. [173,176,178]

All Raman mappings were performed on a LabRAM Aramis (Horiba Jobin Yvon) confocal Raman spectrometer, which is equipped with a Czerny Turner spectrograph enabling a spectral resolution < 1 cm⁻¹, three different Laser sources (532 nm (~2.33 eV) Nd:YAG, 632 nm (~1.96 eV) HeNe, 785 nm (~1.58 eV) laser diodes), four holographic gratings (600, 1200, 1800, 2400 grooves mm⁻¹), and a thermoelectrically cooled CCD detector. The Nd YAG laser was used as a radiation source by focusing the laser beam onto the graphene sample through a 100× objective (N.A. 0.95). A piezo stage (PI) mounted on top of a motorized

Märzhäuser stage enables precise Raman mappings across the graphene sample. The Raman measurements were performed by S. Wundrack from the Optical Analysis Lab in the Chemical Physics department at the PTB Braunschweig.

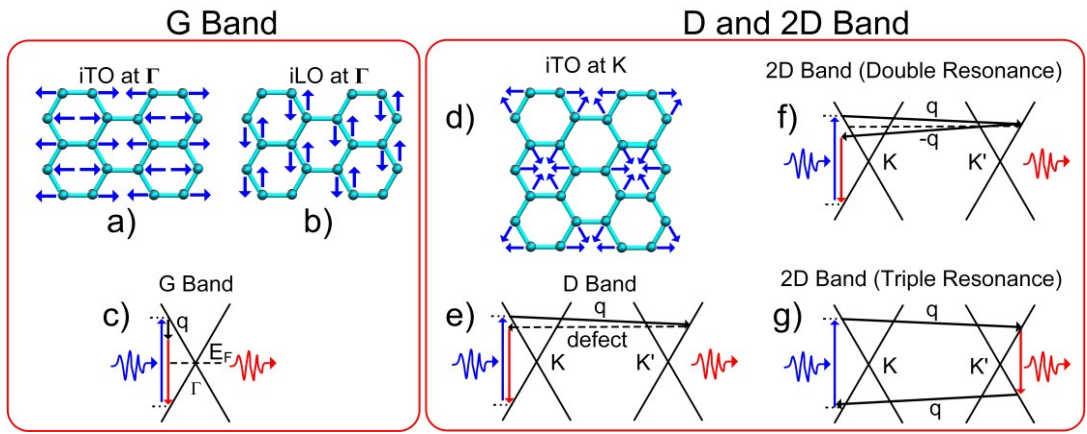

**Figure 3.6. Phonon modes in Raman spectroscopy of graphene.**
Phonon modes iTO (a) and iLO (b) at Γ-point and the sketch of Raman process in graphene for G band (c). (d) Schematic representation iTO at K-point. (e) D band resonant process which involves a scattering from defects (horizontal dotted line). 2D band Raman process occurs through a second-order process, which is either double resonant (f) or triple resonant (g). [178]

## 3.4. Scanning electron microscopy

Scanning Electron Microscope (SEM) is used to scan the substrates by exposing a focused electron beam and generating images, including information about the substrate's topography and composition. As depicted in **Figure 3.7**, the SEM instrument comprises three major sections: the specimen chamber, the electron column, and the electronic controls (PC). The latter provides control knobs and switches that allow for instrument adjustments such as filament current, accelerating voltage, focus, magnification, brightness, and contrast. The electron column is where the electron beam is generated (in a vacuum), focused to a small diameter (via condenser lenses), and scanned across a specimen's surface by electromagnetic deflection coils. The specimen chamber is located at the lower side of the column, wherein the specimens are mounted and secured onto the stage. **Figure 3.7** shows the main components of the SEM. The free electrons are generated via the electron gun located at the top of the column. Electrons are primarily accelerated toward an anode that is adjustable by an applied voltage.

**Figure 3.7. Schematic of the working principle of a scanning electron microscope.**

Adapted from ref. [182]

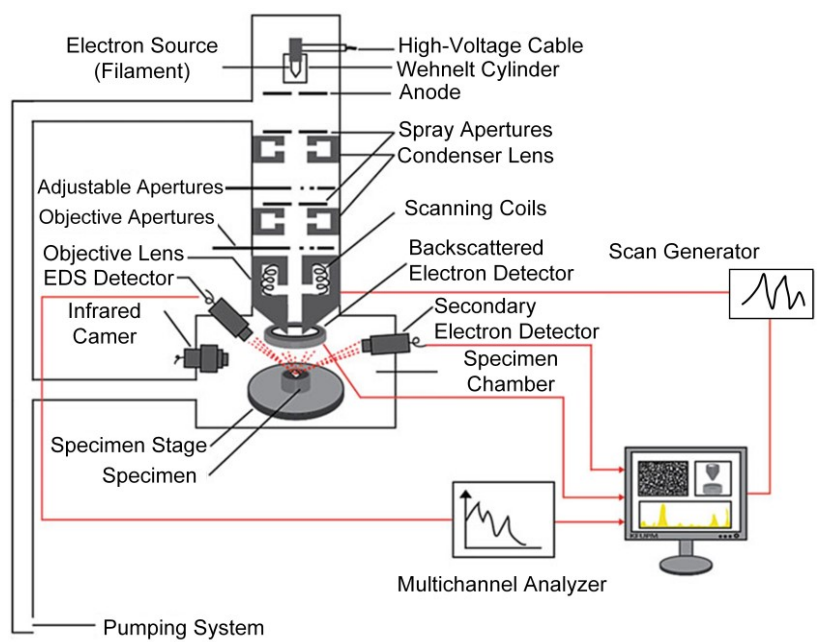

The electron beam will be focused after passing the anode and being influenced by two condenser lenses that cause the beam to converge and pass through a focal point. [182] The electrons in the beam penetrate the surface (a few microns) of a bulk sample, interact with its atoms and generate a variety of signals such as secondary and backscattered electrons (which are used to form images) and X-rays, which are used to obtain elemental constitution of the specimen material. The ultimate lateral resolution of the image obtained in the SEM corresponds to the electron beam's diameter. Advances in the lens and electron gun design yield very fine probe diameters giving image resolutions between 1nm to 20 nm, and depending on the instrument, can reach a point resolution <1 nm. [182]

The nature and type of contrast in SEM image depend on the type of specimen and its interaction with the electron beam and the number of electrons emitted from the specimen based on the operating conditions employed during microscopy. The difference in the signal between the two points may arise due to many factors, including change in specimen topography, the difference in composition, crystal orientation, magnetic or electric domains, surface potential, and electrical conductivity. However, most images depict a combination of contrast mechanisms. For example, a material (the term phase is more appropriate) with a higher atomic number will appear relatively brighter (due to a larger number of backscatter electrons being ejected out of this phase) while a phase with a low atomic number will appear relatively dark. [182–184] These types of contrasts are important and required to be adequately noticed for a correct interpretation of the SEM images, particularly in epitaxial graphene wherein the SEM contrast could arise from various sources, e.g., buffer layer,

SiC, MLG, BLG (material contrast), topography contrast (SiC step edges), or crystal orientation.

The SEM measurements on the samples in this study have been mostly performed at PTB Braunschweig by Peter Hinze and Kathrin Stör from Nanostructuring Lab using a "SUPRA 40" field emission scanning electron microscope (FESEM) from Carl Zeiss SMT that allows acceleration voltages between 0.1 keV and 30 keV and enables magnification up to 1000000x.

## 3.5. Scanning tunneling: microscopy, potentiometry, spectroscopy

The invention and discovery of scanning tunneling microscopy (STM) by Binnig and Rohrer in 1981 [185] has enabled an atomically resolved image of almost any conducting surface. In this work, STM was used to characterize the quality of the graphene with atomic resolution. In addition to STM, which enables us to scrutinize the graphene morphology, also scanning tunneling potentiometry (STP) was used, which facilitates investigating the local charge transport on the atomic scale. Experiments were conducted in a UHV chamber at room- or low-temperatures with a pre-cleaning of the samples in UHV.

The STM is based on the quantum mechanical tunneling effect. The basic idea behind STM is illustrated in **Figure 3.8a**. When the sharp metal tip is brought close enough to the sample surface (height control via piezo-drives ($P_Z$)), electrons can tunnel through the vacuum barrier between tip and sample. By applying a bias voltage on the sample, a tunneling current can be measured through the tip, which is exponentially sensitive to the distance between the tip and the surface. Therefore, the tip is usually < 1nm away from the sample. [186] By using two other piezo-drives ($P_X$ and $P_Y$), the tip scans in two lateral dimensions. A feedback controller is employed to adjust the height of the tip to keep the tunneling current constant. During the tip scanning over the surface, the height of the tip (the voltage supplied to $P_Z$) is recorded as an STM image, which represents the topography of the surface. Therefore, the topography does not show the real surface structure but the contours of a constant integrated LDOS at a certain distance from the sample. [184,186] This operation mode of STM is called "constant current" mode. Constant current mode is mostly used in STM topography imaging. It is safe to use the mode on rough surfaces since the distance between the tip and sample is adjusted by the feedback circuit. [184]

On a smooth surface, it is also possible to keep the tip height constant above the surface; then, the tunneling current's variation reflects the small atomic corrugation of the surface. This "constant height" mode has no fundamental difference to the "constant current" mode. However, the tip could be crashed if the surface corrugation is big. On the other hand, in this mode, the STM can scan very fast to research the surface dynamic processes. [184]

Scanning tunneling potentiometry (STP) is a versatile tool that provides valuable interrelated information about the local potential and spatial topography of the sample on the nanoscale. The main idea of the STP is to measure the electrochemical potential (ECP) locally with the resolution of an STM. The basic principle of STP is shown in **Figure 3.8b**. In the STP setup, the samples are ex-situ contacted with two gold contacts in a shadow mask procedure such that additional bias voltage $V_{Transport}$ can be applied, inducing an electric current ($j$) in the sample [187], as shown in **Figure 3.8b**. The STP measurements were performed at every image point by adjusting the ECP at the tip at a fixed tip-sample distance. The applied bias voltage is switched off for STP, while only the transport potential ($V_{Transport}$) across the sample remains. The potential at the tip is adjusted in a way that the tunneling current $I_T = 0$. Subsequently, the voltage $V_{STP}(x, y) \big|_{I_T=0}$ necessary to compensate the net tunnel current is recorded in a map, that is the STP map. This voltage $V_{STP} = \frac{\mu_{ECP}}{e}$ has been referred to as the local ECP, which is inherently defined by the STP method. [187–189]

The measurements are made at different values of the electron current in the sample plane, especially at zero and forward and reversed current as defined by the potential applied to the sample contacts. The STP measurements were performed to evaluate the charge transport properties in graphene samples, particularly considering the resistance anisotropy on the local scale, as are discussed in Chapter **6**.

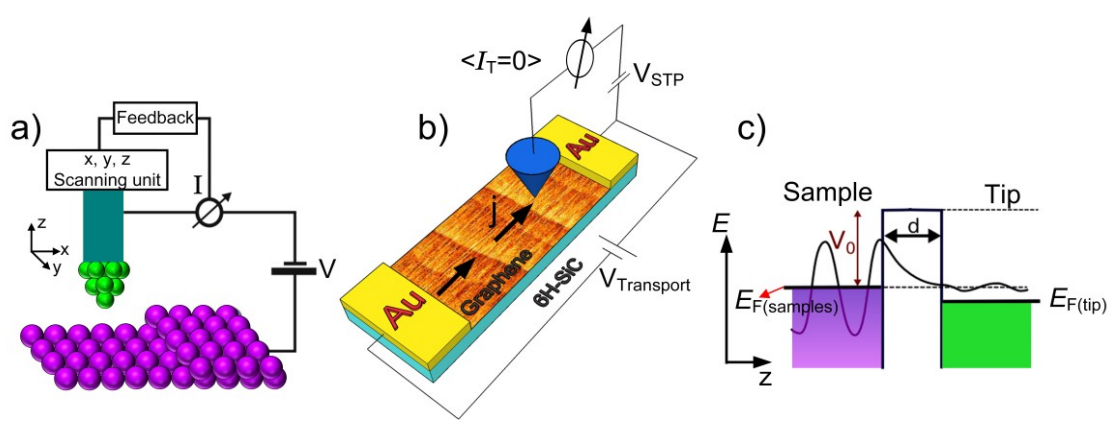

**Figure 3.8. Operating principle of scanning tunneling microscope.**
(a) Sketch of the basic setup of an STM (b) Schematic of scanning tunneling potentiometry. (c) illustration of the quantum tunneling effect in STS measurement. Adapted from refs. [39,190,191].

The scanning tunneling spectroscopy (STS) is a complementary technique that enables probing the LDOS of the sample. This is accomplished by measuring the current-to-voltage characteristic of the tunneling junction, see **Figure 3.8c**. Practically, an STS spectrum (differential conductance $(dI_t/dV_t)$) is obtained using the tip, which is kept over a constant height above a distinct position on the surface. Then the feedback loop is disabled (leaving the tip-sample distance constant during spectroscopy), and the tunneling current $I_t$ is recorded as a function of the tunneling voltage $V_t$ signal. Thus, in the simplest approximation, the differential conductance is approximately proportional to the energy-dependent local density of states of the sample ($dI_t/dV_t \propto D_{\text{sample}}(eV)$), where $D_{\text{sample}}$ denotes the LDOS. In this approximation, the differential conductance measures the sample density of states at the energy eV relative to the Fermi energy of the sample. [190]

The STM measurements in this thesis were performed by T. T. N. Nguyen from the University of Chemnitz within the group of Prof. C. Tegenkamp as well as A. Sinterhauf, G. A. Traeger, and P. Willke from the Georg-August-University of Göttingen within the group of Prof. M. Wenderoth.

## 3.6. Low-energy electron diffraction

As a non-destructive technique, low energy electron diffraction (LEED) is a powerful widely-used method for surface structural analysis. The history of LEED dated back to about 1924, when by derivation from Einstein's well-known matter and energy equation $E = mc^2$, and Planck's theory $E = h\nu$, de Broglie introduced his equation as follow:

$$\lambda = \frac{h}{p}(\text{Å}) = h/\sqrt{2mE\ (eV)} \approx \sqrt{150.4/E(eV)} \qquad (3\text{-}5)$$

for the wavelength ($\lambda$) of an electron ($e$) with mass ($m$), momentum (p), energy ($E$), and $h$ the Planck's constant. The de Broglie's postulation was experimentally observed by Davisson and Germer (1927), confirming that the intensity distribution of low energy electron backscattered from a surface obeys diffraction law. They studied the scattering of electrons from Ni (111) and found that the maximum in the reflected intensity of the elastically scattered electrons at any angle satisfied the plane grating formula (Bragg condition):

$$n\lambda = \alpha \sin\theta = d \qquad (3\text{-}6)$$

where $\alpha$ is the spacing between adjacent rows of atoms and $\lambda$ is given by de Broglie relationship, and n is an integer (see **Figure 3.9b**). [192] Accordingly, the LEED experiment principle is not very complex, as schematically is shown in **Figure 3.9a**. A primary narrow beam of monoenergetic electrons with a low energy range between 25 eV up to 600 eV (correspond to de Broglie wavelength of 0.5 to 2.5 Å, i.e., it is of the order of interatomic distances in a solid) is directed onto a planar single crystal surface at a given angle as shown in **Figure 3.9**. Several diffracted beams of electrons with the same energy as the incident (elastic) beam are produced in the backward direction. [193] The inelastically backscattered electrons are suppressed by the grids. The spatial distribution of these beams and their intensities as a function of the incident beam's angle and energy (elastic) provides information that can be used to analyze the surface structure. Moreover, because the mean free path of low energy electrons (as mentioned above) in a crystal is only a few angstroms, only the first few atomic layers will engage in the diffraction. This leads to no diffraction in the direction perpendicular to the sample surface. Thus, the reciprocal lattice of the surface is a 2D lattice with rods extending perpendicular from each lattice point, as shown in **Figure 3.9a**.

**Figure 3.9. Working principle of low energy electron diffraction.**

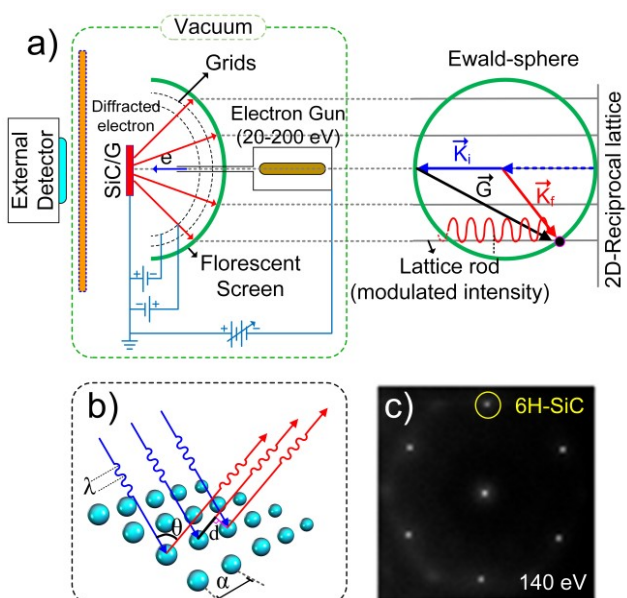

a) Schematic of the working principle of a typical LEED system with a fluorescent screen and hemispherical grids. The elastically backscattered beams hit a luminescent screen after having passed an additional accelerating voltage applied between a transparent grid near the screen. Additional grids make the space between the sample and the near screen region field free to allow for free-electron traveling and filter out the inelastically scattered electrons. Image of the reciprocal lattice with the incident wave vector $k_i$, the diffracted wave vector $k_f$, the 3D scattering vector G. b) The principle of diffraction of electrons at a surface with the symmetrically separated 2D chain of atoms. c) The LEED pattern of a pure 6H-SiC.

This can kinematically be described by the Laue condition relating the wave vector of an incident electron $\vec{K_t}$ and a scattered one $\vec{K_f}$ as follow:

$$\vec{K_t} - \vec{K_f} = \vec{G} \tag{3-7}$$

where $\vec{G}$ is a reciprocal lattice vector, and as mentioned, the perpendicular component is absent and has only a parallel component.

In the so-called spot profile analysis low-energy electron diffraction (SPA-LEED), the detection of diffracted electrons is done by a channeltron detector, unlike the standard LEED with a phosphorous screen. The SPA-LEED has the advantage of a larger accessible reciprocal space in contrast to the conventional LEED. This is due to electrostatic deflection plates that are used to continuously vary the angle of incidence of the electron beam in all directions at the sample position. Subsequently, scanning the incident angle of the electron beam results in a simultaneous variation of the angle under which diffracted electrons from the surface are recorded. This variation of both the incident and the exit angle of the electrons results in a very special scanning mode in reciprocal space, in which the angle between the incident and final scattering vector stays constant while the incident angle is changed. So, the Ewald sphere is rotated around the origin of reciprocal space, resulting in the recorded diffraction pattern not following the Ewald sphere (unlike the conventional LEED) but a sphere with twice as the diameter of the Ewald sphere. [194] The SPA-LEED images are akin to the conventional LEED images but contain quantitative information of the diffraction spot intensities (e.g., peak profiles) and much higher k-space resolution.

The LEED measurements in this work were carried out by J. Aprojanz from the Leibniz University Hannover within the group of Prof. C. Tegenkamp.

## 3.7. X-ray photoelectron spectroscopy

X-ray photoelectron spectroscopy (XPS) is a versatile quantitative spectroscopic technique that enables measuring the elemental composition, empirical formula, chemical state, and electronic state of the elements that exist within a material. XPS spectra are obtained by irradiating a material with a monochromatic photon beam of X-rays, causing the sample to emit electrons (the so-called photoelectrons) while simultaneously measuring the kinetic energy ($E_K$) and the number of electrons that escape from approximately top 10 nm of the material being analyzed. A basic XPS setup consists of an X-ray source, an electron analyzer to measure the kinetic energy of the photoelectrons, a detector to count the number of electrons, data acquisition, and a processing system. [195] **Figure 3.10** illustrates the XPS working principle in which an X-ray photon interacts with a core level electron transferring its photon energy and causing electron emission by the photoelectric effect.

The kinetic energy of the photoelectron ($E_k$) is the difference between the X-ray photon energy $h\nu$ ($h$ is the Plank's constant and $\nu$ is the X-ray frequency) and the binding energy ($E_b$) of the core-level electron. Since the X-ray energy is known and the $E_k$ of the photoelectron can be experimentally determined, the $E_b$ of the emitted electron is given by the following equation:

$$E_b = h\nu - E_k - \phi_{spec} \qquad (3\text{-}8)$$

where $\phi_{spec}$ is the work function of the spectrometer. The $E_k$ is measured in the spectrometer concerning the $E_{vac}$ of the spectrometer.

The XPS measurement and surface analysis need to be performed in an ultra-high vacuum environment (< $10^{-9}$ Torr) to increase the mean free path of the photons and electrons, remove adsorbed gases from the sample and prevent adsorption of contaminants on the surface.

The XPS measurements presented in this work are mainly carried out on so-called intercalated buffer layer samples. This reveals valuable information about the sample and origin of intercalation, as is discussed in detail in Chapter **5**. The XPS measurements were performed by Philip Schädlich from the University of Chemnitz within the group of Prof. T. Seyller. The system operates with a monochromatic Al K$_\alpha$ radiation (1486.6 eV) sample spot size of about 2.5 mm² and a hemispherical Phoibos 150 MCD-9 analyzer (SPECS). Before XPS, samples were degassed at a maximum temperature of ~370 °C for 1 hour.

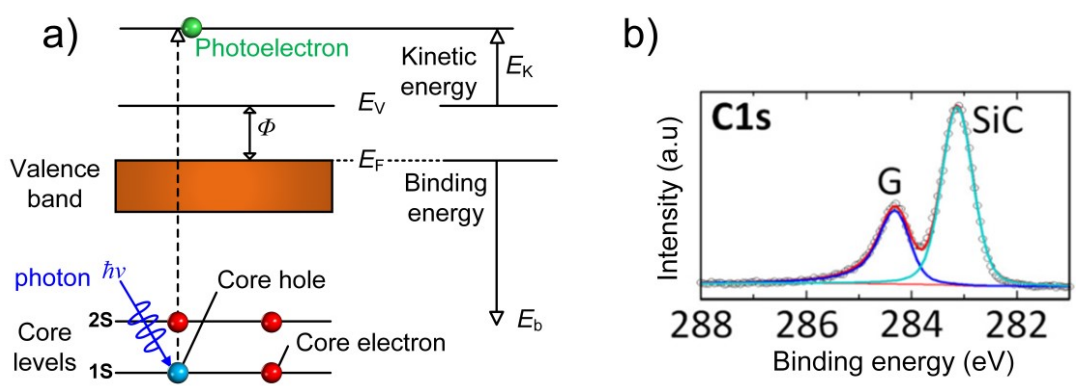

**Figure 3.10. Schematic drawing of the photoelectric effect in XPS.**
a) The process involves electron from the core state via a high energy X-ray irradiation and propagate through the crystal and finally emits into the vacuum. Red (blue) circles represent electrons (holes) within different orbitals (1s, 2s). The XPS spectra of the C 1$S$ core levels of so-called quasi-freestanding monolayer graphene on SiC.

## 3.8. Low-energy electron microscopy

The low energy electron microscopy (LEEM) is a technique based on the principle of LEED. [196] The LEEM measurement relies on elastically reflected electrons created by a low-energy electron beam up to about 100 eV (frequently with less than 10 eV) acceleration voltage to image the sample. Due to the low energy of the electron, LEEM is a highly surface sensitive characterization technique. Therefore, it is a powerful tool to study SiC/G particularly, providing both insights into the local graphene coverage and thickness using reflectivity spectra [23,197] as well as giving information about the local SiC stacking order through dark-field measurement. [78,198–200] The latter is obtainable in the LEEM instrument since it is capable of imaging the diffraction pattern through an intermediate lens, thereby the LEED and dark field images can be captured.

The schematic of a basic LEEM setup is shown in **Figure 3.11**. A focused electron beam with high energy (20 keV) is injected through the objective lens along its optical axis, which then shortly before reaching the sample is decelerated (in the cathode lens) to the desired low energy of a few eV. To be able to produce an image, this incident beam has to be separated from the reflected beam by a magnetic field beam divider (separator). The elastically back-scatted electrons are again accelerated (~20 keV) and deflected into the imaging column (similar to an optical microscope) via the magnetic field. [201] A phosphorous screen is used as a projection plane for LEEM imaging. A so-called LEEM bright-field (LEEM-BF) image is obtained using all back-reflected electrons for imaging.

Moreover, LEEM enables studying an aerial contribution of a certain phase. This can be achieved by aligning an aperture in the diffraction spot, allowing electrons reflection only from the desired phase. Through this so-called dark-field imaging (LEEM-DF) technique, an intensity image of the selected diffraction spot is achieved.

The same experimental setup can be used for X-ray photoemission electron microscopy (XPEEM), which directly measures the work function variation on the surface. [202] In this mode, the sample is illuminated by light (X-ray) at a fixed energy level ($h\nu$), thereby the electrons from the atomic core level of the sample are excited, and they escape from the sample. These photoelectrons with certain kinetic energy equal to $E_{kin} = h\nu - E_{bin} - \phi$ are selected by the hemispherical energy analyzer to form an XPEEM image, where $E_{bin}$ is the core level binding energy and $\phi$ the work function. The XPEEM measurement in this work was performed by Alexei A. Zakharov at the MAXIV laboratory. More information about the measurement setup can be found in ref. [203].

Also, from the low-energy electron reflectivity (LEEM-IV), also known as LEEM-IV of graphene samples, the information about the layer thickness, as well as interlayer distance, can be extracted. This can be perceived from the LEEM-IV spectra, which exhibit a dip (local minima) whenever the electrons can be absorbed by an unoccupied state of the sample. This happens because the confinement of electrons in thin films can create quantum well (QW) bound states. The QW states at discrete energy levels produce peaks in the photoemission energy spectrum or reflectivity of the low-energy electrons. The energy levels of the QW states change with the film thickness. Therefore, the photoemission intensity shows an oscillatory behavior as a function of the electron energy and film thickness, in which the number of dips in spectra corresponds to the number of graphene layers. For a detailed explanation, see refs. [23,197,204,205]. The LEEM-IV investigations of the graphene samples are discussed in detail in Chapter **7**.

The LEEM measurements in this work were carried out by P. Schädlich from the university of Chemnitz within the group of Prof. T. Seyller as well as by Alexei A. Zakharov at the MAXIV laboratory.

**Figure 3.11. Schematic of the LEEM instrument.**
For the details, see the text and also ref. [201,206]

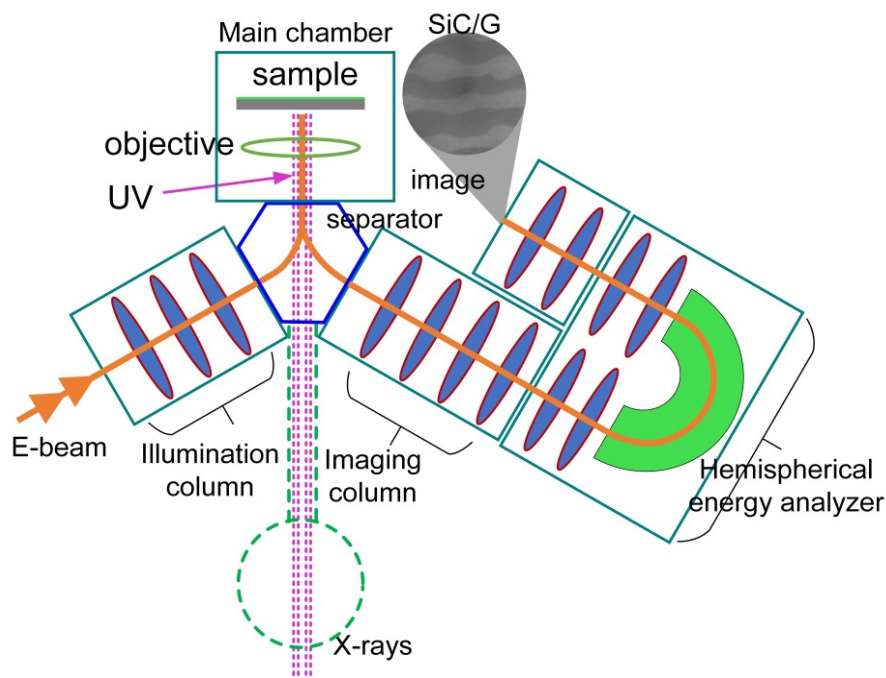

**Figure 3.12. Nano-four-point-probe measurement setup.**

Schematic of the 4-tip transport setup featuring a Keithley source-meter (bottom) and a sketch of STM tips on a graphene sample for the rotational N4PP measurement with respect to the SiC step and terraces (top), Adapted from refs. [191], and [39], respectively.

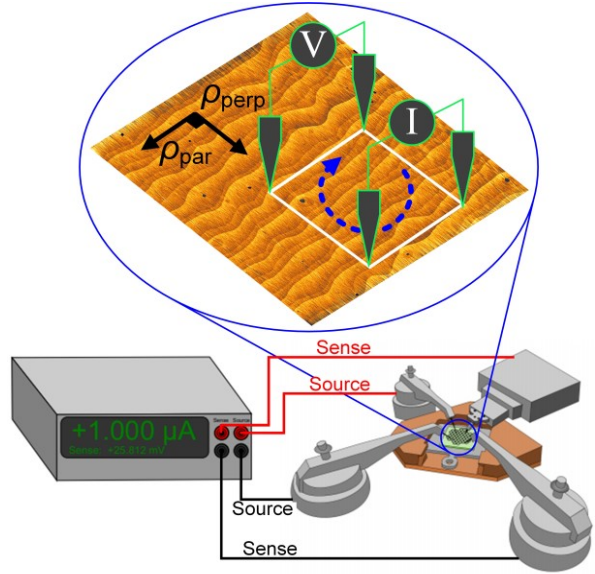

## 3.9. Square nano-four-point-probe measurement

Angle-dependent Nano-Four-Point Probe (N4PP) electronic transport measurement in a square configuration is used in this work to investigate a so-called extrinsic resistance anisotropy of epitaxial SiC/G samples. This method, which enables electronic transport investigations on micrometer scales, also gives valuable information about the sheet resistance (intrinsic resistivity) of the sample. The measurement configuration of the N4PP on the graphene sample is shown in **Figure 3.12**. The setup includes four STM tips in a square arrangement, which are equally-spaced with a certain distance between the probes (in our case 100 μm). Two probes conduct the current through the material, and the two others, measure the voltage difference. The N4PP measurements were carried out for different angles between the direction of the current probes and the step edges. The angles of 0° and 180° (90°) correspond to the current flow parallel (perpendicular) to the steps (see **Figure 3.12**).

The measured resistances $R_\theta$ for a given angle $\theta$ are adequately described by:

$$R_\theta = \frac{1}{2\pi\sqrt{\sigma_\parallel \sigma_\perp}} \times ln \sqrt{\frac{(\sigma_\parallel/\sigma_\perp + 1)^2 - 4\cos^2\theta\, sic^2\,\theta\,(\sigma_\parallel/\sigma_\perp - 1)^2}{(sin^2\theta + \sigma_\parallel/\sigma_\perp \cos^2\theta)^2}} \qquad (3\text{-}9)$$

where $\sigma_\parallel$ and $\sigma_\perp$ denote the conductivities measured parallel and perpendicular to the step direction, respectively, assuming an anisotropic 2D sheet with different conductivities in x- and y-direction. From the fitting procedure, finally the resistivity values perpendicular ($\rho_{perp} = \sigma_\perp^{-1}$) and parallel ($\rho_{par} = \sigma_\parallel^{-1}$) to the step edges are obtained, and the anisotropy ratio is calculated as $A = \rho_{\text{perp}}/\rho_{\text{par}}$. Because the current flow via the semi-insulating SiC substrate and the buffer layer is negligible, the measured resistance is related to the 2D graphene sheet on top. [207,208]

The angle-dependent nano four-point probe (N4PP) measurements were taken in an Omicron UHV nanoprobe system. [209] The samples were kept in UHV at room temperature after a thermal cleaning procedure by heating up to 300 °C. The position adjustment of the tips on the sample surface was carried out by helping an SEM, positioned on the right top of the STM tips. The N4PP Measurements were performed by J. Aprojanz and J. Baringhaus, J. P. Stöckmann, at the Leibniz University Hannover within the group of Prof. C. Tegenkamp.

## 3.10. Van der Pauw measurement

In addition to the above-discussed N4PP method, which enabled measurement on micrometer-scales, Van der Pauw (VdP) [210,211] is another four-point probe technique used in this work which allows transport measurements on large-scale graphene samples. Using the VdP, the sheet resistance ($R_{\text{sh}}$) can be measured. For a proper VdP measurement, the samples must have homogeneous resistivity (no anisotropy), and the contacts should be small and located at the periphery of the sample that could be arbitrarily shaped sample but with a uniform thickness. Implementing the VdP measurement in a cryostat facilitates measuring charge carrier concentration ($n_s$) and mobility ($\mu$) of the sample as a function of temperature.

The resistivity can be derived from a total of eight measurements configuration (see **Figure 3.13c**) that are made around the edges of the sample. For each measurement, the current is applied on two adjacent contacts, and voltage is read off from the two other opposite remaining contact pair, and the corresponding resistance can be simply calculated using Ohm's law $R_{ij,kl} = \frac{V_{kl}}{I_{ij}}$. From the entire measurements, two following values are derived:

$$\rho_A = \frac{\pi}{\ln 2} \, f_A t_A \frac{(V_1 - V_2 + V_3 - V_4)}{4I} = \frac{\pi}{\ln 2} \, f_A t_A \left(\frac{R_{12} + R_{34}}{2}\right) \qquad (3\text{-}10)$$

$$\rho_B = \frac{\pi}{\ln 2} \, f_B t_B \frac{(V_5 - V_6 + V_7 - V_8)}{4I} = \frac{\pi}{\ln 2} \, f_A t_A \left(\frac{R_{56} + R_{78}}{2}\right) \qquad (3\text{-}11)$$

where: $\rho_A$ and $\rho_B$ are the resistivities in ohm-cm; $t_s$ is the sample thickness in cm; $I$ is the current through the sample in amperes; $f_A$ and $f_B$ are geometrical factors based on sample symmetry related to the two resistance ratios $A_A$ and $A_B$. The relation between A and $f$ (where A= $(A_A+A_B)/2$ and $f = (f_A + f_B)/2$)), is shown in the following equations ($f_A = f_B = 1$ for a perfect square symmetry):

$$A_A = \frac{(V_1 - V_2)}{(V_3 - V_4)} \tag{3-12}$$

$$A_B = \frac{(V_5 - V_6)}{(V_7 - V_8)} \tag{3-13}$$

$$\frac{A-1}{A+1} = \frac{f}{ln\,2}\; arc\cosh(\frac{1}{2} exp\, \frac{ln\,2}{f}) \tag{3-14}$$

The average resistivity value is given by

$$\rho_{AVG} = (\rho_A + \rho_B)/2 \tag{3-15}$$

The effect of the sample's thickness is important. The sheet resistance $R_{Sh}$ is the resistivity divided by the thickness of the sample, and the sheet carrier density $n_S$ is the doping level multiplied by the thickness, therefore, the sheet resistance in a 2D system like the graphene sample can be calculated as follow:

$$R_{Sh} = (\rho_A + \rho_B)/2 \tag{3-16}$$

The Hall Effect can be studied on the samples, giving information about the conductivity type, carrier density, and mobility in the graphene samples. By applying a magnetic field ($B$) on the sample with the measurement configuration shown in **Figure 3.13d**, the Hall voltage can be measured. As shown in **Figure 3.13d**, while the magnetic field switched on, the current is applied across one diagonal, and the voltage is measured along the other diagonal. Then, the polarity of the magnetic field is inverted, and a second measurement is carried out. Thus, each measurement is made twice with opposite polarities of the magnetic field and then subtracted to eliminate magnetoresistance contributions. Reversing the current-flow polarity in the sample allows the removal of offset and thermoelectrical voltages [212]. The Hall coefficient is given by:

$$R_H = \frac{t_s V_H}{BI} \tag{3-17}$$

where $V_H$ is the average Hall voltage (RMS value), $I$ is the magnitude of the applied current (RMS value), and $t_s$ is the thickness. Following the contact numbering in **Figure 3.13b**, we have the configurations shown in **Figure 3.13c** for the Hall coefficient measurement. For the abovementioned measurement configuration, eight Hall voltage measurements are obtained, and the average Hall coefficient can be calculated as follows:

$$R_H = \frac{t_s(V_{4-2+} - V_{2-4+} + V_{2-4-} - V_{4-2-} + V_{3-1+} - V_{1-3+} + V_{1-3-} + V_{3-1-})}{8BI} \quad (3\text{-}18)$$

From the resistivity equation **(3-15)** and Hall coefficient in equation **(3-18)**, the mobility ($\mu$) can be calculated:

$$\mu = \frac{\mid R_H \mid}{\rho_{AVG}} \quad (3\text{-}19)$$

Correspondingly, the carrier density ($n_s$) can be derived as follow:

$$n_s = \frac{IB}{q \mid V_H \mid} = \frac{1}{\mu q R_{sh}} \quad (3\text{-}20)$$

**Figure 3.13** sketches the VdP setup equipped with a LabView-programmed measurement system to characterize different types of graphene or other stacking-materials samples (see Appendix **A1** and **A10**) within this thesis. The measurements were all performed at the PTB Braunschweig. The system includes a helium-4 ($^4$He) continuous flow cryostat, as shown in **Figure 3.13a**, which enables measurements at temperatures between ~1.5K and ~350K under a homogenous magnetic field between $(0-250\,\text{mT})$ provided by an electromagnet. For the VdP measurement, first a test is performed to check the functionality of all the contacts. The proper functionality of the contacts can be inferred from their ohmic ($I/V$) features. Moreover, in the case of measuring the SiC/G samples, it should be noticed that the graphene needs to be isolated from the graphene on the sidewalls and backside of the sample to assure the transport only in the top graphene layer, as shown in **Figure 3.13b**.

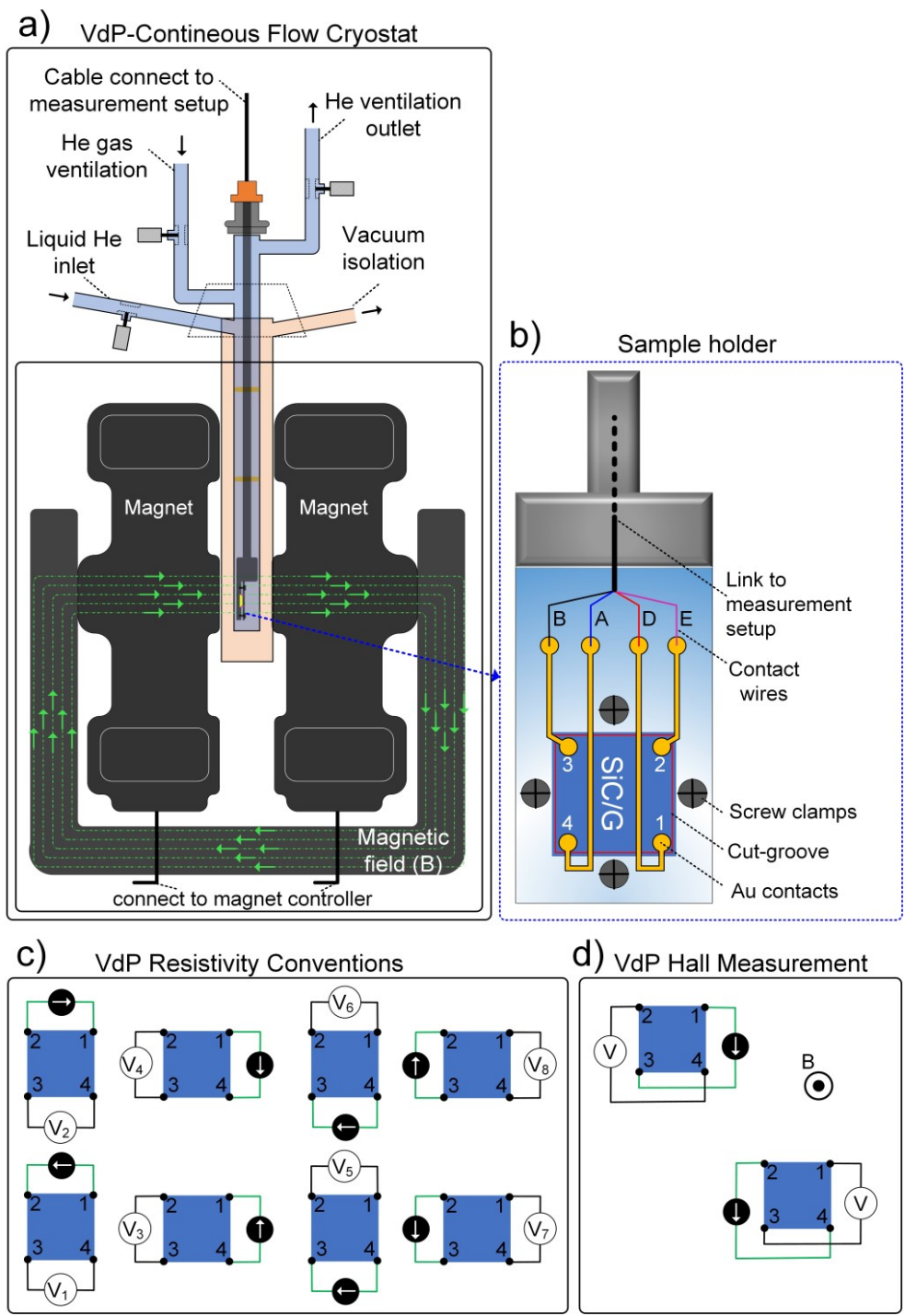

**Figure 3.13. Van der Pauw measurement setup.**

(a) Schematic of VdP cryostat system enabling measurements within the temperature and magnetic field ranges of ~1.5K up to ~350K and 0 – 250 mT, respectively. (b) The sample holder's top view for fast characterization of graphene samples on the mm-scale (5 × 5 mm²). Before the measurements, the graphene sample is isolated from the graphene on sidewalls and the backside by cut-grooves close to each edge (indicated as red lines). (c) VdP resistivity and (d) Hall voltage measurement configurations.

# 4. Fabrication process

**Abstract**

*T*his chapter presents the methods, recipes, and setups used in this work for sample preparation, growth, and device fabrications.

## 4.1. Reactor setup

**Figure 4.1** shows the horizontal inductively heated quartz-tube reactor used for graphene fabrication in this study. The specimens are placed in a susceptor that has a beveled pocket for keeping the samples, as shown in **Figure 4.1**. In the pocket, the primary samples are surrounded by other dummy SiC samples for avoiding the possible influence of the sidewall of the susceptor at high-temperature annealing. The inductive heater can heat the susceptor to 2000 °C. The susceptor's temperature is continuously measured through a window with a pyrometer placed at one end of the reactor within a distance of approximately 30 cm. An insulation layer of graphite surrounds the hot susceptor and reduces the thermal stress of the quartz glass cylinder, which is actively cooled with ambient air from the outside by a radial fan. The quartz tube is connected from one end to a turbopump and a scroll pump, which can evacuate the reactor down to a pressure of $\sim 4 \times 10^{-7}$ mbar at room temperature. Argon, nitrogen, and a mixture of argon/hydrogen (95% Ar and 5% $H_2$) are three separate gases that can be led via different inlets to the chamber. Except for the flow rate of nitrogen gas, which is manually adjusted (usually for venting the system), the flow rate of two other gasses can be controlled automatically through the mass flow controllers. This setup was also used for the hydrogen intercalation process. The setup is automated and can be monitored and controlled by the user via a PC based LabVIEW program. This reactor is based on a non-industrial design presented in ref. [213].

## 4.2. Silicon carbide wafer specifications

The epitaxial growth was performed using four-inch 4H- and 6H-SiC wafers from II-VI GmbH Deutschland. The wafers with a thickness of 500 $\pm$25μm were mostly semi-insulating (vanadium compensating dopants, $R > 10^9$ Ω-cm), classified as prime grade (micro-pipe density <10 cm$^{-2}$), with epi-ready chemical mechanical polishing (CMP) on the silicon face and optical polishing on the carbon face. The growth was carried out on the silicon face. Wafers with various miscut angles towards different crystal orientations were used for the growth. For graphene growth, the SiC wafers were cut into samples typically with the sizes of 5 × 10 mm$^2$ and 10 × 10 mm$^2$ (see **Figure 4.2**) using an automatic dicing saw (DAD3220_DISCO) in the cleanroom at the Physikalisch-Technische Bundesanstalt (PTB). The dicing trenches were aligned up parallel to the primary and secondary flats of the wafers. This adjustment was made while mounting the wafer on the spindle.

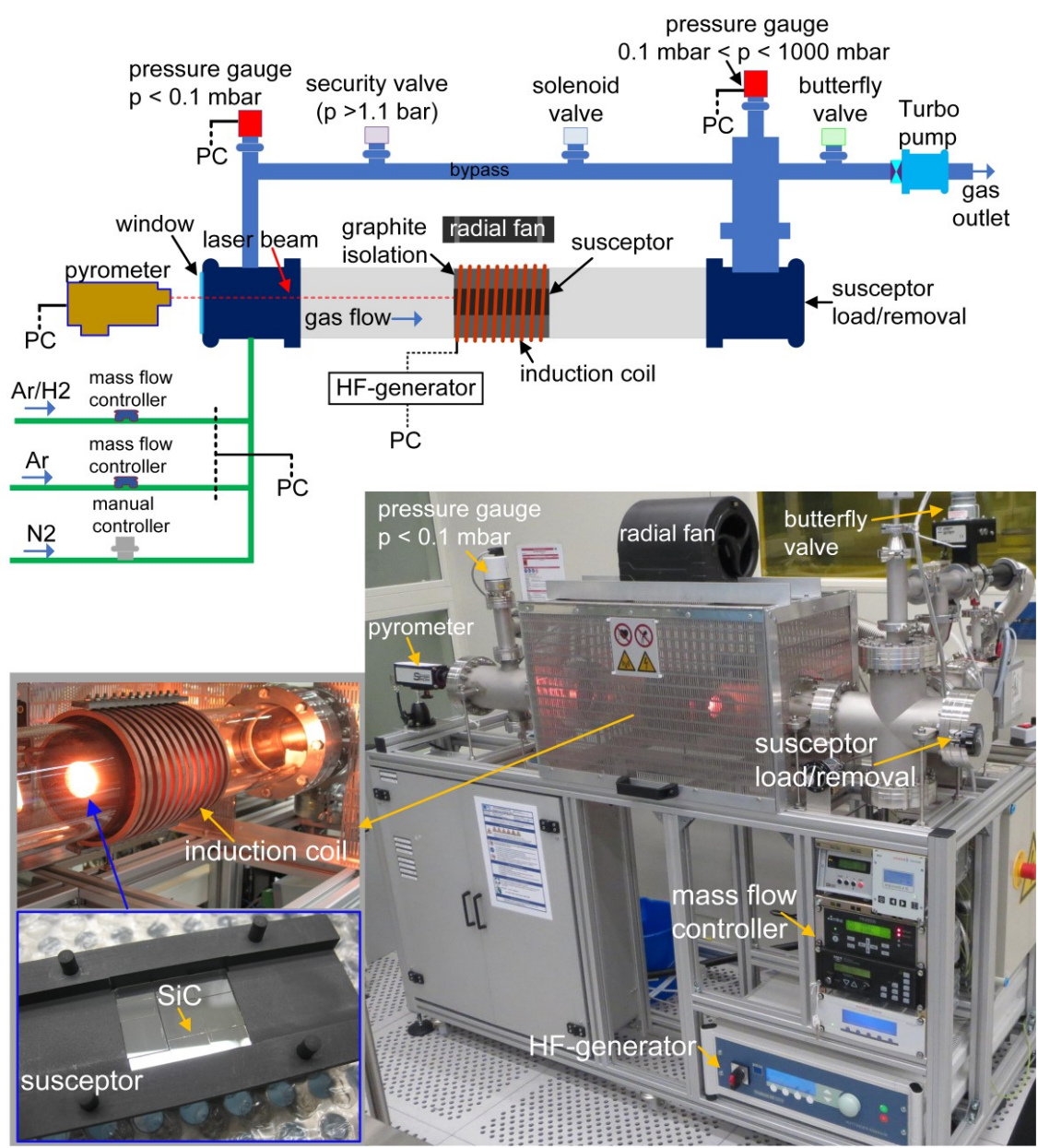

**Figure 4.1. The epigraphene reactor.**

The sketch on top shows the different components of the horizontal inductively heated quartz-tube reactor designed for graphene growth. On the bottom right, the image of the actual reactor shows the oven during functioning at a high temperature. The inset on the left side gives a closer look at the place where the susceptor is located. An open susceptor is shown (left bottom side), which has a pocket for holding the sample. The SiC sample is usually placed in the center of the susceptor surrounded by SiC dummy samples. After mounting the sample, the upper part of the susceptor is attached to the top.

**Figure 4.2. Silicon carbide wafer.**

Top view sketch of a SiC wafer (carbon-face up) used in this study. The primary and secondary flats are towards [11$\bar{2}$0] and [1$\bar{1}$00] crystal directions. The wafers usually have a miscut angle of $\sim -0.06°$ towards [1$\bar{1}$00]. The SiC samples with standard sizes of 5 × 10 mm² or 5 × 10 mm² are cut and split from the wafers using a dicing machine. Each sample is identified with a letter and number (e.g., A6), and a letter ''L'' which the sides of the ''L'' help to quickly realize the crystal orientations for every single sample.

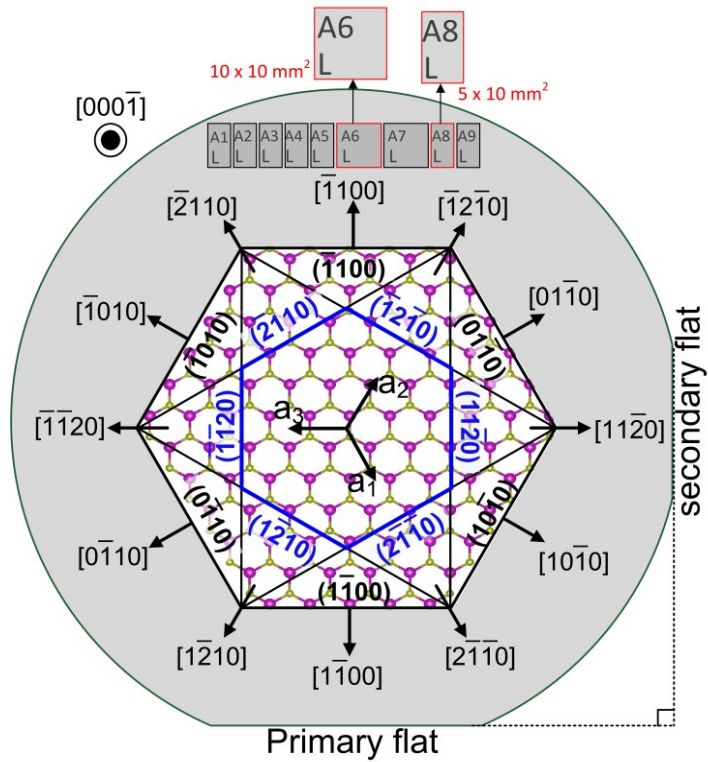

The wafer flats are standardized marks to identify the crystal orientation of the wafer. The sketch in **Figure 4.2** illustrates a top view of a SiC wafer (carbon face up) indicating the standard wafer flats, crystal-planes, and directions.

The larger primary flat is towards the [1$\bar{1}$00] crystal direction and the shorter secondary flat points to the [11$\bar{2}$0] direction. The mainly used wafers had a small ($\sim -0.06°$) misorientation (or miscut angle) towards the [1$\bar{1}$00] direction. This is shown to be one of the most critical parameters for high growth quality. The Si-face of the wafers was covered by an adhesive protection foil to protect it during the dicing process. The foil coverage was done carefully to avoid any trapping air bubbles between the foil and the wafer. The dicing was carried out on the C-face. The dicing parameters were chosen with appropriate care for avoiding any damage to the expensive wafers. The cutting depth was adjusted to be about 60 percent of the wafer's thickness, facilitating easy sample splitting.

After finishing the dicing procedure, the samples could be easily split using tweezers. To identify the samples, they were named by a letter and a number, e.g., A7 and A8 shown in **Figure 4.2**, written on the backside (carbon-side or C-side). Also, the letter ''L'' was scribed on the lower-left edge of each sample on its C-face, which helps to identify the crystal orientation, i.e., long and short sides of the ''L'' is parallel to the secondary flat (11$\bar{2}$0) or primary flat (1$\bar{1}$00) directions, respectively. The scribes were written by a diamond pen.

## 4.3. Hydrogen etching

The surface quality of SiC is decisive to achieve large-area homogenous epitaxial graphene. Hydrogen etching or cleaning is a technique for providing flat SiC surfaces. High cleaning efficiency is an advantage of this technique, which is the main reason to be applied as a pre-treatment step for epitaxial film growth. Hydrogen etching was formerly often used to remove scratches and damages arising from the cutting and polishing processes. Development in SiC surface preparation has led to producing wafers with almost no scratches. However, it is still beneficial for surface cleaning, removing contaminations, and obtaining atomically flat surfaces.

**Figure 4.3. Hydrogen etching of silicon carbide samples.**

(a) The diagram illustrates the hydrogen etching process as a function of time, pressure, temperature, Ar/H2 gas concentration, and gas flow rate. Varying each of the parameters can substantially alter the surface restructuring, especially the step-bunching.

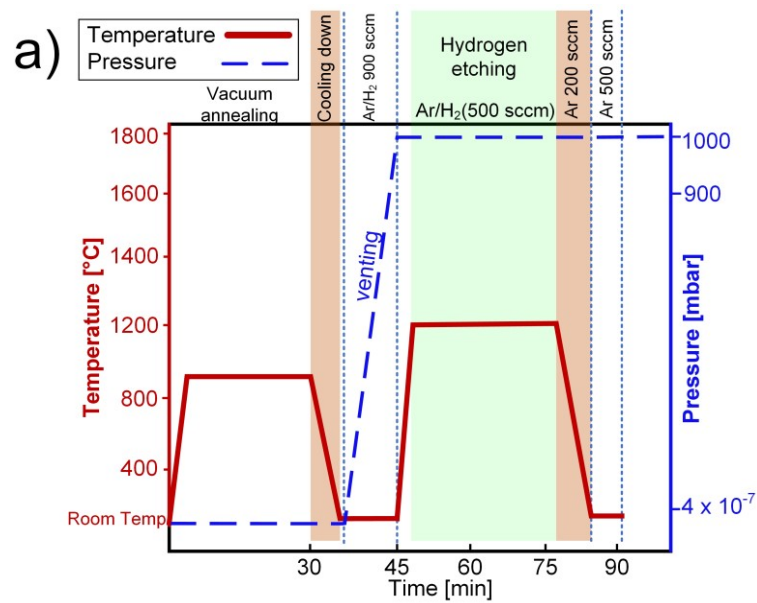

(b) Low–temperature hydrogen–etching applied on a 6H–SiC (0001) sample resulting in a smooth surface with regular terrace and steps. The bright droplet-like spots are silicon droplets, which are typically seen after the H-etching on the SiC surface. For more details, see ref. [37].

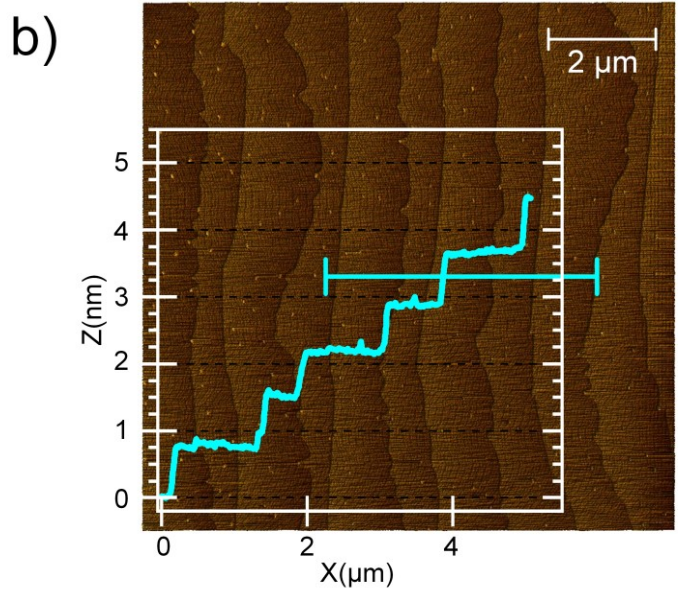

For example, the SiC specimens after dicing (described in section **4.2**) usually contain considerable contaminations, especially when for the protection a durable adhesive foil or photoresist coating is used. Such contaminations stick to the surface, and it is challenging to be removed by standard cleaning procedures, i.e., acetone cleaning. Here, hydrogen etching plays a significant role in surface cleaning. The hydrogen flow-rate and process temperature are two critical parameters in the etching. Depending on the etching parameters, often well-shaped regular but strongly step-bunched terraces are obtained. Heavily step-bunched surfaces are detrimental to graphene growth.

However, it is possible to combine the advantages of the hydrogen-etching if certain parameters in the process are considered. Hydrogen etching processes were intensively studied in previous works and can be found in refs. [37,148]. The diagram shown in **Figure 4.3a** demonstrates the typical hydrogen etching process leading to smooth surfaces, as seen in **Figure 4.3b**. Moreover, a modified process, named hydrogen-cleaning, was used to clean the graphene oven before the main graphene growth. While hydrogen etching is a short process to obtain smooth SiC surfaces, the hydrogen-cleaning is applied to clean the oven. This is critical in the reproducible fabrication of graphene, especially if the reactor is contaminated.

## 4.4. Polymer-assisted sublimation growth

An advanced, so-called polymer-assisted sublimation growth (PASG) method was used to produce graphene on SiC. This technique could be regarded as a modification of standard conventional SiC sublimation growth combined with an additional carbon source. Accordingly, the growth is not based only upon the graphenization of the SiC through thermal treatment and subsequent Si sublimation but also is supported from a foreign carbon-rich source via polymer adsorbates added to the samples. The PASG effectively suppresses the inherent but unfavorable formation of high SiC surface terrace steps during high-temperature sublimation growth. This happens through a rapid formation of the graphene buffer layer, which stabilizes the SiC surface. The growth has gone through multiple optimizations that will be discussed in this work as also have been addressed in refs. [36–39]. In the following, the sample preparation and PASG growth are described.

The sample preparation comprises two main successive treatments. First, the SiC specimens require to be cleaned from contaminations and then be polymerized for the PASG. The applied cleaning process has two sub-steps: firstly storing the SiC specimens in an acetone beaker (Si-face up) for at least 48 hours, and secondly immersing the samples in a fresh mixture of isopropanol and acetone (1:3) and introducing to an ultrasonic bath (USB) for 15 minutes (at 40 °C). A USB equipped with the ability of temperature and power adjustments is indeed highly favorable. For the cleaning, ultra-pure isopropanol and acetone were used. It worth mentioning that the cleaning was investigated using different methods and chemicals, e.g., hydrogen fluoride (HF), piranha, RCA (SC1/SC2) method, and other known techniques introduced in different studies. [214–216] However, the experiments showed that the abovementioned simple cleaning process was sufficiently efficient for obtaining clean samples.

After cleaning, the polymerization process was achieved by an easy-to-use so-called liquid-phase deposition (LPD) technique. [37] The LPD polymerization was applied using a mixture of pure AZ5214E photoresist diluted in isopropanol (5ml/25 ml). The LPD process with three sub-steps is depicted in **Figure 4.4a**. For the LPD, the samples are first immersed in a beaker containing diluted AZ5214E photoresist and introduced to the USB for 15 min at 40 °C. This step is followed by rinsing the sample for ~45 seconds using an isopropanol wash bottle to remove the excess polymer from the surface. Right after that, the samples are dried by spin-drying at a speed of ~6000 rpm (30 seconds), which could be supported by a simultaneous nitrogen gas blow on the sample. All the processes were carried out in the cleanroom conditions in the yellow area. The LPD results in uniform distribution of nano-sized polymer adsorbate on the surface, which is crucial for the growth homogeneity. The polymer could also be applied directly on the surface by spin-coating; however, it often leads to a non-uniform distribution of adsorbates resulting in irreproducible and inhomogeneous graphene growth with bilayer inclusions. [217,218]

For the prepared specimens, the growth procedure begins with a cleaned and evacuated ($P \leq 1.0 \times 10^{-6}$ mbar) reactor. The oven cleaning usually includes several successive steps, e.g., hydrogen cleaning, vacuum cleaning, and cleaning in argon ambient. Each of the steps has several temperature windows with varying the applied gas flow. The samples are mounted into the susceptor (**Figure 4.1b**), then it is loaded into the oven. The diagram in **Figure 4.4b** shows the growth process for the graphene monolayer fabrication. The complex growth has three initial annealing steps at lower temperatures of 900 °C (vacuum, 30 min), 1200 °C (Ar atmosphere, 900 mbar, 10 min), and 1400 °C (Ar atmosphere, 900 mbar, 2 min) before the graphene growth at 1750 °C (Ar atmosphere, 900 mbar, 6 min). During the temperature ramp, an intermediate interruption of the growth process was performed by cooling the system to room temperature after initial annealing in a vacuum ($P \leq 1.0 \times 10^{-7}$ mbar, 900°C, 30min). The system then was vented by introducing argon gas to change the pressure to ~900 mbar.

**Figure 4.4. Polymer-assisted sublimation growth.**

(a) Sample preparation for the PASG process by liquid phase deposition (LPD) of polymer (AZ5214E) adsorbate onto the SiC substrate. The LPD includes three successive steps: (1) the specimens are cleaned in acetone, isopropanol, and an ultrasonic bath. (2) The samples are immersed in an ultrasonic bath of diluted polymer/isopropanol in a beaker. (3) Rinsing the sample by isopropanol and spin-drying leading to the remaining nano-sized polymer adsorbates uniformly distributed on the sample.

(b) The growth diagrams of bilayer-free epitaxial monolayer graphene. The growth results in sequential step-patterns are discussed in Chapters 5–7. Changing the buffer layer (cyan) and graphene layer (violet) temperature windows can alter the coverage, thickness, surface restructuring. See text and ref. [39] for more details.

(c) The growth diagrams of the graphene-free epitaxial buffer layer. It indicates implementing the Ar flow as a growth parameter. The recipe results in a high-quality buffer layer with remarkable suppression of surface step-bunching. [38] This is an alternative growth. Similar results were attained by different temperature/time treatments.

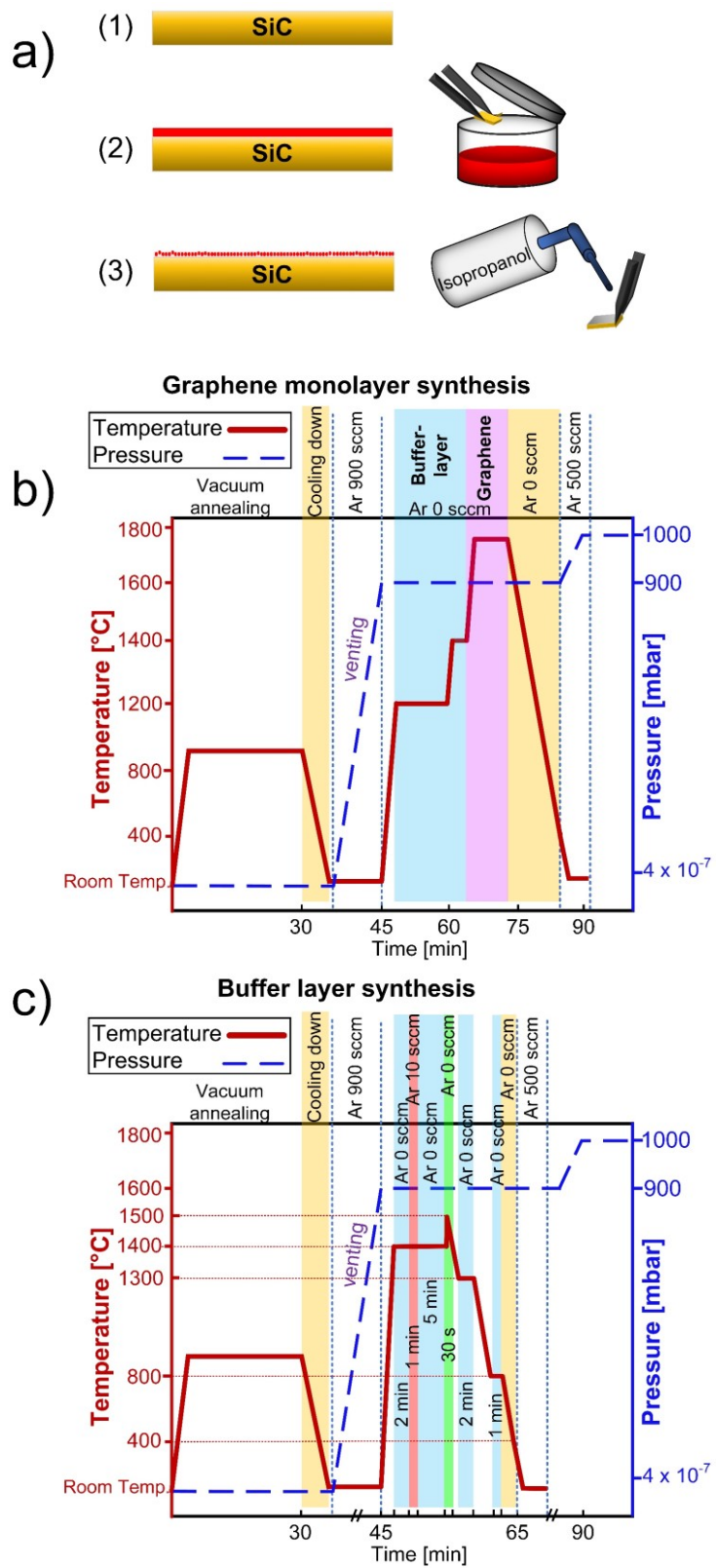

This additional cooling step was performed for two main reasons: (i) avoiding the possible influence of the argon flow-rate on the sample during pressure change (vacuum to ~900 mbar), (ii) increasing the carbon condensation and nucleation sites on the SiC surface for accelerated buffer layer growth. [37] The process was followed by intermediate annealing at 1200 °C and 1400 °C for 10 and 2 minutes, respectively. Afterward, the samples were heated directly up to 1750 °C and annealed (6 min) while the argon flow-rate was kept at zero sccm. All the temperature ramps were applied at the same heating rate of ~7 °C/s. Finally, the heater was switched off, and the samples were allowed to cool down to ~400 °C (no Ar flow), then to room temperature under Ar flow of 500 sccm.

Furthermore, by proper thermal treatments, homogenous bilayer graphene can be achieved, although the graphene growth on SiC(0001) is known to be self-limiting. This is acquired by the same recipe shown in **Figure 4.4b**, but the growth at 1750 °C is extended to about one hour and several temperature windows up to 2000 °C. The results will be discussed in Chapter **5**.

**Figure 4.4c** shows the diagram of a buffer layer growth. This process results in outstanding buffer layer coverage appealing for intercalation purposes to achieve coherent QFMLG layers or fabrication of other sub-dimensional materials. This growth is special since it implements the Ar flow-rate as a growth utensil in combination with other parameters. It is an alternative growth recipe, and similar results were achieved by manipulating and optimizing other growth parameters. The results will be discussed in Chapter **5** and can also be found in refs. [38].

The growth optimization was carried out on more than 1200 samples types, which statistically underline the successful reproducibility of graphene syntheses. The presented fabrication method and optimization enables the growth of ultra-smooth bilayer-free graphene sheets with unprecedented reproducibility, a prerequisite for the wafer-scale fabrication of high-quality graphene-based electronic devices.

## 4.5. Device fabrication

In this section, the general lithography process applied to pattern graphene into an electronic device is described. Here, the primary fabrication technique is based on electron beam lithography (EBL) to define the patterns on epitaxial graphene. Alternatively, photolithography is also possible, but because of its lower resolution is not preferred in this study and thus is not used. The lithography process begins when the graphene sample has passed initial inspection using an optical microscope, AFM, and sometimes Raman spectroscopy. The fabrication produce is technically composed of three successive steps, as illustrated in **Figure 4.5a**.

In the first step, the contacts and markers are structured. To this end, the graphene is covered with an EBL compatible photoresist using spin coating. The bottom layer, closest to the graphene, consists of a poly(methyl-methacrylate) (PMMA) based copolymer, and the top layer consists of ARP 630-670 series (e.g., P672.06). These are both positive resists, which become soluble in specific developers upon exposure to the electron beams. Furthermore, these two resists are sensitive with different ratios to the developers, which facilitates control over the resist profile. Additionally, a 20 nm chromium layer is evaporated on the sample. This thin reflective chromium layer helps to overcome the EBL's focusing problem on the transparent SiC/epitaxial graphene. After EBL illumination, the chromium layer is removed. Then, the exposed areas are developed. Next, the graphene is structured by AC-plasma using an oxygen/argon gas mixture. Since graphene adherence to any deposited metal (e.g., Au) is weak, therefore, in this step, areas where metal bond pads (anchors) are placed will also be etched to allow bonding to the SiC substrate. The first step also includes the etching of the markers (at the corners) to help precise alignment of future layers. After the plasma etching, the PMMA mask is lifted.

The structured graphene is then contacted by Ti/Au metal contacts. The thin titanium sticks very well to the etched areas (SiC) and is sealed by the gold layer to avoid oxidation. These two layers act as a coupling agent between the final gold overlap layer and the SiC substrate. Titanium does not show proper contact with graphene, so it is not directly deposited onto the graphene. In the second step, the graphene Hall-bar is structured by EBL illumination on two-layers positive photoresists A-RP 630-670 series and atop SX AR-PC 5000/90.2 (Electra 92) layer. The latter resist is electrically conductive and is required for an efficient EBL illumination. Afterward, the Electra 92 is developed under flowing distilled water, and the exposed areas of the photoresist are developed by the same procedure described in the first step.

In the third step, the electronic contact with the graphene is realized. The EBL is applied to the sample covered with the same two photoresists (as described in the first step) to open a window (by a developer) for the metallization of the gold contacts. With a thickness of 50 nm, this gold layer overlaps the previously deposited Ti/Au contacts and the nearby uncovered graphene. **Figure 4.5** depicts the graphene Hall bar with two main sizes of either ($100 \times 400$ μm²) or ($200 \times 800$ μm²) designed and used for the QHE measurements presented in Chapter **8**. For the measurements, the samples are glued to a chip carrier, e.g., TO8, as shown in **Figure 4.5**, bonded using an aluminum wire wedge bonder from West

Bond Inc. The lithography process of the graphene devices was performed mainly based on an optimized approach developed by A. Müller. [219]

For the Van der Pauw measurements, the samples are mainly measured without any prior lithography, as explained in Chapter 3. However, for better contacting, which enables multiple measurements (without damaging the graphene in VdP setup) and avoiding durable typical EBL procedures, an easy and innovative so-called pen-patterning method was used to generate fast contacting to the graphene samples. This technique is briefly explained in Appendix **A1**.

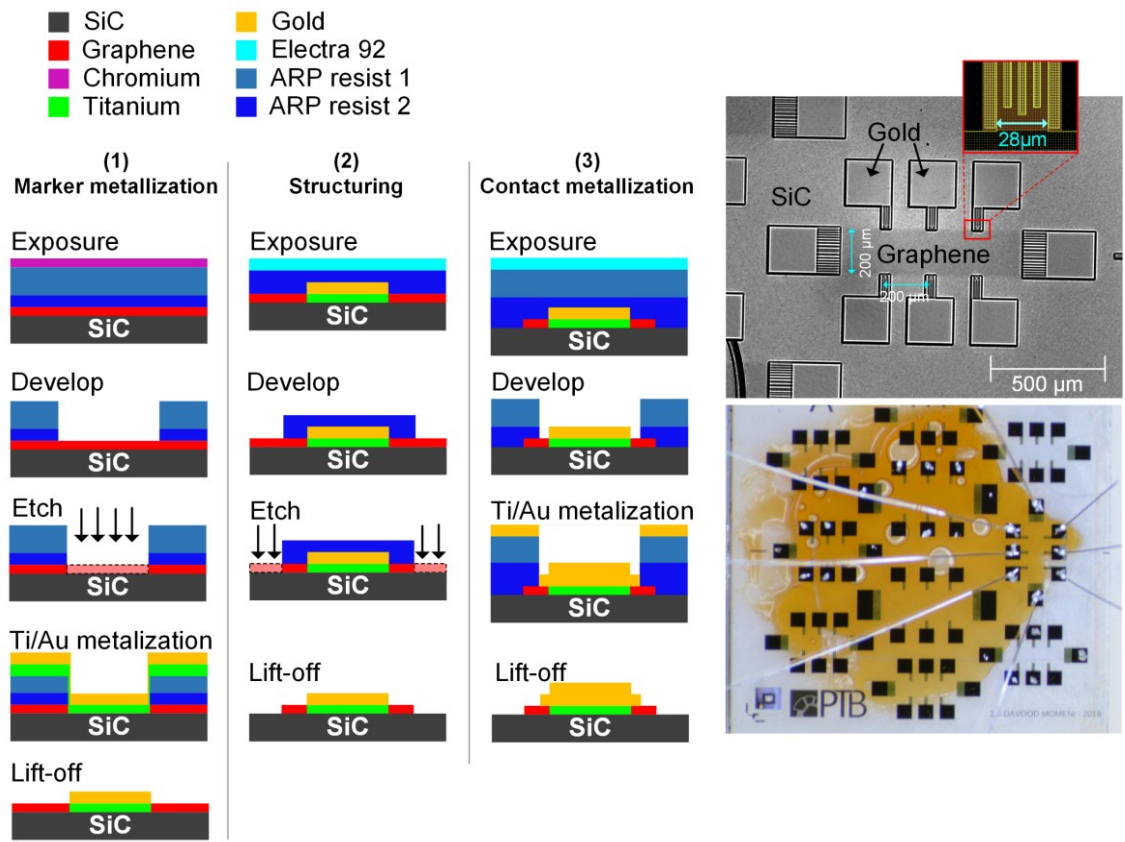

**Figure 4.5. The fabrication process of the graphene Hall-bars.**
Left: Schematic representation of three EBL steps, (1) Marker metalization, (2) structuring, and (3) contact metallization. Graphene is removed using oxygen/argon plasma. Right-top: optical microscope image of a finished graphene Hall-bar device with a size of $200 \times 800$ μm². Right-bottom: an optical image of a ready-to-measure Hall-bar mounted and bonded on a TO8 chip carrier. The sample includes several Hall-bars designed with different sizes. For more detail, see the text.

## 4.6. Charge carrier tuning

Notwithstanding several existing methods, precise control over the charge carrier density in SiC/G is still a delicate task to accomplish. For the QHE applications, it is desirable to tune the graphene close to the charge neutrality condition ($E_F = E_D$) that enables reaching high electron mobility and quantization at low magnetic fields. This section presents two of those methods to tune the graphene Hall-bars.

The first method, the so-called photochemical gating, is based on the chlorinated photosensitive polymer (methyl styrene-co-chloromethyl acrylate), commercially available as ZEP520. The photo-gateable heterostructure is formed by spin-coating the sample with 55 nm PMMA as a neutral spacer layer followed by a deposition of 300 nm ZEP520, as shown in **Figure 4.6a**. A deep ultraviolet light (DUV) illumination with a wavelength of ≤ 254 nm activates the ZEP520A layer and generates an electric field above the graphene, which leads to the gating effect. Through the DUV exposure, the chemical bonding in the ZEP520 polymers is changed, which results in the formation of Cl radicals acting as effective electron acceptors. [220] These acceptors can take the electrons from the lower-lying graphene layer. By varying the dose (increasing illumination time), the density of acceptors increases, and the electron concentration in the graphene can be gradually reduced. By this technique, the electron density in the graphene can be reduced from several $10^{12}$ cm$^{-2}$ to about $10^{10}$ cm$^{-2}$. After the measurements, the sample can be illuminated again, or the carrier concentration can be restored to its original value by heating the device above the polymer glass transition temperature of $T_g \approx 170$ °C. [220]

Alternatively, a high electrostatic potential gating with ions can be produced by corona discharge for a reversible tuning of the SiC/G carrier density. To this end, the sample first needs to be covered with a dielectric PMMA layer as a host material for the ions generated by the corona discharge gun. This can be regarded as similar to the metallic gate of a field-effect transistor (FET) that is replaced by the ions deposited on the dielectric layer, which induce a surface charge density on the underlying semiconductor.

Corona discharge can be created by applying a high voltage to a sharp tip or wire to generate an electrical discharge that ionizes the surrounding gas. If performed in air, the predominant ionic species are $H_3O^+$ and $CO_3^{-2}$ which move along the lines of the electric field. When a negative voltage is applied, positive ions drift towards the discharge source, while negative species propagate away from it and are deposited on a target substrate and vice versa. [221]

The carrier density control of SiC/G samples was performed through a piezo-activated antistatic gun (Zerostat) similar to ref. [221] with two distinctions in the measurement setup and ambient condition. **Figure 4.6b** depicts the measurement configuration in which half of the contacts are connected to a positive DC voltage (red color cabling), and the other half are grounded (blue color cabling) while the current is measured. This, in addition to replacing the DC current source with a DC voltage source, assures the safety of the device during the measurements and charge spikes. Furthermore, the tuning was carried out in the presence of a gentle nitrogen flow (5N) that experimentally turned out to be resulting in higher stability of charge tuning compared to when performing the tuning in air. The corona discharge gun produces a 1 to 2-sec long pulse of positive or negative ions with the polarity depending on the compression or expansion of the piezo-crystal. The variation in the electronic properties of SiC/G during exposure to the corona ions is monitored continuously by measuring the current flow in the sample. **Figure 4.6c** demonstrates the resistance change of a SiC/G sample under investigation. By each pulse of ions, the carrier concentration of SiC/G, which is initially $n$-doped, can be tuned close to the Dirac point or be doped towards hole doping, depending on the number of ion pulses. By comparing the experimental data of doping in SiC/G, the observed maximum in the measured resistances corresponds to the Dirac point, indicating the cross-over from $n$- to $p$-doping. As will be discussed in Chapter **8**, the aim is to keep the electron doping with fine adjustment close to the Dirac point for QHE measurements.

Also, it is worthwhile to mention that a so-called post-treatment such as hydrogen-treatment or air-annealing (see Appendix **A2**) is applied to the samples to modify the doping level and will also be discussed in Chapter **8**.

### Figure 4.6. Charge carrier tuning of epitaxial graphene.

(a) Photochemical gating. The layout of SiC/ graphene/ polymer heterostructure consists of a PMMA spacer layer (red) on top of the graphene and an EP520A layer (blue) deposited on top, which is activated via UV-light (λ = 254 nm) exposure.

(b) Electrostatic gating. The graphene Hall-bar is covered with a 55 nm PMMA layer as a host material for the ions generated by the zerostat. The measurement setup sketch shows a constant DC voltage source is applied to half of the contacts, and the rest of the contacts are grounded. Corona discharge tuning is performed while a very gentle flow of nitrogen gas is applied to the sample.

(c) Monitoring variation of electronic properties of SiC/G upon each corona discharge pulse (occur at spike regions). The carrier density can be reversibly changed from initial electron-doping to hole-doped graphene.

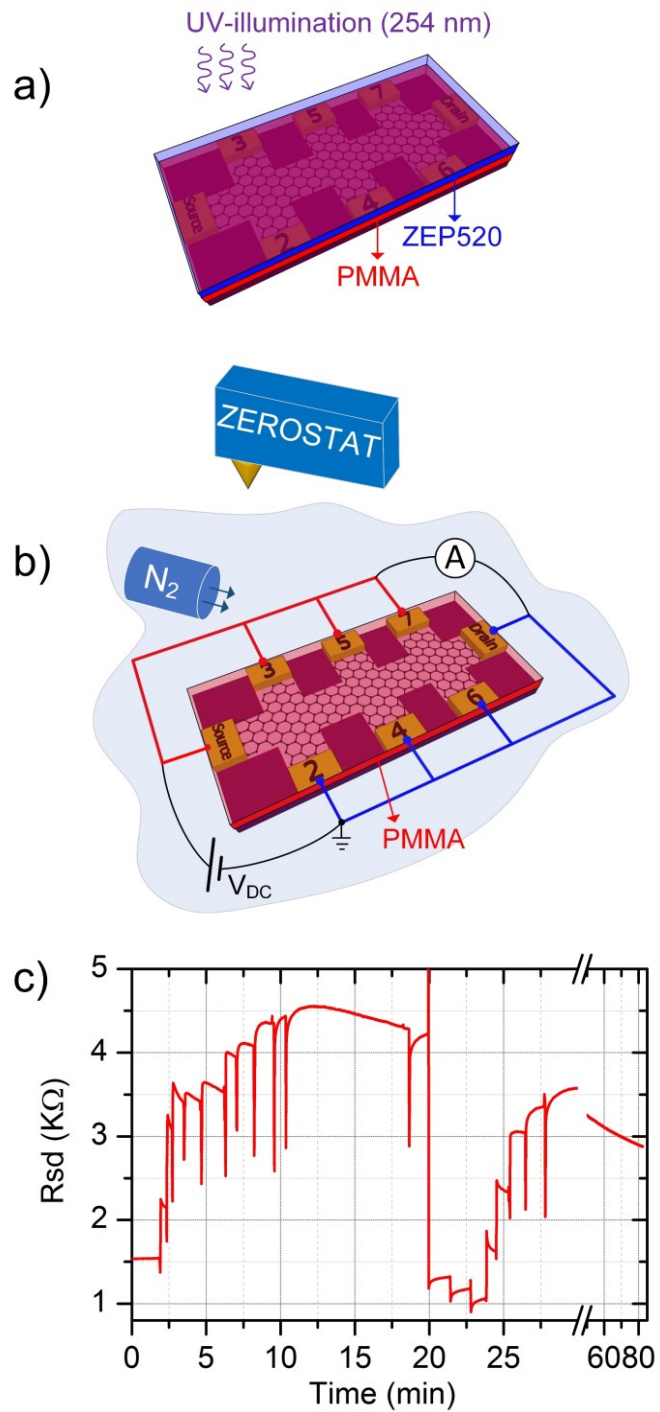

# 5

## 5. High growth control of epitaxial graphene on SiC

### Abstract

*T*his chapter focuses on the fabrication of graphene on the silicon-face of hexagonal silicon carbide. Over 1200 samples of different types were synthesized to optimize the growth and its reproducibility. Here the most important results are discussed. This chapter gives a comprehensive overview of the growth of a so-called buffer layer (BFL), monolayer graphene (MLG), bilayer graphene (BLG), and quasi-freestanding monolayer/ bilayer (QFMLG/ QFBLG) layers. The high quality of these layers helps to better understand the growth kinetics and mechanism, a so-called step-bunching, and surface restructuring and recrystallization. The large-scale homogeneity of the samples enables both local- and macro-scale studies on different sample types resulting in several salient features in epigraphene that each thematically will be discussed separately in the following chapters (i.e., 5, 6, and 7). In this chapter, the optimization of the growth considering an influential but so-far neglected parameter, the "argon gas flow-rate," is addressed. The conditions for the fabrication of epitaxial BFL are studied. Subsequently, the main challenges in BFL growth: the poor or excessive- "coverage" at the SiC step-edge regions, which both dramatically degrade the quality of the BFL sample, are discussed. In addition to high-quality MLG, the growth conditions to achieve coherent and scalable epitaxial BLG are presented. Also, the challenges and efficiency of the intercalation technique concerning the intercalant purity and intercalation conditions are studied. Several characterization techniques such as AFM, Raman spectroscopy, XPS, LEED, VdP, and STM are used to examine the samples. The presented results are partly published in refs. [36–39,118,148].

## 5.1. Introduction

Inert gas (e.g., argon or nitrogen) counter-pressure has been used for many years to improve silicon carbide sublimation growth, which prevents unwanted crystal growth before reaching the optimal growth temperature. [222,223] The growth of epitaxial graphene in an argon atmosphere of elevated pressure of about 1bar was a breakthrough in the progress of obtaining high-quality large-area graphene layers. [16,17] However, little attention has been paid to the gas flow velocity of the ambient process gas. However, the presented results in this study show that the SiC decomposition rate can be controlled through the Ar flow-rate without varying the total pressure and substrate temperature. Herein the focus is initially drawn on the impact of argon flow-rate on optimization and improvement in growing epitaxial buffer- and graphene layers. By taking into account the influence of the argon flow-rate, optimization of growth parameters, and modification of a so-called polymer-assisted growth technique is achieved, and ultra-smooth bilayer-free graphene layers are produced. [36–39,118,148]

The QFMLG can be fabricated by decoupling an epitaxially grown buffer layer from the underlying SiC substrate, e.g., via hydrogen intercalation. [18,108,224] The hydrogen intercalation allows the fabrication of $p$-type monolayer graphene combined with the advantage of the large-scale graphene epitaxial growth directly on semi-insulating SiC substrates with reduced influence on the atop graphene layer. [16,108,119,121] Hence, this approach offers a versatile platform for potential applications as an alternative to epitaxial graphene (EG) with $n$-type charge carriers in the pristine state. This has prompted various interesting experimental intercalation studies by applying different elements. [18,96,99,101,109,225–227]

State-of-the-art QFMLG can be fabricated with high quality proven by low defect-related D-peak intensities in local Raman measurements and high charge carrier mobilities in transport measurements of micrometer-sized Hall bars. [108,228,229] However, it is quite challenging to obtain homogenous QFMLG over mm or cm areas, as can be obtained with EG. [36,37,39,148] An important reason is the lower temperature used for buffer layer growth (about 1400 °C, ~1 bar) compared to graphene growth (> 1600 °C, ~1 bar) which limits the carbon supply and surface mass transport and thus, the formation of a coherent large-area buffer layer. Such problems are often observed at the SiC step-edge region where the sublimation rate is strongly enhanced. [230] Moreover, hexagonal SiC shows terraces with inequivalent surface energies and decomposition velocities [133,141], which complicates the epitaxial growth concerning thickness control and coverage. Due to these facts, either incomplete buffer layer coverage (at low growth temperatures) [36] or additional graphene-layer formation (at elevated growth temperatures) [119,120,231] are common ramifications at step edges. Practically, such defected buffer layers prevent a reproducible fabrication of large-area homogeneous QFMLG, which is unfavorable regarding electronic device fabrication, e.g., for achieving superior transistor performance with high cut-off frequency [117], quantum Hall metrology applications [19,36,39,232] and even beyond that for growing other 2D materials. [43,105]

This chapter is organized as follows: first, the influence of the mass flow-rate of argon, which is used as an inert atmosphere for the epitaxial buffer layer and graphene growth processes, is investigated. Subsequently, the ultra-smooth buffer layer and graphene monolayer fabrication by taking into account the effect of the Ar flow-rate in combination with other determining growth parameters (e.g., $T$, $p$) is presented. These are supported by atomic force microscopy (AFM) and Raman spectroscopy investigations, which prove that optimized Ar mass flow conditions lead to the formation of highly homogenous buffer or graphene layers, and after intercalation to high-quality large-area QFMLG and QFBLG, respectively. The intercalation conditions regarding the gas impurity, time, and temperature are studied and discussed. The STM measurements demonstrate the freestanding graphene layers smoothly bridge over the SiC steps on the adjacent terraces. This is further supported by mm-scale Van der Pauw (VdP) as well as μm-scale nano-four-point probe (N4PP) measurements of millimeter-sized samples with high charge carrier mobilities up to 1300 and 3300 (cm²/Vs) for QFMLG and QFBLG, respectively, at room temperature. [38,233] Finally, the graphene growth is pursued, aiming at homogenous epitaxial bilayer graphene fabrication. Accordingly, the growth conditions are discussed, and the quality of the samples is scrutinized using AFM, Raman, and STM measurements.

## 5.2. Sample preparation

The experiments (Exp.1-Exp.7) were performed on the Si-face of the samples (5 × 10 mm²) cut from a semi-insulating 4H- and 6H-SiC wafer with a nominal miscut of about −0.06° towards [1$\bar{1}$00]. The substrates were prepared by liquid phase deposition of polymer adsorbates on the surface as described for the polymer assisted sublimation growth (PASG) technique described in Chapter **4** and refs. [36–39]

The influence of Ar mass flow-rate on the buffer layer growth (Exp.1) is exemplary shown on three samples $S_0$, $S_{100}$, and $S_{1000}$. After vacuum annealing at 900 °C, the buffer layer was grown at 1400 °C (900 mbar Ar atmosphere, 30 min) under Ar mass flow rates of 0 (zero), 100, and 1000 sccm, respectively. The process diagram is shown in **Figure 5.2c**. Since the surface diffusion is highly temperature-dependent, studying the influence of argon flux on surface morphology can be more effectively investigated for the opted low temperature (1400°C) than graphene growth temperatures (above 1500°C, 1000 mbar). This is demonstrated in section **5.3.1**.

The impact of the Ar mass flow-rate on the graphene growth (Exp.2) was demonstrated on two samples $G_0$ and $G_{20}$ grown at 1750 °C (Ar atm., 900 mbar, 6 min), under zero, and 20 sccm Ar mass flow, respectively. All other parameters were kept constant. (See section **5.3.2**)

The optimized buffer layer sample BFL1 (Exp.3) was grown under 0 sccm Ar flow (900 mbar) by an annealing procedure with temperature steps at 1400 °C (8 min), 1500 °C (5 sec), 1300 °C (2 min), and 800 °C (1 min). The detailed process diagram can be seen in **Figure 4.4b**. The BFL2 represents, in contrast, a weakly covered buffer layer sample, which was grown at 1400 °C for 5 min (900 mbar). Comparing the quality of the BFL1 and BFL2 explicitly demonstrates the significance of growth optimization, especially when homogenous freestanding monolayer graphene is desired. This is compared in Exp.4, which includes the application of H-intercalation on two buffer layer samples with different quality (BFL1 and BFL2) which are afterward named QFMLG1 and QFMLG2, respectively. The hydrogen intercalation was conducted at 900 °C (60 min). This is discussed in section **5.4**.

Next, it is shown that the buffer layer samples can be surprisingly intercalated in a nitrogen environment with 99.999% purity. This is demonstrated in Exp5., which includes stepwise buffer layer intercalation processes in nitrogen (5N) and hydrogen environments separately and subsequent investigations using Raman spectroscopy. For this purpose, the intercalation of the buffer layer samples was performed from ~300 to ~1000 °C with temperature increments of 100 °C. This is discussed in section **5.5.3**.

Similarly, quasi-freestanding bilayer graphene (QFBLG) was shown (Exp.6) by H-intercalation on $G_0$ (epigraphene) at 1050 °C (2 hours). The intercalation was done in hydrogen (5%) and argon (95%) gas mixture (1000 mbar). The optimal temperature was determined by Raman spectroscopy and large-scale VdP measurements.

Finally, the epitaxial bilayer graphene (Exp.7) was grown at several temperature windows of 1750 °C (10 min), 1850 °C (30 min), and 1900 °C (10 sec) at 900 mbar in an argon atmosphere. All the samples in this study are listed in **Table 5.1**, including the main growth parameters. More details can be found in Chapter **4**.

| | Sample | Polytype | Aim | Ar (sccm) | H2 (sccm) | $P$ (mbar) | $T$ Growth (C°) | Time (min) |
|---|---|---|---|---|---|---|---|---|
| **Exp.1** | S$_0$ | 6H | Inspecting Ar flow-rate in BFL growth | 0 | - | 900 | 1400 | 30 |
| | S$_{100}$ | 6H | | 100 | - | 900 | 1400 | 30 |
| | S$_{1000}$ | 6H | | 1000 | - | 900 | 1400 | 30 |
| **Exp.2** | G$_0$ | 6H | Inspecting Ar flow-rate in epigraphene growth | 0 | - | 900 | 1750 | 6 |
| | G$_{20}$ | 6H | | 20 | - | 900 | 1750 | 6 |
| **Exp.3** | BFL1 | 6H | High-quality BFL1 vs. low-quality BFL2 | 20 | - | 900 | 1300-1400 | 10 |
| | BFL2 | 6H | | 0 | - | 900 | 1300-1400 | 5 |
| **Exp.4** | QFMLG1 | 6H | High-quality QFMLG1 vs. low-quality QFMLG2 | - | 100 | 900 | 900 | 60 |
| | QFMLG2 | 6H | | - | 100 | 900 | 900 | 60 |
| **Exp.5** | QFMLG3 | 6H | Nitrogen (5N)-intercalation (QFMLG3) vs. H- intercalation (QFMLG4) | - | - | - | 0-1000 | see text |
| | QFMLG4 | 6H | | - | - | - | 0-1000 | see text |
| **Exp.6** | QFBLG | 6H | Homogenous QFBLG | - | 100 | 900 | 1050 | 120 |
| **Exp.7** | epi-BLG | 6H | Homogenous BLG | 0 | - | 900 | 1800/ 1900 | 5/ 60 |

**Table 5.1. List of the samples and experiments in this study.**
The aim and main growth parameters of each experiment are briefly described. For detailed information, see text and refs. [36–39].

## 5.3. Influence of argon flow-rate

### 5.3.1. Ar flow-rate in buffer layer growth

The surface morphology of the buffer layer samples grown under different Ar mass flow rates (S$_0$, S$_{100}$, and S$_{1000}$) are plotted in **Figure 5.1** (see the process diagram in **Figure 5.2**). **Figure 5.1a and b** show for sample S$_0$ (zero Argon flow) a smooth surface with regular terraces and step heights of ~0.75 nm. The clear $(6\sqrt{3} \times 6\sqrt{3})R30°$ spot profile analysis low-energy electron diffraction (SPA-LEED) pattern in **Figure 5.1c** indicates the formation of the buffer layer which homogeneously covers the terraces as shown by the even phase contrast. The homogenous buffer layer growth is attributed to the PASG growth, which favors buffer layer nucleation over the entire terrace. [36,37]

The different phase contrasts (lighter colors) along the step edges (see inset in **Figure 5.1b**) are ascribed to two different effects. The bright-line originates from the local phase shift induced by the topographical difference in height. The narrow stripes of light contrast around the step edges are attributed to material contrast which could originate from uncovered SiC areas or already graphene domains. Since graphene growth is rather unlikely at the low growth temperature of 1400 °C [36], an inferior buffer layer growth at the step edges is assumed. The missing buffer layer coverage along the step edges indicates an insufficient carbon supply in these areas, which is attributed to carbon diffusion and preferred buffer layer nucleation on the terraces. These line defects separate the buffer layer areas on neighboring terraces, which is unsuitable for the fabrication of large-area QFMLG through intercalation.

For 100 sccm Ar flow, the surface morphology changes. Although the $(6\sqrt{3} \times 6\sqrt{3})R30°$ LEED pattern indicates the formation of the buffer layer on the terraces, **Figure 5.1f**, the AFM images of sample $S_{100}$ in **Figure 5.1d**, and e show that the smooth terraces are interrupted by canyon-like defects which erode into the SiC terraces and terminate at the following terrace step. These canyon-defects are known to form at gaps in the buffer layer. [234,235] Here, the increased Ar flow rate alters locally the thickness and homogeneity of the near-surface layer of species (Knudsen layer) [236,237] during the growth, where it causes a faster local SiC decomposition and surface mass diffusion, leading to the canyon-defects before a continuous buffer layer has formed on the terraces.

For much higher Ar flows (**Figure 5.1g and h**), the accelerated SiC decomposition induces an etching of the SiC surface. No buffer layer can be formed under these conditions, as indicated by the (1×1) LEED pattern of the bare SiC surface, **Figure 5.1i**. The AFM image of $S_{1000}$ in **Figure 5.1g** shows wide terraces and pronounced terrace broadening and step bunching with step heights of ~2.5 nm. Nanometer-sized islands with a triangular-shaped basal plane and heights about ~5.5 nm are frequently observed on the surface. [38] A similar Ar flow dependence behavior was observed for typical sublimation growth (SG) without using the PASG technique. [37,38]

The investigations show that with increasing Ar mass flow, the SiC surface decomposition is enhanced while the Ar pressure in the reactor is kept constant. This can be understood in a model in which a quasi-thermal equilibrium exists between Si and C species in a surface layer and those in the adjoining gas phase.[236,237] For higher Ar flow, the species in the gas phase are increasingly "blown away" by collision processes with the Ar atoms. This perturbation enforces enhanced SiC decomposition to maintain the equilibrium. The decomposition process competes with the buffer layer growth since the C-rich surface reconstruction is known to stabilize the SiC surface by the covalent bonds in-between. [36,235] The final state of the surface is determined by the rates of the involved processes.

For zero and small Ar flows, the slow SiC decomposition is self-limiting by the generated carbon for buffer layer growth. For a high Ar flow, when a fast SiC decomposition rate exceeds the nucleation and growth rate of the buffer layer, an etching of the SiC surface is the consequence. Both extreme cases are displayed by the samples S₀ and S₁₀₀₀, respectively. For moderate Ar flow both, etching and buffer layer growth can appear simultaneously but spatially separated as seen for S₁₀₀. The control of the SiC decomposition by the Ar mass flow without changing process temperature or the Ar background pressure opens a new parameter range for improved epitaxial graphene growth.

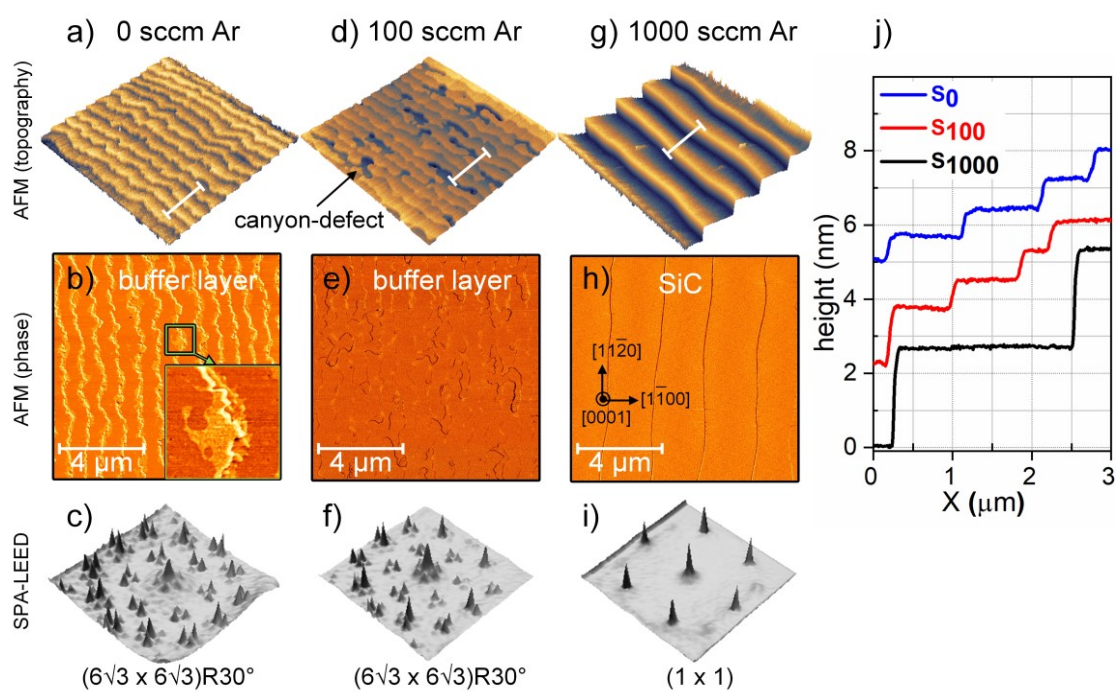

**Figure 5.1. Influence of Ar flow-rate on graphitization of 6H-SiC (0001).**

Experiments performed (1400 °C, 1 bar Ar atmosphere, 30 min) under three different argon mass flows. The AFM topography and phase images are plotted (a), and (b) for $S_0$ (0 sccm Ar), (d) and (e) for $S_{100}$ (100 sccm Ar), and (g), and (h) for $S_{1000}$ (1000 sccm Ar). The inset (1.2 × 1.2 μm²) in (b) shows a close-up of a line defect of sample $S_0$: The lighter narrow stripes are discontinuities in the buffer layer located around the terrace step edge. The step edge itself appears as a very bright line. The dark spots in (d) show canyon defects in the buffer layer on the terraces of sample $S_{100}$.

(j) The step-height profiles, extracted from the indicated line in the AFM topography images, indicate the giant step bunching under 1000 sccm Ar flow. The LEED image of each sample shows typical patterns: A $(6\sqrt{3} \times 6\sqrt{3})R30°$ reconstruction for buffer layer on $S_0$ (c) and $S_{100}$ (f) and a $(1 \times 1)$ SiC crystal structure for $S_{1000}$ (i), acquired at 140 eV.

This situation can be better perceived in the thermodynamic diagram of the process shown in **Figure 5.2c**. The idea of using an inert gas like argon in epitaxial graphenization of SiC to tackle the problem of high Si sublimation under vacuum is almost similar to that of what Langmuir [238] and Fonda [239] considered about the thermodynamics concerning increasing the lifespan of the tungsten filament by replacing the vacuum by argon gas instead. [240] Similarly, it can be assumed that the evaporation from the SiC substrate in a gas at atmospheric pressure is a diffusion phenomenon within a certain limited range of evaporating species (interface layer) adjoining the substrate's surface.

The heat transfer mechanism under different gas flows, inferred from the model described in the literature [241], is shown in **Figure 5.2c**. The heat convection resulting from the random molecular motion (diffusion) dominates near the surface (interface layer) where the fluid velocity is low. When the gas flow is zero, this heat convection by diffusion extends in the boundary layer due to the buoyancy force, causing density change and the thermal gradient, which creates a thermal current. This leads to high partial pressure (or concentration) at the interface layer, increasing the surface's supersaturation rate. Under such conditions, the resulted extra pressure is compensated by a pressure sensor that opens the outlet to retain the $P \approx 1$ bar.

By increasing the Ar flux, the heat is transported mainly due to bulk fluid (advection) in the boundary layer (which in the interface layer is still due to the diffusion), however, the flux reduces the thickness of the interface layer. This leads to an increase of sublimation rate, compared to very low or zero gas flow, presuming a constant $Ts$.

At intensive gas flow, the thermal gradient and thermal current toward the outlet increases. Moreover, the irregular flux in the boundary layer and its lower $T$ leads to condensation and random reflection of some of the specimens back to the substrate surface, which is probably the reason for the formation of the aggregating specimens as well as triangular-like structures on the surface. Moreover, as a consequence of the intensive gas flow, no significant supersaturation on the surface occurs, which is why no buffer layer is observable in its LEED pattern in **Figure 5.1i**.

The formation of two-dimensional nuclei is a susceptible function of the supersaturation, which is negligible below a critical supersaturation and rapidly increases above it. [242,243] The supersaturation is far lower in the case of increased gas flow compared to low gas flow. In the absence of gas flow, the displacement means of mobile molecules in the interface layer are almost uniform and are led by a driving force coming from terraces surface energies (step flow). This uniformity is perturbed by increasing the gas flow, causing the appearance of canyon-like structures at 100 sccm. This shows, however, a mass diffusion on the surface as a result of moderate gas flow. This distortion is mainly due to the irregular flow of gas, causing a local change in the thickness of the gas film and, indeed an incoherent shift in surface diffusion, which results in the formation of canyon-like structures because of higher Si sublimation.

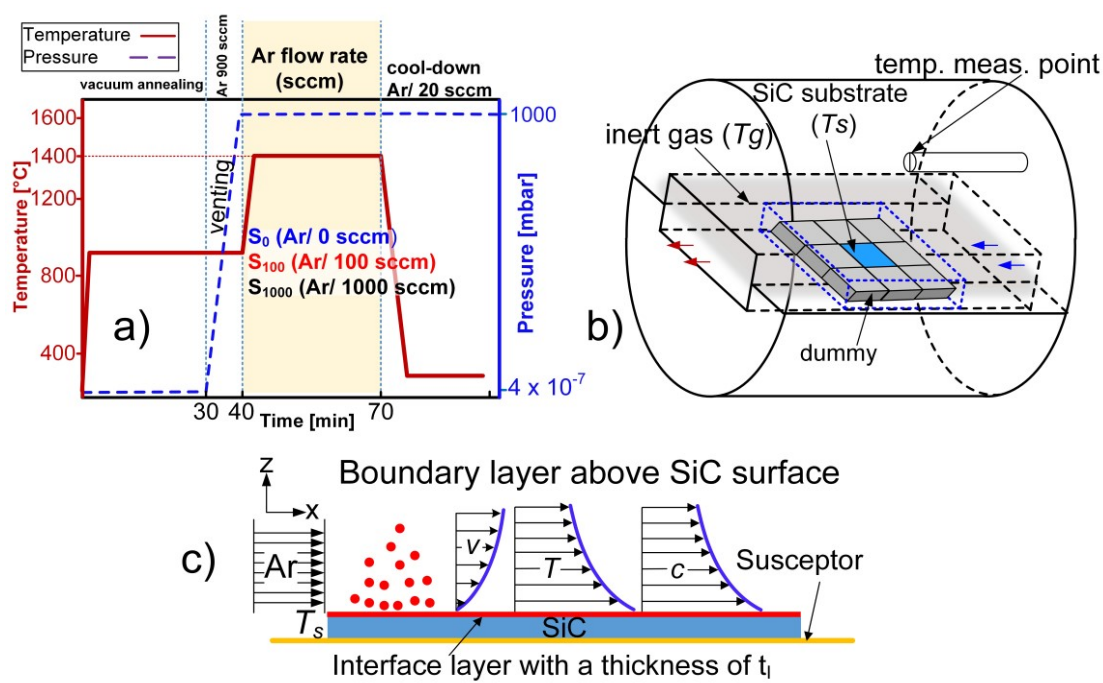

**Figure 5.2. Schematic of argon flow-rate experiments.**
(a) Diagram of buffer layer growth under the same condition but different Ar flow-rate of 0, 100, and 1000 sccm for three samples named $S_0$, $S_{100}$, $S_{1000}$, respectively. (b) Schematic of susceptor and samples during growth in graphene reactor, for more detail, see the text and Chapter **4**. (c) Drawing of the thermodynamics condition demonstrating the change in the growth kinetics (velocity ($v$), temperature ($T$), concentration ($c$)) at the surface of the sample at the presence of different Ar flow rates, see text for more detail.

## 5.3.2. Ar flow-rate in epigraphene growth

In the following, it is demonstrated that the rate of the Ar mass flow also has a substantial impact on the surface morphology of epitaxial monolayer graphene. Two exemplary graphene samples $G_0$ and $G_{20}$ are compared, grown at 1750°C under zero and 20 sccm Ar gas flow, respectively. The AFM images for both samples (**Figure 5.3a and b**) reveal smooth and regular terraced surfaces covered with monolayer graphene (see Raman spectrum in **Figure 5.5e**).

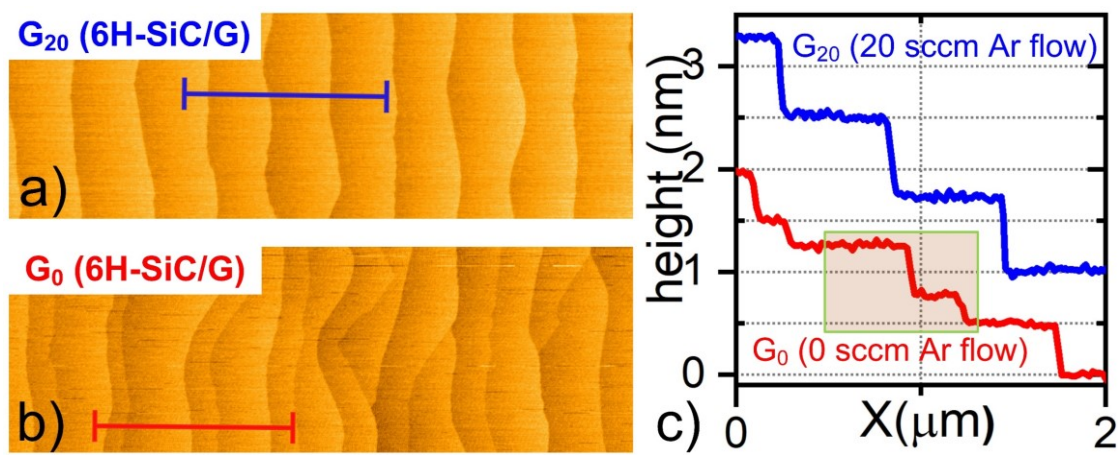

**Figure 5.3. Effect of Ar flow-rate on epitaxial graphene growth on 6H-SiC(0001).**
(a),(b) AFM topography of two graphene samples, $G_{20}$ and $G_0$, grown under 20 sccm and zero argon flow, respectively, at 1750° and 900 mbar Ar pressure. (c) Comparison of AFM height profiles of both samples, $G_0$ and $G_{20}$. The lower step heights and the step pairs of ~0.25/~0.5 nm (indicated rectangle) exhibit a slower step retraction velocity for growth under zero Ar flow.

Under the slightly increased Ar mass flow of 20 sccm, a homogenous step height of ~0.75 nm is observed (corresponding to 3 Si-C layers), see cross-section in **Figure 5.3a**. For zero Ar flow, the step height is further reduced, and a sequential pattern of step pairs with heights of ~0.25 nm and ~0.5 nm is observed, which consequently leads to narrower terrace widths. This result confirms the model explained above. For the higher Ar flow, a faster decomposition of the SiC layers leads to a step height of ~0.75 nm corresponding to half of a 6H-SiC unit cell. The slower SiC decomposition rate of zero Ar flow results in gradually retracting SiC layers. The formation of the observed step pairs of one and two SiC layers is attributed to different retraction velocities of the SiC layers related to the inequivalent surface energy of the specific SiC layer sequence of the 6H polytype. [133,141]

The retraction process stops when large-area buffer layer coverage on the terraces stabilizes the SiC surface. Once the SiC surface morphology is stabilized by the buffer layer, this structure is frozen and remains stable even when the temperature is further increased for graphene growth to 1750 °C. Accordingly, the SiC morphology is not significantly altered in the subsequent graphene formation process, which is regarded as the formation of the second buffer layer and the detachment and conversion of the first buffer layer into monolayer graphene. The high quality of such ultra-smooth graphene layers was already shown by Raman measurements and nearly isotropic resistivity. [36,37,39]

The experimental results of the influence of Ar flux on 4H-SiC samples are very similar to 6H-SiC samples. Similar experiments on 4H- and 6H-SiC samples without PASG preparation which yielded the same results. [37] It was also found that at a high argon gas flow, noticeable triangular-shaped structures on the surface appear. The shape of such structures could be due to the hexagon-triangle at the surface of SiC forming as a result of stress and condensation of evaporating species back on the substrate. Increasing the gas flow and also leaving out the polymer preparation cause rather the appearance of such triangles. [37,38] (See Appendix **A4**).

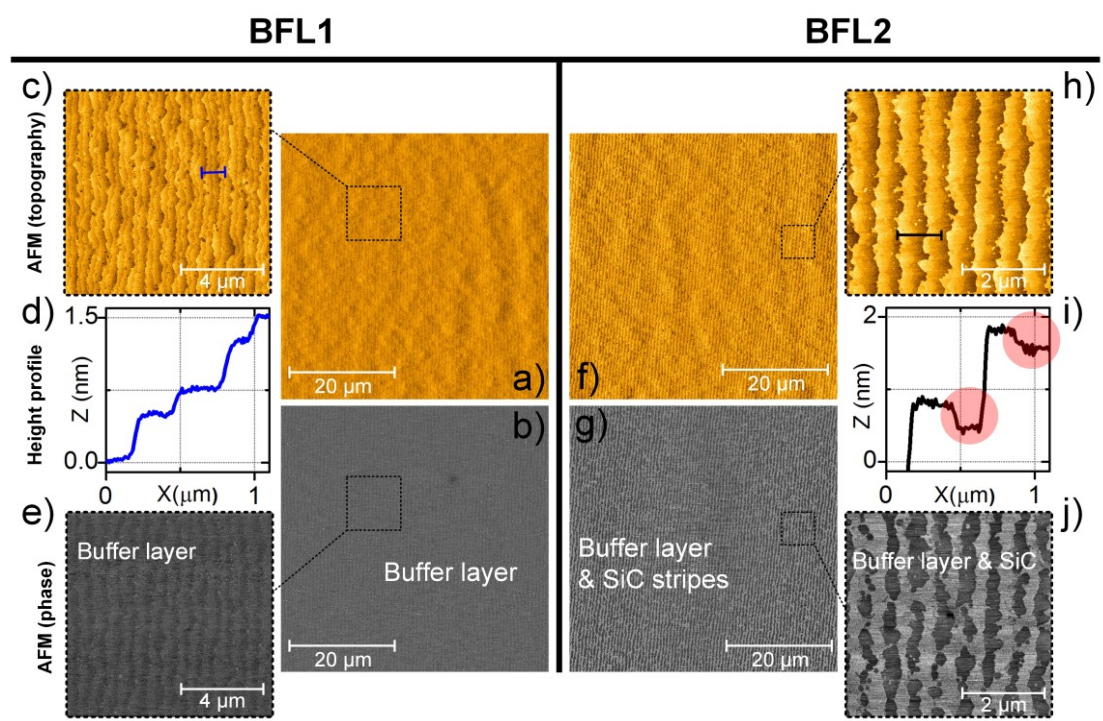

**Figure 5.4. Comparing optimized and non-optimized buffer-layer samples.**
(a) AFM topography and phase (b) images of the optimized buffer layer (BFL1) sample. A closer look at the cut-out AFM images (c) and height profile (d) reveals a highly smooth surface with sequential steps of ∼0.25 nm and ∼0.5 nm on BFL1. (f) Large-scale AFM topography and (g) phase of the non-optimized buffer layer sample (BFL2). Zoom-in AFM topography (h) shows the step defects and lack of complete buffer layer coverage on the terraces, as shown in the height-profile (i) of the BFL2 sample. The AFM phase images of BFL1 (e) and BFL2 (j) demonstrate alternating phase color contras. The phase-contrast on the BFL2 is traced back to the different materials, i.e., the buffer layer and SiC stripes. However, the regular phase-contrast pattern on the terraces of BFL1 is attributed to surface energy difference originating from the different underlying SiC layer sequences. This simple measurement reveals a fascinating interaction between the SiC terminations and atop carbon layers. See refs [38,78] and Chapter **7** for more detail.

## 5.4. Optimization of buffer layer growth

Comparing the samples $S_{100}$ and $S_0$ illustrate that the presence of argon flow-rate causes a change in surface mass diffusion; therefore, it may be implemented in the growth to overcome buffer layer low coverage. Herein, although almost similar steps and terraces are observed on $S_0$ and $S_{100}$, the latter shows very small straggly uncovered SiC area with brighter phase-contrast inferred from the AFM phase **Figure 5.1b**, which for $S_0$ looks rather like semi-ordered weakly covered regions along with the steps (AFM phase in **Figure 5.1c)**. The lack of proper coverage of the buffer layer at step regions is a critical problem limiting obtaining a large-size QFMG after intercalation. **Figure 5.4a-e** exhibits the buffer-layer sample (BFL1) grown on 6H-SiC substrates by applying a very mild Ar flux (20 sccm) for several minutes in between the two zero Ar-flux annealings at 1400°C. For comparison, a buffer layer sample (BFL2) with low coverage is shown in **Figure 5.4 f-j**.

The buffer layer improvement was started with the growth condition of $S_0$ but with an optimized time and annealing protocol as given in Chapter **4**. The AFM topography of this optimized sample BFL1 in **Figure 5.4a** shows a very smooth buffer layer with step heights below ~0.75 nm (**Figure 5.4d**) and a repeating pattern of step pairs of ~0.25 and ~0.5 nm which indicates a reduced step retraction compared to the samples $S_0$ and $S_{100}$. No canyon defects appear in this sample, and the corresponding phase image in **Figure 5.4b and e** also shows no line defects, which indicates a continuous buffer layer that spans over the terrace edges. The small step heights are supposed to be additionally beneficial for the linking process of the buffer layer on neighboring terraces.

Although the buffer layer (BFL1) thoroughly covers the surface (see **Figure 5.5a and b),** an appreciable phase contrast between both alternating terraces is observed, which is attributed to the influence of the underlying SiC layers. This contrast and its origins are extensively discussed in Chapter **7**. In the following, the quality of the buffer layers (BFL1 & BFL2) will be further examined by applying hydrogen intercalation and Raman investigations.

## 5.5. Optimization of quasi-freestanding monolayer graphene

### 5.5.1. Hydrogen intercalation

So far, the result of an optimized buffer layer sample (BFL1) and a low-quality buffer layer sample (BFL2) were presented in section **5.4**. Here the goal is to achieve a high-quality QFMLG layer by hydrogen intercalation technique. By comparing the results of hydrogen intercalation on the BFL1 and BFL2 samples, the importance of the BFL quality in the uniformity of the QFMLG is further demonstrated. The structural homogeneity and lateral coverage of the BFL1 and BFL2 samples are studied by Raman measurements. The spectrum of BFL1 shows broad features, between 1200 and 1700 cm$^{-1}$ (upper spectrum in **Figure 5.5a**) which are related to the vibrational density of states (vDOS) of the $(6\sqrt{3} \times 6\sqrt{3})R30°$ surface reconstruction. [244] The integrated intensity area of these broad Raman bands is regarded as a measure for lateral coverage in the Raman mapping of the buffer layer, and it is plotted in **Figure 5.5b** for an area scan of $20 \times 20$ μm². Additionally, no Raman spectral changes were observed in the vDOS bands during the Raman mapping indicating a homogenous distribution of the buffer layer across the investigated area. The nearly monochrome green colored area visualizes that a continuous and homogenous buffer layer has formed. This becomes obvious when the Raman map is compared to that of a BFL2 sample, which was grown under non-optimized conditions, see **Figure 5.5e and h**. There, the spatial variation of the integrated buffer layer intensity displays a considerable non-homogenous coverage and the partial lack of the buffer layer.

It should be noticed that here the resolution of the Raman measurement is ~1 μm, which is about the terrace width and thus higher than the uncovered areas seen in the AFM image in **Figure 5.4f-j**. Therefore, the mapping is an average and superposition of the backscattered Raman spectra from all scanning, including both covered and uncovered areas. Nevertheless, the inhomogeneity can be inferred from the mapping in **Figure 5.5f and g** but less evident in the FWHM mapping of the 2D peak in **Figure 5.5h**. The upper spectrum in **Figure 5.5e** shows no graphene-typical 2D peak (at ~2700 cm$^{-1}$), which indicates the absence of EG domains on top of the buffer layer but shows a large D peak in the bottom spectra after H-intercalation indicating a significant defect density. Raman mappings have been evaluated regarding the characteristic peak parameters such as peak position and peak width using a non-linear curve fitting algorithm.

Furthermore, the mean values of these quantities and the standard deviations were calculated from these data. The appearance of the 2D peak in the Raman spectra of the QFMLG1 sample (lower spectrum in **Figure 5.5a**) proves that quasi-freestanding monolayer graphene has been produced, in agreement with LEED shown in **Figure 5.6e**. The relaxation of the graphene layer is indicated by the redshift of the 2D peak position at $2669 \pm 2.7$ cm$^{-1}$ compared to the 2D peak position at ~$2731 \pm 1.5$ cm$^{-1}$ of monolayer epitaxial SiC/G van-der-Waals bonded to the buffer layer. From the small FWHM value of $27 \pm 2.2$ cm$^{-1}$, a high carrier mobility value is expected. [245]

The Raman spectrum of the QFMLG sample in **Figure 5.5a** (lower spectrum) shows a well pronounced G peak (1587 $\pm$ 1.1 cm$^{-1}$) with an FWHM of 9.6 $\pm$ 1.9 cm$^{-1}$. A very small D peak at ~1339 cm$^{-1}$ in the Raman spectrum of QFMLG1 can be attributed to a small density of remaining defects in the graphene lattice. The small $I_D/I_G$ (peak maxima) ratio of about 0.1 is comparable to that of other high-quality graphene samples. [246] From the $I_D/I_G$ ratio, a defect density of $n_{def.} = (3.3 \pm 0.7) \times 10^{10}$ cm$^{-2}$ is estimated. [176,246] This extraordinarily high graphene quality was found over the entire area of 20 × 20 µm$^2$ in the mapping of the $I_D/I_G$ (peak values) ratio in **Figure 5.5c**.

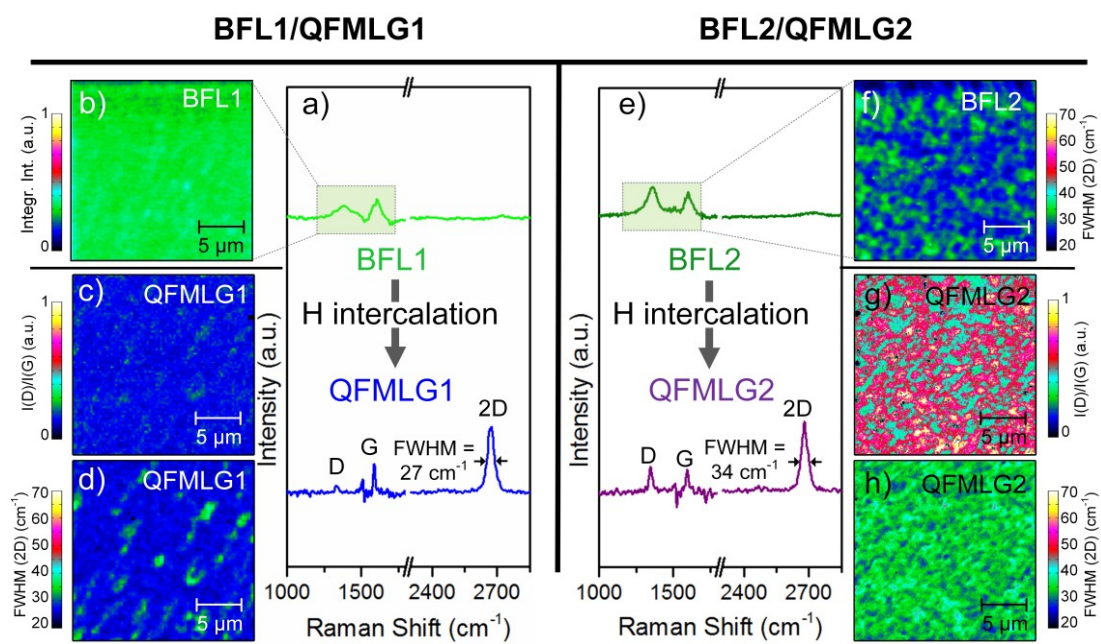

**Figure 5.5. Raman spectroscopy of BFL samples before and after hydrogen intercalation.**

Two samples, BFL1 (optimized growth) and BFL2 (low-quality) are compared before and after the H-intercalation. a) shows the Raman spectra of an optimized BFL1 (upper spectrum) and the resulting QFMLG1 (lower spectrum) after hydrogen intercalation. In (b), the 20 × 20 µm$^2$ map of the integrated intensity of the buffer layer Raman band is plotted. In (c) the intensity ratio of D- and G-peak (peak values) and in (d) the linewidths (FWHM) of the 2D peak of the QFMLG1 sample are displayed in areal maps.
(e) Shows Raman spectra of BFL2 and the resulting QFMLG2 obtained after H-intercalation. In (f) the 20 × 20 µm$^2$ map of the integrated intensity of the BFL2 Raman band is plotted. (g) Areal maps of the intensity ratio of D- and G-peak (peak values) and (h) linewidths (FWHM) of the 2D peak of the QFMLG2 sample. For the analysis of the data, see the text.

Measurements at different positions and on other optimized buffer layer samples suggest a similar low defect density over the whole sample surface. The origin of the defects is correlated with the SiC crystal imperfections that induce defects in the buffer layer during growth and the intercalation deficiency like partial intercalation or etching which could have reasonably been enhanced by a low concentration of the applied hydrogen. [247–249]

In **Figure 5.5a and e**, the Raman spectra were subtracted from a reference Raman spectrum of pure SiC to remove the spectral overtones related to SiC. A spectral artifact (wiggles) appears (in blue and red spectra) at ~1500 cm$^{-1}$ and ~1700 cm$^{-1}$, which is due to a slight mismatch between the Raman spectrum of the samples and the reference spectrum of SiC. The Raman mapping of the FWHM of the 2D peak in **Figure 5.5d** shows predominantly blue marked regions that are related to 2D peak widths of ~27 cm$^{-1}$, which give evidence of a very high homogeneity QFMLG1. Bilayer formation, in this case, is excluded since it would result in much larger FWHM values > 45 cm$^{-1}$. [245] The small green-colored areas show increased FWHM values slightly above 30 cm$^{-1}$ which could arise from low strain variations at the nanoscale, leading to a superposition of slightly different 2D peak positions within the Raman laser spot and thus exhibiting an artificial broadening of the 2D peak width in the acquired Raman spectrum and mapping.

### 5.5.2. Uniformity investigation at the atomic scale

The QFMLG was produced by hydrogen intercalation of the optimized buffer layer (BFL1), and the STM and LEED scrutiny of the surface are shown in **Figure 5.6c-e**. The large area detachment of the buffer layer is proven by the typical LEED pattern in **Figure 5.6e** of quasi-freestanding graphene (QFMLG1), giving further support to the Raman measurements in **Figure 5.5a-d**. [18]

The $(6\sqrt{3} \times 6\sqrt{3})R30°$ buffer layer pattern has disappeared since the correlation of the buffer layer superstructure to the underlying SiC surface lattice is lost. Additionally, the atomic structure was investigated using an Omicron low-temperature STM at 77 K with a tungsten tip. The detailed topography of the QFMLG1 sheet near ~0.25 nm high, ~0.5 nm, and ~0.75 nm high step edges are displayed by high-resolution STM images in **Figure 5.6a-d**. The graphene lattice is seen on both the upper and lower terraces. The observation of the hexagonal crystal structure (lattice constants of 2.46 Å) proves that the SiC in this area is completely covered with graphene, and line defects are absent.

Furthermore, dislocations and domain boundaries are not observed. The 4 × 4 nm$^2$ STM images taken across the step edges reveal a coherent graphene layer that spans smoothly over the step edge from one terrace to the next.

A larger two-dimensional image in **Figure 5.6a** shows the unchanged lattice orientation over the step. This finding is similar to the case of monolayer graphene covered steps. [248,249] This is also very similar to the case of QFMLG on the higher step of ~0.75 nm, which randomly was observed on the surface of this sample, see **Figure 5.6d**. However, in this QFMLG, the warp up (down) of the graphene sheet at the upper (lower) terrace, as observed for EG [248,249], is not observed.

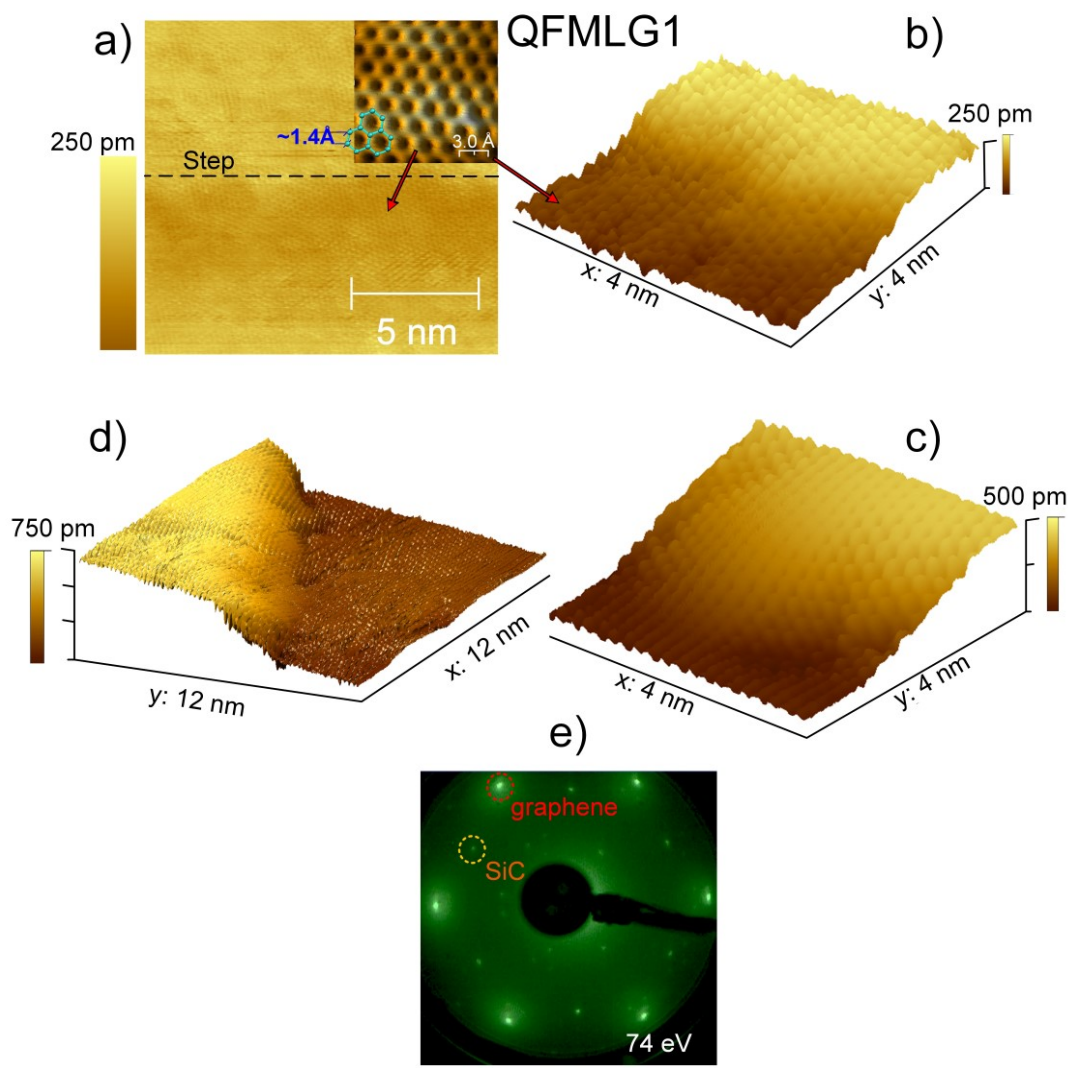

**Figure 5.6. STM measurements of the optimized QFMLG1 sample grown on 6H-SiC.**

After H-intercalation, a coherent sheet of QFMLG is obtained. (a) STM inspection (4 × 4 nm$^2$) of the QFMLG near terrace steps of minimum feasible step-height of ~0.25 nm (a, b), ~0.5 nm (c), and ~0.75 nm (12 × 12 nm$^2$) (d) show perfect coverage with the single freestanding graphene layer. (e) The atomic resolution topography of the QFMLG in the marked square terrace area is obtained by constant-current STM (0.2 nA, −0.5 V). Three hexagonal carbon rings (cyan) are indicated (inset a). (e) Shows the typical LEED pattern of quasi-freestanding graphene.

### 5.5.3. Impact of impurities in intercalation agent

Aside from the above-discussed successful intercalation of buffer layers by 5% hydrogen (95% argon), it was striking to observe much the same effect for the BFL samples that were annealed (600- 1000 °C) in 5N nitrogen (99.999% purity) environment. After this process, the samples were not semi-insulator anymore but instead turned into semi-metallic freestanding graphene monolayers. Due to the almost inert-like nature of molecular nitrogen with its triple covalent bonds, a direct influence of nitrogen in intercalation is unlikely since its dissociation is a strongly endothermic process. However, it cannot be considered as impossible because the dissociation of nitrogen may occur at lower energies under certain circumstances, e.g., in the vicinity of a metal (e.g., ruthenium, lithium, or iron) [250–255] due to a catalytic effect, or the gas concentration (e.g., diluted in argon) [256], or presence of magnetic field [257]. Moreover, graphene growth under nitrogen ambient was shown to result in low nitrogen-doped graphene. [258] Also, theoretical studies expect obtainable charge neutrality on epigraphene via nitrogen intercalation. [259]

Since the abovementioned cases could plausibly happen in our graphene reactor, herein, a systematic study was conducted to figure out the origin of the intercalation. All sources of impurities, e.g., leakage in oven and gas impurities, were checked. Finally, it turned out that this does not occur using nitrogen ambient with 6N purity. From this examination, two interrelated conclusions are inferred: (i) the leakage in the graphene oven can be excluded, and (ii) the origin of the intercalation must be explored instead in the gas impurity itself. From the gas manufacturer, it is known that for standard 5N nitrogen, an impurity inclusion below 5 ppm ($H_2O \le 3$, $O_2 \le 3$) is expected, which for 6N nitrogen is smaller than 1 ppm ($H_2O \le 0.5$, $O_2 \le 0.5$). This is the reason for the intercalation and is further supported by XPS measurements, which showed the appearance of an ultrathin silicon oxide layer located between the SiC substrate and quasi-freestanding graphene, by considering the silicon bonding states in the Si 2p core-level spectrum, see the XPS results in ref. [118]. Moreover, no oxygen species bound to carbon atoms were derived from the XPS data that would reveal a defective graphene layer, and thus graphene oxidation. These results show that such a low concentration of oxygen impurities in nitrogen could be adequate to turn the BFL into freestanding graphene layers. This is further shown with step-by-step Raman measurement on two samples named QFMLG3 and QFMLG4, which were intercalated on very similar buffer layer samples (like optimized BFL1 sample) with hydrogen and nitrogen (5N), respectively. The results are illustrated in **Figure 5.7**. For all the process steps, except the intercalation agent, all other parameters (e.g., $t$, $T$, and $P$) were kept the same.

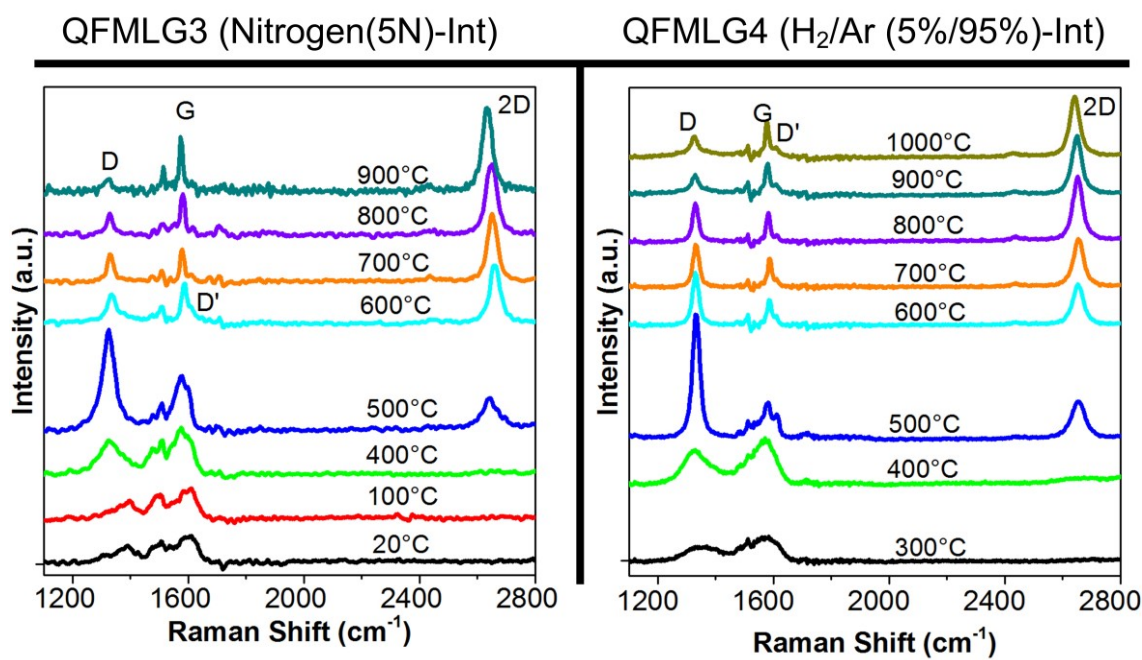

**Figure 5.7. Comparative temperature-dependence intercalation of BFL in N₂ (99.999%) and 5% H₂ (95% Ar) ambient by Raman spectroscopy.**

Two similar quality BFL samples (like BFL1 in **Figure 5.4a-e** and **Figure 5.5a-b**) were used, which after intercalation using nitrogen (5N) and hydrogen are named QFMLG3 (left-side) and QFMLG4 (right-side), respectively. Both samples show similar behavior by increasing temperature despite the different served intercalant agents. Up to 400 °C, yet none of the two samples show intercalation.

For temperature above 400°C, the D peak starts to increase which is accompanied by the appearance of the 2D peak, initiating the transition into graphene. By increasing the temperature on both samples, the 2D peak increases, and D peak declines further, demonstrating more effective intercalation and reduction of defect density. However, the D peaks do not vanishes implying yet existence of crystal defects in both samples.

**Figure 5.7** demonstrates the Raman spectrum of the buffer layers (black color spectrum on the bottom of both experiments) and their transition into QFMLG. For both intercalations, the Raman characteristic of the buffer layer does not change at low-temperature annealing. Raising the annealing temperature to 400 °C changes noticeably the Raman spectrum, which is predominantly indicated by a spectral softening of almost all phonon bands of the buffer layer, thus implying slight structural changes of the buffer layer lattice, which could probably be attributed to the gradual disappearance of the phonon-dispersion behavior of the buffer layer resulting from the steady conversion into QFMLG. Moreover, the rise of the D and G peaks in the Raman spectrum of the buffer layer at 400 °C, indicating the start of the lattice transformation from the buffer layer into QFMLG. The further increase to an annealing temperature of 500 °C results in an increased D peak intensity and the simultaneous appearance of the 2D peak, emphasizing the spectral superposition of QFMLG and buffer layer phonon modes in the Raman spectrum.

Upon further annealing at higher temperatures, the D peak decreases, and the 2D peak increases, revealing a complete transformation of the BFL to QFMLG on both sets of samples. The similarity of the transition on both experiments is fascinating and might indicate the decisive role of temperature than the intercalant. This seems reasonable since the intercalation is known to be obtainable through vast material choices. [38,43,101,226,228,260]

Additionally, the electronic transport on the intercalated large-sized sample ($5 \times 5$ cm$^2$) using the Van der Pauw method at room temperature underlines the high quality of the samples. The VdP measurement on QFMLG3 (treated in 5N nitrogen) revealed a hole doping with $p \approx 2.9 \times 10^{13}$ cm$^{-2}$ and improved charge-carrier mobility of $\mu \approx 620$ cm$^2$V$^{-1}$s$^{-1}$ compared to other studies ($\mu \approx 420$ cm$^2$V$^{-1}$s$^{-1}$). [114] Compared to QFMLG4 (H-intercalation), lower mobility was achieved (see **Table 6-3**), which could be due to a lower concentration of the intercalant but also a higher carrier density in QFMLG3.

It is worth to be mentioned that, contrary to the above experiment on buffer layers, the intercalation of epitaxial monolayer for the fabrication of QFBLG in the 5N nitrogen ambient was not successful as in hydrogen/argon (5%, 95%) atmosphere. This gives additional evidence that the intercalant concentration is a crucial parameter and should be high enough for proceeding efficient intercalation.

## 5.6. Optimization of quasi-freestanding bilayer graphene

The intercalation can also be carried out on epitaxial monolayer graphene to convert it into two freestanding graphene layers on top of each other. This happens with the same mechanism as before through the separation of the BFL, and thereby a QFBLG is obtained. Here the QFBLG sample was fabricated via hydrogen intercalation of an optimized monolayer graphene sample (like G$_0$, see **Table 5.1**). **Figure 5.8** illustrates the Raman investigation of the QFBLG sample. The upper Raman spectrum in **Figure 5.8a** shows the typical fingerprint of epitaxial graphene, indicating the G peak at 1601 cm$^{-1}$ and 2D peak at 2731 cm$^{-1}$, whereas broad phonon bands from the buffer layer arise in the range of 1200 and 1700 cm$^{-1}$. **Figure 5.8b** shows a homogenous distribution of the 2D peak width of epitaxial graphene over an area of $20 \times 20$ μm$^2$ with an averaged 2D peak width of ($33 \pm 1.5$ cm$^{-1}$).

After H-intercalation (lower spectrum in **Figure 5.8a**), the broad buffer layer related Raman band around the D peak disappears since the detached buffer layer is transformed into the second free-standing graphene layer. Therefore, the 2D peak becomes broader (FWHM around 59 cm$^{-1}$), and the line shape becomes asymmetric (see also ref. [38]), which indicates the formation of bilayer graphene. The FWHM map of the 2D peak in **Figure 5.8d** reveals the uniform distribution of bilayer graphene.

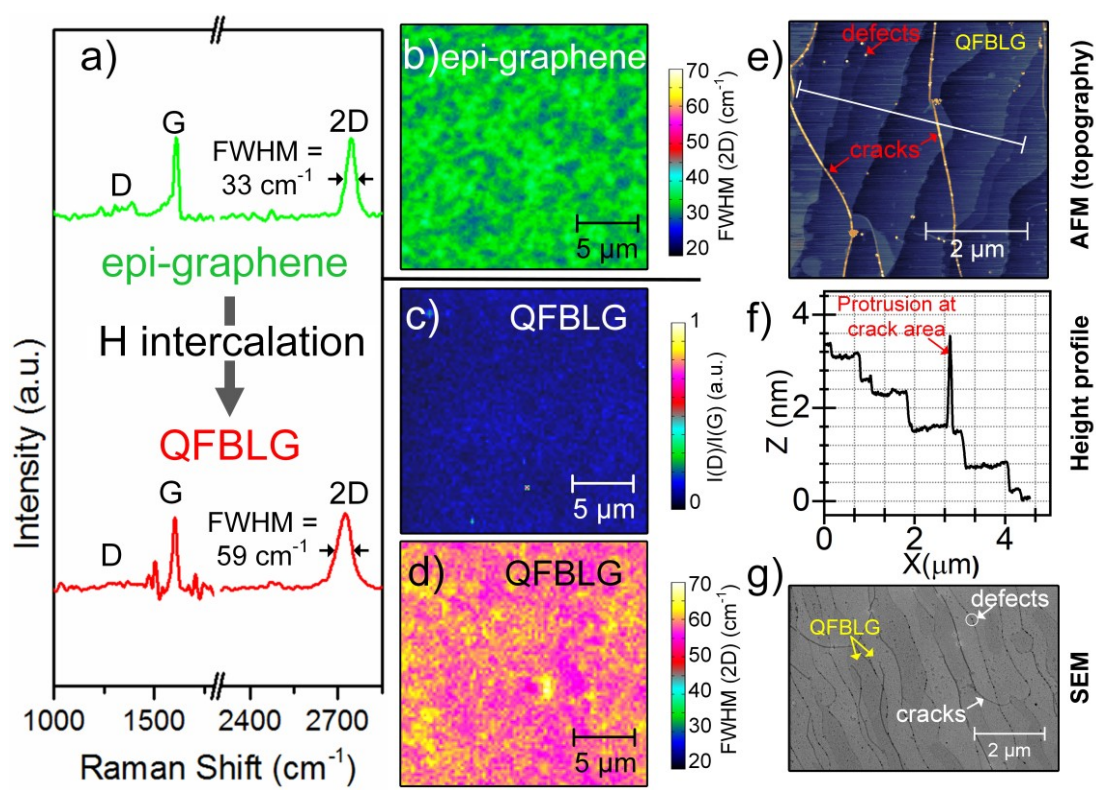

**Figure 5.8. Micro-Raman spectroscopy of graphene MLG sample before and after hydrogen intercalation.**

(a) Raman spectra of epitaxial monolayer graphene and the resulting QFBLG obtained after hydrogen intercalation.

(b) The line widths (FWHM) of the 2D line of the epitaxial graphene layer show homogeneous monolayer graphene without bilayer inclusions.

(c) Areal maps of the intensity ratio of the D and G peak (peak values) and (d) line widths (FWHM) of the 2D peak of the QFBLG sample.

(e) AFM image of the QFBLG sample indicating defects and cracks which appear on the surface after the H-Intercalation. The crack areas do not necessarily appear at the steps regions but also on the terraces, as can be seen in height profile (f) and SEM image (g). SEM image shows a reflectivity contrast on the neighboring terraces that is entangled with the bottom SiC terraces (see Chapter **7** for more detail).

The slightly increased FWHM values (yellow areas) could again be caused by local strain variations. The quality and homogeneity of the QFBLG are further underlined by the low values and even distribution of $I_D/I_G < 0.1$ (peak maxima) ratios, which indicate a low defect density of about $n_{def.} < 2.0 \times 10^{10}$ cm$^{-2}$, see **Figure 5.8c**. The QFBLG samples in this study show compellingly high mobilities up to 3500 cm$^2$/Vs at room temperature. A detailed comparison of transport properties and scattering mechanisms of different discussed samples is reported in Chapter **6**.

Here, in connection with the discussion in section **5.5.3**, in the following, the critical concern about the applied hydrogen (5%) diluted in argon (95%) for the intercalation is concisely discussed. Although the main idea of intercalation of BFL/SiC or MLG/SiC is taking advantage of the nearly freestanding nature of graphene with reduced impact of SiC substrate (e.g., for de higher carrier mobility) [117], however, the technique itself may cause structural defects in the graphene as well. Some of these defects are already reported in the literature [247,249,261]. Here, the low concentration of the used hydrogen (5% in Ar ambient) is technically counterproductive in the intercalation procedure. The used gas concentration, which hitherto must be followed based on safety regulations at the PTB, propel inefficient intercalation. Thereby, partial intercalation is highly plausible, as is demonstrated in **Figure 5.9**.

The STM measurements in **Figure 5.9a and b** clearly show such local intercalated QFBLG areas on a sample which experiences inefficient hydrogen intercalation. **Figure 5.9c and d** show the atomic resolution of QFBLG and MLG, which in the latter, the corrugation of the buffer layer below the graphene is discernable. The height profile in **Figure 5.9e** confirms a displacement of layers after the intercalation process. A distance of ~1.6 Å was measured between the QFBLG and MLG layers, which is in good agreement with other studies. [262–265] The QFBLG and MLG areas were distinguished from the scanning tunneling spectroscopy (STS) measurements (not shown), indicating similar typical STS features as in other studies. [249,266–271] As was shown in **Figure 5.8**, an alternative for improving the intercalation could be extending the processing time, however it may also generate additional defects. Aside from that, interestingly, gentle intercalation was reported to be applicable for improving the carrier mobility in MLG epigraphene for QHE measurements. This, which happens through the saturation of Si dangling bonds on the SiC surface [272], will be further discussed in Chapter **8** regarding the graphene functionalization and magneto-transport.

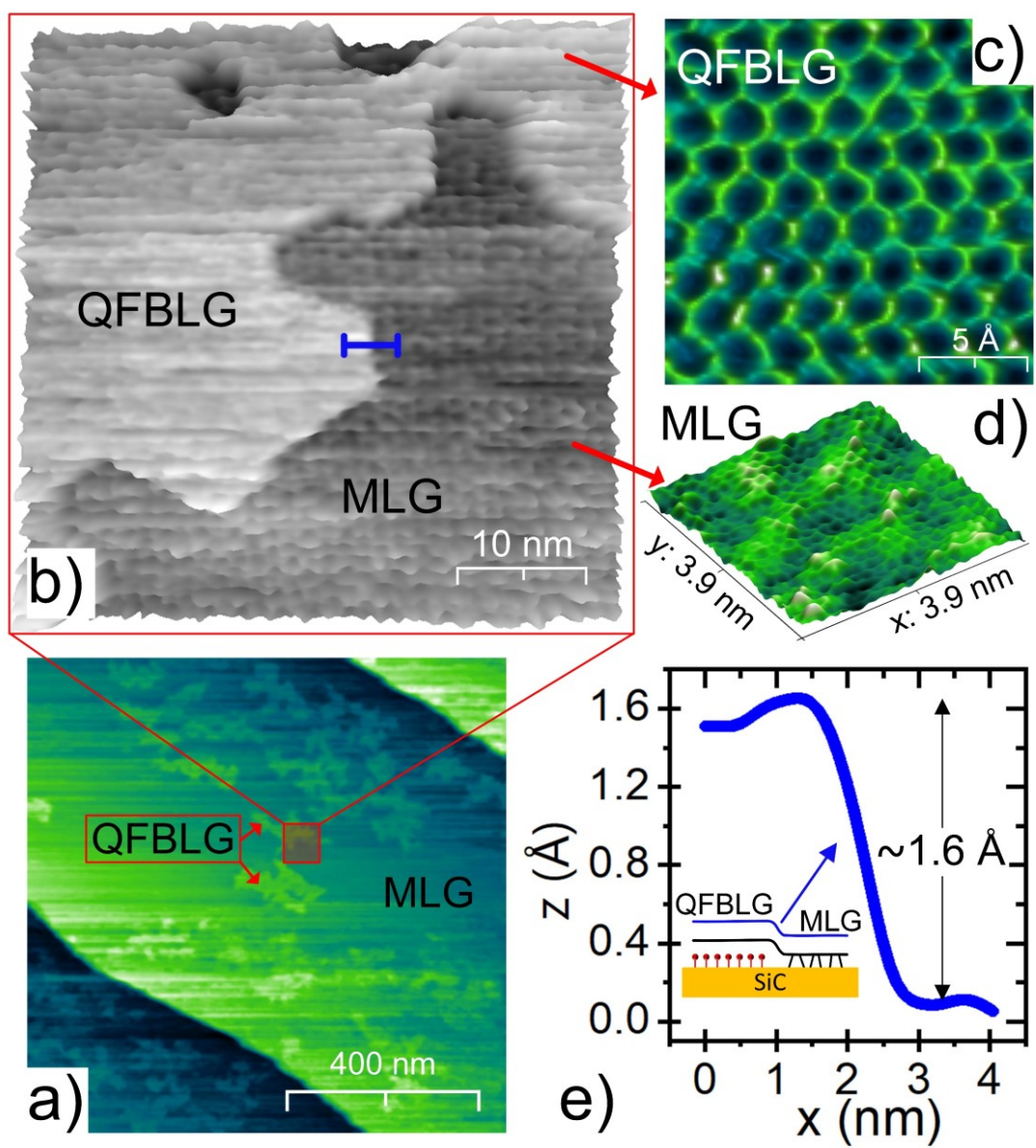

**Figure 5.9. Incomplete hydrogen intercalation of epigraphene.**
(a) STM measurements reveal partially intercalated areas epigraphene(−0.1 A, +2V).
(b) zoom-in STM image shows the coexistence of QFBLG and MLG layers (−0.4 A, +2V).
(c) Atomic resolution of QFBLG (−0.4 A, +2V) (d) and MLG (−0.8 A, +0.3V). A corrugation which arises from the underneath buffer layer is observed for MLG in (d). (e) The cross-sectional plot capture along the blue line in (b) shows a height difference of about 1.6 Å between the QFBLG and MLG layers.

## 5.7. Epitaxial growth of bilayer graphene

Taking the process one step further, the growth of epitaxial bilayer graphene (epi-BLG) can be obtained through an extended annealing process. This is achieved by a similar epi-MLG growth process, but at higher temperatures and adequate growth time, i.e., 1800 °C to 1900 °C, for 5 to 60 min, respectively. Several groups have already reported bilayer graphene growth using different techniques. [13,91,273,274] However, since the graphene synthesis on SiC (0001) is known to be self-limiting, the growth of a homogeneous large-area bilayer (or multilayer) is challenging. A decrease in growth rate due to the increase in the graphene layer thickness entails the self-limitation of the process. The reason is believed to be a result of restricted Si sublimation after the formation of the first epigraphene and buffer layers, which confines the space on top of the SiC substrate for further Si loss, and thus further graphenization. [87,154,275] This is reasonable, since even after the formation of the first BFL (at ~1400 °C, ~1 bar argon), higher energy (e.g., by increasing temperature to ~1800 °C) is required to convert it to the graphene layer by the formation of a second BFL under the first one. Proceeding the growth for extra graphene layers, i.e., BLG, or few-layer graphene (FLG), would be accompanied by further Si depletion, which must ensue through defects in the top carbon layers and thus can lead to additional defects as well.

Following the high-quality epi-MLG and BFL growth using our PASG and the so far discussed optimization techniques, it is interesting to see this method to fabricate epi-BLG. The epi-BLG synthesis was carried out using multiple time and temperature windows, as briefly shown in **Table 5.1**. **Figure 5.10** shows AFM inspections of the epi-BLG sample in this study. The morphology and step-height profile of the sample shows still almost homogenous terraces, however, with larger step-heights compared to those for epi-MLG samples in **Figure 5.3**. No sequential pattern of step-terraces was formed as was shown for epi-MLG in **Figure 5.3b and c**. The higher step-heights result from further step-flow and SiC usage (Si sublimation) for providing enough carbon for the epi-BLG production. The Raman mapping of the FWHM (2D) peak in **Figure 5.10d** verifies a true formation of epi-BLG [274], including small MLG and trilayer inclusions, which emphasize the above-discussed challenge facing a coherent epi-BLG growth. It worth mentioning that in contrast to the *p*-type nature of the QFBLG and QFMLG obtained by H-intercalation, the epi-BLG is electron-doped. [38,276–278] As for the same reason in epi-MLG, the presence of the buffer layer and the electric dipole that exists between the graphene and SiC interface (BFL) induces an electrostatic asymmetry between the layers, which results that the Dirac point to be located below the Fermi energy in the epi-BLG band structure. [22,279,280]

The presented results demonstrate significant progress in the growth of a coherent and scalable epitaxial bilayer graphene fabrication as an excellent platform for further investigations, e.g., intercalation, electronic and magneto-transport, and superconductivity effect. [43,94,278,281–283]

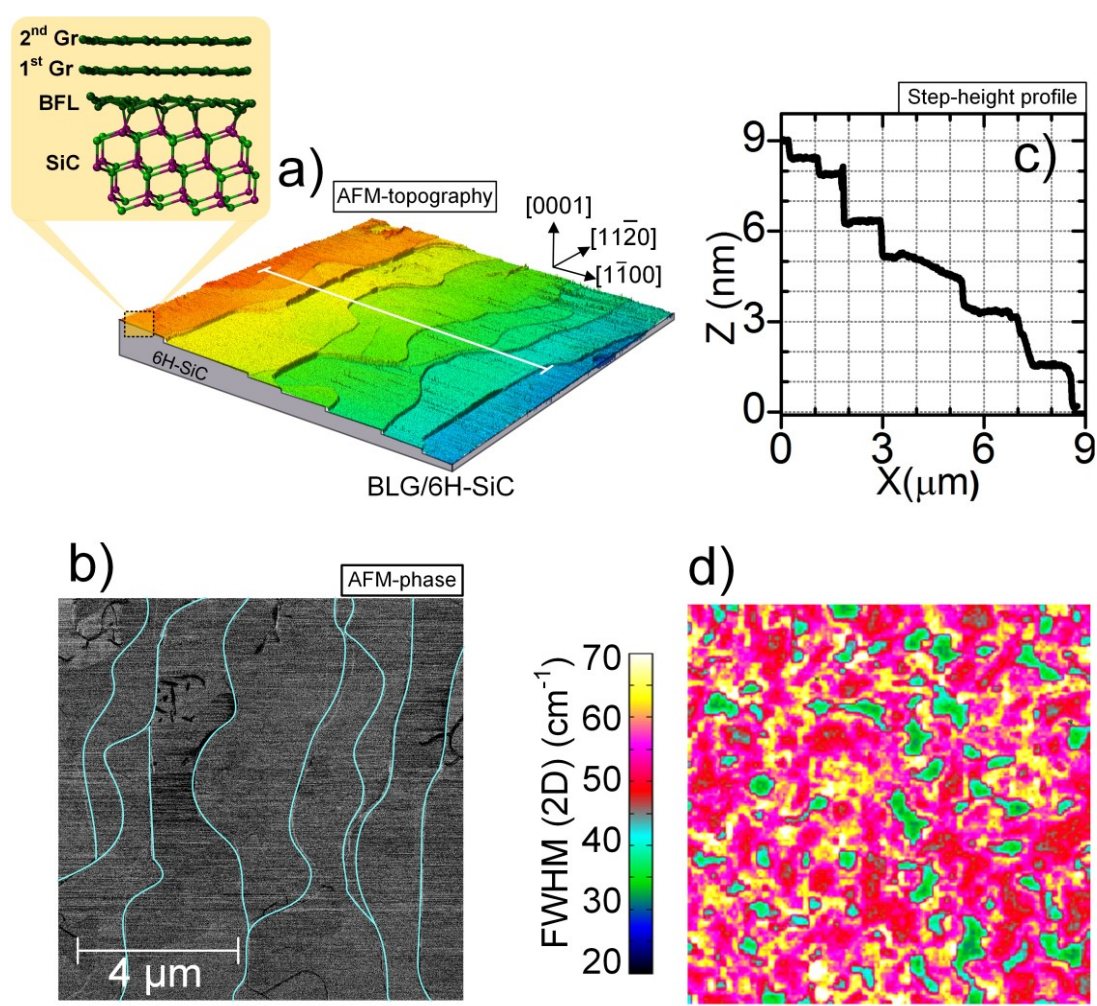

**Figure 5.10. Epitaxial bilayer graphene on 6H-SiC (0001) substrate.**
(a) AFM topography of epi-BLG grown sample on 6H-SiC (0001). Schematic of an epi-BLG structure illustrates two stacking graphene layers on top of a buffer layer, partially bonded to the bottom SiC substrate.
(b) AFM-phase image and (c) corresponding height profile taken from the white line cross-section shown in (a).
(d) Areal map (20 × 20 μm²) of line widths (FWHM) of the 2D peak verify the creation of epi-BLG; nevertheless, monolayer and trilayer graphene patches can also be seen.

## 5.8. Conclusion

In summary, this chapter presented AFM, STM, LEED, and Raman measurements, which indicate the strong influence of the argon mass flow-rate on the formation of the buffer layer and the graphene growth. For a given temperature and constant Ar pressure, the Ar mass flow rate controls the SiC decomposition rate, which can be qualitatively understood by thermal equilibrium considerations. This new finding has the potential to improve the graphene quality by avoiding accelerated step bunching at higher temperatures and graphene roughening for lower Ar pressures, respectively. By properly chosen growth parameters, it is possible to prevent structural defects (canyon defects and step defects) and obtain a continuous, large-area buffer layer without graphene inclusions as well as bilayer-free graphene monolayer.

Optimization of epitaxial monolayer growth resulted in highly homogenous and sizeable layers without bilayer-inclusions. The further characterizations in the following chapters (i.e., **6**, **7**, and **8**) will highlight the quality of such graphene samples.

The QFMLG and QFBLG produced by hydrogen intercalation exhibit excellent homogeneity and very small resistance anisotropy over areas in the millimeter range. This indicates the presence of coherent quasi-freestanding graphene layers over large areas. Moreover, it was shown that surprisingly, the ultra-low concentration of oxygen impurities in 5N-nitrogen is sufficient to intercalate a buffer layer. This was shown in a framework of Raman spectroscopy comparative study of stepwise temperature-dependence intercalation of BFL samples in both hydrogen and 5N-nitrogen environments. The decisive role of intercalant purity was discussed, and it was shown that, e.g., the applied 5% H (95% Ar) gas is practically ineffective in intercalation. This study suggests that implementing pure hydrogen intercalation of, e.g., BFL sample can be obtained at lower temperature and shorter time, which reduce the possible process-induced defects.

Also, a uniform synthesis of epitaxial bilayer graphene was presented. These results demonstrate a significant improvement in achieving sizeable epi-BLG.

This study supports the promising application potential of epi-graphene on SiC for quantum Hall metrology applications and QFMLG and QFBLG for superior transistor performances and extends the capability of epi-MLG, BFL, or epi-BLG to be implemented as a platform for growing other 2D or sub-dimensional materials or metamaterials.

# 6. Resistance anisotropy of epitaxial graphene

## Abstract

*T*his chapter aims to study a so-called extrinsic resistance anisotropy in the family of epitaxial graphene on SiC. It has been already known that monolayer epigraphene on SiC (0001) exhibits the chirality of an ideal graphene sheet. [284] However, in-plane defects are known as the dominant scattering sources by distorting graphene's lattice symmetry. [285] Herein, the prominent purpose of obtaining graphene on SiC with shallow step heights that ensure better electronic transport properties is often thwarted by step bunching of the SiC surface during the sublimation growth. The resistance anisotropy of epigraphene, which highly depends on the substrate and graphene uniformity, is known to be severely degrading the epitaxial graphene-based electronic device performances, particularly in metrological applications, as the primary purpose of this thesis. This chapter explores the origin of the resistance anisotropy in diverse sample types with different quality (e.g., SiC step heights, graphene thickness). It is shown that the resistance anisotropy highly rises by the change in the graphene thickness, i.e., adlayer inclusion. [286–290]. Finally, for the first time, epigraphene with nearly no directional dependence of the resistance is presented. The combinations of nm-, $\mu$m-, and mm-scale measurements convincingly show that the remaining minimal anisotropy (~3%) is only correlated with the substrate's unavoidable step-structures. By this, the minimum resistance anisotropy in epigraphene is reached. Often discussed, other sources for extrinsic anisotropy play no role in the produced bilayer-free monolayer graphene, and it proves on the hand graphene's intrinsic isotropy. Part of this chapter is published and can be found in refs. [37–39,291].

## 6.1. Introduction

Being a sheet of carbon just a single atom thick makes graphene an excellent material with many novel physical properties such as its outstanding carrier mobility. However, it is challenging to save high mobility when the graphene is transferred or grown on a substrate. For example, at room temperature, a $\mu = 15k$ cm$^2$/Vs was measured for exfoliated graphene on SiO$_2$ substrate [12,13,55], while it was reported to be ~140k cm$^2$/Vs on boron nitride [292] and 25k cm$^2$/Vs (at low temperature 200k cm$^2$/Vs) for suspended graphene. [233,293,294] This implies that graphene experiences ranges of scattering mechanisms that substantially degrade its transport properties. Therefore, understanding and possibly manipulating the interaction between substrate and top graphene layers is crucial both from a scientific perspective and for obtaining practical applications.

Graphene fabrication, through SiC sublimation growth, the so-called epitaxial growth on SiC substrates, has the potential to be used as a basis for future electronics applications. [8]–[11],[18]–[21] The method is capable of wafer-scale graphene manufacturing directly on the insulating SiC substrate, and both are highly favorable for device fabrication. The performance of electronic devices highly depends on the graphene's crystal quality and its size demanding coherent electronic properties over large areas. This is, however, challenging for epitaxial growth. The morphology of the substrate, in particular, SiC terrace steps are known to strongly deteriorate the performance of graphene-based electronics, e.g., by limiting the geometry of devices, lowering the cut-off frequency in high-speed electronics [299], degrading carrier mobility [300] in FET devices, [287,301] or leading to anisotropies in the quantum Hall effect (QHE). [272,289]

Rotational square probe measurements have quantified a conductance anisotropy of about 70% for epitaxial graphene layers grown on the Si-face of 6H-SiC. [286] Other four-terminal electronic transport measurements showed a pronounced resistance anisotropy of approximately 60% and even more than 100% for epitaxial graphene produced by sublimation growth (SG) methods and chemical vapor deposition (CVD), respectively. [233,301] In all cases, higher resistance values were observed for transport perpendicular to the SiC surface terraces, which indicates a correlation with the terrace step edges of the SiC substrate.

The impact of individual step edges of the substrate on the electrical resistance of the epitaxial graphene layer was investigated by various local scanning tunneling potentiometry (STP) studies, which revealed an additional step-induced resistance contribution for charge carrier transport in monolayer graphene across the step edges. [302–304] Various physical scattering sources were discussed, e.g., detachment from the underlying substrate leading to a potential barrier induced by a doping variation. [70,304] Also, local scattering by charge built up, graphene defects, as well as local strain at step edges, were addressed as potential origins. [286,305,306] For example, a high density of grouped steps on the substrate is entangled in lower carrier mobility than the exfoliated graphene on SiO$_2$. [16,73,245,307] Moreover, a ten times lower electron mobility in step regions than terrace regions was reported for the graphene on the vicinal SiC substrate. [290]

It is also commonly reported that step density [144], step height [308], and step-bunching [288] increase the graphene's resistance. Another more considerable contribution arises from the transition region between mono- and bilayer (ML-BL) graphene due to a quantum mechanical wave function mismatch. [188,302,309,310] In particular, an ML-BL transition at a SiC step edge causes a significant increase in the local resistance. Moreover, magneto-transport measurements in bilayer-patched monolayer graphene showed that bilayers could cause anomalies in the quantum Hall effect. [311] The influence of bilayer regions on charge magneto-transport also depends on the bilayer position and its carrier density, which latter determines the metallic or insulating behavior of the bilayer. Accordingly, magneto-transport in graphene can be interfered either shunted by the bilayer or constricted through the monolayer graphene regions in case of metallic or insulating bilayer's characteristic, respectively. [312] This suggests that bilayers substantially impact the transport properties of graphene devices, and an impact on the resistance anisotropy is expected. Since the formation of bilayer graphene is often observed at step edges higher than three Si-C bilayers [133,234], it is highly favorable to keep SiC step heights below ~0.75 nm to prevent bilayer formation during epitaxial graphene growth.

This study presents the successful realization of ultra-smooth monolayer graphene sheets on 4H-and 6H-SiC polytype substrates by the so-called polymer-assisted sublimation growth (PASG) technique and several optimizations. [36–39] Rotational square probe measurements of the monolayer graphene reveal nearly vanishing resistance anisotropies of only about 3%. This value is in good agreement with the anisotropy determined from STP measurements at individual terrace steps. Hence, it can be regarded as the ultimate lower limit of resistance anisotropy only given by step-induced resistance contributions. This study shows that nearly perfect resistance isotropy of epitaxial graphene sheets can be achieved by careful control of the growth conditions. It is also shown that the resistance properties in epigraphene are modified depending on the SiC termination underneath. [291]

Moreover, in contrast to such outstanding quality of the sample, three other epigraphene samples grown under different conditions, e.g., sample properties and preparation, are presented, which all together give an insight into the origin of extrinsic resistance anisotropy of epigraphene. Moreover, the resistance anisotropy is also studied for the first time in quasi-freestanding monolayer/bilayer graphene (QFMLG/ QFBLG) samples obtained by hydrogen intercalation technique (see the growth condition in Chapter **5** and ref. [38]). These all are further supported by mm-scale VdP analysis in combination with μm-scale nano-four-point probe (N4PP) measurements of millimeter-sized samples. There is a good agreement between the performed multiscale measurement from local to large-area mm ranges, remarking the significant improvement of the epitaxial graphene growth quality.

## 6.2. Sample preparation

This study is conducted on several samples, including SiC/G, QFMLG, and QFBLG, each with distinctive features (e.g., morphological, electronic properties) to investigate the resistance anisotropy and its origins. The standard sample preparation (discussed in detail in Chapters **4** and **5**) is briefly explained in the following. The growth of epitaxial graphene was performed on the Si-terminated face of SiC substrates ($5 \times 10$ mm²) cut from semi-insulating 6H and 4H polytype wafers (nominally -0.06° towards[$1\bar{1}00$]), in the following referred to as sample S1 and S2, respectively. The epi-ready surface conditioning allows high-quality epitaxial growth without hydrogen pre-etching. [37] For S1 and S2, a particular growth procedure was applied, including the polymer-assisted sublimation growth (PASG) technique and special temperature ramps, as described in Chapters **4** and **5** as well as the ref. [39]. The subsequent high-temperature growth process was identically carried out on both polytype substrates in a horizontal inductively heated furnace. [213] For graphene growth, no argon gas flow was applied. For more details, see refs. [37,38].

For comparison, five other graphene samples, including three typical monolayer epitaxial graphene named (S3-S5) as well as one QFMLG and one QFBLG sample are used in this study, listed in **Table 6-1**. Graphene sample S3 was grown by conventional sublimation growth (SG) after pre-annealing in Ar atmosphere (1000 mbar) on a small miscut 6H-SiC substrate. [234] S4 is a PASG graphene sample on a 6H-SiC substrate with a large miscut angle of ~0.37° [36].

| Sample | SiC-polytype | $[1\bar{1}00]$ | $[11\bar{2}0]$ | Process | AFM | |
|--------|--------------|------------|------------|---------|-----|-----|
| | | | | | $h_{step}$ (nm) | bilayer |
| S1 | 6H | - 0.06° | 0.00° | PASG | ~0.25- ~0.75 | no |
| S2 | 4H | - 0.06° | 0.00° | PASG | ~0.25- ~1.0 | no |
| S3 | 6H | - 0.01° | - 0.01° | SG | ~0.75 | small |
| S4 | 6H | - 0.01° | 0.37° | PASG | ~0.75- ~3.5 | scattered |
| S5 | 6H | - 0.01° | 0.37° | H2/SG | ~3- ~15 | extended |
| QFMLG | 6H | - 0.06° | 0.00° | PASG/H- | ~0.25- ~0.75 | no |
| QFBLG | 6H | - 0.06° | 0.00° | PASG/H- | ~0.25- ~0.75 | yes |

**Table 6-1. List of the samples used in this study.**

The graphene of S5 was fabricated by sublimation growth on a hydrogen pre-etched 6H-SiC substrate. [234] The main parameters (1750°C, ~1 bar Ar atmosphere, 6 min) of the graphene growth were kept the same for all samples. The growth process of the QFMLG and QFBLG sample is described in detail in Chapters **4** and **5**, as well as in ref. [38].

## 6.3. Surface morphology

The AFM topography images of the graphene monolayers grown on 6H- and 4H-SiC substrates, samples S1 and S2, are shown in **Figure 6.1a and d**. The exceptionally smooth and homogeneous surface morphology is a typical result and can be found on the entire surface of the samples. This is confirmed by multiple AFM measurements at different positions in the center and near the edges of the samples, as well as by optical microscopy inspection throughout the surface. The corresponding histograms in **Figure 6.1c and f** are the results of AFM inspection of about 200 steps collected from 9 different positions on the substrates, including edge regions. For most of the terrace steps on both polytypes, we found heights below ~0.75 nm.

A closer inspection of the topography in **Figure 6.1a** reveals a regular and alternating terraces sequence with a ~0.25 nm high step in front of a terrace with ~0.5 nm step-height for the 6H-SiC sample. This situation is depicted in the height profile of **Figure 6.1b**. The clear majority of the terrace steps (~90%) exhibit such a sequential pattern, and only occasionally (10%) steps with ~0.75 nm height are observed, see the histogram in **Figure 6.1c**. Higher steps were not found, which confirms that the PASG technique has effectively suppressed the step bunching.

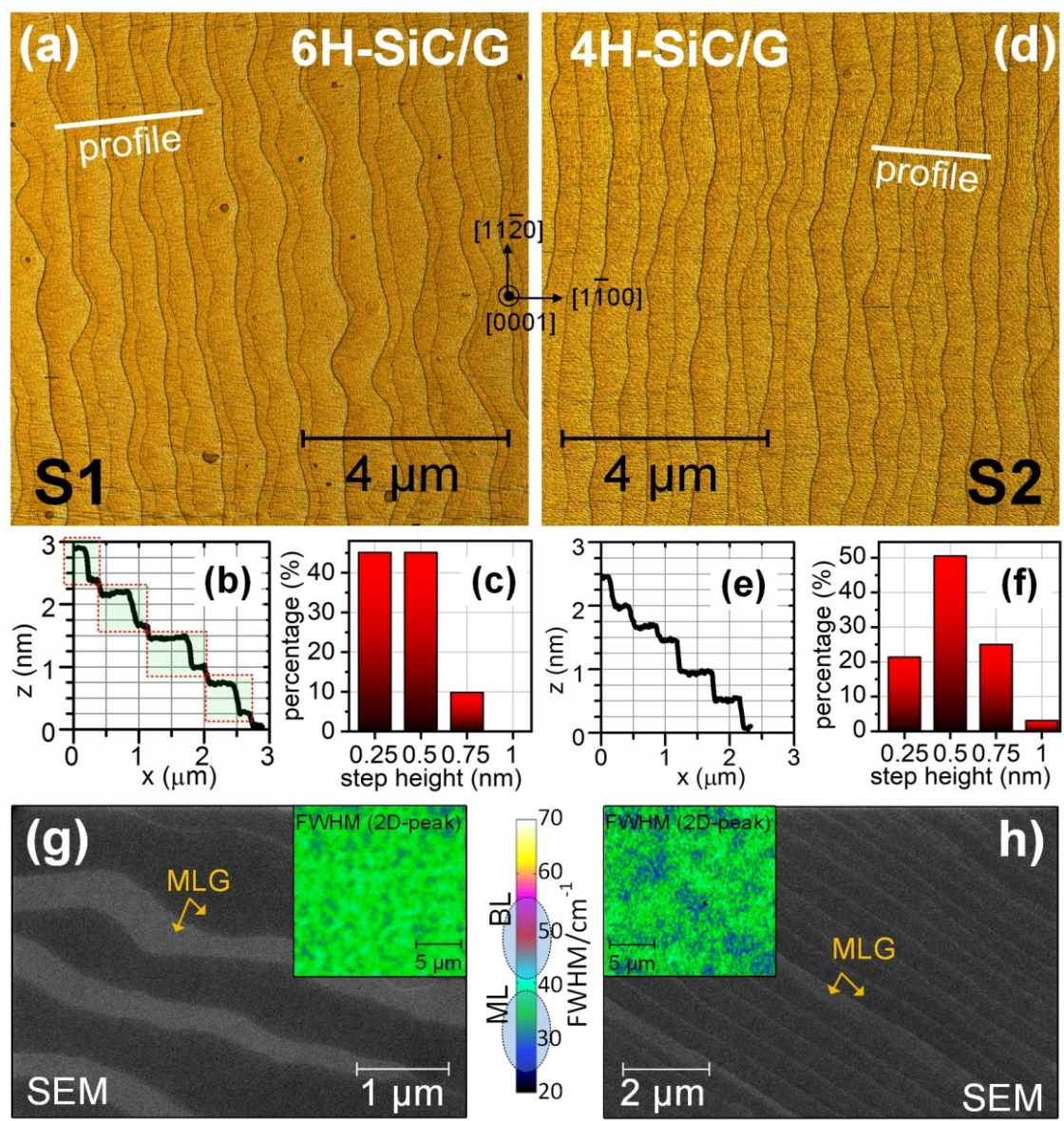

**Figure 6.1. Ultra-smooth bilayer-free monolayer epigraphene.**

AFM measurements of monolayer graphene grown by the PASG method on 6H-SiC (sample S1) and 4H-SiC (sample S2). (a) Surface topography of S1. (b) Height profile along the profile line in (a) showing the pairwise sequence of ~0.25 and ~0.5 nm steps (marked by red dotted rectangles) typical when using 6H-SiC substrates. (c) Statistical evaluation of nine AFM images from the center, edges, and corners of the sample indicating the remarkable homogeneity all over the sample. (d) Surface topography of S2 using 4H-SiC substrates as well as (e) the corresponding height profile and (f) the step height distribution. The scanning electron microscopy images ($E$ = 1 kV) (g) and (h) of both samples show a contrast (backscattered electrons) in graphene. Since from the Raman mapping of FWHM(2D-peak) (inset), monolayer graphene on both S1 and S2 is identified, therefore the BSE contrast in SEM images cannot be due to graphene thickness variations (e.g., bilayer inclusions). This is attributed to the bottom SiC termination. (See Chapter 7 for more detail)

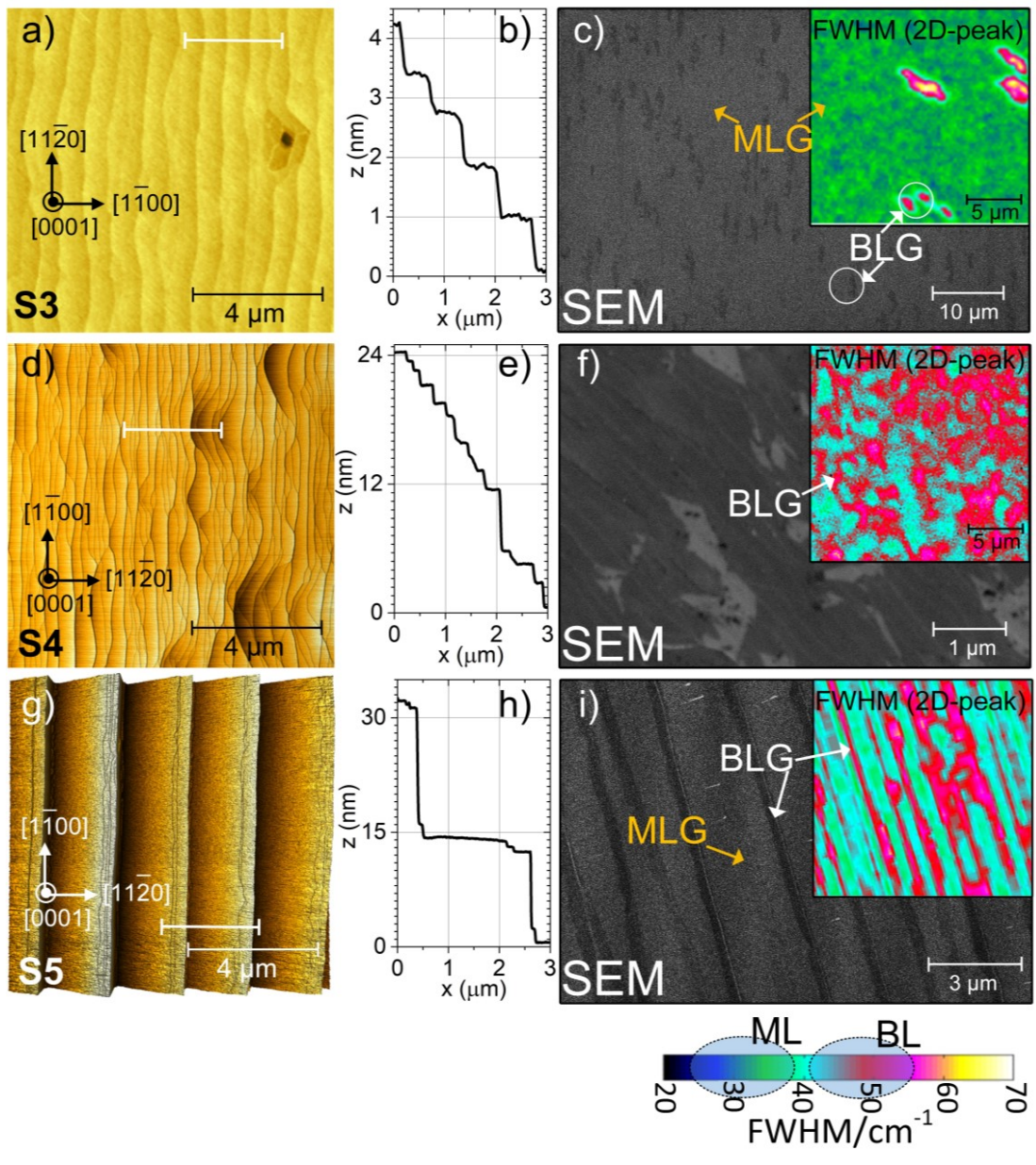

**Figure 6.2. Morphology inspection of the graphene samples.**

Three samples were grown under different growth circumstances. AFM topography of S3 (a) grown by standard sublimation growth (SG) on 6H-SiC with small miscut angle ([1$\bar{1}$00]). (b) Height profile along the profile line in (a) indicating steps ~0.75 nm. (c) SEM image and Raman mapping of FWHM (2D-peak) illustrating scattered BLG inclusions. Surface topography S4 (d) and S5 (g) with PASG and SG growth, respectively, on two identical 6H-SiC samples with a high miscut angle of 0.37° ([11$\bar{2}$0]). The height profiles illustrate that the PASG method results in significant suppression of step bunching on S4 (e) while on the contrary wide terrace and giant steps appear on S5 (h). SEM ($E$=15 kV) image (f) shows that S4 contains noticeable BLG, which are scattered on the surface, as are seen in Raman mapping of its FWHM (2D-peak) (inset). SEM ($E$ = 15 kV) image of S5 (i) and Raman mapping of FWHM(2D-peak) (inset) indicates BLG formation mainly at step areas where the substantial step bunching during the growth lead to giant steps.

Interestingly, for the S1 with the sequential terrace-step pattern, the SEM imaging (**Figure 6.1g**) indicates an alternating contrast associated with the backscattered electrons (BSE). Any bilayer attribution is excluded since the Raman mapping of the FWHM of the 2D-peak shows pure single-layer graphene. This is different on S5, whereon the stripes in the SEM image are related to MLG and BLG coexistence, as seen in **Figure 6.2i**.

For the graphene on the 4H-SiC polytype, also very regular terraces and steps are observed, **Figure 6.1d and e**. The step height histogram in **Figure 6.1f** shows a different and somewhat wider height distribution than the 6H polytype. Although the majority (50%) of steps are ~0.5 nm high as before, a smaller percentage (20%) of ~0.25 nm steps and a higher proportion (25%) of ~0.75 nm steps are measured. Here, a small portion (~3%) of ~1 nm high steps is observed. Nevertheless, the high percentage (70%) of low steps with heights of ~0.25 and ~0.5 nm is remarkable and exceeds the results for conventional sublimation growth on 4H-SiC. [133,313]

The surprising BSE contrast in the SEM image of the 4H-SiC/G sample (S2) is also observable, but not in a sequential manner as for S1. Again, the Raman spectroscopy (**Figure 6.1h**) verifies a pure MLG coverage on S2. This SEM contrast, which is entangled with the SiC substrate termination, is studied in detail in Chapter **7**. However, from such BSE contrast in the SEM images, a modification of graphene resistance is plausible, which motivates local transport measurements by STP and will be discussed in section **6.5**.

The ultra-smooth graphene layers found on both SiC polytypes are a unique feature of the PASG technique. A second typical property of PASG graphene layers is the suppression of graphene bilayer formation, which can be regarded as a result of the very low SiC step heights ≤ ~0.75 nm in agreement with Raman mappings. [36,37] The observed formation of the ~0.25/~0.5 nm step-pairs on the 6H-SiC substrate is related to the specific surface-energy sequence of the SiC bilayer planes of the 6H polytype. This is demonstrated in a step-retraction model in detail in Chapter **7**, Appendix **A5**, and **A6**.

For the S2, the surface arrangements are not developed like sequential terraces on S1, since the 4H-SiC unit cell has only two distinct terrace energies per unit cell. [133,141] However, it is evident that an overall reduction of the step heights is achieved by the PASG technique compared to SG growth on 4H-SiC substrates. [133,313] This is more obvious considering three other samples, named S3, S4, and S5, grown under different conditions. The S3 is a result of the standard sublimation growth without the PASG method. Although under argon ambient with intermediate annealing, the growth was improved (compared to the growth in a vacuum) regarding the sample's morphology (**Figure 6.2a,b**) [16,39,234], still bilayer patches can be noticeably found throughout the sample, see SEM and Raman images in **Figure 6.2c**. Also, from the general morphology comparison between the samples, it is noticed that the direction of the miscut angle for S4 and S5 samples (towards [11$\bar{2}$0])) boost the step-bunching compared to S1-S3 samples (miscut direction [1$\bar{1}$00]). The morphological properties of the QFMLG and QFBLG samples (which were achieved through H-intercalation on optimized grown BFL and epi-MLG samples) can be found in Chapter **5** and ref. [38]. In the following, the resistance anisotropy of these samples will be analyzed.

## 6.4. Resistance anisotropy on μm-scales

The electronic properties of the graphene samples were investigated by angle-dependent nano four-point probe (N4PP) measurements in an Omicron UHV nanoprobe system. [314] The samples were kept in UHV at room temperature after a thermal cleaning procedure by heating up to 300 °C. The STM tips were placed in a square arrangement with 100 μm spacing, and electrical current was flowing between two adjacent tips while the voltage drop was measured between the two opposite ones, **Figure 6.3i**. From the ohmic *I-V*-curves, which were measured in the current range from −10 μA to +10 μA, the absolute resistance values $R$ were calculated. See section **3.9** for more detail. The N4PP measurements were carried out for different angles between the direction of the current probes and the step edges. The angles of 0° and 180° (90°) correspond to current flow parallel (perpendicular) to the steps, and $R_0$ denotes the averaged absolute resistance from the parallel (0° and 180°) measurements, see **Table 6-2**. The measured resistances $R_\theta$ for a given angle θ are adequately described by equation **(3-9)**, which includes the $\sigma_\parallel$ and $\sigma_\perp$ denoting the conductivities measured parallel and perpendicular to the step direction, respectively, assuming an anisotropic 2D sheet with different conductivities in x- and y-direction. [209] From the fitting procedure, finally, the resistivity values perpendicular ($\rho_{perp} = \sigma_\perp{}^{-1}$) and parallel ($\rho_{par} = \sigma_\parallel{}^{-1}$) to the step edges are obtained [208], and the anisotropy ratio is calculated as $A = \rho_{perp} / \rho_{par}$, see **Table 6-2**.

Since the current flow via the semi-insulating SiC substrate and the buffer-layer is negligible, the measured resistance is related to the 2D graphene sheet on top. For the applied rotational square method, it was shown that it is sensitive to both a possible intrinsic anisotropy of the graphene and additional superimposed effects (extrinsic anisotropy), e.g., step edges. [209] Due to the isotropic dispersion of the density of states near the Fermi level, an isotropic resistivity for graphene is expected. [6,286] Any measured anisotropy is, therefore, related to extrinsic effects.

**Figure 6.3a** shows the anisotropy related resistance contribution $(R-R_0)$ as a function of the rotation angle of all samples S1 to S5 and two free-standing samples (QFMLG and QFBLG). The calculated curves and the experimental data agree very well except for S5, where higher resistance values for angles > 110° are probably due to tip-induced defects. The $(R-R_0)$ curves in **Figure 6.3b** show for samples S3, S4, S5 QFMLG, and QFBLG an apparent maximum at an angle of 90°, which corresponds to transport perpendicular to the step edges. This indicates that step related sources are responsible for the extrinsic anisotropy in these epitaxial graphene layers. The resistance anisotropy increases to $A_{S3} = 1.17$, $A_{S4} = 1.79$ and $A_{S5} = 1.66$, $A_{QFMLG} = 1.2$, and $A_{QFBLG} = 1.42$, respectively. Thus, the values $A_{S1} = 1.03$ and $A_{S2} = 1.02$ of the PASG samples S1 and S2 can be regarded as practically isotropic, which verifies the assumption of intrinsic isotropy of the graphene monolayer. This also demonstrates that extrinsic effects can be reduced to a level where they practically play no role when advanced graphene growth procedures are applied as the presented PASG method on low miscut 4H- and 6H-SiC substrates. The N4PP measurements also show that the resistivity on the terraces is significantly reduced by the PASG method, which is demonstrated by the lower values of $R_0$ and $\rho_{par}$ for S1, S2, and S4 compared to the other samples. Hall measurements show that this is due to increased electron mobility, see refs. [36,37].

| Sample | N4PP | | | | STP | |
|---|---|---|---|---|---|---|
| | $R_0$ ($\Omega$) | $\rho_{par}$ ($\Omega$/sq) | $\rho_{perp}$ ($\Omega$/sq) | Anisotropy ratio | Anisotropy ratio | $\rho_{sheet}$ ($\Omega$/sq) |
| S1 | 68 | $629 \pm 1$ | $647 \pm 1$ | $1.03 \pm 0.002$ | $1.03 \pm 0.02$ | $570 \pm 20$ |
| S2 | 67 | $611 \pm 2$ | $620 \pm 2$ | $1.02 \pm 0.005$ | $1.04 \pm 0.02$ | $615 \pm 20$ |
| S3 | 184 | $1755 \pm 13$ | $2046 \pm 15$ | $1.17 \pm 0.01$ | | |
| S4 | 112 | $1339 \pm 39$ | $2397 \pm 56$ | $1.79 \pm 0.04$ | | |
| S5 | 202 | $2121 \pm 48$ | $3531 \pm 54$ | $1.66 \pm 0.03$ | | |
| QFMLG | 140 | $1339 \pm 38$ | $1609 \pm 42$ | $1.20 \pm 0.01$ | | |
| QFBLG | 52 | $546 \pm 1$ | $780 \pm 2$ | $1.42 \pm 0.03$ | | |

**Table 6-2. Results of resistance anisotropy measurements.**
Samples used in this study and the results from the angle-dependent nano four-point probe (N4PP) measurements and scanning tunneling potentiometry (STP).

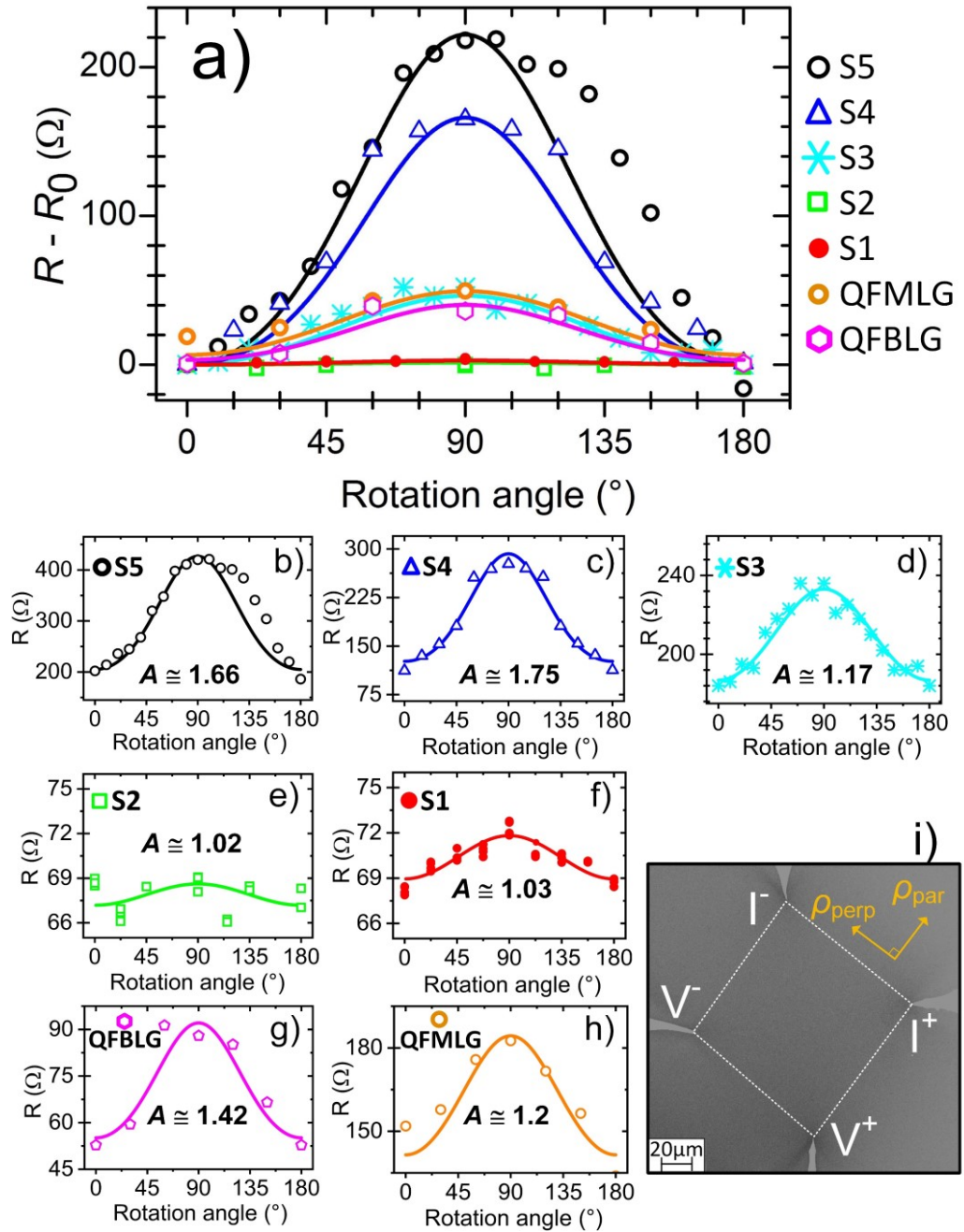

**Figure 6.3. Resistance anisotropy test using N4PP measurements.**

Resistance measurements by rotational four-point probe measurements ($100 \times 100$ μm²) as a function of rotation angle for five epitaxial graphene samples produced under different growth conditions as well as one QFMLG and one QFBLG sample, see **Table 6-1**. A rotation angle of 0° corresponds to transport parallel to the terraces and at 90° perpendicular to the step edges. The anisotropy values $A$ are given as calculated from fit curves of $\rho_{\text{perp}}$ / $\rho_{\text{par}}$. (a) Anisotropy related resistance contribution ($R$-$R_0$) as a function of the rotation angle of all five graphene samples S1 to S5, QFMLG, and QFBLG. The fitted curves (solid lines in a-h) are calculated using a model for anisotropic 2D sheets, as explained in the literature. [209] (b-h) Resistance variation as a function of the rotation angle for each sample. (i) Schematic diagram of the rotational N4PP method. The SEM image shows the STM tips on a graphene sample for the N4PP measurement at a rotation angle of 90°.

The rotational four-point probe measurements also highlight the homogeneity of the QFMLG and the QFBLG samples. The lowest resistance values for each sample are measured for transport parallel to the terraces, whereas the maximum value is obtained at an angle of about 90°, which corresponds to a current direction perpendicular to the step edges. [39,286] The calculated anisotropy values ($A = \rho_{perp} / \rho_{par}$) are 1.2 and 1.42 for QFMLG and QFBLG, respectively. Despite missing comparative values in the literature, these values are regarded as a sufficient low anisotropy indicating a good homogeneity. However, they are larger compared to the optimized ultra-smooth epitaxial monolayer graphene with nearly unity isotropy values ($A = 1.03$). [39] It is supposed that the observed anisotropy could stem from the intercalation related defects and local strain variation as, e.g., observed in the Raman spectrum of the QFMLG sample, see **Figure 5.5 a-d**. The low concentration of the intercalating gas agent (5% hydrogen) could have also intensified the defects and result in inhomogeneous or partial intercalation. [38,268,315]

**Figure 6.3b-h** shows the measured resistance $R$ as a function of the rotation angle for each sample in this study. For S1 and S2 epitaxial graphene samples, a very slight resistance increase of a few Ohm is observed at angles around 90°, which indicates that step related effects are noticeable also from these very flat surfaces. However, they are of minimal impact on the resistance anisotropy, which is expressed by the obtained very small values of $A_{S1} = 1.03$ and $A_{S2} = 1.02$. This is underlined by the comparison to anisotropies of about 1.7 for epitaxial graphene growth in vacuum using H-etched SiC substrates. [286] Comparing these results on these different types in the epitaxial graphene family allows us to have a better understanding of the contributors, e.g., the impact of the substrate, preparation, and bilayer inclusions in the extrinsic resistance anisotropy.

## 6.5. Resistance anisotropy on nm-scales

### 6.5.1. Local transport on steps

The assignment of the very small resistance anisotropies of the PASG samples S1 and S2 to step related effects is further investigated by scanning tunneling microscopy (STM) and scanning tunneling potentiometry (STP) measurements at room temperature, which gives an insight into the local sheet resistance and the defect resistance induced by substrate steps. [302,304,316] **Figure 6.4a** shows an example of a monolayer graphene sheet crossing a substrate step with a height of ~0.5 nm, which is located in the center (x = 0 nm) of the STM topography image taken in an area of 200 nm × 50 nm. The accompanied potential-jump is clearly visible at the same position in the simultaneously acquired potential map plotted in **Figure 6.4b**. **Figure 6.4c** shows the averaged potential across the flat graphene monolayer regions and the substrate step, from which the local electric field $E_{\text{sheet,x}}$ in the x-direction as well as the voltage drop $\Delta V$ caused by the step are deduced.

Using the macroscopic current density, an almost linear increase in resistances with step heights is found: $\varrho_1 = 4 \pm 2$ $\Omega\mu$m, $\varrho_2 = 10 \pm 2$ $\Omega\mu$m and $\varrho_3 = 13 \pm 2$ $\Omega\mu$m for monolayer graphene crossing a substrate step with heights of 0.25 nm, 0.5 nm or 0.75 nm, respectively, which is in good agreement with literature values. [302,304,316] The step resistance values are independent of the overall crystal morphology of the 4H- and 6H-SiC surface. The STP results can be compared with the N4PP measurements by setting the additional voltage drop at steps and their relative frequency $c_i$ [#/$\mu$m] in relation to the electric field $< E_{sheet} >$ on the terraces, accordingly, $A_{\text{STP}} = (< E_{sheet} > + \sum c_i < \Delta V_i >)/< E_{sheet} >$ resulting in an anisotropy of $1.03 \pm 0.02$ for S1 and $1.04 \pm 0.02$ for S2. The excellent agreement with the anisotropy value close to 1.0 from the N4PP measurements demonstrates that we reached the ultimate lower limit where the resistance contribution of the substrate steps is the sole cause for the measured anisotropy.

Two implications follow from the linear relation between the step height and the local defect resistance at the step. When using SiC substrates with the same miscut angle, a similar step related resistance anisotropy value is expected because step density and step height can commensurate with each other during the surface restructuring processes. A more significant anisotropy is expected for larger substrate miscut angles, which increase the number of steps, its height, or both. These conclusions are valid if only step related resistances in monolayer graphene are considered. Additional extrinsic effects can cause higher resistances and larger anisotropies. An important source for the resistance anisotropies of our samples S3-S5 is attributed to graphene bilayer domains. Local STP measurements have found that the electronic transition from monolayer to bilayer graphene results in an elevated resistance value, which approximately corresponds to that of monolayer graphene over a ~0.75 nm high SiC step. [188,302–304] Moreover, when the ML-BL transition is accompanied by a topographic height change, the resistance again drastically increases, e.g., by a factor of 4 at a 1 nm substrate step. [302]

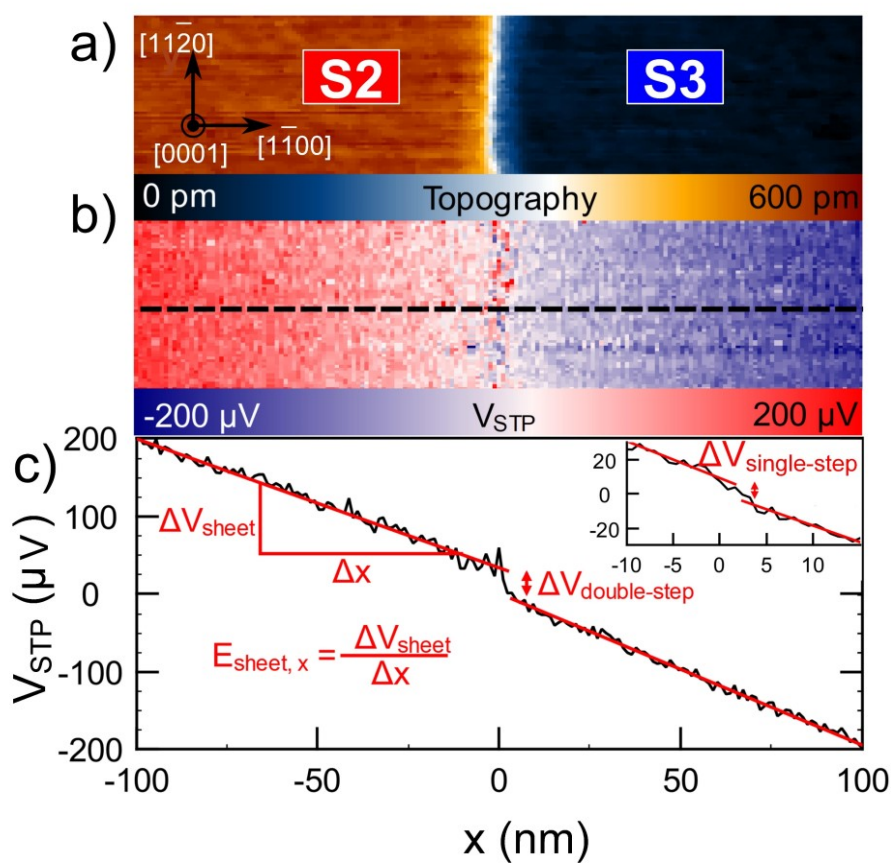

**Figure 6.4. Scanning tunneling potentiometry inspection of step-induced resistance in epitaxial graphene on SiC.**

(a) Constant current topography of monolayer graphene sheet with a ∼0.5 nm step in the center, (tunnel conditions: $I$ = 150 pA, $V_{bias}$ = 30 mV).

(b) The simultaneously acquired potential map with an average current density of $j$ = 3.6 A/m.

(c) The cross-section along the line in (b), averaged over all potential values perpendicular to the dashed line in (b). The local electric field component in x-direction $E_{sheet,x}$ is calculated from linear fits to the monolayer area (solid red lines in (c)). The step causes an additional, local voltage drop $\Delta V \approx 36\ \mu V$. The inset represents the equivalent situation of a monolayer graphene sheet covering a single SiC-bilayer substrate step with a height of ∼0.25 nm.

(d) Schematic of the STP experiment setup, see Chapter **3** and ref.[39,188,291] for more details.

These bilayers related local resistance increase can lead to a macroscopic resistance directional dependency, according to the shape and distribution of the bilayer inclusions. Since the bilayer inclusions are not symmetric but show an elongated shape at terraces and are very often embodied as bilayer stripes along the terrace step edges, their presence results in a higher resistance for current flow perpendicular to the terrace step edges compared to current flow on the terraces parallel to the step edges.

For the graphene sample S3, a larger anisotropy ($A_{S3} = 1.17$) was obtained compared to S1 and S2 ($A_{S1, S2}$ ~1), although all were grown on low miscut substrates. As discussed above, this discrepancy is not clear if only step related contributions are considered. The additional resistance anisotropy is attributed to the scattered, micrometer-sized, asymmetric bilayer spots, which are located mainly at the terraces edges of the sample S3, see **Figure 6.1a,b** in ref. [234]. This comparison clearly shows that the nearly vanishing resistance anisotropy of the PASG samples S1 and S2 is related to the absence of bilayer graphene.

The significantly increased resistance anisotropy of the samples S4 and S5 compared to S1, S2, and S3 is expected because of the 6-times larger SiC miscut angle. Under the assumption of a linear correlation between step-height and step resistance, according to the abovementioned STP anisotropy equation, one can estimate for pure monolayer graphene an anisotropy of $A_{STP} \approx 1.2$. The measured anisotropy values of $A_{S4} = 1.79$ and $A_{S5} = 1.66$ are much higher and are attributed again to bilayer graphene on the terraces. Both samples show more significant bilayer coverages compared to S3, and by taking into account the much higher step concentration in S4 and the giant step edges in S5, respectively, this should drastically increase $\rho_{perp}$ and the anisotropy. On the other hand, transport along the terraces can vary, e.g., caused by local planar ML-BL transitions. This is reflected by the higher $\rho_{par}$ value of S5 compared to S4, which results in a smaller anisotropy value, $A_{S5} < A_{S4}$, although $\rho_{perp}$ of S5 shows the highest value of all samples. This is probably due to the very high terrace steps in S5, which cause extensive graphene bilayer stripes along the upper side of the step edges.

### 6.5.2. Local transport on terraces

The BSE contrast observed in the SEM images of S1 and S2 samples in **Figure 6.1c, g,** brings to the mind a possible variation of the electronic properties, i.e., the resistance of graphene correlated with underneath the SiC terraces. This contrast, which first might naively be regarded as typical material contrast (e.g., BLG, BFL, or SiC) as shown for instance, in samples S3, S4, and S5 in **Figure 6.2c, f, and i,** was verified to be appearing in pure MLG samples using Raman spectroscopy in **Figure 6.1g, h (inset)**. To explore the origin(s) of this effect, which was addressed first in ref. [37], various characterization techniques are employed. Here the transport properties in the 6H-SiC/G sample (like S1 in **Table 6-1**) using the STP measurement are concisely discussed.

**Figure 6.5. Scanning tunneling potentiometry investigation on local sheet resistance at terrace regions at room temperature.**

Using the macroscopic ohmic resistance, step resistances of $\sim$6 $\Omega\mu$m, $\sim$12 $\Omega\mu$m, $\sim$18 $\Omega\mu$m for single, double, and triple steps, respectively. This verifies the previous measurements (within the uncertainty range) in **Figure 6.4**.

Measured voltage drop along the graphene layer induced by the cross voltage $V_{cross}$ when crossing a single step (a), double step (b), and triple-step (c). Dashed lines represent the slope of the voltage drop (shifted for clarity). The figure was edited from ref. [291].

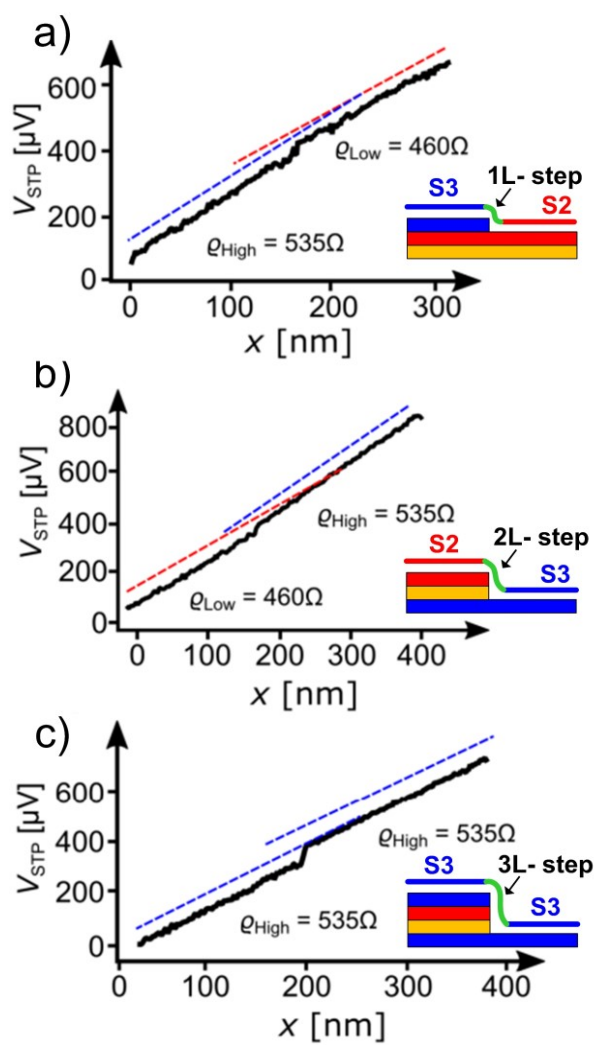

The STP measurement was carried out on terraces with different step heights, i.e., one Si-C bilayer heights (1L-step), 2L, or 3L, as shown in **Figure 6.5a-c**. As already shown in **Figure 6.4**, for each terrace to terrace transport in **Figure 6.5**, a step-resistance is seen too, which increases almost linearly corresponding to the terrace-step height. However, the interesting point is regarding the variation of sheet resistance ($R_{sheet}$) of graphene on non-identical terraces (i.e., S2, and S3).

The identification of SiC surface terminations beneath the graphene (and BFL) is discussed in detail within a framework of a step-flow model in Chapter **7** and ref. [78]. So far, let us admit that through the step restructuring of the SiC during the growth S1/S1* terminations disappear from the surface, and the remaining S2/S2* and S3/S3* terminations shape the surfaces below the atop carbon layers. These SiC surface configurations below the graphene are schematically shown in the inset of each measurement in **Figure 6.5a-c**. For these two terraces, two distinct sheet resistances were extracted from the STP measurements, which are referred to as $\varrho_{High}$ and $\varrho_{Low}$. These resistance variations verify the expected idea of electronic properties variation from the BSE contrast in SEM images (**Figure 6.1g and h**). [37] The mean $\varrho_{High}$ = 535$\Omega$ and the mean $\varrho_{Low}$ = 460$\Omega$. The mean $\varrho_{High}$ and $\varrho_{Low}$ deviate by (14 $\pm$ 1)% from each other at room temperature. [291] Notably, examining the local sheet resistance at low temperatures revealed a significant variation of up to 270%. The resistance variation measured by STP was attributed to the distance change of the graphene on the terraces, i.e., the larger distance results in reduced sheet resistance. [291] However, whether the distance of graphene on either S2 or S3 is always larger or smaller remains yet unanswered [291] and requires future investigation.

It is worthwhile to mention that from the STP sheet resistance output [291], a mobility value of about ~2,000 cm²/Vs (at $T$ = 8K) is extracted, which one may interpret it as a low-quality graphene sample. However, it should be noticed that graphene is extremely sensitive to environmental influences. [317] For the STP study, the samples were exposed to air (despite a UHV degassing process- 400 °C, 30 min) and were seen stencil lithography process, which indeed their invasive impacts on graphene's transport properties cannot and should not be excluded. In addition to the informative STP results, the origin of this resistance variation and BSE contrasts in SEM images and AFM phase-contrasts [37,38] of the graphene layer will be further discussed in Chapter **7**.

## 6.6. Resistance anisotropy in mm-scales

The microscopic anisotropy test is further investigated using macroscopic VdP measurements in a helium flow cryostat in a magnetic field up to 250 mT. For the measurement, the samples were first to cut into square-shape of 5 mm × 5 mm, and the graphene on top of the sample was isolated from the graphene on the side and the back of the substrate by scribing cut-grooves on each side close to the edge of the sample (~0.1 mm from the edge), as shown in the schematic of the VdP configuration in **Figure 6.6**. For more details about the VdP measurement, see section **3.10**. The samples were cleaned using isopropanol and acetone in an ultrasonic bath for 10 minutes before the VdP measurement. Four gold pins in the square configuration were pressed firmly onto the surface to contact the graphene at the corners. The ohmic characteristic and the linearity of the Hall ramps were tested before the measurements. The VdP measurements were carried out at room temperature (295 K) and 2.2 K, see **Table 6-3**.

As expected, both QFMLG and QFBLG samples show the typical high $p$-type carrier density of about 6.7 × 10¹² and 6.3 × 10¹² cm⁻², respectively. [18] For the QFMLG, we obtain mobility of ~1160 cm²/Vs at room temperature and for $n$- type epitaxial graphene (like S1) ~1110 cm²/Vs. However, at 2.2 K, the epitaxial graphene shows considerably higher mobility of 2459 cm²/Vs, whereas the mobility of QFMLG remained almost constant at $\mu \approx 1170$ cm²/Vs.

This is due to different scattering mechanisms in these two types of epitaxial graphene monolayers. While the temperature dependence of mobility in the epitaxial graphene (S1) is due to longitudinal acoustic phonons in graphene and SiC, the dominant scattering in QFMLG is attributed to Coulomb scattering induced by charged impurities. [229,318] Although higher values were already reported for micrometer-sized QFMLG confined Hall bars on the single terrace [108,119,229,231,319], this is still a remarkable result concerning the low concentration of hydrogen and the produced large-size sample. In contrast to QFMLG, the measurements on QFBLG (after intercalating a sample like S1) revealed noticeable higher mobility values of ~2700 cm²/Vs (RT) and ~3350 cm²/Vs (2.2 K). This temperature dependence of the carrier mobility in QFBLG is referred to as an interplay of different scattering mechanisms, temperature-dependent Coulomb scattering, and the charge impurity density. Note that the higher mobility compared to QFMLG could stem from the screening of Coulomb scatterers in the substrate and bilayer graphene. [229,320,321]

It is worth mentioning that investigation of QFMLG and QFBLG was carried out on a considerable number of sample sets showing much higher mobilities of ~1300 cm²/Vs, ~2000 cm²/Vs, and ~3300 cm²/Vs for QFMLG, epigraphene, and QFBLG at room temperature, respectively. However, for consistency, only the results of the samples with similar treatment and the same growth and intercalation processes are presented.

The homogeneity of the samples can also be derived from the VdP sheet resistances $R_{par}$ and $R_{perp}$ measured in two orthogonal directions, parallel and perpendicular to the step edges. As before, higher resistance values were obtained for perpendicular transport. The QFMLG and QFBLG samples show anisotropy values ($R_{perp}$ / $R_{par}$) values of 1.18 and 1.37 (at room temperature), which are in excellent agreement with the microscopic four-point probe measurements. This result indicates the very good electronic homogeneity over mm scales of our QFMLG and QFBLG samples in excellent agreement with the AFM and Raman data. For QFBLG, a comparable VdP study [233] revealed a much stronger anisotropy of about 200%, which was attributed to high step edges (~10 nm) and multilayer graphene along the step edges. The absence of multilayer graphene in the produced QFMLG and QFBLG and the low step heights can thus be regarded as highly beneficial for homogeneous electronic properties. This result gives evidence of the high quality of the PASG method and the possibility of optimization of the growth parameters.

**Figure 6.6. Sketch of the VdP configuration.**

The graphene (green) on top of the sample ($5 \times 5$ mm²) was isolated from graphene on sidewalls and backside by cut-grooves close to each edge (indicated as red lines). See section **3.10** for more detail.

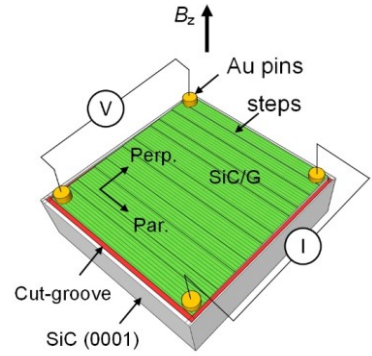

| sample | $T$ (K) | $R_{sheet}$ ($\Omega$sq) | $R_{par}$ ($\Omega$) | $R_{perp}$ ($\Omega$) | $A = \frac{R_{perp}}{R_{par}}$ | $p, n$ ($10^{12}$ cm$^{-2}$) | $\mu$ (cm²/Vs) |
|---|---|---|---|---|---|---|---|
| QFMLG | 295 | 812 | 164 | 194 | 1.18 | $p$ = 6.7 | 1160 |
|  | 2.2 | 837 | 174 | 195 | 1.16 | $p$ = 6.4 | 1170 |
| epi-MLG | 295 | 979 | 209 | 219 | 1.05 | $n$ = 6.8 | 1110 |
|  | 2.2 | 365 | 79 | 82 | 1.04 | $n$ = 6.9 | 2460 |
| QFBLG | 295 | 364 | 68 | 93 | 1.37 | $p$ = 6.3 | 2670 |
|  | 2.2 | 345 | 64 | 90 | 1.4 | $p$ = 5.4 | 3350 |

**Table 6-3. Millimeter-scale Van der Pauw measurement results.**

The results of VdP measurements on 5 mm × 5 mm large samples of QFMLG, epitaxial graphene, and QFBLG. (see **Figure 6.6** and ref. [38]).

## 6.7. Conclusion

In summary, a comprehensive resistance anisotropy study in multiscale nm, μm, and mm geometries were performed. Various samples, including epitaxial graphene grown by different sample preparation and growth methods on 4H- and 6H-SiC(0001) substrates with small and large miscut angles as well as QFMLG/QFBLG samples by H-intercalation were grown and investigated. In agreement with STP measurements, the rotational square probe measurements reveal very small resistance anisotropies of ~3% for graphene layers grown by PASG on SiC substrates with a small miscut angle. This anisotropy value is traced back to the step resistances of the monolayer graphene across the SiC steps measured by STP on the nano-scale. The main reason for the vanishing small resistance anisotropy was identified to be the absence of bilayer domains, while the specific step resistances are similar to other graphene.

The PASG and fine growth optimization methods allow the uniform fabrication of ultra-smooth graphene with most of the terrace step edges being ~0.5 nm or lower, which prevents the formation of graphene bilayer domains. In particular, on the 6H-SiC substrate, a very high percentage of 90% is achieved with a typical pattern of alternating steps of ~0.25 nm and ~0.5 nm in height, which is related to the SiC layer sequence in this polytype. This study shows that graphene growth using the PASG method and fine-tuning of the growth parameters bears the potential to reduce the terrace step heights down to an ultimate level of a single Si-C bilayer. Since SiC substrate steps cannot be entirely avoided, it is impossible to achieve perfect resistance isotropy for epitaxial graphene. However, for the produced bilayer-free graphene on ultra-low terraces negligible small deviations from isotropy can be obtained.

The QFMLG and QFBLG produced by hydrogen intercalation exhibit excellent homogeneity and very small resistance anisotropy over areas in the millimeter range. This indicates the presence of coherent quasi-free-standing graphene layers over large areas and remarks on the significance of the optimizations discussed in Chapter **5**.

Surprisingly, the SEM inspections showed a BSE contrast in pure monolayer graphene on SiC (non-identical) terraces, which give rise to a plausible electronic property difference on these terraces. This was verified by the STP measurements revealing a resistance variation of graphene on the substrate terraces. The investigation of this effect and possible explanations will be discussed in further detail in Chapter **7**.

This study supports the promising application potential of epigraphene on SiC for quantum Hall metrology applications as well as QFMLG and QFBLG for superior transistor performances and extends the capability of epigraphene or the buffer layer to be implemented as a platform for growing other 2D materials or metamaterials. This study also highlights the importance of bilayer-free graphene growth for all kinds of epitaxial growth techniques whenever isotropic properties are demanded for perfect device performance. It makes the device orientation independent of step direction and improves the freedom to design device layouts, thereby promoting the potential for future device applications of epitaxial graphene. Finally, the applied optimization enabled us to synthesize graphene on non-identical SiC terraces. Thanks to this, the mesoscopic interaction of bottom SiC terminations and top carbon layers could systematically be investigated.

# 7. SiC stacking-order-induced doping variation in epitaxial graphene

**Abstract**

*T*he excellent quality of the fabricated graphene samples enables us to study at sub-nanometric scales the interaction between substrate and atop carbon layers. Thanks to the advanced growth control, the graphene can be fabricated on non-identical SiC terminations, whereon the "proximity effect" significantly changes, in contrast to general assumptions. Accordingly, the graphene's work function depends on its exact position on the underlying SiC termination. This is attributed to the spontaneous polarization doping of the hexagonal SiC substrate. The effect is elaborated using several characterization analyses. It is experimentally shown for the first time that the hexagonal SiC polarization doping (usually known as a bulk concomitant) is also surface dependent. This agrees with theoretical predictions. Thereby, a sequential doping variation is observed in top C-layers on the pairwise terraces-steps of 6H-SiC. This finding opens a new approach of nano-scale doping-engineering based upon the substrate properties, not merely in hexagonal SiC but also in other dielectric polar crystals. The main results are also published in ref. [78].*

## 7.1. Introduction

Epitaxial graphene growth provides wafer-scale graphene fabrication for electronic devices for a wide range of potential applications. [11,295,296,322] The most commonly used SiC substrates are the hexagonal 4H and 6H polytypes, which exhibit a spontaneous polarization induced by the hexagonal stacking sequences in the SiC unit cell. (see section **2.4.4** for more detail) The spontaneous polarization of the hexagonal SiC polytypes leads to a phenomenon called polarization doping, i.e., a *p*-type doping of the order of 6-9 $\times$ 10$^{12}$ cm$^{-2}$ in so-called quasi free-standing graphene on hydrogen saturated SiC(0001). [30,31] On the other hand, epitaxial graphene residing on the buffer layer shows an *n*-type conductivity with a charge carrier density of the order of 10$^{13}$ cm$^{-2}$. This is attributed to an overcompensation of the polarization doping by electron transfer from a donor-like buffer layer and interface states to the graphene layer. In general, the graphene properties on the SiC terraces are assumed to be uniform. These terraces, considering the stacking order in the unit cell of hexagonal SiC, result in inequivalent surface "terminations." Given that, some doubt is raised because theoretical investigations estimate a certain dependence of the doping on the stacking sequence and surface terminations of 4H- and 6H-SiC. [142,161–164] Moreover, a very recent nano-scale transport study reported a 270% variation of the sheet resistivity for epitaxial monolayer graphene on two different terminated 6H-SiC terraces. [291]

The polarization doping, as a doping effect without using any impurity dopants, has the potential to be used for engineering localized electric field not merely in epigraphene but also other dielectric materials, e.g., pyroelectric wurtzitic III-nitrides. This also can be an excellent platform for other experimental and theoretical studies of defects in SiC (e.g., vacancy or divacancy) [323–331]. In this work, it is experimentally shown that the stacking terminations of the hexagonal SiC substrate have a biasing effect on the surface potential and the doping level of the overlying epitaxial monolayer graphene. To this end, using the so-called polymer-assisted sublimation growth (PASG) technique, monolayer graphene on identical and non-identical SiC terraces is fabricated. In this chapter, various surface-sensitive measurement techniques, namely AFM, STM, LEEM, KPFM, and XPEEM, indicate different electronic properties of graphene on the inequivalent SiC surface terraces types in association with their cubic and hexagonality nature. These SiC terraces can be clearly assigned to the specific stacking terminations within the framework of an extended SiC step retraction model. This model includes a joint hexagonality and cubicity considering the position of each single atom within the SiC unit cell.

## 7.2. Graphene and buffer layers on identical/non-identical SiC terminations

The epitaxial buffer and graphene layers in this work were grown on semi-insulating 6H-SiC samples with a nominal miscut of about −0.06° toward [1$\bar{1}$00] (from II−VI Inc.). Epitaxial growth was carried out in a horizontal inductively heated furnace. [213] The buffer layer and graphene samples were grown by the PASG technique in an argon atmosphere (∼900 mbar) at 1400 °C and 1750 °C, respectively. [36,38,39] The control of the surface morphology was attained by taking into account the influence of Ar flux during the sublimation growth. [38] The details of growth and optimization can be found in Chapter **4** and refs. [36–39].

AFM images in **Figure 7.1 (a-d) and (g-j)** show the epitaxial monolayer graphene and buffer layer, respectively, with two types of surface morphologies. The origin of the different surface morphology will be explained in the next section. The AFM topography images of the graphene samples in **Figure 7.1a and b** show that one sample exhibits regular ∼0.75 nm step heights while the other one displays a step pattern consisting of alternating ∼0.25 nm and ∼0.5 nm high steps. The phase images of these samples in **Figure 7.1c and d** show a very interesting behavior. For the ∼0.75 nm stepped surfaces, the same phase is observed on all terraces. Only step edges appear as narrow regions with increased phase. On the other hand, ∼0.25 nm/∼0.5 nm stepped surface clearly shows an alternating phase (**Figure 7.1d**), which changes from one terrace to the next. As for the other sample, step edges appear as narrow regions with increased phase. The observation of two different phase values indicates different material properties of the graphene layer on neighboring SiC terraces.

This phase contrast is not caused by a different number of graphene layers. The integrated Raman spectra (areal scan over  20 × 20 μm$^2$ ) in **Figure 7.1e and f** reveal a similar spectrum for both samples and a typical 2D-peak (at ∼2724 cm$^{-1}$ and full-width-half-maximum of about 33 cm$^{-1}$) which proves that both samples are uniformly covered with monolayer graphene. [38,274] Similar contrast can also be seen in the scanning electron microscopy of the sample with sequential terrace-steps, see Chapter **6**.

A very similar result is deduced from the AFM investigation of the buffer layer samples (**Figure 7.1g-j**). Again, the regular ∼0.75 nm stepped SiC terraces (see **Figure 7.1i**) show no phase contrast, and only for the binary ∼0.25 nm/∼0.5 nm stepped terraces an alternating AFM phase contrast is observed (see **Figure 7.1j**). The integrated Raman spectra in **Figure 7.1k and l**) show a broad buffer layer related vibrational band, which indicates a homogenous buffer layer coverage. [244]

The agreement of the phase contrast of graphene and the buffer layer samples for the same substrate step structure clearly indicates a substrate-related effect modifying the properties of the overlying layer, whether it be the epitaxial graphene (which includes the buffer layer) or the buffer layer alone. An intrinsic effect of the graphene layer itself thus can be ruled out. In AFM experiments, the phase image contrast arises from local variations of the energy dissipation in the tip-surface interaction, which results in damping and shift of the tip's oscillation frequency giving information about the chemical/mechanical/electrical heterogeneity of a surface. [332] Thus, the observed phase contrast on non-identical terraces is associated with a change of the surface properties originating from the different SiC stacking terminations of the corresponding terraces below. Therefore, in the following, the formation and nature of the SiC surface terraces are explored.

**Figure 7.1. Epitaxial monolayer graphene and buffer layer on 6H-SiC with identical/non-identical terraces.**

Atomic force microscopy (AFM) images of (a- d) epitaxial monolayer graphene and (g-j) buffer layer on 6H-SiC substrates with different terrace step heights of ∼0.75 nm and sequential pairs of ∼0.25 nm/ ∼0.5 nm, respectively.

The cross-sections in the inset of (a, b, g, h) are taken along the blue line. (c, i) The phase images of the homogeneously stepped (∼0.75 nm) samples show no phase contrast except for an increased phase at the step edges. (d, j) Only the phase images of the samples with step pairs (∼0.25/∼0.5 nm) show a sequential contrast on the terraces.

(e, f, k, i) Raman spectrum of each sample. The displayed spectra were integrated over 14000 single spectra from an area of 20 × 20 μm². (e, f) The narrow 2D line widths of around 30 cm⁻¹ indicate that the graphene sample is thoroughly covered with monolayer graphene. (k, l) The broad vibrational DOS distribution indicates the existence of buffer layer graphene.

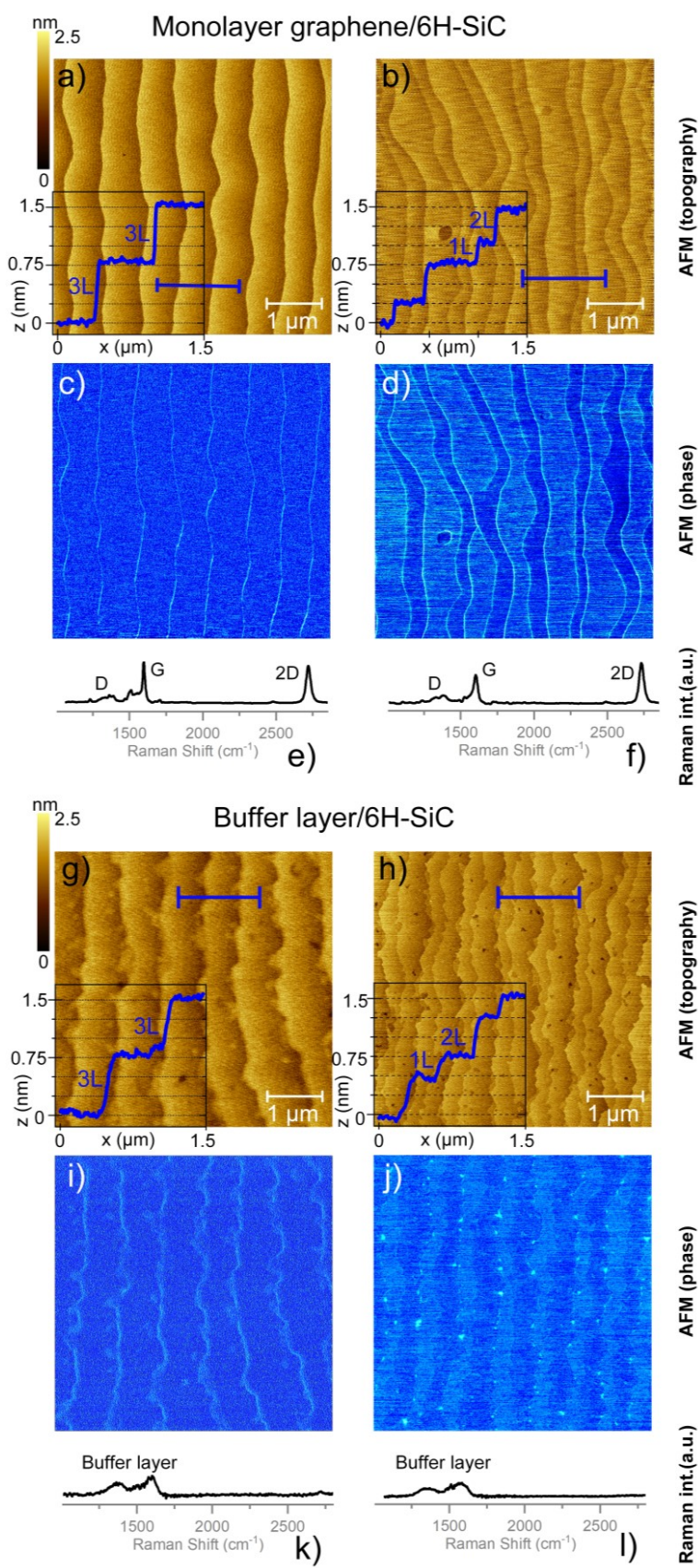

## 7.3. Step retraction model of 6H-SiC/G

To further study the interaction of individual SiC surfaces and the graphene layers, as it was initially inferred from the AFM investigations, it is required to identify and map the substrate terminations. The creation of the SiC terraces during graphene can be understood in the framework of the SiC step retraction model applied to the 6H-SiC(0001) substrate. **Figure 7.2a** illustrates the unit cell of 6H-SiC consisting of 6 Si−C bilayers called A, B, C, A, C, B (from bottom to top). For this investigation, it is more instructive to focus on the six resulting Si-terminated surface-terraces which differ in three inequivalent stacking sequences of the underlying Si-C bilayers, i.e., S1, S2, and S3, and another three stacking sequences, S1*, S2*, S3*, which are equivalent to the first ones but rotated by 60°, see **Figure 7.2b**. [140] The number gives the number of SiC bilayers between the surface and the first hexagonal stacking arrangement. For simplicity, they are named S1, S2, and S3 if the rotation can be neglected. The eclipsed and staggered orientation of subsequent Si−C tetrahedra are called hexagonal (*h*) and cubic (*k*), respectively, which leads to discrete *h, k,* and *k* stacking orders of A, B, and C. In a more detailed model, one can assign to each atomic layer a hexagonal or cubic orientation, which is sketched in **Figure 7.2a**. [326] For the on-bonds in axial configuration ([0001] direction), this results in *hh, kk,* and *kk* stacking for the layers A, B, and C, respectively. The off-bonds (basal configuration) are described by *kk* (for S2) but also by mixed *hk* (for S3) and *kh* (for S1) stacking orders. The hexagonality of each surface terrace can be considered as the joint cubicity-hexagonality of the corresponding on- and off-bonds. Hereafter, this step-retraction approach is referred to as a so-called joint cubicity-hexagonality (JCH) approach. For the polytype of interest here (6H-SiC), the layers are stacked in cubic (*k*) (four layers) and hexagonal (*h*) (two layers) order. The position of each atom is depicted in **Figure 7.2a**. According to the JCH approach, the S2 in both on- and off-bonds has a cubic nature (100% *k*), while S3 and S1 have ~66% *k* (~33% *h*) and ~33% *k* (~66% *h*) nature, respectively.

As discussed in section **2.4.2**, the step-bunching mechanism, e.g., during graphitization or hydrogen etching of the SiC surfaces, has been interpreted in two main step-flow models [137,138]. Both models agree that the hexagonal stacking layers (S1/S1*) are energetically less stable than the cubic layers. However, they seem to disagree on the second and third decomposition velocities for (S2/S2*) and (S3/S3*) layers in 6H-SiC (see **Figure 2.7**). This discrepancy is explained in this study by the JCH step-flow model, which is based on the experimental results that include the step-height and terrace-width analysis. The presented step-retraction model considers the position of the atoms in each termination within a joint hexagonality in on- and off-bonds.

Before the growth, the terrace structures have single Si-C steps (1L) of ~0.25 nm high and can be regarded as the initial surface of the used low-miscut angle substrates, as sketched in **Figure 7.2b and c**. During graphene growth at high temperatures, a restructuring of the SiC surface takes place, which can be described by the retraction of individual Si–C bilayers with different velocity in a step retraction model. [133,137,138] The step retraction is driven by the minimization of the surface energies, which depend on the surface hexagonality of the individual terraces. [133,162] This step retraction velocity is indicated by the length of the horizontal arrows in **Figure 7.2b**.

The high retraction velocity of the S1 surface can be attributed to the strongest hexagonality (*hh* on-bond) of this layer in agreement with refs. [137,138]. Thus, the corresponding S1 surface disappears at first. S2 and S3 remain, which results in periodically stepped surfaces with ~0.25 nm (1L) and ~0.5 nm (2L) step-heights, as observed in the AFM image in **Figure 7.1b and h**. This situation is sketched as an intermediate state in **Figure 7.2c**. From this model, we can clearly assign the S3 terrace being above a 1L step and an S2 terrace above a 2L step. It is further assumed that S2 is the most stable surface because of its least hexagonality (*kk* on-bond and *kk* off-bond) and, therefore, the width of the S3 terrace (*kk* on-bond but *hk* off-bond) is decreasing faster than S2, which agrees with the model in ref. [133,137] but not ref. [138]. The width of the initially wider S3 terrace (see buffer layer AFM image in **Figure 7.1h** and ref. [38]) decreases, and for advanced step retraction, the S3 terraces become narrower than S2. This situation is found for all graphene samples, see **Figure 7.1b**, Appendix **A5**, and in refs. [36,38,39]. (Such a pattern could not be primarily formed if S2 would retract faster than S3.) This can be easily examined. For instance, a 2L step following an S3 termination in the downstream direction would lead to an S1 termination. This can be excluded since S1 has the lowest stability. Finally, only the most stable S2 terraces remain with step heights of ~0.75 nm (3L), as sketched as the final state in **Figure 7.2c**. This situation is observed in the AFM images in **Figure 7.1a and g**.

The preparation of a ~0.25/~0.5 nm stepped surface with alternating S2 and S3 terraces of nearly 100% efficiency is a specific advantage of the PASG method. [39] The rapid formation of the buffer layer by the additional polymer-related carbon supply stabilizes the SiC surface and reduces the step bunching velocity. As a result, the terrace structure can be "frozen in" at the intermediate state with S2 and S3 terraces before the final state with S2 surfaces only is reached, leading to highly isotropic graphene transport properties as discussed in Chapter **6**. [36–39] During all the surface restructuring, the additional carbon from polymer and growth optimization provides a uniform growth and effectively minimize the step-bunching.

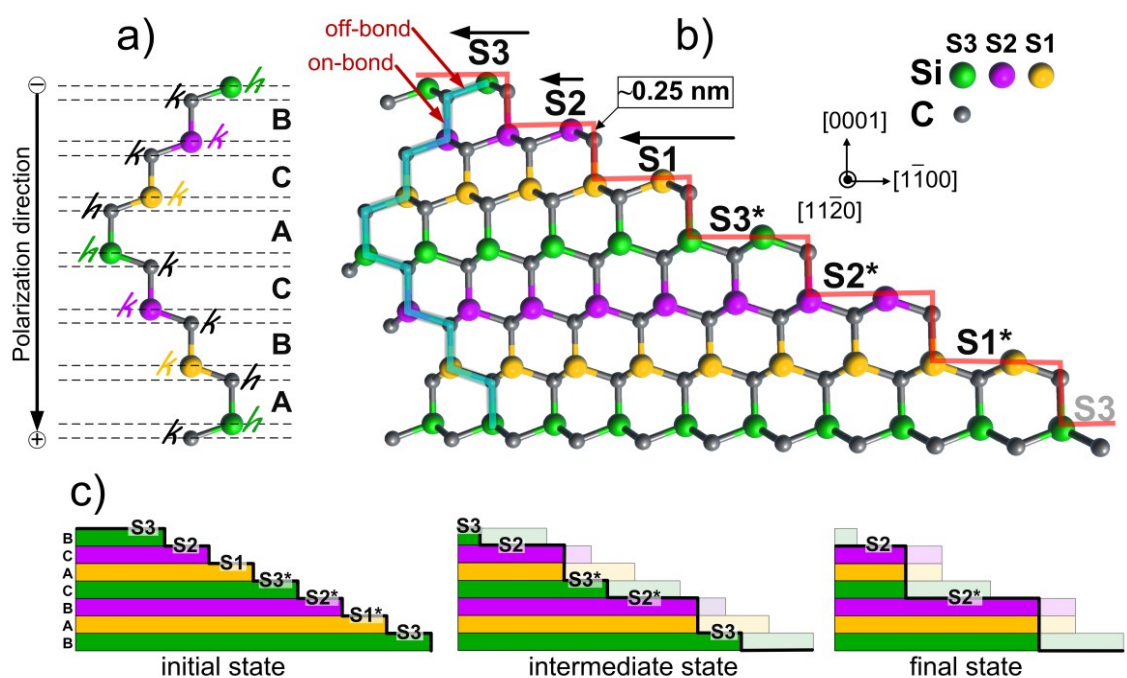

**Figure 7.2. Structural model of the 6H-SiC (0001) substrate and schematics of the corresponding step patterns in the joint-cubic-hexagonal (JCH) step retraction model.**

(a) Layer sequence of the Si-C bilayers in the 6H unit cell denoted as BCACBA. For the 12 atoms in the unit cell of 6H polytype, 3 Si and 3C atoms are non-equivalent regarding their positions hexagonal ($h$) or cubic ($k$). The position of each atom is shown in the unit cell. The hexagonal layer A is characterized by ($hh$) on-bonds. B and C are either completely cubic ($kk$) or one-half ($hk$, or $kh$), depending on the position of Si and C atoms in the neighboring off-bonds.

(b) Schematic side-view of 6H−SiC (0001) projected in (11$\bar{2}$0) plane with the six possible surface terminations S3, S2, S1, S3*, S2*, S1*. The terrace widths are strongly reduced. A terrace width of ~240 nm is estimated for a miscut angle of 0.06° consistent with experimental results. The surfaces Sn and Sn* (n = 1-3) are energetically similar but rotated by 60° related to each other. For simplicity, they are treated as similar if the rotation is neglectable. The arrows mark the different retraction velocities of the Si-C layers in the step retraction model, which are related to the individual surface energy and the surface hexagonality.

(c) Basic terrace step patterns of the 6H-SiC surface developing in the step retraction model. In the initial state, the individual S1, S2, and S3 terraces are separated by single SiC monolayer steps (1L) of ~0.25 nm in height. In the intermediate state, the S1 surface with the fastest retraction velocity has disappeared, and an alternating sequence of S2 and S3 surfaces remain with steps of ~0.25 nm (1L) above S2 and ~0.5 nm (2L) above S3. After advanced step retraction S2 becomes wider than the S3 terrace since the step velocity of S3 is faster than S2, which is depicted here. In the final state, the most stable S2 terraces remain with ~0.75 nm (3ML) in height.

## 7.4. Identification of SiC terminations below the graphene

In addition to the AFM phase images on the buffer and graphene layers (in **Figure 7.1**), a systematic LEEM measurement is also performed on a monolayer 6H-SiC/G sample similar to **Figure 7.1b**. In LEEM, the elastically backscattered low-energy electron beams are imaged, revealing information about the electronic and structural properties of the sample (see section **3.8**). LEEM is a powerful tool to study SiC/G, providing both insights into the local graphene coverage and thickness using reflectivity spectra [23,197] as well as giving information about the local stacking order through dark field measurement. [198–200]

Bright-field (BF) LEEM images are mainly governed by the reflectivity contrast, which is related to the density of states for wave vectors perpendicular to the surface. A bright-field (BF) LEEM image of the ~0.25/~0.5 nm (1L/2 L) stepped graphene sample is displayed in **Figure 7.3a**.

In the upper part of this image, a regular contrast pattern of alternating narrower brighter and wider darker stripes is observed. From the similarity to the AFM pattern **Figure 7.1b** (and scanning electron microscopy **Figure 6.1g, and h**), the BF-LEEM stripes are identified as the terraces S2 (wide) and S3 (narrow). From the correlation between terrace width and step height (narrower S3 being above a 1L and wider S2 being above 2L step), we can ascribe the corresponding step-heights to the boundaries between areas of different brightness, which is visualized by the top blue profile in **Figure 7.3a**. The underlying regular SiC step structure is sketched in part (i) of **Figure 7.4c** (For clarity, the covalently bonded buffer layer and the overlying graphene layer are left out in this sketch). The bright-field image depicts the reflected intensity of the $0^{th}$ order low-energy electron diffraction (LEED) spot. The usual attribution of the BF contrast to a graphene thickness variation [68,199] is not valid here since the sample is unambiguously covered only with monolayer graphene. As will be discussed further below, the different brightness in the BF-LEEM images can be related to a small variation of the surface potential, which is induced by the stacking of the SiC crystal underneath.

It is worthwhile to mention that the SiC step edges can be discerned in the LEEM images (See **Figure 7.3a**) as dark lines on both sides of the terraces, which is due to interference effects at step edge areas. [23] Also, note that exclusively steps with a height of 3L or ~0.75 nm do not result in a change of the bright field contrast, likely due to similar energies. These findings become more evident when we compare the bright field image with the dark field image of this sample area (**Figure 7.3b**).

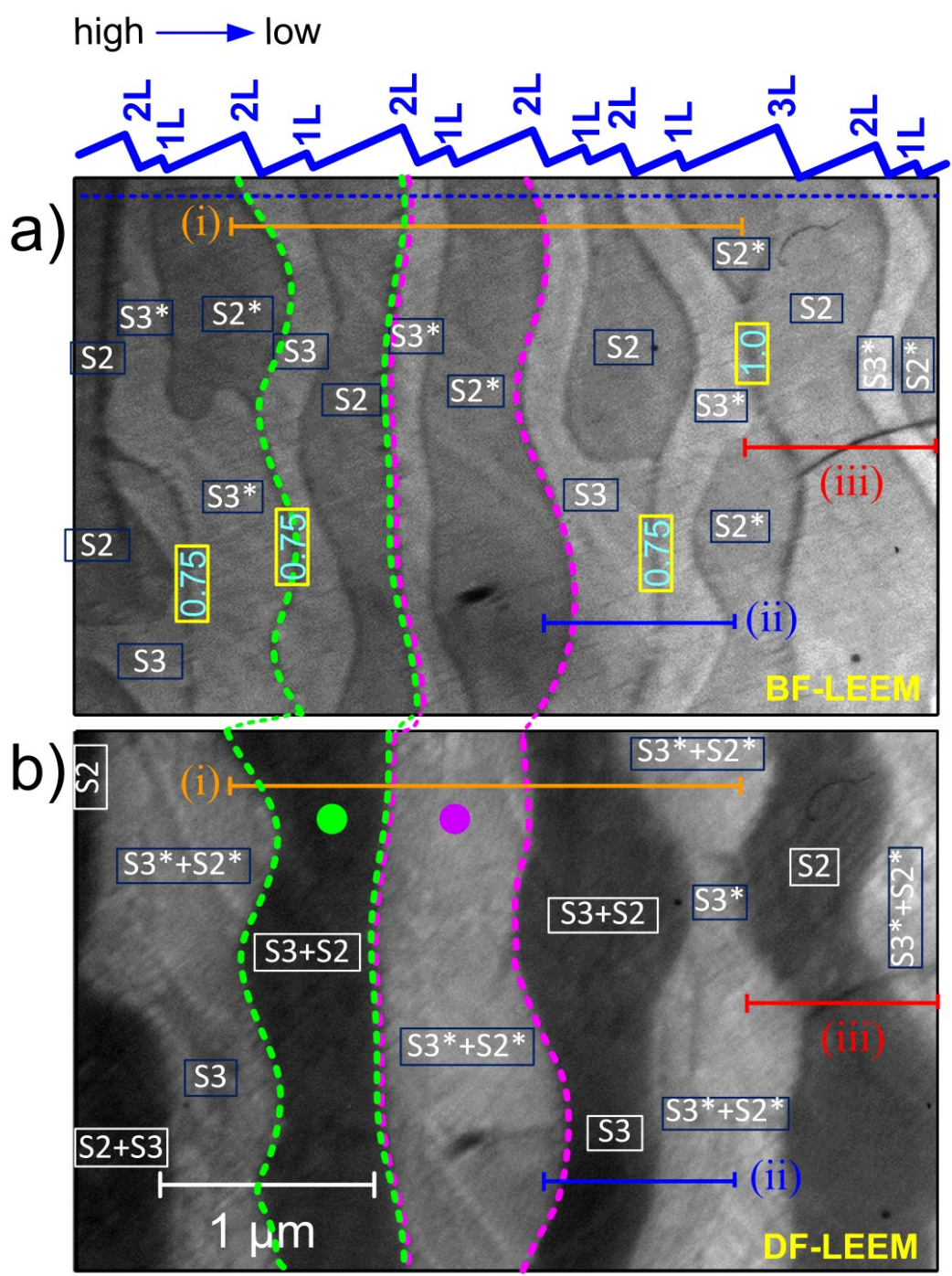

**Figure 7.3. LEEM investigation of PASG on 6H-SiC (0001).**

(a) BF- LEEM ($E$ = 4.4 eV) image of PASG monolayer graphene on 6H-SiC(0001) showing a stacking-related reflectivity contrast. The terraces can be distinguished by their reflectivity and are labeled S2, S2*, S3, S3*, as explained in the text. From this, a height profile (top blue line) is deduced for the upper part of the LEEM image (along the dotted blue line).

(b) Dark-field LEEM ($E$ = 11 eV) image of the same surface area using the diffraction spot marked with a black circle in **Figure 7.4b**. The contrast is caused by the 60° crystal rotation of the terminating SiC layers of the substrate. Areas of the same brightness are labeled by the SiC terraces as deduced from the BF-LEEM image.

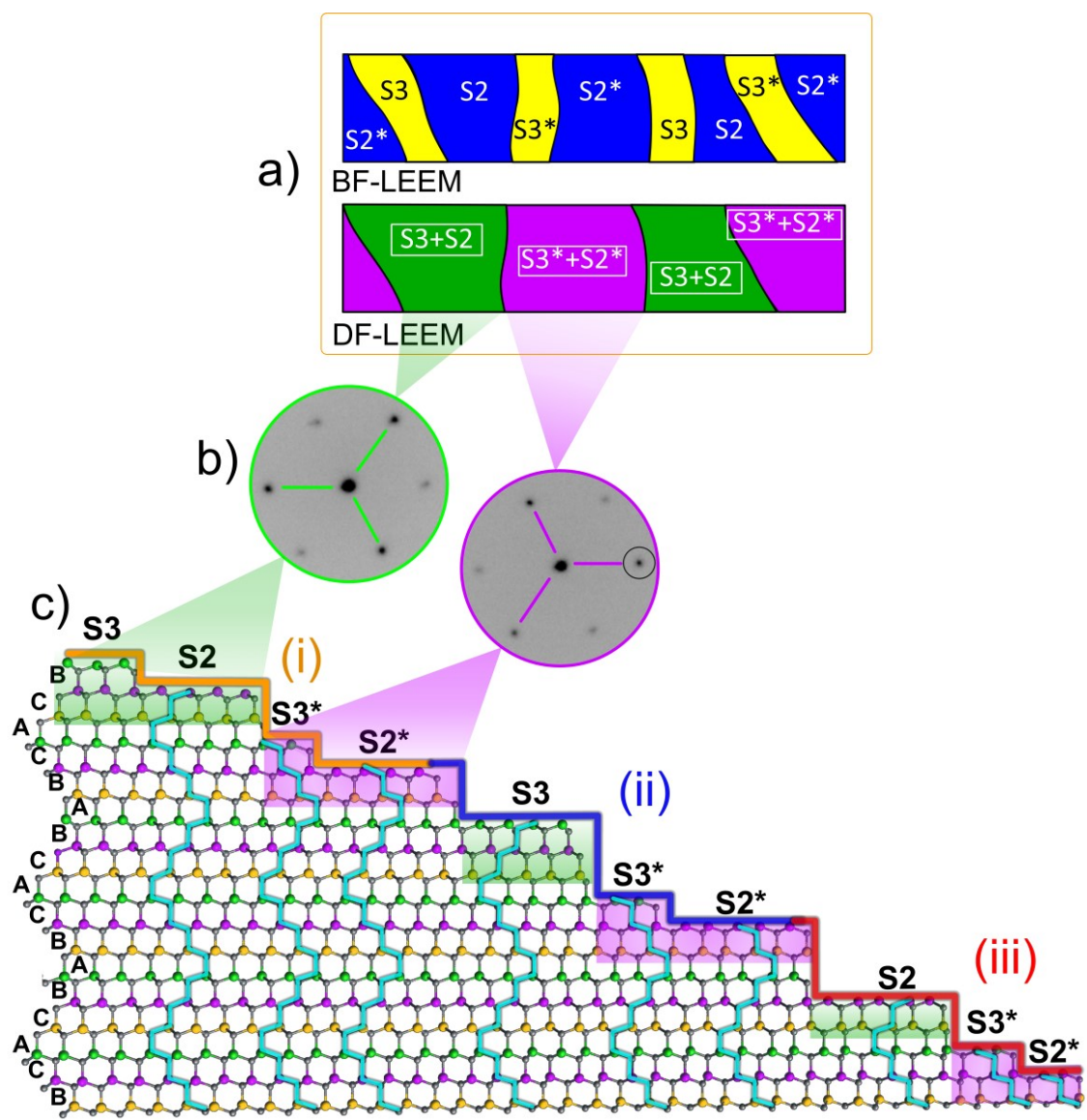

**Figure 7.4. Surface restructuring model of PASG on 6H-SiC (0001) supported by LEEM investigation.**

(a) Sketch of the BF- and DF-LEEM image for a selected area (orange line (i)) in **Figure 7.3a and b**, which represents a region with a regular 1L/2L step pattern.

(b) Two µ-LEED ($E$ = 37 eV) patterns from neighboring areas with different dark-field LEEM contrast (marked by the green and violet dot in **Figure 7.3b**). The 60° rotation of the satellite spots indicates the corresponding SiC crystal rotation.

(c) Schematic step restructuring model of the 6H-SiC substrate during epigraphene growth for three typical step patterns observed in the LEEM images **Figure 7.3a and b**. Buffer and graphene layers are not shown for clarity. Area (i) demonstrates the characteristic regular pattern of 1L/2L step pairs with the S3/S2 and S3*/S2* terrace sequence along the line (i) in both LEEM images in **Figure 7.3a and b**. The cyan-colored line-tracks indicate the different SiC crystal rotation below the terrace pairs S3/S2 and S3*/S2*, which give rise to the DF-LEEM contrast. Areas (ii) and (iii) with an irregular step sequence including ~0.75 nm (3L) steps fully explain the observed BF-and DF-LEEM contrast patterns along the line (ii) and (iii) in **Figure 7.3a and b**.

Next to the very regular S2/S3 terrace pattern in the upper part of this BF-LEEM image, there also irregular terrace configurations are observed in the lower part of **Figure 7.3a**. As an example, the blue line (ii) crosses a bright/bright transition and the red line (iii) include a dark/dark transition. The corresponding step structure is sketched in parts (ii) and (iii) in **Figure 7.4c** and can be explained by a ~0.75 nm (3L) terrace step in both cases. Such terrace configurations occur since the terraces of the starting SiC substrate are not entirely equal in length (and width). On both sides of the ~0.75 nm steps, the same terrace type (both are either S2 or S3) is found, which results in the same LEED reflection intensity, i.e., no contrast in the BF-LEEM image is observed. This interpretation allows a consistent explanation of the bright and dark areas of the complete BF-LEEM image by the corresponding terraces types (i), (ii), and (iii), as marked in **Figure 7.4c.**

To test the surface structure model, also the dark field (DF) LEEM image was recorded using the moiré diffraction spot marked in **Figure 7.4b**. The moiré diffraction spots arise from multiple scattering at the graphene and SiC lattices and thus carry information about the SiC lattice orientation. The resulting DF-LEEM image is shown in **Figure 7.3b**. It also exhibits a sequential binary contrast pattern, but interestingly, the width of both stripes is wider and not congruent with the stripes observed in the corresponding BF image, **Figure 7.3a**. For the regular stepped upper part of the LEEM images (see the line (i) in **Figure 7.3a and b**), we can deduce that two terraces (one bright and one dark stripe in BF) comprise one stripe in the DF image.

This situation is depicted in **Figure 7.4a**. A single SiC terrace has a three-fold rather than a six-fold symmetry [333], also indicated by the selected area LEED images in **Figure 7.4b** acquired from neighboring stripes in **Figure 7.3b** (marked with violet and green spots). In the LEED pattern, three of the six satellite spots are prominent on one stripe while on the neighboring stripe, thus it is found that the pattern is rotated by 60°. The origin of the DF contrast is the 60° rotation between the equivalent surface terminations Sn/Sn*, which causes a 60° rotation of the respective threefold diffraction spots patterns (see **Figure 7.4b**) of the terrace, which is sketched as green and purple areas in **Figure 7.3b**. Thus, when crossing a step from Sn to an Sn* (or vice versa) terminated terrace (S2-S3*, S2*-S3, S2*- S2, S3*-S3) the LEED pattern is rotated by 60° and a change of the DF contrast appears. On the other hand, no contrast inversion occurs when crossing a step from S3 to S2 or from S3* to S2*, since the direction of the underlying SiC is preserved.

Using this interpretation also the irregular regions of the LEEM image can be explained, which is sketched in **Figure 7.4c** for the lines (ii) and (iii). With the described model, the different contrast patterns of the dark- and bright-field LEEM images complement each other and result in a consistent view of the underlying SiC terrace structure. With the input of the AFM results and the presented step retraction model, an unambiguous assignment of the step heights is possible, as are labeled in **Figure 7.3a-b** (white color texts) in addition to the steps which are ≥ 0.75 nm (yellow color). All the possible terrace-step types are shown in **Figure 7.4c** and marked with colors (type(i) orange, type(ii) blue, and type(iii) red) as also identified in the actual LEEM measurements in **Figure 7.3a and b**. Also, the surface terminations on the 6H-SiC underneath the graphene can be easily followed, considering the cyan color track-lines.

## 7.5. Verification of the SiC terrace type at nm-scales

So far, the step-flow model and LEEM inspections revealed that the graphene lies on SiC domains, which are rotated by 60° crystal rotation. This finding, however, raises a question about the crystal structure of the buffer layer, which is partially bonded to these SiC surfaces.

A confirmation of the presented surface terrace model and a visualization of the SiC surface crystal of the graphene sample orientation is provided by an STM investigation performed at 77 K with a tungsten tip STM (For details about the STM measurement, see Chapter **3**). **Figure 7.5a** shows a large-scale topography of the irregular stepped area (iii) of **Figure 7.3a and b**. From the measured step heights (3L, 2L, and 1L), an assignment of the underlying SiC surfaces next to each terrace step is possible. This is shown in the histogram in **Figure 7.5b** created from the black dashed line in **Figure 7.5a**. From the histogram step-heights of 1L = (2.48 ±0.13) Å, 2L = (4.92 ±0.2) Å, and 3L = (7.49 ±0.13) Å, are extracted. (Uncertainty values are calculated from the FWHM of the Gaussian fit) For the applied STM energies (-0.4 nA, 1.7 V), the graphene appears transparent, and a high-resolution image of the buffer layer structure is resolved with the characteristic $(6\sqrt{3} \times 6\sqrt{3})R30°$ superlattice. [334,335] High-resolution images of the areas in the vicinity of the three consecutive terrace steps are shown in the insets of **Figure 7.6a-c**. They show a smooth buffer layer formation.

On all four terraces, S2*, S2, S3*, and S2*, see **Figure 7.6a-c** triangular-shaped structures are identified, partly marked as red/yellow triangles. (Note that such pyramids are different from the growth-induced triangular-structure in ref. [38] and Appendix **A4**) Such corrugation structures are known to form during growth and surface reconstruction. [80,336–338]

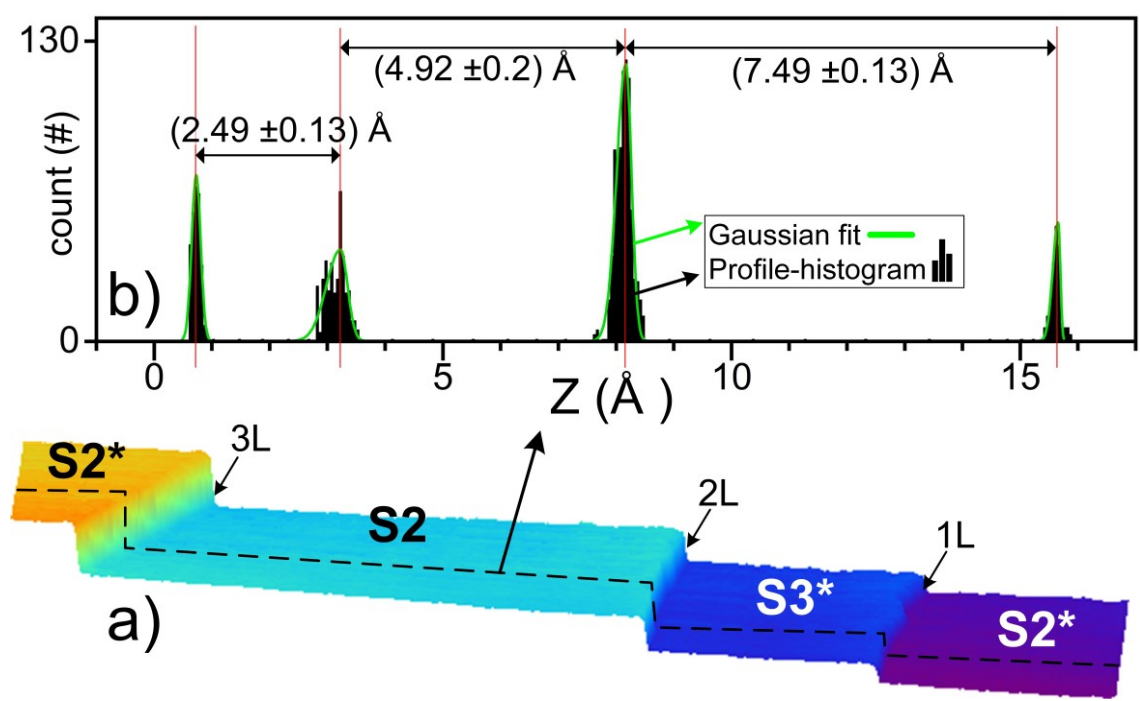

**Figure 7.5. Scanning tunneling microscopy of graphene on 6H-SiC with non-identical surface terminations.**

(a) STM image (-0.3 nA, 0.1 V) of monolayer 6H-SiC/G with non-identical surface terminations measured in the area along with the line type (iii) in the LEEM image of **Figure 7.3a and b**. The assignment of the surfaces (S2*, S2, S3*, and S2*) and the corresponding step heights are a direct result of the interpretation of the LEED images within the step retraction model and are schematically displayed in the area (iii) of **Figure 7.4c**.

(b) The Histogram (black columns) and the corresponding Gaussian fit (green curves) of step-heights plotted from the black dashed line in (a). Atomic resolution STM images of the graphene buffer layer on neighboring terraces around the 3L, 2L, and 1L step edges, respectively, shown in (c-e).

A close inspection of the STM images in **Figure 7.6a-c** reveals that the orientation of the triangles above and below a terrace edge is different for the three step-heights. (For clarity, the orientation and the rotation angle are depicted above each image.) For the consecutive terraces S2* to S2 (3L step of ~0.75nm) in **Figure 7.6a** and S2 to S3* (2L step of ~0.5 nm) in **Figure 7.6b** the triangle orientation rotates by 60°. The same rotation is observed for the (6 × 6) quasi corrugation (diamonds sketched in blue) and the $(6\sqrt{3} \times 6\sqrt{3})R30°$ diamonds (sketched in yellow) of the buffer layer. Only for the 1L (~0.25 nm) step crossing from S3* to S2* terrace the triangle orientation does not change as well as the direction of the diamonds. Note, the 120° (or −60°) rotation of the diamonds of the first and the final S2* terrace, in **Figure 7.6a and c**, is a result of the 3-fold rotational symmetry. Both terraces are otherwise equivalent. The observed rotations of the SiC surface lattice are in excellent agreement with the SiC terrace model depicted in the corresponding area (iii) of **Figure 7.4c** and it is consistent with the step-flow model, the μ-LEED, and DF-LEEM results, which showed that for the S to S* crossings the underlying SiC surfaces are accompanied by a 60° rotation, but the crystal orientation is retained for all Sn* to Sn* or Sn to Sn crossings. The tight correlation of the $(6 \times 6)$ and the $(6\sqrt{3} \times 6\sqrt{3})R30°$ diamond rotations across the terrace steps show very instructively that the buffer layer strictly follows the rotation of the underlying SiC lattice.

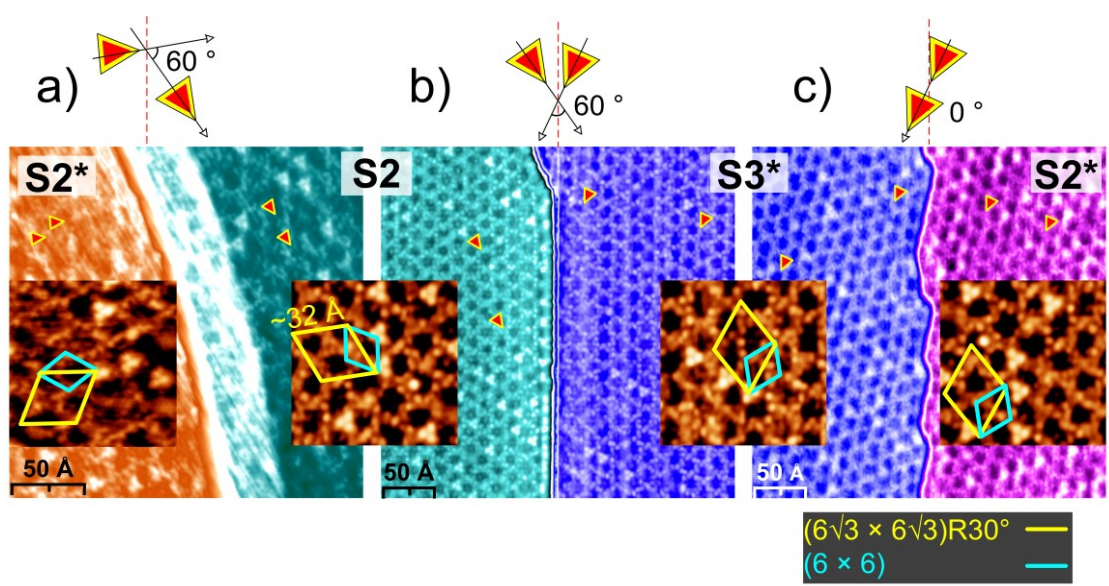

**Figure 7.6. Atomic resolution STM inspection of graphene on 6H-SiC with non-identical surface terminations.**

(a-c) STM images (-0.4 nA, 1.7 V) illustrate the transition between identical and non-identical 6H-SiC terminations (below the graphene) from the areas shown in **Figure 7.5a**.

(a) shows the transition area from terrace S2* to S2, which is correlated with a 60° rotation of the SiC substrate, see the area (iii) in **Figure 7.3a** and **Figure 7.4c**. This rotation manifests itself in a 60° rotation of triangular-shaped structures partly marked by the red/yellow triangles. For clarity, the directions of the triangles are sketched above each image. These triangular structures span a $(6 \times 6)$ nanomesh, which also rotates by 60° indicated by the blue diamond in the high-resolution insets. The buffer layer characteristic $(6\sqrt{3} \times 6\sqrt{3})R30°$ super-lattice is indicated by a yellow diamond, and it follows the 60° rotation of the SiC surface.

(b) The transition from an S2 to an S3* terrace is also characterized by a 60° rotation of the Si clusters (triangles), the $(6 \times 6)$ nanomesh, and the buffer layer superlattice.

(c) For the S3* and S2* transition, no rotation of the triangles and, therefore, of the buffer layer is observed in agreement with the missing SiC crystal rotation.

## 7.6. Work function of graphene on non-identical terraces

So far, AFM phase images reveal that the surface properties of graphene monolayers are different depending on the stacking termination of the underlying SiC terrace, which is in good agreement with the observed reflectivity contrast in the BF-LEEM images. In this section, additional methods are used to quantify the energy difference between the terraces, and the possible origins are discussed.

First, ambient KPFM-AM is used to measure the surface potential ($\Delta V$) and the work function $\phi$ of the graphene layer. [172,339] The AFM topography of a binary 1L/2L stepped monolayer graphene sample is shown in **Figure 7.7a**. The measured step heights allow the assignment of the terraces to the underlying SiC surfaces S2 and S3. The KPFM surface potential maps of the same area are displayed in **Figure 7.7b** and a section enlargement in **Figure 7.7d**. The weak binary contrast pattern (dark and light grey) of neighboring terraces is very similar to the AFM phase image in **Figure 7.1d**. The potential values of two neighboring terraces are calculated by taking the median values from an area of $100 \times 600$ nm$^2$ (dashed rectangles in **Figure 7.7d**). The corresponding histogram of the potential values in **Figure 7.7e** clearly shows the difference between the S2 and S3 surfaces. The potential difference of monolayer graphene on both terraces S2 and S3 results in $V_{S2} - V_{S3} = 9 \pm 2$ mV (in the air) which corresponds to a work function difference of about $\Delta\phi_{S2-S3} \approx -10$ mV. ($\Delta\phi_{S2-S3} = (\phi_{probe} - eV_{S2}) - (\phi_{probe} - eV_{S3})$). [68,340]

The KPFM map in **Figure 7.7b** also shows the elevated surface potential of bilayer graphene (BLG) spots (red areas), which have formed around a substrate defect on an S3 terrace. The homogenous potential of the bilayer graphene is also used to measure the relative difference to the monolayer graphene on both types of terraces. The local potential variation along the line scans correlated with an S2 and S3 terrace (magenta and green line in **Figure 7.7b**) is measured, and the values are plotted in the histogram in **Figure 7.7c**.

A clear difference between the S2 and the S3 related potential is observed with a value of $V_{S2} - V_{S3} = 12 \pm 2$ mV in reasonable agreement with the previous value from areal integration. The potential difference between monolayer and bilayer graphene is much larger, with a value of ~60 mV, which is consistent with literature data but smaller than the reported values for vacuum measurements. [119,172,339–341] It is known that the absolute surface potential value is reduced by moisture and atmospheric adsorbates on the surface and thus makes it difficult to precisely assess the surface potential difference. [172,340,342] It is worthwhile to mention that a potential difference of 10-20 meV for monolayer graphene on different terraces can also be seen in AM-KPFM (in air) measurement in ref. [339], which was not discussed.

As a second method, low-energy electron reflectivity (LEEM-IV) measurements are used to deduce the graphene specific energy from the energy dispersion of the reflected electron beam in BF μ-LEED geometry. [78,343] The LEEM-IV measurements in **Figure 7.8a** were performed in the vacuum with thermally cleaned surfaces on two neighboring terraces, S2 and S3. All LEEM-IV spectra show one prominent minimum, which is the signature of monolayer graphene in agreement with the Raman measurements in **Figure 7.1f**. [23,344] The lateral distribution of the minimum energies in the LEEM-IV map in **Figure 7.8b** taken at the same position as the BF-LEEM images in **Figure 7.3a** shows an identical contrast pattern (compare the upper part of **Figure 7.3a**). This shows that next to the reflected intensity of BF-LEEM, also the energy of the minimum is correlated with the underlying SiC surface termination.

For the neighboring terraces S2 and S3, an energy difference of ($E_{S3} - E_{S2}$) equals to (60 ±10) meV is estimated from the minimum energy histogram in the inset of **Figure 7.8b**, which indicates a distinct difference in the graphene properties on both terraces. It worthwhile to mention that nearly the same value was measured for the graphene on 4H-SiC with non-identical terraces (Appendix **A6**), however this polytype does not show a sequential pattern of terraces like 6H-SiC. [39] Although, in an early publication, the minimum energy was related to the graphene work function under the assumption that the reflectivity spectrum is related to discrete energy levels in the conduction band along the Γ- A direction of the graphene band structure, [23,345] a recent study suggests, that the graphene interlayer bands play the decisive role. Following this model, the minimum position depends on the distance between the buffer and graphene layer and the corresponding correlation-exchange potential. [204,343] To explain the shift of the LEEM-IV minima, at least one or both parameters should be different for graphene on S2 and S3 terraces. Since High-resolution STM measurements [291] on comparable 1L/2L stepped PASG graphene samples revealed fluctuating step heights variations (with smaller as well as larger on both 1L and 2L steps), thus systematic step heights differences can be excluded. Also, the presented high-resolution STM measurements shown in **Figure 7.5a-b** support this idea. This suggests a considerable difference in the correlation-exchange potential for both terraces, which is sensitive to the different charge densities. More detailed model calculations are necessary to verify this assumption. [78]

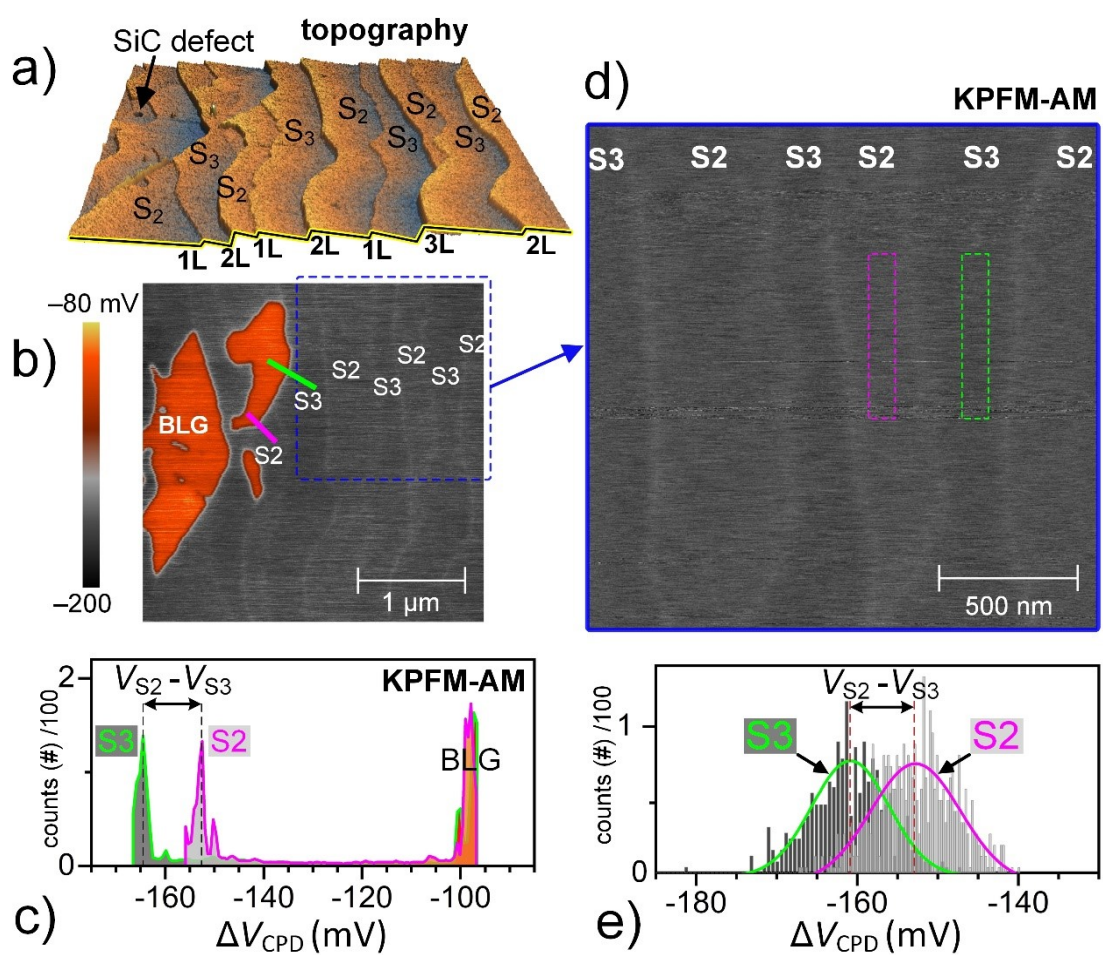

**Figure 7.7. KPFM-AM measurements of surface potential and work function of monolayer graphene on non-identical 6H-SiC terraces.**

(a) The AFM topography image allows an unambiguous assignment of the surface terraces S2 and S3 based on the step height sequence, as explained in the text.

(b) Surface potential map from KPFM-AM measurement of the same surface area. The red areas mark bilayer graphene (BLG) spots, which have formed at a SiC defect. Marked are the positions of two line scans (green and magenta-colored lines) correlated with an S3 and an S2 terrace, respectively, for the histogram evaluation.

(c) Histogram of the surface potential differences along the two line scans S2 (magenta) and S3 (green) indicated in the inset of (b). A clear difference of the potential values for the S2 and S3 surface is observed with $V_{S2} - V_{S3} = 12 \pm 2$ mV. A separation of ~60 mV to the potential values of the BLG is clearly visible.

(d) Section of KPFM image in (b) (blue square) shows the potential contrast of monolayer graphene on the terraces S2 and S3. The green and magenta-colored rectangles ($100 \times 600$ nm²) indicate the area for the calculation of the potential values.

(e) Histogram of the surface potential difference (median values) extracted from the green (S3 terrace) and magenta (S2 terrace) rectangles shown in (d). A potential difference $V_{S2} - V_{S3} = 9 \pm 2$ mV is estimated.

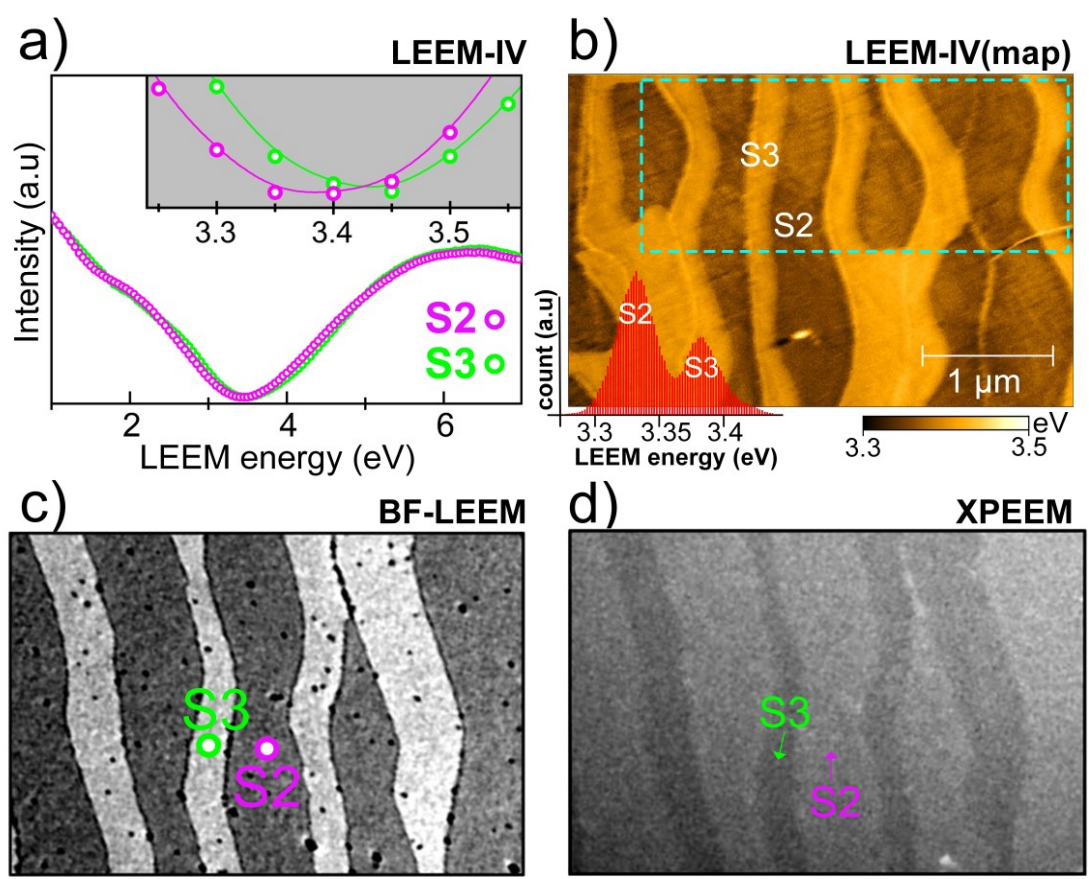

**Figure 7.8. Graphene on non-identical 6H-SiC terminations: surface analysis by LEEM, LEEM-IV, and XPEEM.**

(a) Two LEEM-IV spectra taken at two different terraces S2 and S3, as indicated in (c). The inset shows the minimum position and a difference in energy of ~60 meV between the S2 and S3 curve.

(b) The LEEM-IV map shows the lateral distribution of the minimum energy from the LEEM-IV spectra in an area of $4 \times 3$ $\mu m^2$. This LEEM-IV map is taken from the same area as the BF-LEEM image in **Figure 7.3a**, and the comparison reveals a congruent contrast pattern related to the SiC terraces underneath. The histogram (inset) shows the distribution of LEEM-IV minimum energies taken from the area marked with the dashed-cyan rectangle. The histogram clearly indicates an energy difference of 60 $\pm$10 meV between graphene on the S2 and S3 terminations.

(c) BF-LEEM image from another 1L/2L stepped monolayer graphene sample. The indicated dots S2 and S3 mark the position where the LEEM-IV spectra in (a) were measured. The incident electron energy was 2.7 eV, and the field of view (FoV) is 10 $\mu$m.

(d) XPEEM image taken at the same positions as the BF-LEEM image in (c). The FoV is 10 $\mu$m, X-ray excitation of 80 eV, and detection at 1 eV. Note the inverse contrast. The XPEEM contrast is related to the work function of the graphene. The higher work function terraces show a darker contrast. The white lines stem from 3L step edges between terraces of the same SiC surface termination.

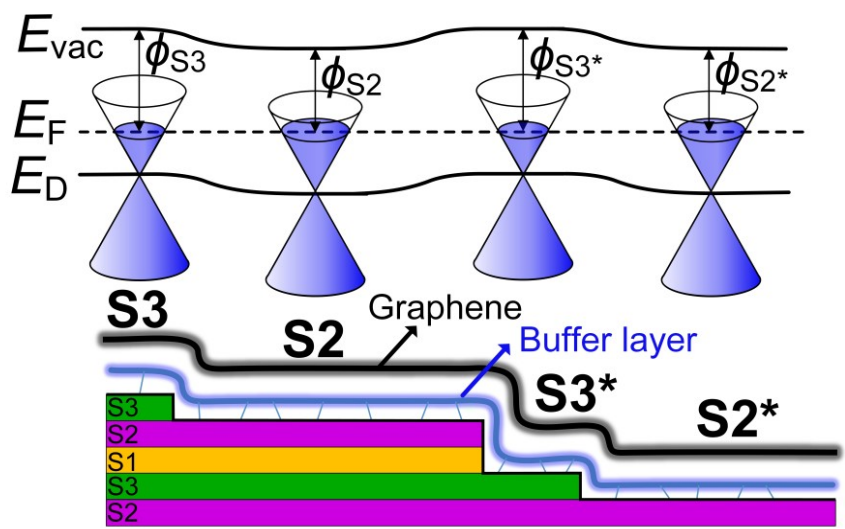

**Figure 7.9. Schematic representation of stacking-order-induced doping in epitaxial graphene on hexagonal SiC.**
Schematic energy diagram of epitaxial monolayer graphene on the 6H-SiC terraces S2 and S3 as derived from XPEEM and KPFM measurements. The variation of the work functions $\phi$ at S2 and S3 terraces indicate a shift of the Dirac cones by the same amount (Dirac energy $E_D$). This results in a spatial modulation of the graphene surface potential.

As the third approach, X-ray photoemission electron microscopy (XPEEM) was applied, which directly visualizes the work function variation on the surface (see section **3.8** for more details). [202] From the broad secondary electron energy distribution generated by X-ray excitation (photon energy of $h\nu = 80$ eV), those with a low kinetic energy of 1 eV are selected and measured. Therefore, the intensity map visualizes spatial variations of the work function. The XPEEM image of a 1L/2L stepped graphene sample is displayed in **Figure 7.8d**, and it shows the mentioned contrast pattern of the alternating surface terraces: narrow (S3) and wider stripes (S2). The XPEEM image taken at the same position as the BF-LEEM image (**Figure 7.8c**) shows a congruent pattern which verifies this assignment. In contrast to BF-LEEM images where an arbitrary contrast is obtained, the intensity in the XPEEM image is unambiguously correlated with the magnitude of the work function, namely, the layer with the lower work function generates a brighter contrast. [202]

Thus, the S2 related graphene terraces (light grey stripes) have a lower work function compared to S3 terraces (dark grey stripes) in agreement with the KPFM result. This results in a consecutive substrate-related doping engineering in top carbon layers that is schematically represented in **Figure 7.9**. This work function difference and the related potential difference are regarded as the reason for the contrasts observed in the AFM phase and BF-LEEM images. The different interaction of the AFM tip and the varying potential on the non-identical terraces results in damping and a phase shift of the tip's frequency. In the BF-LEEM experiments, the slightly different surface potential and charge state, respectively, varies the reflection behavior on non-identical terraces and results in a variation of the reflected intensity of the $0^{th}$ order electron beam.

The KPFM and XPEEM reveal in good agreement a lower work function for monolayer graphene on S2 compared to S3 terraces, and also, the LEEM-IV result support it. A reliable value for energy difference of ~10 meV was measured by KPFM, knowing that this value might depend on the environmental conditions. [342] The strong XPEEM contrast points to a higher energy difference, but beam damage effects prevent a more precise estimate of the energy difference. More detailed studies are necessary to clarify the absolute value of the energy difference.

Thus far, other experimental results are lacking. The local measurements of the sheet resistivity by high-resolution scanning-tunneling potentiometry (STP) on similar 1L/2L stepped samples (thermally cleaned) have indicated a difference of the average graphene sheet resistance by ~14% at room temperature and ~270% at low temperature (8 K) on both terraces S2 and S3. For terraces connected by a 3L step, a smaller variation of < 3% is measured. [291] The analogy to the results presented here supports the idea of a strong impact of the SiC terrace termination on the graphene properties. The scanning-tunneling spectroscopy (STS) analysis suggests that the doping variation cannot result in such substantial resistance variation on S2 and S3, therefore, it was attributed to the distance variation between graphene and buffer layer and substrate. [291] A question remains open considering the atomic distance between STS tip and surface (~1 nm) and the in-between tunneling current as well as a probably tip-induced band bending at the surface which all could influence the STS measurement.

The presented measurements convincingly show a correlation of the SiC substrate termination with the electronic properties and the work function of epitaxial monolayer graphene. The graphene's $n$-type doping level is mainly determined by two effects, namely, an overcompensation of the SiC bulk polarization doping by donor-like buffer layer and interface states. [30,31] The strong intrinsic bulk polarization in hexagonal SiC substrates produces a negative pseudo-charge at the SiC surface, which induces positive charges in the graphene to account for overall neutrality. This polarization doping effect shifts the Fermi energy in freestanding monolayer graphene far below the charge neutrality point to $E_D - E_F = -0.3$ eV. [30]

A specific terrace related polarization effect can be deduced from ab-initio pseudo charge calculations, which show a different valence band charge density for cubic and hexagonal SiC layers in the 6H polytype and a different charge density depending on the distance to the next underlying hexagonal layer. [163,164] In a recent publication, the total polarization doping effect of SiC surface and bulk was investigated by standard density functional theory calculations, which show a SiC surface termination dependent doping variation of $2 \times 10^{12}$ cm$^{-2}$ for free-standing monolayer graphene. [161] However, the corresponding shift of the Dirac point of ~100 meV is larger than the KPFM value. For the exact modeling of the investigated epitaxial graphene samples, the impact of the intermediate buffer layer and interface states must be taken into account. A variation of the graphene work function induced only by a different distribution of buffer layer related donor-like states is excluded by the observation of AFM and BF-LEEM contrast patterns in buffer-layer free, *p*-type H-intercalated, quasi-freestanding monolayer [38] and also quasi-freestanding bilayer graphene. [to be published]

The observation of an AFM phase contrast of the insulating buffer layer sample, see **Figure 7.1j** indicates that the SiC surface-related polarization effect also exists in the absence of graphene and the corresponding donor-like states, therefore it bears no relation to an interplay between the buffer layer and graphene.

This discussion shows that the assumption of a SiC stacking termination dependent polarization doping can explain the presented experimental results. Other effects, e.g., an influence of terrace dependent stress is less probable by the observation of strong contrast patterns also in more relaxed free-standing mono and bilayer graphene. [38] [to be published] A partial influence of different defects at both SiC terraces types can, however, not be ruled out. An example of a possible defect state in hexagonal SiC is the basal vacancy/divacancy, which can occur in a quasi-cubic or quasi-hexagonal position (shown in **Figure 7.2a**) with an energy difference of 0.03 eV. [324] Both quasi-positions are typical for the cubic S2 (0% hexagonality) and S3 (50% hexagonality) surfaces. To which extend such an effect plays a role requires further studies. [78]

## 7.7. Summary

In conclusion, by using various measurement techniques (AFM, STM, μ-LEED, LEEM, LEEM-IV, KPFM, XPEEM), it is for the first time evidenced a direct dependence of electronic properties of epitaxial graphene on the underlying SiC stacking termination. This was realized by employing advanced epitaxial growth techniques, including the PASG method. A periodic sequence of two different SiC terraces with the distinction in cubic and hexagonal nature of the surfaces were prepared, which develop during the high-temperature graphene synthesis. The terraces in the stacking orders are unambiguously identified as S2 and S3, as scrutinized in AFM and BF-LEEM and supported by additional DF-LEEM and STM measurements. The formation of the observed terraces is successfully interpreted in a so-called JCH step-retraction model in which the step retraction velocity increases with the hexagonality of the SiC surface layer. The KPFM and XPEEM result explicitly indicate an alternating work function of the graphene on periodic SiC surface terraces, which confirms for the first time a theoretical prediction in which the graphene doping depends not only on the bulk polarization but also on a SiC termination dependent polarization doping effect. A value of about 10 meV was estimated for the work function difference of monolayer graphene on S2 and S3 terraces from KPFM measurements. The periodically modulated graphene energies self-ordered by the underlying SiC terraces could act as a template for further graphene functionalization schemes on sub-micron scale structures. Moreover, these findings are applicable to other polar dielectric substrates and other sub-dimensional systems. Note that for graphene on 4H-SiC(0001) substrates with only two different starting surfaces, we found no periodical variation of the surface work function (see Appendix **A6**), since step retraction results in the formation of equivalent S2 and S2* terrace terminations, confirming the above interpretation.

# 8

# 8. Magneto-transport in epitaxial graphene

## Abstract

*T*his chapter is devoted to experimental results of quantum Hall effect measurements in epitaxial graphene to test the quality of the samples. Two techniques, namely electrostatic-gating and chemical-gating, are used to adjust the charge carrier density in the graphene sample close to the charge neutrality point. The stability and efficiency of the employed methods are analyzed, considering their suitability for the metrological applications. Furthermore, the challenges and considerations for device microfabrication, in particular concerning quasi-freestanding layers, are discussed.

## 8.1. Introduction

Quantum resistance metrology based only upon the fundamental constants, the electron's charge and Planck's constant, utilizes a precise redefinition of the ohm unit in two-dimensional electron systems through the von Klitzing's constant $R_k = h/e^2 = 25812.807557\ (18)\ \Omega$, the resistance quantum. [59,346] This aim has so far been implemented using GaAs/AlGaAs heterostructure devices with a high level of precision. [63,347,348] Graphene, with broadly spaced discrete electron energy levels in a magnetic field (Landau levels), enables the quantum Hall effect under rather relaxed conditions, i.e., at higher temperatures and lower magnetic field. [65,322,349] This can also be realized in epitaxial graphene on SiC with the advantage of sizeable and reproducible graphene production directly on an insulator substrate. [11,19,36] The drawback is, however, its strong electron doping [66,67] which requires a fine controllable, robust and long-lasting charge tuning close to the Dirac point, as otherwise, it would demand too high magnetic flux densities for available magnets to reach the onset of the plateau $v = 2$. [63,350,351] This problem is represented in two exemplary samples shown in **Figure 8.1**, including an epigraphene and a QFMLG sample without any charge density modification. Thus, a gateless carrier density control of about $10^{11}$ cm$^{-2}$ in epigraphene is crucial to obtain a robust resistance plateau ($v = 2$) at low magnetic fields (desirably $< 5$ T). There have been vast and ongoing efforts to accomplish this delicate task. Various charge tuning techniques with distinct aspects, e.g., simplicity, efficiency, reversibility, toxicity, time-stability, and cost, have been reported.

Aqua regia (HNO3 : HCl : H$_2$O) is a material example of this approach, but the toxicity, as well as fine-control tuning, are the main problems. [232,352] Another recent study on molecular doping using tetrafluoro-tetracyano-quino-dimethane (F$_4$TCNQ) molecules mixed in a liquid solution with poly (methyl-methacry-late) (PMMA) indicates air-stable and durable doping with the capability of fine-tuning, however, it lacks the reversibility of the process, i.e., initial doping cannot be recovered. [353–357] Besides, molecular doping via fluorinated fullerene (C$_{60}$F$_{48}$) with its high electron affinity is an effective surface acceptor for graphene that has the advantage of fine doping even toward $p$-type functionalization. [358]

Although not meeting the criterion for resistance metrology, poly [para-xylylene], commonly known as Parylene can be used to passivate the graphene surface and modify the doping. [359,360] Moreover, molecular adsorption using NO$_2$ [361] as well as low-temperature hydrogen annealing [272] or air/nitrogen-annealing (see Appendix **A2**) are other approaches for doping modulation in epigraphene. The latter is also employed in this study to saturate the silicon dangling bonds on the SiC substrate beneath the graphene and thus improving the carrier mobility. Also, electrostatic-gating [221] seems to be a reversible, fast, and very easy-to-apply technique, however, the stability over time, doping homogeneity, as well as the area of coverage (for sizeable samples) are the major cons. Photochemical gating [220], using ZEP-520A, is another alternative to tune the charge carrier concentration, but the reversibility of the process is limited, considering the number of activation cycles applied to the sample. The hydrogen annealing two methods are applied and discussed in this chapter.

**Figure 8.1. QHE in epigraphene without charge carrier tuning.**

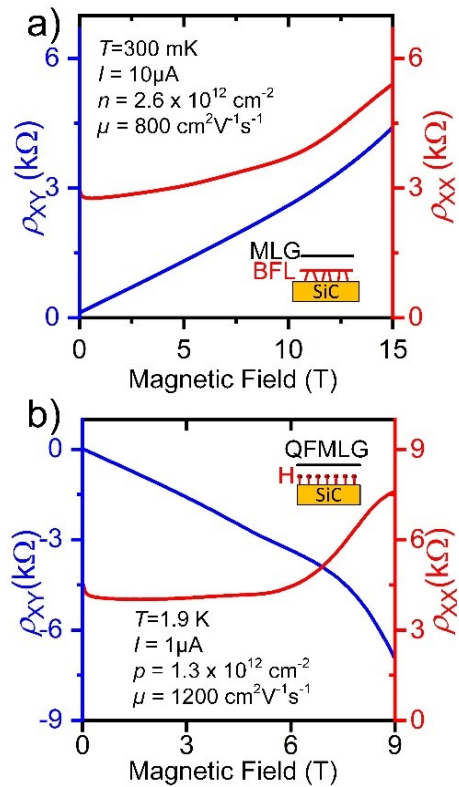

(a) shows QHE measurements in a bare epigraphene device without any encapsulation. The sample exhibits low carrier mobility due to both high charge density and environmental influences.

(b) A similar QHE measurement conducted on an intrinsically hole-doped (inferred from the sign of $\rho_{XY}$ ramp) quasi-freestanding monolayer graphene obtained by H-intercalation of the epitaxial buffer layer. Here the sample is protected by a thin PMMA layer (55 nm), however, it reveals a high charge carrier concentration.

Charge density tuning close to the Dirac point and encapsulation of epigraphene are two essentials for the realization of a QHR plateau ($\nu = 2$) at low magnetic fields.

## 8.2. Sample preparation

The epitaxial graphene layers were grown on semi-insulating 6H–SiC and 4H–SiC samples with a nominal miscut of about −0.06° toward [1$\bar{1}$00]. The QFMLG and QFBLG samples were prepared by hydrogen intercalation of the buffer layer (900 °C, 60 min, ∼1mbar) and monolayer graphene (1050 °C, 120 min, ∼1mbar), respectively. The growth methods and morphological properties of the samples were described in chapters 4–7. The quantum Hall effect (QHE) measurements were performed on graphene Hall bars with different sizes of either $100 \times 400$ μm², $200 \times 800$ μm², or $400 \times 1200$ μm², fabricated by conventional electron beam lithography (see section **4.5**). The charge carrier tuning was achieved employing either electrostatic-grating or chemical-gating methods. These techniques were explained in section **4.6**. For the electrostatic-grating tuned sample, a hydrogen annealing at 500 °C (60 min, 1 mbar) was previously applied, aiming to improve the mobility.

## 8.3. Results and discussion

In the following, the quality of the samples and the restrains in the resistance quantization following two charge-tuning methods are investigated by low-temperature magneto-transport measurements on the lithographically patterned graphene Hall bars.

### 8.3.1. Electrostatic-gating

**Figure 8.2** illustrates the magneto-transport on a graphene Hall bar before and after applying the electrostatic-gating using corona discharge. **Figure 8.2a** shows the QHE on the Hall bar that was encapsulated with 55 nm PMMA, which acts as a host for the later corona ionization. The sample exhibits quantization and reaches plateau ν = 2 with a Shubnikov-de-Haas oscillation at ν = 6. A carrier concentration of about $6 \times 10^{11}$ cm⁻² with $\mu = 3500$ cm²/Vs was measured on this sample. This doping value is lower than that of typically expected for epitaxial graphene being in a range $10^{12}$–$10^{13}$ cm⁻². It was often noticed that for the samples that were kept while in nitrogen shelves, a shift of the carrier density towards hole doping happened. [218] This effect occurs due to the surface physisorption and can be excluded here since such adsorbates are removed by sample cleaning before coating with PMMA. The measured lower doping is correlated with the annealing in the presence of hydrogen, which does not intercalate the samples (safely lower temperature < 600 °C) [38,118], but very likely saturate the Si dangling bonds on the substrate below the graphene and buffer layers. This result is in good agreement with the literature. [272]

The sample was afterward taken out of the cryostat and imposed for a charge tuning by corona discharge. The explanation of this method can be found in Chapter **4**. **Figure 8.2b** plots the change of the resistance that almost doubles after two successive corona pulses. The positive ions on the PMMA layer gate the electrons out of the graphene, thus corresponding to this carrier depletion, the resistance increases. From the curves, at pulse areas, it can be inferred that the sample is still $n$-type. Please note that the resistance ($R_{sd}$) in **Figure 8.2b** is indeed the resistance measured between one-half of the connected contacts and the rest linked contacts, as shown in the circuit configuration in **Figure 4.6b**. The resistance value was adjusted based on experiences on similar devices and expected to decrease the doping down to desirably low ranges. Slightly drift from the adjusted doping happens during the chip transfer onto the probe-stick and loading into the cryostat.

**Figure 8.2c** demonstrates the QHE measurements on the corona-tuned sample within the range of $-15$ to $15$ T. The sample is well quantized by reaching the plateau ($\nu = 2$) at a low magnetic field of about 2 T. The low carrier concentration of $2.5 \times 10^{10}$ cm$^{-2}$ indicates a successful performance of the electrostatic-gating technique. The sample shows the mobility of about ~19000 cm²/Vs that implies the high quality of the produced graphene sample. As it is seen, the electrostatic-gating is a fast, reversible, and easy-to-apply method, however, it bears several technical issues. Checking the resistance values ($R_{sd}$) after removing the sample from the cryostat indicated that the tuned samples keep the adjusted carrier density while they are cooled down inside the cryostat. However, the tuning is not stable in ambient conditions.

Moreover, there are two other crucial concerns: (i) the electrostatic sparks during tuning and (ii) the homogeneity of the doping. The case (i) is observed during the tuning in which corona charges leading to discharges at areas like metal bonds and contact pads, which can cause damages to the device. This effect was more harmful when in the circuit configuration for corona tuning, a constant current was fed to the system instead of constant voltage. Even taking into account the considerations mentioned in (i), the case (ii) makes the electrostatic-gating by corona-discharge inappropriate for a robust QHR. This is because the doping adjustment is not uniform throughout the sample, inherited from the technique itself. Perhaps when the sample size is small, it would be more efficient, nevertheless for the sample size in this study, inhomogeneous doping is inferred. Such nonuniformities were stronger when the samples were tuned very close to the Dirac point compared to higher doping levels. Therefore, this technique is deemed improper for a reliable QHR metrology.

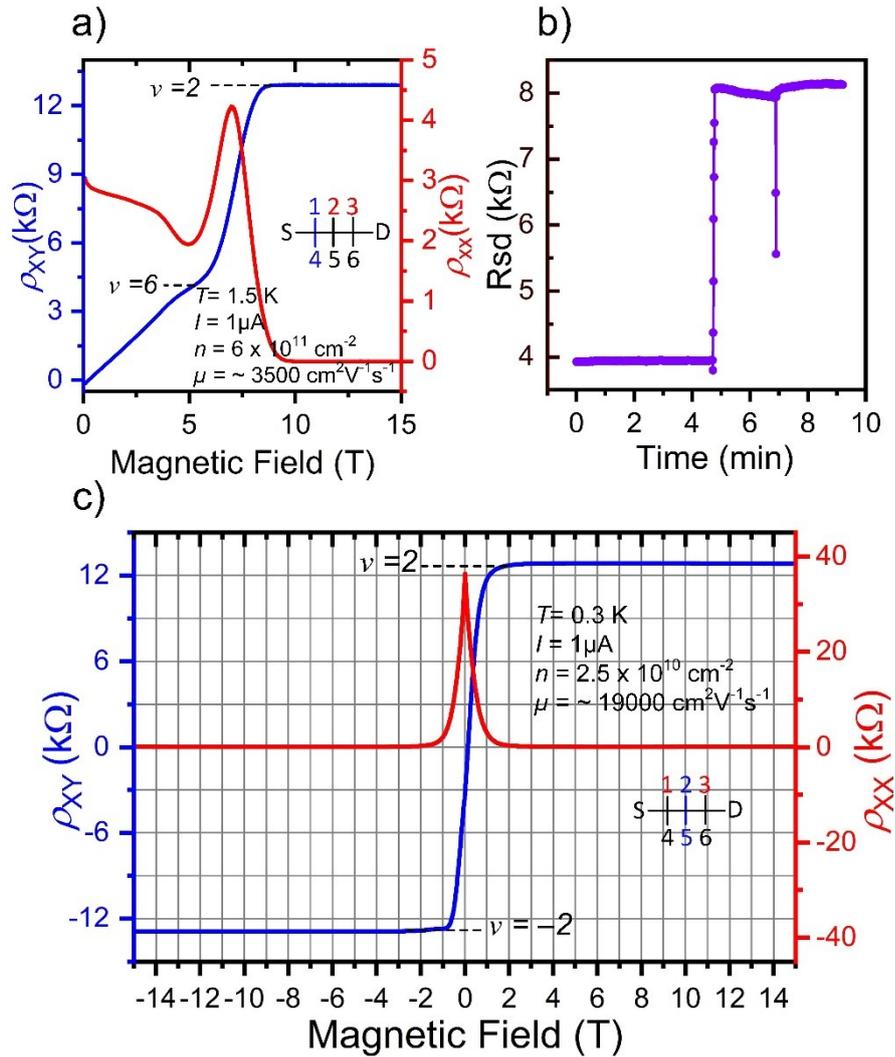

**Figure 8.2. QHE measurement in epigraphene tuned by electrostatic-gating.**

(a) QHE measurement on graphene Hall bar (400 × 1200 μm²). A hydrogen annealing (500 °C, 60 min) was applied to the sample before being coated with 55 nm PMMA. The Hall resistance $\rho_{XY}$ as a function of the magnetic field exhibits a good quantization.

(b) The electron densities adjusted by electrostatic-gating using corona-discharge. Upon two successive pulses, the measured resistance in the sample increases as a result of electron depletion in the graphene. (details can be found in Chapter **4**).

(c) QHE measurement on the same sample shows a successful carrier density reduction down to $2.5 \times 10^{10}$ cm⁻² with substantially high mobility of $\mu = \sim 19$k cm²V⁻¹s⁻¹. This underlines the high quality of the graphene sample, and also the significance of charge carrier density adjustment. In the inset of each QHE measurement, the measured contact pairs are sketched (red for $\rho_{XX}$ and blue for $\rho_{XY}$ contact).

The carrier density $n$ and mobility $\mu$ are determined from Hall measurements as $n = 1/eA$ and $\mu = A/\rho_{XX}$, with $e$ the elementary charge, the Hall coefficient $A = dR_{XX}/dB$. The longitudinal sheet resistance $\rho_{XX} = R_{XX}W/L$, and $R_{XY}$ the transversal resistance.

## 8.3.2. Chemical-gating

Next, the doping results from the chemical-gating using ZEP520 polymer are discussed in comparison with the previous electrostatic-gating method. **Figure 8.3** demonstrates the magneto-transport on two graphene samples grown on 4H– and 6H–SiC polytypes. In the inset of **Figure 8.3a, and b**, the 8-terminal graphene Hall bars are sketched, and the corresponding contact pairs for longitudinal (red color) and vertical (blue color) resistances are marked. The samples were tuned by the photochemical-gating technique [220], resulting in a $2.2 \times 10^{11}$ cm$^{-2}$ for 4H–SiC/G and $5.4 \times 10^{10}$ cm$^{-2}$ for the 6H–SiC/G sample. The different doping levels are mainly due to the difference in the period of the applied UV-illumination on the samples, already indicating that a fine-tuning by this method is challenging to control. However, the doping level in both samples was successfully reduced, leading to wide resistance plateaus at ~12.9 k$\Omega$ (corresponding to $R_K/2$), and simultaneously $\rho_{xx}$ approaches zero ohms, indicating a good and homogenous quantization in both samples. The 4H–SiC/G samples with higher $n$-doping reached $\nu = 2$ at ~4 T and the 6H–SiC/G sample at $B > 1.5$ T. For the 4H–SiC/G sample, mobility of ~3700 cm$^2$/Vs is estimated, and the 6H–SiC/G exhibits $\mu = 14$k cm$^2$/Vs.

Again, the QHE experiments verify the high quality of the fabricated graphene samples on both polytypes, as was also deduced from the experiments using electrostatic-gating by corona-discharge.

By chemical-gating using ZEP520, the electron density in the graphene can be reduced by UV-illumination and also restored to its original value by subsequent heating of the device above the polymer glass transition temperature of $T_g \approx$ 170 °C. [220] The doping remains constant after cooling down in the cryostat, but at room temperature deviates from the adjusted values, as similarly observed for the corona-discharge method. The formation of cracks in the polymer heterostructure was reported on the samples that were cooled downed to low temperatures. [362] This problem, which limits the reusability of the samples, is not observed when a thicker PMMA spacer layer (> 100 nm) was used.

So far, from this comparative study, the high quality of produced graphene samples is verified. On the other hand, it is concluded that the corona-discharge and electrostatic-gating by ZEP520 both are not suitable for a reliable and robust QHR meteorology. An alternative to this problem was suggested by a recent study using F$_4$TCNQ incorporated in PMMA, resulting in an air-stable, tunable, and reliable doping of SiC/G. [358] However, this technique was beyond the subject of this thesis.

## Figure 8.3. QHE measurement on epigraphene tuned by photochemical-gating.

Magneto-transport measurements on PASG graphene Hall bars with a size of 100 × 400 μm². The photochemical-gating method was applied to adjust the carrier density in the samples. [220]

(a) QHE measurement result of 4H–SiC/G sample shows a good quantization and reaching plateau $\nu = 2$ at ~4 T. A carrier density of $n = 2.2 \times 10^{11}$ cm⁻² with mobility of $\mu = 3700$ cm²V⁻¹s⁻¹ was measured on this sample. Very similar to the measurement result shown in **Figure 8.2a**.

(b) QHE result of the 6H–SiC/G indicates an electron concentration of $5.4 \times 10^{10}$ cm⁻² and a high mobility of $\mu = 14$k cm²V⁻¹s⁻¹. Schematic of 8–terminal resistance measurement and contact pairs are shown in the insets. The quantum Hall resistance exhibits a broad plateau at filling factor $\nu = 2$ with a value of $\rho_{XY} \approx 12.9$ kΩ ($R_K/2$). The longitudinal resistivity $\rho_{XX}$ approaches zero Ohm at about $B = 2$T.

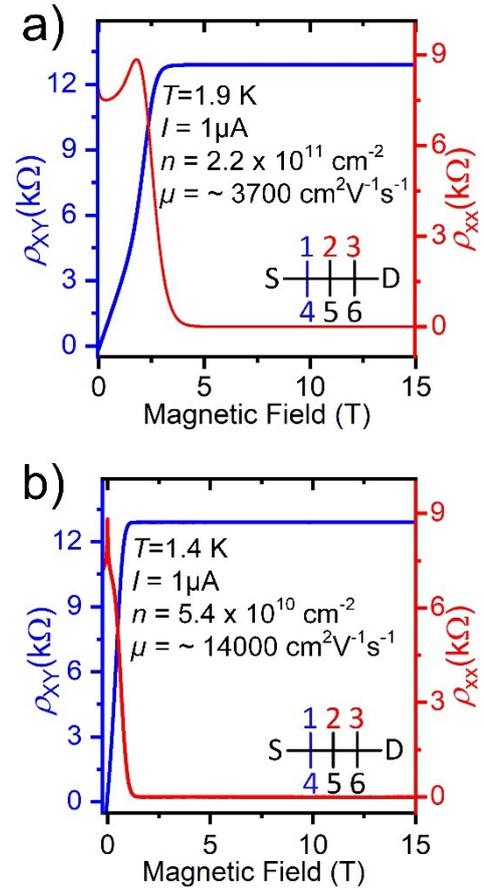

Aside from the doping, a challenge is still regarding the standard lithography process for patterning graphene-based device fabrication. This can be understood, for example, by comparing the potential mobility values in suspended graphene on SiO₂ and that of in SiC/G. [363] In addition to the strong doping, the interaction between substrate and graphene, the buffer layer, and different scatterings, which all are degrading the potential mobilities in SiC/G, also the disorder induced by the lithography process should be considered. [364]

This seems to be even more deteriorating in the case of quasi-freestanding graphene layers. **Figure 8.4** demonstrates the device fabrication on a QFBLG sample at the initial step of the lithography process. While from the optical microscope image of the sample in **Figure 8.4a**, a good graphene quality is inferred, after a simple standard cleaning in isopropanol/acetone baker in an ultrasonic bath, the graphene was severely damaged, see **Figure 8.4b**. This can be clearly seen in the AFM topography and phase images in **Figure 8.4c and d**, which show that graphene sheets were fractured into flakes of graphene on the surface. This may indicate a strong strain effect in this material system. It may initially sound contrary to the expectations of rather relaxed Van der Waals sheets for the quasi-freestanding graphene layers.

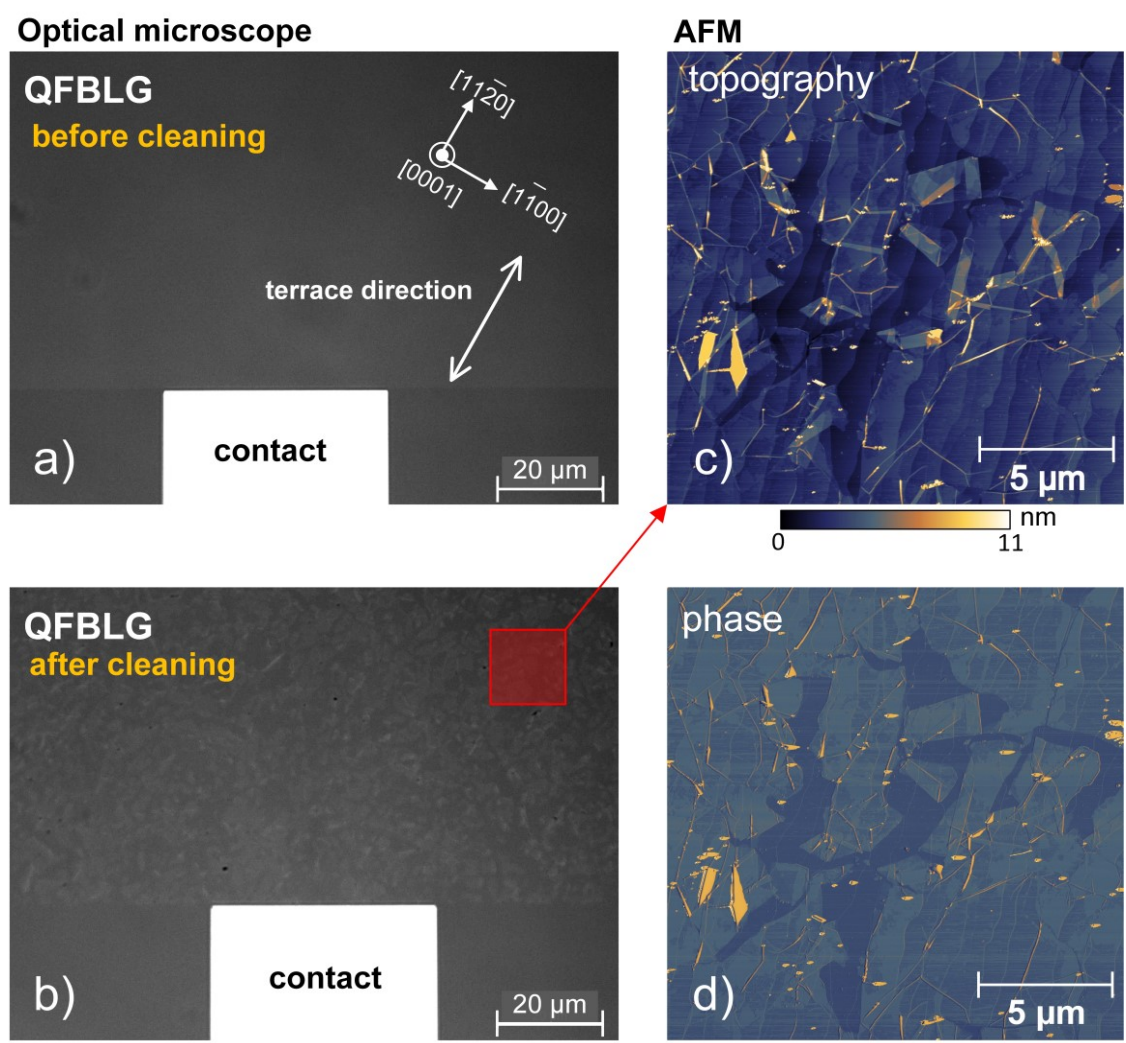

**Figure 8.4. Microfabrication on quasi-freestanding graphene needs special care.**

Optical microscopy of a quasi-freestanding bilayer graphene sample before (a) and after (b) the cleaning in a mixture of isopropanol/acetone and an ultrasonic bath. Although the QFBLG sample shows a good quality with well-ordered terrace-steps, however after the cleaning process is heavily damaged. AFM topography (c) and phase (d) inspections show that the graphene sheet is torn out to the graphene flakes on the surface. Similar examples with probably the same cause can be seen in Appendix **A9**.

However, there are some pieces of evidence that an interaction between the layers may induce such a strong strain-like effect. [200] Herein, three other related observations are shown in Appendix **A9**. Further investigation is required to understand the origin of this effect. This significant disorder indicates that special care is required in the device fabrications process using quasi-freestanding graphene. One approach to mitigate such degradation is the encapsulation of graphene by deposition of, e.g., a dielectric (like $Al_2O_3$) or metal (like Pd) on the fresh fabricated and cleaned graphene samples. Two examples can be seen in Appendix **A7**. Through this encapsulation, the graphene is saved from environmental influences as well as lithography induced defects. By a proper etching, the graphene can then be easily patterned.

## 8.4. Conclusion

In summary, magneto-transport in epitaxial graphene layers was conducted using electrostatic-gating by corona-discharge and photochemical-gating by ZEP-520 for charge tuning. The QHE measurements demonstrated high-quality sample types demonstrating high mobility values up to 19000 $cm^2/Vs$. Although the employed methods led to successful charge carrier reduction in the graphene samples, both suffered either from air-instability or weak controllability. It was observed that performing the corona-discharge doping method in nitrogen ambient instead of the air atmosphere results in further stability of surface ions by reduction of water and oxygen concentration.

In general, monitoring the studies on the functionalization of graphene, it is seen that so far, several approaches have been reported, each with benefits and drawbacks. This indicates that there is still much room for technological improvements. Future works need to be concentrating on further effective techniques to acquire robust and stable charge carrier tuning in SiC/G. Moreover, the lithography process needs to be compatible with the specific graphene type, as was exemplary shown for the QFBLG sample. This suggests that the graphene samples must be first covered, preferably after the fabrication to be protected from the environmental influences as well as lithography processes.

# 9

## 9. Summary and outlook

## 9.1. Summary

This thesis explored the synthesis of epitaxial graphene on silicon carbide for the fabrication of high-quality samples required, particularly in metrological applications. To meet this aim, several optimizations were imposed to gain reproducible and robust processes. Various characterization techniques were employed to scrutinize the structural and electronic properties of the samples. These have led to a better understanding of the complex fabrication processes, influence of multiple contributing parameters, considerations in device fabrications, as well as substrate and graphene layers interplay. The experimental results of this thesis can be categorized into four detailed studies that the highlights are briefly summarized in the following.

● **T**he first experimental section included a detailed study on the synthesis of five different sample types, i.e., epitaxial buffer layer, epitaxial monolayer graphene, epitaxial bilayer graphene, and two others known as quasi-freestanding monolayer and bilayer (QFMLG/QFBLG), each with certain characteristics features. Accordingly, high-quality ultra-smooth bilayer-free monolayer graphene with unprecedented reproducibility was successfully grown. [36–39] Similarly, coherent graphene-free buffer layer sheets were fabricated. [38] By this, two practical challenges in buffer layer growth are overcome, i.e., the lack of a buffer layer or formation of extra carbon layers at step regions. Such high quality of the buffer layer is essential when uniform quasi-freestanding monolayer graphene (QFMLG) is desired.

Moreover, such a homogenous buffer layer helped better understand the complex growth, including surface restructuring and recrystallization mechanisms. It was shown that by properly choosing a set of growth parameters, it is possible to prevent structural defects (canyon defects and step defects) and obtain a continuous, large-area buffer layer without graphene inclusions and a bilayer-free graphene monolayer. To this end, it is important to notice the decisive but less-regarded influence of the argon flow rate in the growth process. It was found that for a given temperature and constant Ar pressure, the Ar mass flow-rate strongly influences the SiC decomposition rate, which can qualitatively be understood by thermal equilibrium considerations. This new finding has the potential to improve the graphene quality by avoiding accelerated step bunching at higher temperatures and graphene roughening for lower Ar pressures, respectively.

Furthermore, the optimization to obtain uniform epitaxial bilayer graphene was presented. It is technically challenging to fabricate epitaxial bilayer graphene on SiC (0001) due to a self-limiting effect on the surface. However, by a proper treatment and growth recipe, homogenous epitaxial bilayer graphene was successfully grown.

In addition, the QFMLG and QFBLG produced by hydrogen intercalation exhibit excellent homogeneity over areas in the millimeter range. Herein, in addition to the major intercalation parameters (pressure, temperature, time), also the intercalant purity is vital in the efficiency of the intercalation process, as was evidenced by two detailed investigations: using (i) 5% Hydrogen (95% Ar) and (ii) low-concentration oxygen impurities in nitrogen 5N (99.999%). From these, two consistent implications were deduced. Firstly, the intercalation occurred consistently in both cases, and secondly, the process is not as efficient as when pure intercalant agents are applied. This low concentration can be somewhat compensated by prolonging the process; however, it cannot be overextended as otherwise induces other defects due to etching effects that occur in graphene layers.

● **In** the second experimental part, the quality of produced types was analyzed, conducting electronic transport measurements from nanometer to millimeter scale. An important challenge confronting the electronic applications of epitaxial graphene is related to its extrinsic resistance anisotropy. Using an angle-dependent nano four-point probe (N4PP), it was demonstrated that monolayer epigraphene can be produced on both 4H- and 6H-SiC(0001) polytypes with a resistance anisotropy as low as only 2%. This study systematically compared various samples with different quality (e.g., step height, thickness variation, miscut angle) concluded that the anisotropy value is traced back to the step resistances of the monolayer graphene across the SiC steps as also verified by scanning tunneling potentiometry (STP) transport on the nanoscale. The main reason for the vanishing small resistance anisotropy was identified to be the absence of bilayer domains. Eventually, the small remaining resistance anisotropy was entirely attributed to the resistance and the number of substrate steps that induce local scattering. Thereby, the data represented the ultimate limit for resistance isotropy of epitaxial graphene on SiC for the given miscut of the substrate. Similar results were deduced from the mm-scale Van der Pauw measurements underlining the excellent uniformity of the samples. Also, the QFMLG and QFBLG produced by hydrogen intercalation exhibit excellent homogeneity and very small resistance anisotropy over areas in the millimeter range. This also indicates the presence of coherent quasi-free-standing graphene layers over large areas.

Because SiC substrate steps cannot be entirely avoided, it is impossible to achieve perfect resistance isotropy for epitaxial graphene. However, for the produced bilayer-free graphene on ultralow terraces, negligible small deviations from isotropy can be obtained. This shows significant progress in the fabrication of uniform epitaxial graphene layers. In general, this study highlights the importance of bilayer-free graphene growth for all kinds of epitaxial growth techniques whenever isotropic properties are demanded for optimum device performance. It makes the device orientation independent of step direction and improves the freedom for designing device layouts, thereby promoting the potential for future device applications of epitaxial graphene.

● **B**ased on the third experimental results, it was shown that the presented advanced growth yields a precise graphene thickness control. Thereby graphene can be fabricated on either identical or non-identical hexagonal SiC terrace terminations. This provides an excellent platform for studying the mesoscopic interaction between the individual SiC terminations and the top carbon layers. Accordingly, a tight correlation between the termination of the SiC stacking and the graphene properties was observed in various characterization techniques. This correlation was attributed to a proximity effect of the SiC termination-dependent polarization doping on the overlying graphene layer. Therefore, it was concluded that the stacking termination of the SiC terraces types in association with their cubic and hexagonality nature has a biasing effect on the surface potential and the doping level of the overlying epitaxial monolayer graphene. The unambiguous identification and assignment of the SiC terraces were described within the framework of an extended SiC step retraction model.

● **F**inally, the fourth experimental study devoted to magneto-transport in the produced graphene types. Two techniques, namely electrostatic-gating and chemical-gating, were employed to tune the graphene samples close to Dirac points to reach the quantization ($v = 2$) at reasonably low magnetic fields (< 5T). This goal was accomplished through these techniques; however, each method has certain drawbacks that make them not entirely suitable for a reliable, robust, and durable system as is demanded in quantum resistance metrology. Therefore, other alternatives should be sought. The measurements, however, underline the excellent quality of the samples. To protect the graphene layers from the possible defects during lithography processes and/or other environmental influences, it was suggested to initially encapsulate them with a depositing layer, e.g., $Al_2O_3$ by atomic layer deposition as shown in Appendix **A7**.

In short, the presented PASG and fine growth optimization methods bear the potential to reduce the terrace step heights down to an ultimate level of a single Si-C bilayer. For the produced bilayer-free graphene on ultralow terraces, negligible small deviations from the resistance isotropy can be obtained. This study supports the promising application potential of epigraphene on SiC for quantum Hall metrology applications as well as QFMLG and QFBLG for superior transistor performances and extends the capability of epigraphene or the buffer layer to be implemented as a platform for growing other 2D materials or metamaterials.

## 9.2. Outlook

From the growth optimization, it was concluded that an early-growth-stage suppression of step-bunching is the key component to improve the quality of epitaxial graphene. From the initial results, it seems yet that the step-bunching may ultimately be frozen. This meets the objective of optimized epitaxial graphene fabrication and motivates further investigations.

Moreover, the quality of epitaxial bilayer graphene synthesis can be even further optimized and improved to gain higher thickness control to yield pure bilayer graphene.

Different defects originating from the SiC substrate or growth bear interesting properties. For instance, it was shown that triangular-shaped structures could form on the SiC surface during graphenization, in which even under special conditions can thoroughly be densely-grown beside each other (see Appendix **A4**). [37,38,298] These structures provide intriguing zero-dimensional platforms for further investigations. For example, recent studies demonstrate a room-temperature strain-induced quantum Hall phase in graphene due to giant pseudomagnetic fields (> 41 T) induced by triangular nanoprisms. [38,365,366] Moreover, the variation of SiC miscut angles on the surface leads to the formation of different domains wherein the polarity may change due to variation in SiC crystal planes. An example can be seen in Appendix **A8**. Also, a deformation of epigraphene in ambient or under different measurement conditions, i.e., SEM and LEEM, was observed, see Appendix **A9**. Further studies are required to understand the origin of these defects.

Although the buffer layer's $(6\sqrt{3} \times 6\sqrt{3})R30°$ structure was already revealed back in 1975 [8], its true structure is still not completely clear. For instance, recent studies predict that rather than a single honeycomb super-lattice, other crystal symmetries might coexist. [21,310,367,368] Additionally, from the SiC crystal rotation and the following buffer layer reported in this thesis (see Chapter **7**), a question arises about the possible formation of buffer layer domains with boundary dislocation on different (non-identical) terraces. Assuming honeycomb crest of the buffer layer superlattice, these domains should not form, but considering probable other coexisting symmetries, it is then likely to have boundary dislocations in the buffer layer structure. Moreover, in some studies, the buffer layer (due to its covalent bonds to the SiC) is considered as being part of the SiC bulk too. [369] However, one may need to consider a probable dissimilar bonding nature. For example, for the 6H-SiC, it was shown in Chapter **7** that the position of Si atoms on S3 (*h* position) and S2 (*k* position) is different therefore the corresponding bonds to the top buffer layer maybe not identical, as it is also the case in the bulk of 6H-SiC. These issues require further theoretical and experimental investigations.

Moreover, by the presented advanced growth, the graphene was formed on non-identical SiC terminations. On such periodic surface configurations, the top graphene layers exhibit an alternating doping variation in association with the underlying SiC surfaces. The effect is attributed to the polarization effect originating from the pyroelectric nature of hexagonal SiC. This finding opens a new approach for a nano-scale doping-engineering on dielectric polar substrates without the use of impurity dopants. This is expected to be a general effect obtainable in other polar dielectric materials like GaN. Additionally, such alternating terrace types in the hexagonal SiC have possible nonequal basal vacancy/divacancy defect states, which can occur in a quasi-cubic or quasi-hexagonal position. [324] Such possible symmetric location is an added value for potential applications and motivates further research.

Among the three main challenges facing in epigraphene-based electronics, i.e., the size restraints, reproducible thickness control, and "as-grown" intense carrier concentration, only the latter is the biggest remaining hurdle which requires appropriate alternatives. When considering the epigraphene-based quantum resistance metrology, a fine-tuning of charge carrier density with long-term stability (ideally constant) with proper graphene-isolation from the environmental influences is essential. These requirements have not yet been fully accomplished by the existing methods and demand proper approaches.

Aside from that, based on a theoretical investigation of the magneto-transport in stepped graphene, a new kind of Aharonov–Bohm interferometers is expected. [370,371] The epigraphene in this study was produced on very regular steps-terraces with 90% in a periodic manner. [38,39] Therefore, it is interesting to investigate the direction dependency of the applied magnetic field in QHE measurements on such samples to test the abovementioned theory.

Epitaxial graphene is also a versatile platform for the fabrication of other sub-dimensional materials. It was shown that the intrinsic SiC micropipes and additional holes induced by the growth process (see Appendix **A7** and **A10**) as "mediation sites" can utilize gallium intercalation. This leads to the fabrication of graphene-metallene heterostacks of Van der Waals combined 2D-layers of bilayer graphene (BLG) and gallium stacking layers on SiC substrate (SiC/Ga/BLG). Initial results obtained from VdP setup and the Hall effect measurements on large-scale samples show robust macroscopic superconductivity at $T < 4$ K. The features observed in the VdP measurement, however, cannot be explained entirely and needs future works. The results, however, are in good agreement with the recent work published in ref. [41].

Finally, it seems that large-area fabrication of 2D/sub-dimensional materials systems is generally still in their infancy. There exist a limited number of manufacturers, and the main clients are still research labs. Industrialization and commercialization of these materials probably will need at least a decade from now. As seen from such a perspective, this would be a "golden period" for start-up companies, which can effectively boost the advancement of technologies in 2D materials synthesis and technologies for being adapted in the conventional semiconductor industry as the next-generation alternatives.

# Appendix

## A1. Pen-patterning

Decreasing the size down to atomic scales requires special consideration for device microfabrication. In particular, two-dimensional material systems are sensitive to conventional lithography, which often leaves significant contaminations on the samples. This is highly degrading to the electronic properties of 2D sheets like graphene.

Here within a framework of an exemplary application, a low-cost, quick, and easy-to-apply technique is presented, which has a potential for patterning contacts to the graphene[*]. As mentioned in Chapter **4**, the VdP setup used in this work causes local damages to the graphene sample, where the fixed equidistant gold pins meet the graphene, see **Figure 3.13b**, and **Figure 6.6**. Therefore, the samples after the first measurement and taking out, most of the time, could not be reused for further investigations. To deal with this problem, as well as avoiding durable lithography processes, which may even cause other surface inhomogeneities (e.g., polymer adsorbates), a so-called pen-patterning is presented. **Figure 9.1** illustrates the simple fabrication process of graphene Van der Pauw structures by handwriting the patterns to be defined using a soft pen, which includes four steps.

The process begins with patterning the sample for etching of graphene at the contact areas. Firstly, the graphene areas not to be etched are covered by ink by pen writing. Then the contact areas (which are free of ink) on the sample ($5 \times 5$ mm$^2$) are etched by an oxygen/argon gas mixture using the ink as a mask. As **Figure 9.1(ii)** shows, the etching was successful, and graphene below the ink remained safe, even applying the exact etching parameters that were used for etching the graphene after the EBL process on PMMA (see Chapter **4**). Then the sample can be easily cleaned using acetone and isopropanol. In the third step, the sample is patterned for Ti/Au (20 nm/50 nm) contacts metallization. Ti sticks well to the etched areas (SiC) and is covered by the gold layer, which all overlap the graphene areas, as shown in **Figure 9.1(iii)**. Finally, the lift-off process is quickly performed by immersing the sample in an acetone/isopropanol beaker and ultrasonic bath for 10 minutes. The following VdP measurements verified the proper electronic contacts, indicating mobility of $\mu \approx 900$ cm$^2$/Vs with an electron density of $n \approx 1.7 \times 10^{13}$ cm$^{-2}$. The entire fabrication process, excluding the time for system evacuation (i.e., Ar/O$_2$ plasma or metallization device), takes about 1 hour, which is significantly timesaving compared to a conventional lithography process. Here, the pen-patterning concept was performed manually using a Staedtler Lumocolor pen. It can be much improved, for example, by implementing available drawing robots in the market to obtain sharp edges for the patterns. Moreover, employing this technique, the graphene could be easily locally etched or intercalated. The demonstrated process is an initial prototype that has the potential for improvements and applications.

---

[*] The author acknowledges valuable discussion with A. Fernandez.

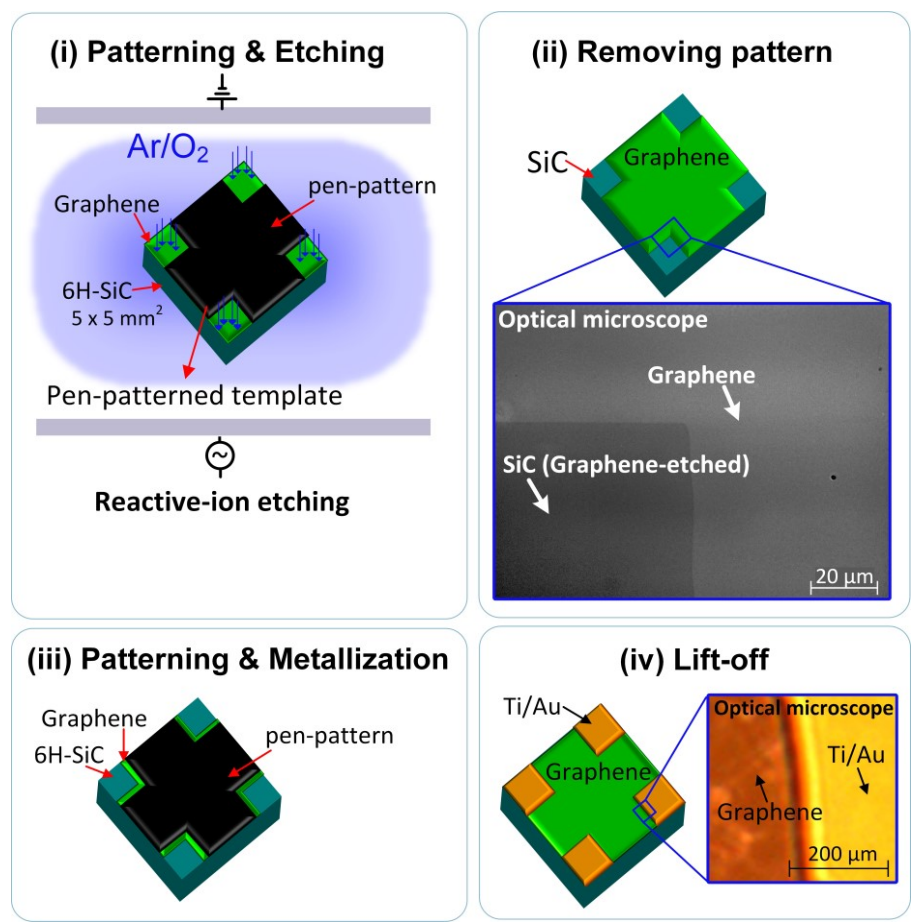

**Figure 9.1. Pen-patterning prototype.**

The pen-patterning procedure for the fabrication of graphene Van der Pauw structures has four successive steps. (i) The pattern is sketched on the sample and the contact areas where the graphene is left without ink-covered are etched away using oxygen/argon RIE plasma. (ii) The pattern is then simply washed away by acetone and isopropanol. The optical microscope image demonstrates successful etching. (iii) A new pattern is drawn for the metallization. Narrow graphene lines (depicted with green color) are not covered with ink to provide graphene/gold contacting overlaps. Metallization is performed by 20nm Ti and 50 nm Au deposition. (iv) A fast lift-off process can be easily done by acetone/isopropanol cleaning in the ultrasonic bath. The optical microscope image demonstrates the graphene and contact regions after the cleaning. The patterning was carried out manually, therefore the edge areas are not sharp. However, it can be substantially improved using available drawing machines in the market.

## A2. Magneto-transport in strong *p*-type epigraphene

A charge carrier drift may occur in unprotected bare epigraphene since the graphene is sensitive to its environment, e.g., the water and oxygen existed in air through surface physisorption alter the doping towards hole-doping. [317,372,373] Here, for a ready epigraphene Hall-bar, which was annealed in a dry-nitrogen oven (at 90 °C for 72 hours, 1 bar) and later covered with 55 nm copolymer, a strong *p*-doping is deduced from the QHE measurement shown in **Figure 9.2**. One could achieve lower doping (i.e., sufficiently close to the Dirac point), by a shorter annealing period.

**Figure 9.2. Magneto-transport in strong hole-doped SiC/G.**

The QHE measurement in a monolayer epigraphene reveals a strong hole-doping with $p = \sim 4.6 \times 10^{12}$ cm$^{-2}$, and $\mu = \sim 350$ cm²/Vs. The graphene-Hall bar was annealed at 90° for 72 hours in a dry-nitrogen oven. The sample was next coated with 55 nm copolymer, bonded on a chip carrier, and without any further treatment was measured at $I = 1\mu$A, $T$=300 mK.

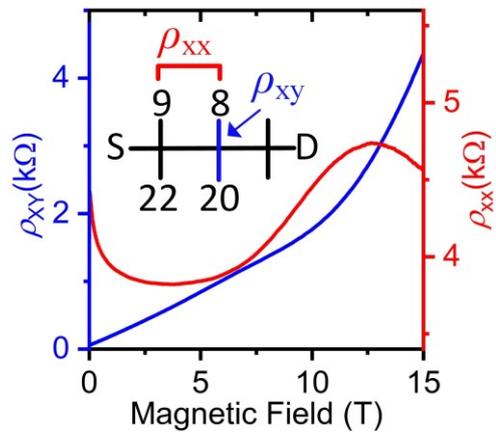

## A3. Influence of Ar flow-rate on conventional epitaxial growth of buffer layer

**Figure 9.3** shows the influence of the argon mass flow rate on buffer layer growth on the samples without polymer preparation that are investigated by AFM and SEM. Three samples S′$_0$, S′$_{100}$, and S′$_{1000}$ are 4H-SiC with a nominal miscut of about -0.06° towards [1$\bar{1}$00] and were processed at 1400 °C (1 bar argon atm., for 30 min) at different Ar flow of 0, 100, and 1000 sccm, respectively. This experiment was carried out under the same conditions like the one in Chapter **5**, and it aims to study the influence of argon flow on samples grown from another SiC polytype (4H-SiC) in the absence of polymer preparation.

For the high argon flow of 1000 sccm, the sample's surface undergoes strong step bunching without any buffer layer growth **Figure 9.3g, h, i**. This is similar to the PASG sample in Chapter **5** (see **Figure 5.1g, h, i**). In both cases, the high Ar flow leads to surface etching. For moderate Ar flow, however, the situation is different. The surface of the sample without polymer preparation shows stripes of the covered buffer layer and bare SiC, **Figure 9.3d, e, f**. This is in contrast to the PASG sample (**Figure 5.1d, e, f**) where the provided carbon species from polymer lead to surface super-saturation and well buffer layer coverage, although the Ar flow caused canyon-like defects. For the case of zero argon flow (**Figure 9.3a, b, c**), the surface looks very good with uniform coverage, while the terraces appear less ordered in comparison with the PASG sample (**Figure 5.1a, b, c**).

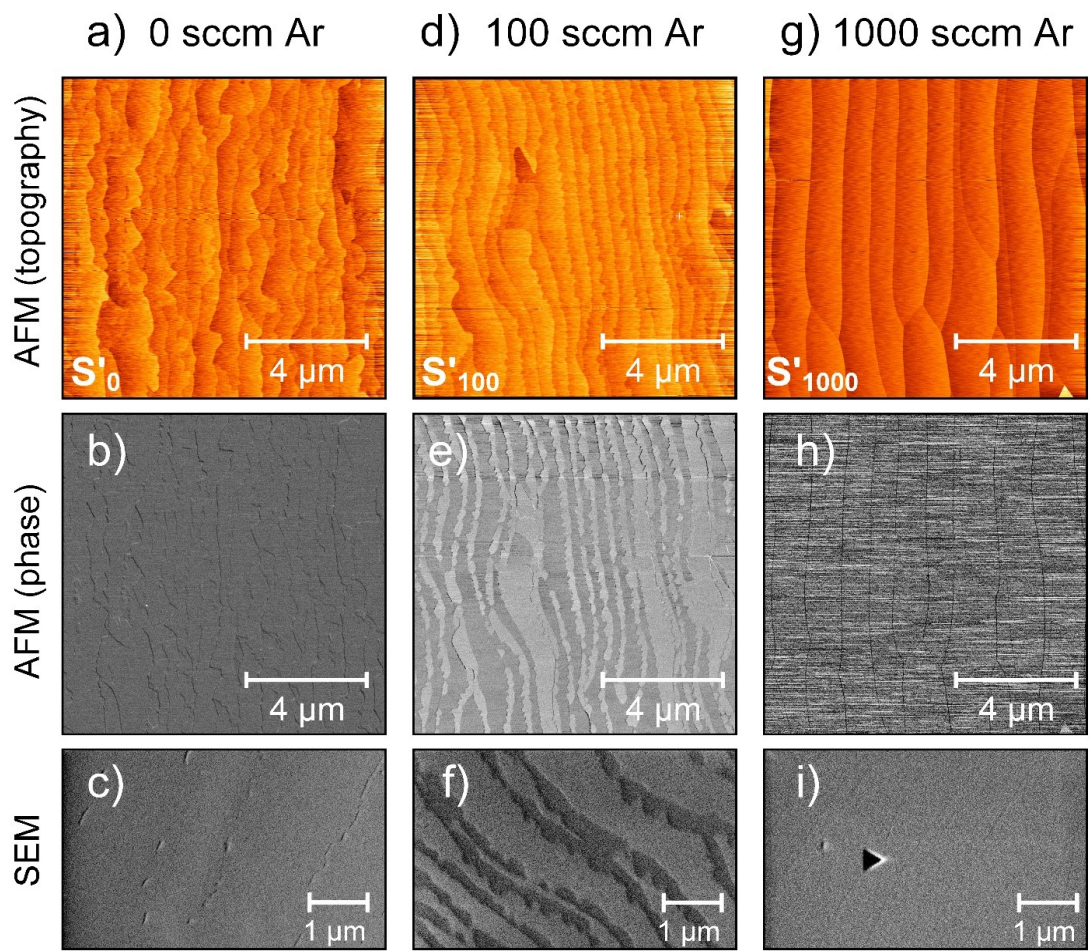

**Figure 9.3. Influence of Ar flow-rate on graphitization of 4H–SiC (0001) without PASG.**

Inspecting the influence of the argon mass flow on graphitization of 4H-SiC(0001) at 1400°C (1bar Ar ambient, 30 min) under three different argon gas flows: (a) $S'_0$ (Ar/ 0 sccm), (b) $S'_{100}$ (Ar/100 sccm), and (c) $S'_{1000}$ (Ar/1000 sccm). The sample was grown by typical sublimation growth without polymer preparation. $S'_0$ processed under no Ar flow representing good buffer layer coverage in AFM phase (b) and scanning electron microscopy SEM (1kV) (c) images. The moderate flow Ar for $S'_{100}$ distorts its surface growth, causing the formation of buffer layer stripes on this sample, as can be seen in the AFM phase (e) and SEM (f) images. The intensive argon flux on $S'_{1000}$ prevents buffer layer formation and results in severe step bunching on this sample (g), (h), (i).

## A4. Formation of triangular-shaped structures

It is observed that the increase of the Ar flow leads to the formation of triangular-like structures. This can be seen in **Figure 9.3g, h, and i,** (and also refs. [37,38]) for the sample processed under 1000 sccm Ar flow. Also, rather increase of the Ar flow escalates the density of such structures, as shown in **Figure 9.4** for the sample processed at 1400 °C (30 min, 1 bar Ar) in the presence of the Ar flow of 2000 sccm. The aggregated mass along with the giant steps and the triangular-like structures is the typical morphology all over the surface of this sample. Although, here, the properties of such triangular-shaped structures have not been further studied, however, they very resemble the cubic SiC grown on other substrates elsewhere. [374,375]

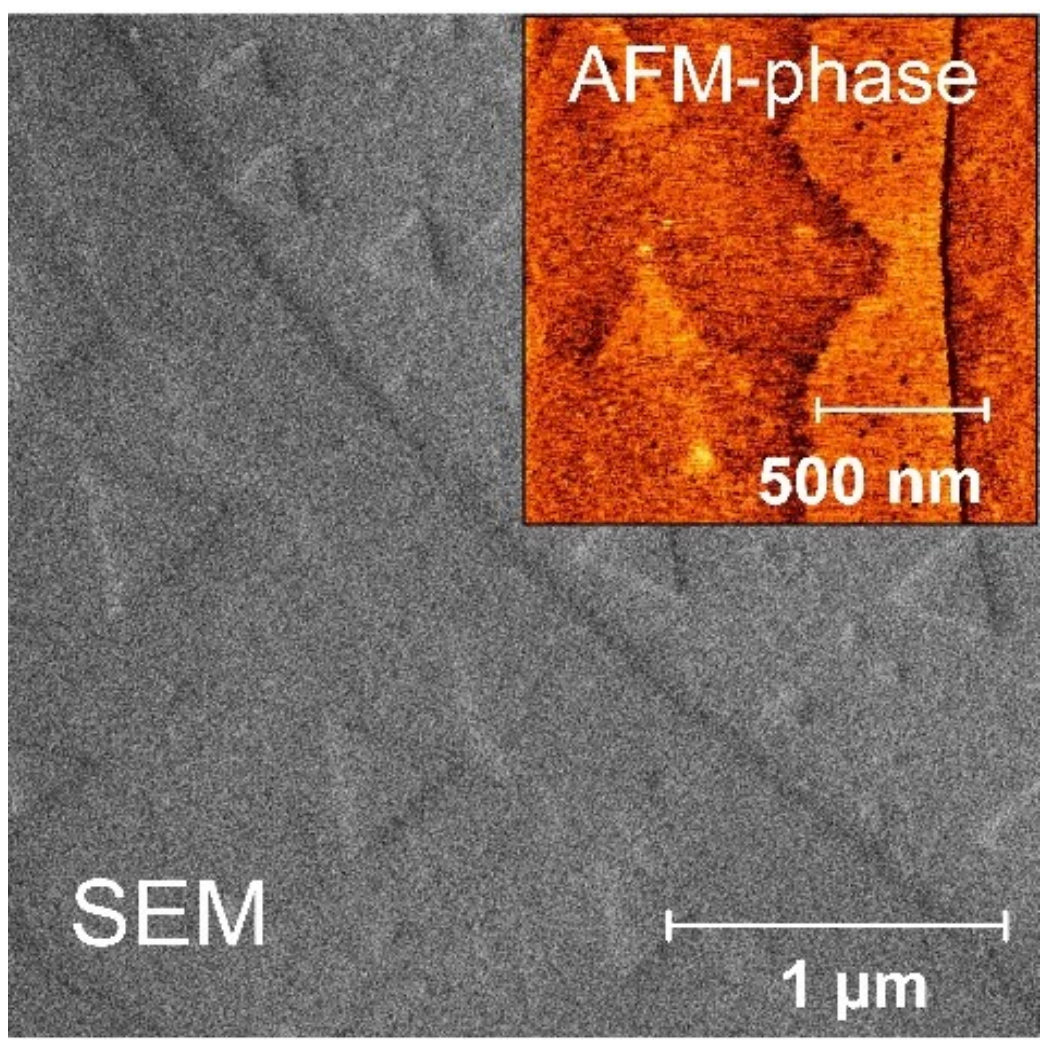

**Figure 9.4. Formation of triangular-like structures at high argon gas flow.**
Scanning electron microscopy (1kV) of a 6H-SiC sample after annealing at 1400°C (1 bar in Ar ambient, 30 min) under 2000 sccm Ar flow-rate. The inset shows the AFM phase image of the same sample.

Moreover, such triangle-shape structures appear not merely under intensive gas flow but also under a lower pressure condition. **Figure 9.5** shows AFM inspection on the surface of a 6H-SiC sample, which is processed at 50 mbar (1400 °C, 30 min). The triangle-like structures appear to cover the substrate entirely.

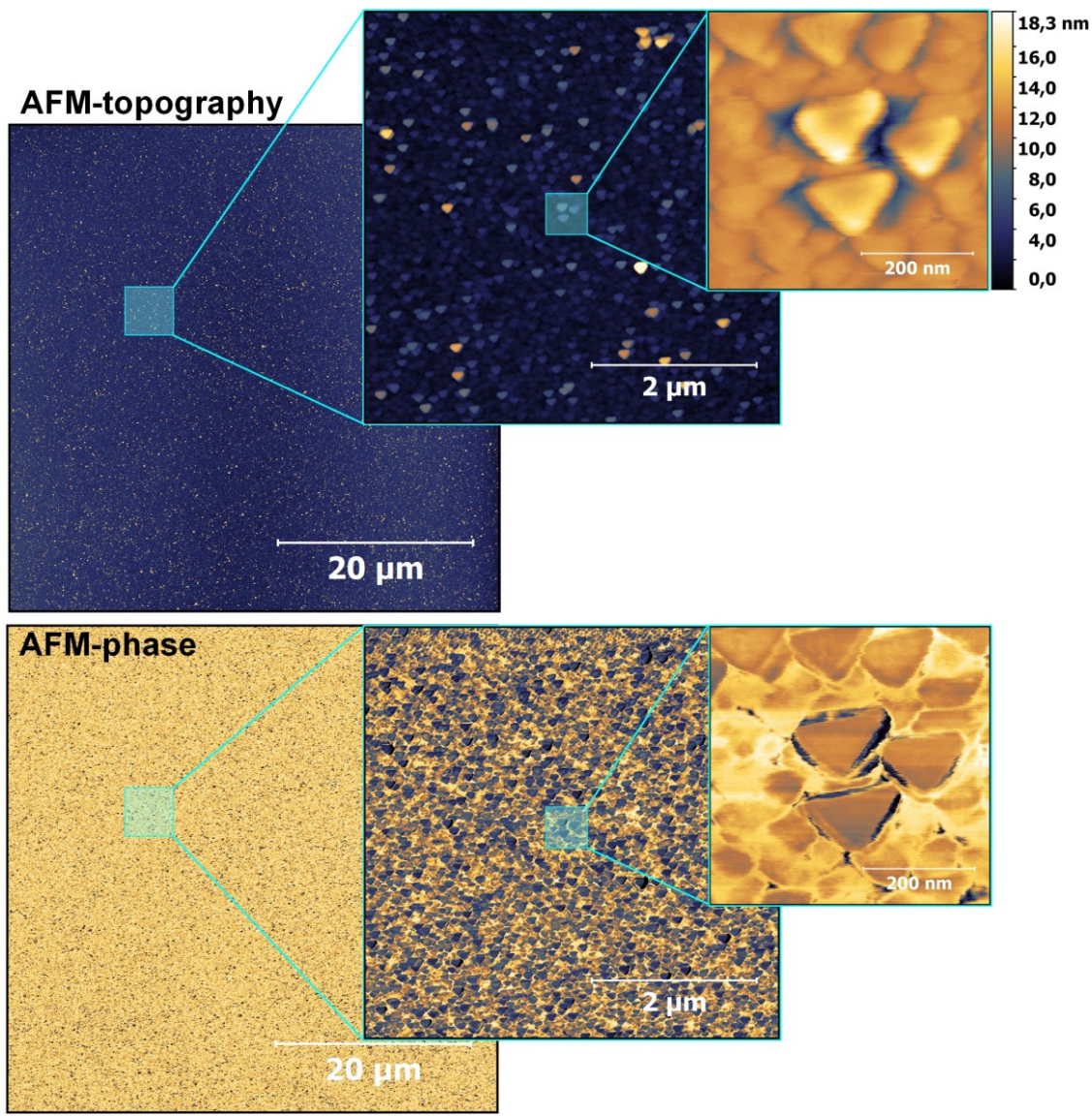

**Figure 9.5. Densely-grown triangular-form structures on 6H–SiC.**
AFM topography and phase images of triangular-form structures. These structures appeared on the surface of a 6H–SiC sample processed at 1400 °C (30 min), but the pressure was set to 50 mbar instead of the standard ~1 bar of argon.

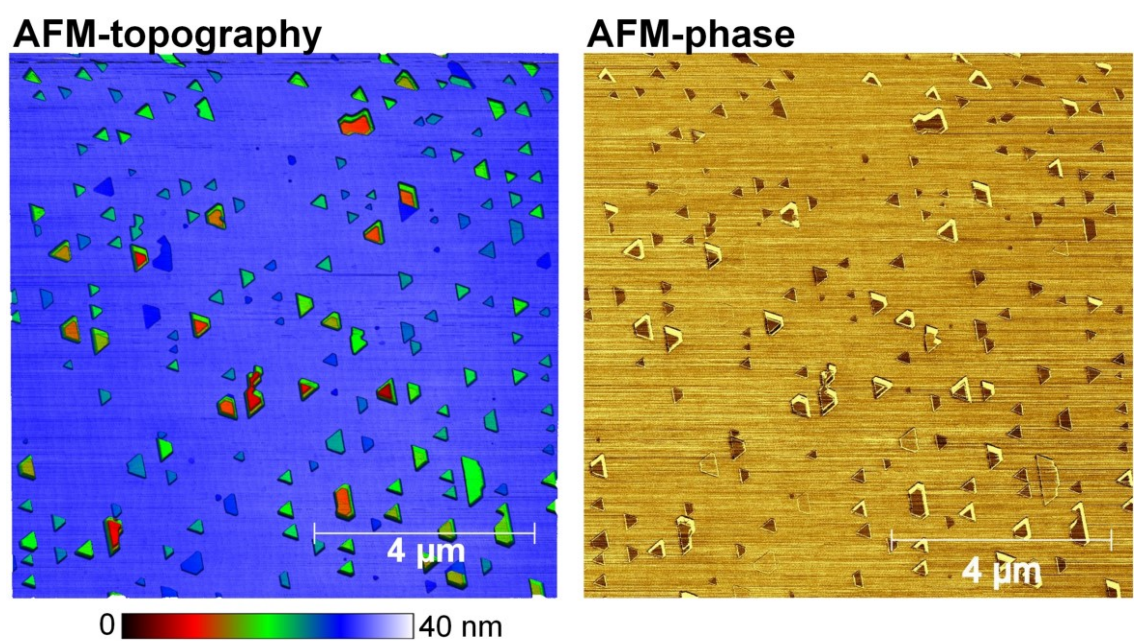

**AFM-topography**

**AFM-phase**

0 [colour scale bar] 40 nm

**Figure 9.6. Triangular-like structures in SiC/G.**
AFM topography and phase images indicate that the triangular-like structures can also be formed in epigraphene samples.

Moreover, triangular-like structures can also be formed during the graphene growth, see **Figure 9.6**. Such so-called nanoprisms in the SiC substrate were recently reported to be able to generate strain-induced uniform fields of ~41 T, enabling the observation of strain-induced Landau levels at room temperature. [38,365,366] Considering the orientation of the triangular structures, it is noticed that they are dispersed on the surface of the sample, interestingly, either at 0°, 120° or 180° with respect to each other, most likely due to the staking fault regions in the bottom substrate.

## A5. Step-retraction model of 6H-SiC/G

The step flow model for 6H-SiC/G was described in Chapter **7** is shown here in more detail. **Figure 9.7** shows the proposed step-flow model, which combines the AFM experiments results, illustrates that the decomposition of 6H-SiC terrace-step leads to three types of step-bunching scenarios: (i) terraces appear in pairwise sequences with a bilayer (1L) and 2L steps terminating to S3/S3* and S2/S2*, respectively, (ii) 3L steps terminating to S3/S3*, or (iii) 3L steps terminating to S2/S2*. Each of these situations represents a step sequence in which the attained termination is accompanied by the lowest surface energy.

### Figure 9.7. Schematic of the step-flow model of PASG on 6H-SiC.

(a) Initial state: three different terraces (step-heights ~0.25 nm) types named S3/S3, S2/S2*, and S1/S1* shape the surface of 6H-SiC before the growth. These terminations have inequivalent surface energies, which lead to different surface decomposition velocities. (b) After intermediate annealing (buffer layer growth) at about 1300-1400 °C, the S1/S1* with the highest surface energy (lower stability) disappears, leaving the surface with two remaining surfaces of S2/S2* and S3/S3* which govern the step bunching for the rest of the growth. As a result, the terrace widths of S2/S2* is smaller than S3/S3*.

(c) Since the decomposition velocity of S3/S3* is faster than S2/S2*, the latter extends in width as the surface undergoes further recrystallization and restructuring. Three terrace-step configurations are identified: orange color (i) sequential pattern of steps with heights of ~0.25/~0.5 nm, blue color (ii) ~0.75 nm steps terminate to S3 and S3* surfaces, and red color (iii) where ~0.75 nm steps terminate to S2 and S2* terraces.

Configuration (ii) and (iii) usually occur while the SiC terraces are not equal in width in the initial state (the exemplary missing terraces are schematically shown with red hachured-squares).

(d) Formation of higher steps of ~0.75 nm by further annealing. This model only illustrates the SiC surfaces, and atop carbon, layers are omitted for the sake of simplicity. (See also Chapter 7).

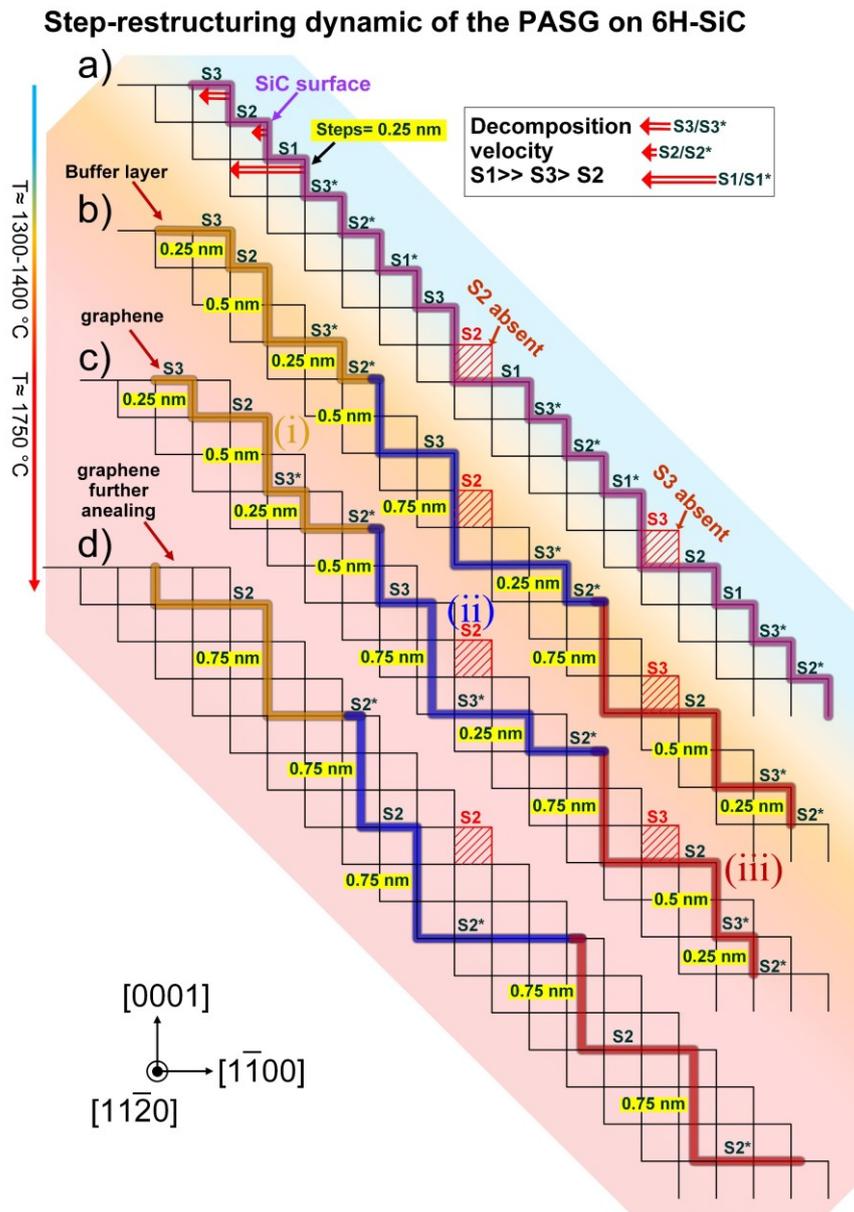

Step-restructuring dynamic of the PASG on 6H-SiC

## A6. Influence of SiC polytype in substrate-induced doping

Considering the stacking order-induced doping variation that was found for 6H-SiC/G, it is interesting to study this effect in the 4H polytype too. As illustrated in **Figure 9.8**, the 4H-SiC is composed of 4 Si−C bilayers, in which two of those are energetically different (i.e., S2/S2* and S1/S1*), thereby leads to two distinct types of decomposition velocities, indicated by horizontal arrows in as shown in **Figure 9.8**. However, similar to the 6H polytype, the Sn and Sn* terminations have the same structure except for 60° crystal rotation with respect to each other. The 4H-SiC, in **Figure 9.8**, has hexagonal symmetry with 50% hexagonality owns an equal number of hexagonal and cubic stacking layers. For the 8 atoms in the unit cell of the 4H polytype, 2 are non-equivalent for both Si and C atoms regarding their positions as of hexagonal (*h*) or cubic (*k*), as each is depicted in **Figure 9.8**. The hexagonality of each surface terrace can be considered as the joint hexagonality of the corresponding on- and off-bonds, shown in **Figure 9.8**.

The identification of the 4H-SiC terminations after graphene growth can be described in a framework of the step-flow model. The concept of the step-flow model for the 4H-SiC/G is similar to the 6H-SiC/G (see Chapter **7** and **Figure 9.7**) with less complexity. Let us see the experimental result of graphene growth on 4H–SiC, as demonstrated in **Figure 9.9**. **Figure 9.9** shows the AFM topography, phase, and step height profile of a 4H-SiC/G sample. At first sight, a sequential steps pattern, like for the 6H-SiC/G (**Figure 7.2**), is not observed. [39] This is due to the nature of the 4H-SiC and the number of stacking layers in its primitive unit cell that register a different shape of step patterns, in comparison to 6H-SiC/G. The sample's topography clearly indicates the uniformity and homogeneity of the graphene growth on this polytype as the terraces-steps regularly shaping the surface as on the 6H-SiC. However, interestingly, phase contrast is randomly observed in the AFM phase image of this sample too, which is highlighted with green color marks in **Figure 9.9**. Alike already proved for the 6H-SiC/G, it can be reasonably expected that the origin of the phase contrast in the AFM phase image of the 4H-SiC/G sample is again rooted in the dissimilarity of the SiC surface terminations. This can be simply examined as follows. By knowing that the S2/S2* terminations are the most stable ones, reasonably, these terminations are most likely the last standing terraces remaining below the graphene and buffer layers after the growth. Accordingly, it can be deduced that the brighter phase contrast in the AFM phase images which statistically have higher distribution are S2/S2*, and the randomly distributed terraces with darker contrasts are S1/S1*. Keeping this in mind, we can subsequently map the surfaces in the step-height profile (terraces-steps aligned downwards), as shown in **Figure 9.9**. To do so, by considering the experimentally measured step heights, we can go downwards from one terrace to the next terrace and identify each one with respect to the stacking order in the unit cell and the distance that we walk downward (1L, 2L, 3L, etc.). This mapping is shown in **Figure 9.9**, wherein each terrace is marked with white color and the step-height with a highlighted yellow color. As can be seen in **Figure 9.9**, the initiation of the mapping started with S2 (it could also be S2*, but it does not change the mapping process drastically because Sn and Sn* are similar except for their 60° crystal rotation with respect to each other).

The mapping of terraces fits captivatingly with the expectations considered and discussed above, and it shows that the phase contrasts are is only observed on dissimilar terraces terminations. Accordingly, the darker and brighter AFM phase contrasts belong to S1/S1* and S2/S2*, respectively. It is worthwhile to mention that this phenomenon is not an accident but observed and examined on many samples, therefore it can be considered as a general fact, as it was already cleared for the 6H-SiC/G.

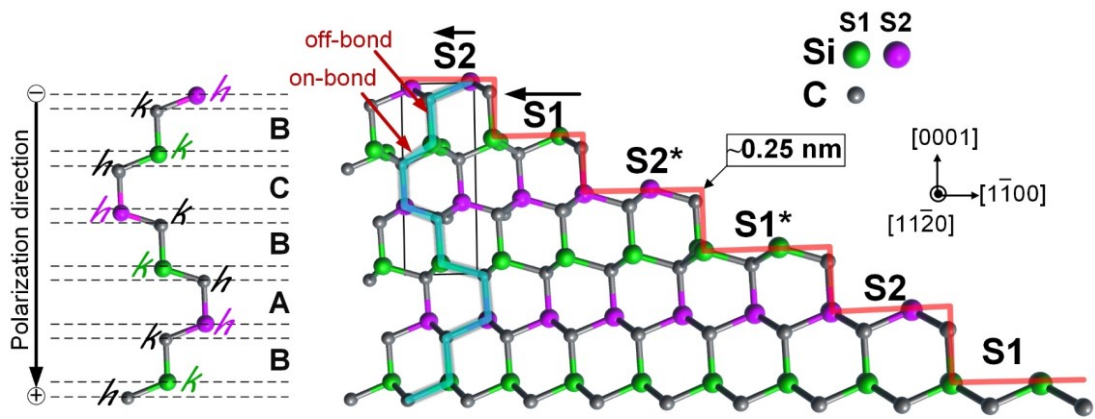

**Figure 9.8. Schematic of the unit cell of 4H-SiC and position of C and Si atoms.**

(Right side) side-view of 4H–SiC (0001) unit cell, which includes four sequential stacks (marked with a black rectangle) of Si–C bilayers projected in (11$\bar{2}$0) plane. The Si–C bilayers are denoted as BCBA, leading to Sn, and Sn* (n=1, 2) terminations, which are energetically similar but are 60° rotated related to each other. For each of Si and C atoms regarding their hexagonal ($h$) and cubic ($k$) positions in the unit cell is shown (left side).

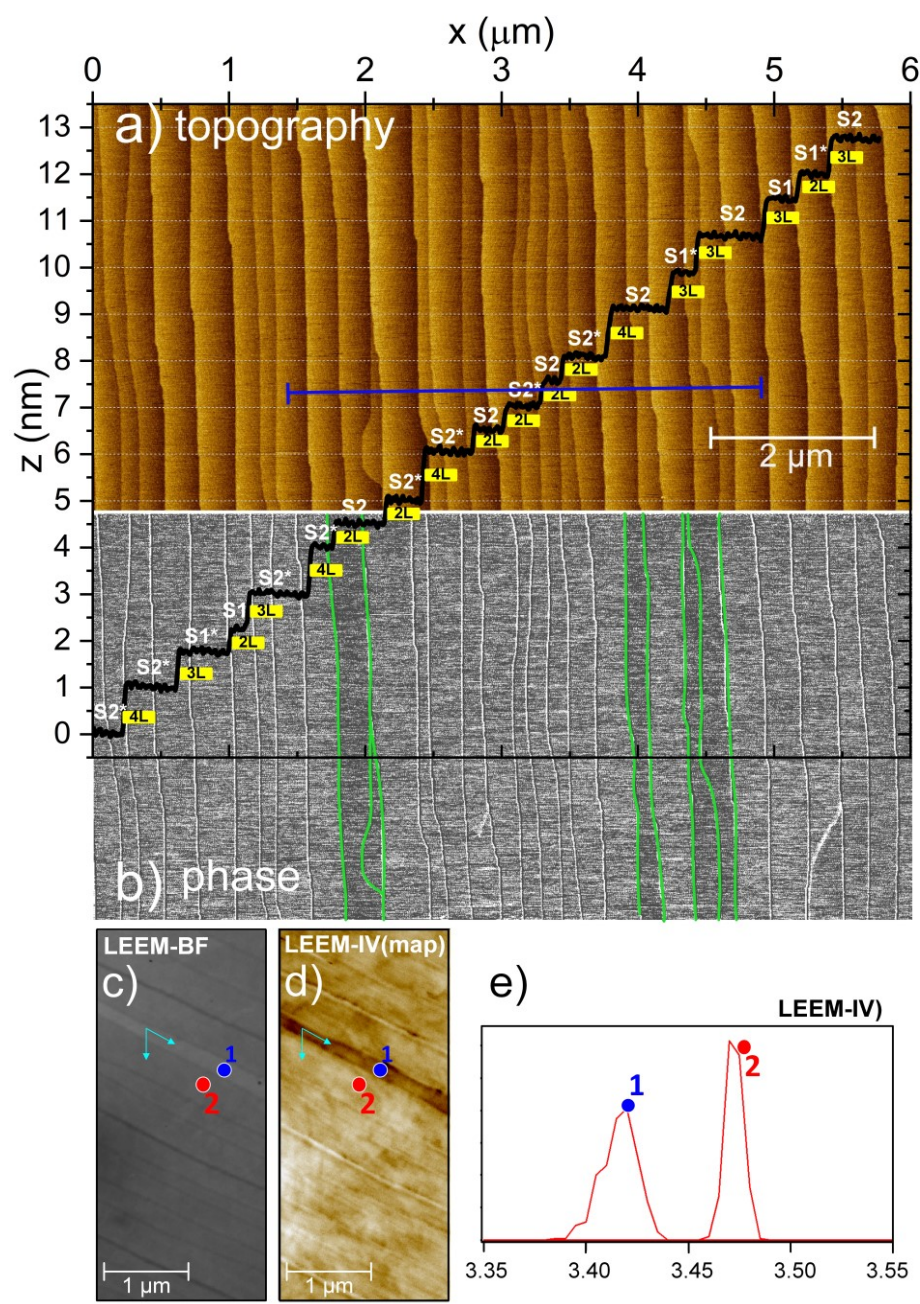

**Figure 9.9. AFM and LEEM-IV studies on monolayer graphene on 4H-SiC(0001) with non-identical surface terminations.**

(a) AFM topography image implies a highly homogenous surface where the terraces-steps shape the surface of the sample in regular order as can be deduced from the step-height profile (inset) from the cross-sectional blue line in (a). (b) AFM phase demonstrates a phase-contrast that appears randomly on the adjacent terraces (marked with green color). Each of the terraces is identified and addressed (white color). The height of each step is highlighted in yellow (1L= 1 Si−C height of ~0.25 nm). (c) LEEM (BF) and (d) LEEM-IV images of monolayer 4H-SiC/G. Similar to 6H-SiC/G in **Figure 7.8**. (e) The Reflectivity curves show two different values with a variation of about ~60 meV for the minimum associated with the graphene interlayer state.

Also, the LEEM investigations on a 4H-SiC/G sample are in agreement with the step-flow model and AFM studies. The LEEM-BF in **Figure 9.9** reveals a reflectivity contrast on the neighboring terraces. More importantly, although the LEEM-IV spectra on these areas indicate a single prominent minimum, which clearly verifies the formation of monolayer graphene on both terrace domains [23,344], however, again the LEEM-IV spectra on the adjacent terraces reveal graphene with an energy difference of about ~60 meV. This is attributed to the two bottom non-equal terminations, i.e., S2/S2* and S1/S1* on 4H-SiC, leading into modulations of electronic properties of top graphene layers. [78]

## A7. Al$_2$O$_3$ encapsulation

It is important to protect the graphene layers from possible defects and adsorbates, e.g., during lithography processes (e.g., see **Figure 8.4**) or environmental influences (e.g., see **Figure 9.12**). A solution could be an initial passivation/encapsulation of a metal or dielectric layer, which can be later structured or removed. **Figure 9.10** demonstrates two examples of this approach by the deposition of a 5nm Al$_2$O$_3$ layer (by ALD method).

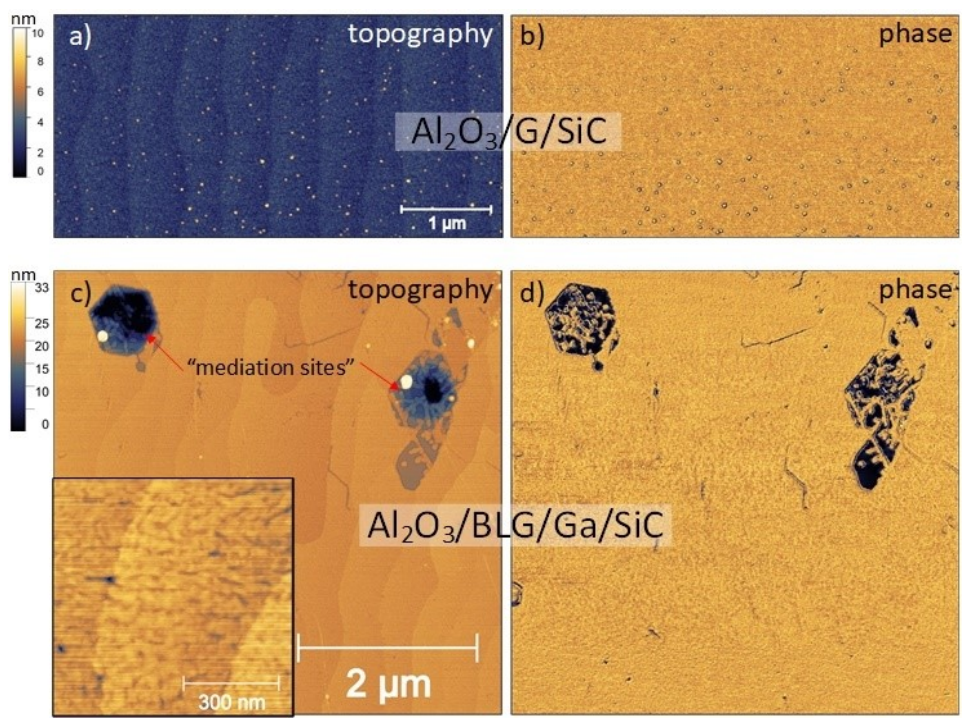

**Figure 9.10. Sample protection by Al$_2$O$_3$ encapsulation.**
Two examples of 2D material systems which are protected by 5nm aluminum oxide. AFM topography (a) and phase (b) images of a graphene sample covered with Al$_2$O$_3$. Similarly, AFM topography (c) and phase (d) images of a heterostack of 2D-gallium (gallenene), bilayer graphene on 6H-SiC isolated with Al$_2$O$_3$. The latter was created by Ga intercalation. (see **A10** for more detail)

# A8. Influence of miscut variation in epigraphene growth

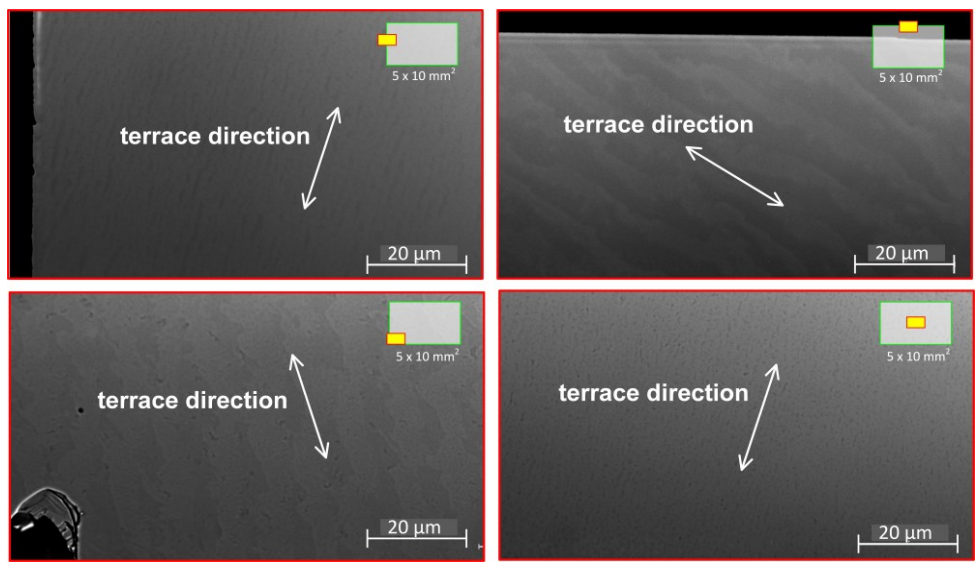

**Figure 9.11. Influence of miscut variation in graphene growth.**
Optical microscope images of a sample with a size of 5 × 10 mm² (sketched as green rectangles in the insets) show intensive miscut variation. Four positions are shown as marked with the yellow rectangles in the inset of each image. From the direction of step-terraces, the variation of the miscut angle can be deduced. While the quality of the sample is good, as seen in the bottom right image, the other images show a substantial deviation from the typical sample's quality. The impact of such miscut angle deviations on, e.g., electronic properties of the graphene layer, requires further investigation.

# A9. Deformation of epitaxial graphene

This section is devoted to an observation regarding graphene deformation under different conditions. This deformation appears as cracks and protrusion areas, as well as dot-like defects on the graphene samples. In the following, this observation by various characterization techniques is demonstrated. **Figure 9.12a-c** shows AFM and Raman investigation on pure monolayer graphene on 6H-SiC with a pairwise step-terrace configuration of ∼0.25/∼0.5 nm. This is named the 1st measurement. In chapters 5-8, different characterization methods verified the high-quality of the samples with almost vanishing resistance anisotropy. [36–39] The 2nd measurement was performed on the same sample after being kept for about 3 months in a sample box in the office under ambient condition. Surprisingly, the AFM topography and phase images in **Figure 9.12d and e** show the appearance of cracks and protrusion areas in the sample, not only at the step regions but also on terraces. The Raman spectroscopy indicates the formation of the bilayer graphene, see **Figure 9.12f**. The depth of cracks and the height of protrusion areas are almost 1 nm, inferred from the AFM height profile and STM image in **Figure 9.12g and h**.

**Figure 9.12. Graphene deformation under ambient condition.**
The 1st measurement frames the AFM (a and b) and Raman (c) measurements indicating a high-quality of single graphene on 6H-SiC, with periodic terraces with steps of ∼0.25/∼0.5 nm height. 2nd measurements belong to the same sample after being kept for 3 months inside a sample box in the office under ambient condition. AFM topography (d) and phase (e) indicate the formation of cracks and protrusion areas on both step and terraces regions. Closer looks are shown in the inset marked as blue circles in (d) and (e). (f) Raman spectroscopy verifies the formation of the bilayer graphene. An about 1 nm depth and height is deduced from the AFM height profile (g) and STM topography (h) for cracks and protrusion areas, respectively.

Almost a similar effect was observed during the SEM and LEEM scanning of the sample. **Figure 9.13a** shows SEM imaging a pure 6H-SiC/G sample, which deforms upon exposure to electron microscopy and cracks dot-like defects appear on the surface. In LEEM imaging, which applies electrons with much lower energy, interestingly more dot-like defects are created, see **Figure 9.13b**.

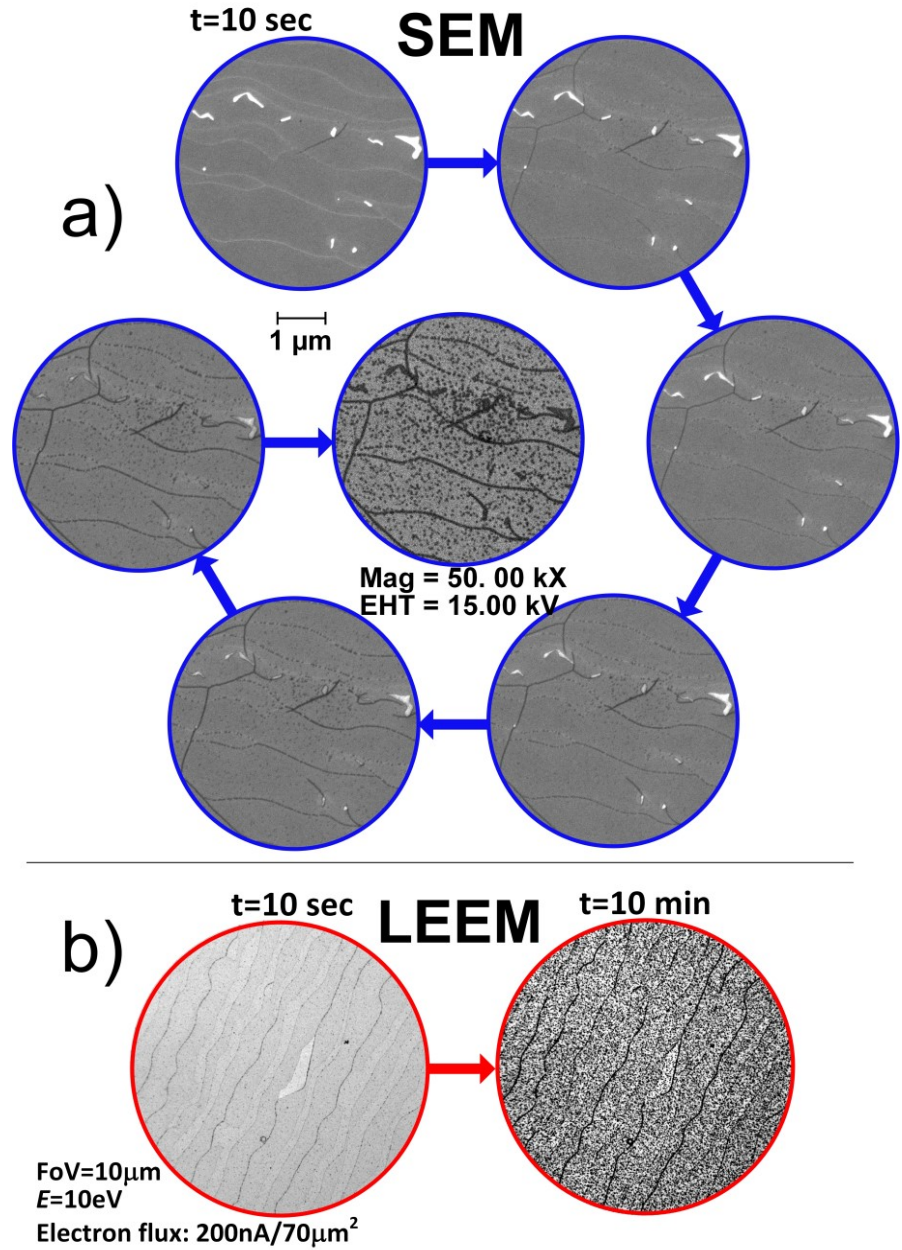

**Figure 9.13. Graphene deformation under electron microscopy.**

(a) Scanning electron microscopy of a monolayer 6H-SiC/G. The entire imaging was done over 35 min. The sample cracks and dot-like defects can be found throughout the scanning area, which develops further over time. When electron energy is lower in LEEM (at MAXIV lab) inspection (b), the surface after 10 min deforms, and plenty of dot-like defects generate on the surface.

A much stronger deformation was observed when the samples were intercalated. Accordingly, applying a simple cleaning in acetone/isopropanol mixture and ultrasonic bath led to severe damages of the quasi-freestanding bilayer graphene sample. **Figure 9.14** shows an optical microscope image of a QFBLG sample after the abovementioned simple cleaning. While on the right side, the graphene is still healthy and well-ordered (terrace-steps can still be identified), on the left side, the graphene is almost heavily destroyed and removed from the surface. The possible origins were shortly discussed in Chapter **9**.

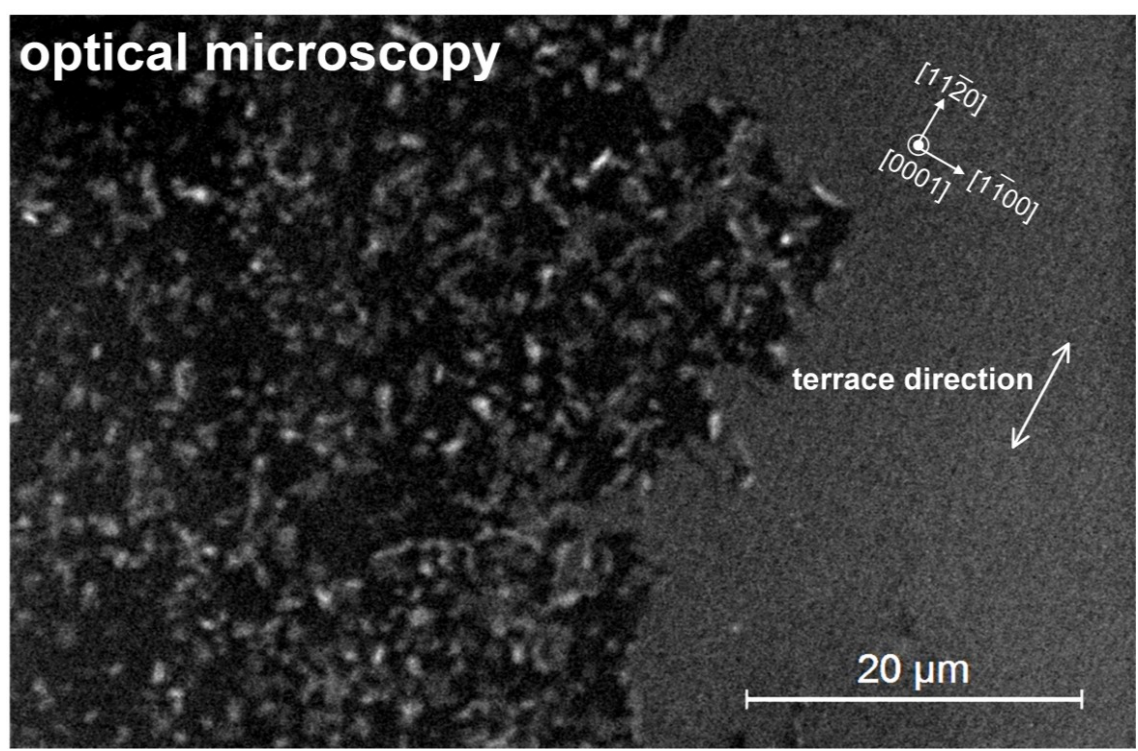

**Figure 9.14. Graphene deformation during cleaning in the ultrasonic bath.**
Optical microscope image of quasi-freestanding bilayer graphene after cleaning in acetone/isopropanol mixture and 15-minutes ultrasonic-bath. On the right side, the graphene is still fine, and terrace-steps can be seen. On the left side, the graphene is strongly damaged and even removed from the surface.

## A10. A VdP study on a large-area BLG/Ga/SiC superconducting heterostack

Epitaxial graphene growth on silicon carbide (SiC/G) is an excellent method for obtaining sizeable graphene layers, a versatile platform for the fabrication of other sub-dimensional materials too. Accordingly, we fabricated a graphene-metallene 2D heterostacks of Van der Waals combined 2D-layers of bilayer graphene (BLG) and gallium stacking layers on SiC substrate (SiC/Ga/BLG). [43]

**Figure 9.15a** shows the optical microscope image demonstrating the propagation of gallium into SiC/G, which lies in between the bottom SiC and top graphene layers. The AFM topography image in **Figure 9.15b** highlights the excellent homogeneity of the samples regarding the small step heights (mostly < 1 nm). The ultra-smooth morphology of the SiC/G as a result of effective suppression of surface step-bunching during the epitaxial growth facilitates a homogenous formation of 2D-Ga layers. [41] The AFM image was captured at a border region where the SiC/G and SiC/Ga/BLG coexist. This material difference is projected in the optical microscopy contrast observable in the AFM phase image in **Figure 9.15c**, as marked by the dashed blue line. The brighter belongs to the SiC/G, and the darker arises from the SiC/Ga/BLG. For more detail, see ref. [43]. Although a precise explanation of the intercalation mechanism is yet needed to be understood [41,43], however, there is experimental evidence that the intrinsic SiC micropipes and additional holes induced by the growth process (**Figure 9.15b,c**) as "mediation sites" utilize the Ga intercalation without the necessity of further defect engineering, e.g., by using oxygen plasma. [41,43]

The VdP measurements were carried out in a cryostat system equipped with helium-4 ($^4$He) continuous flow that enables measurements at temperatures in the range from ~1.5 K to ~350 K under a homogenous magnetic field (B) between 0 to 250 mT provided by the Helmholtz coil (see Chapter **3** for more detail). For the measurement, the samples were first to cut into a square shape of 5 mm × 5 mm. The graphene/gallenene on top of the sample was isolated from the graphene on the side and the back (C-side) of the substrate by scribing cut-grooves on each side close to the edge of the sample (~0.1-0.2 mm from the edge) [38], as can be seen in **Figure 9.16a**. In the VdP setup, the sample was connected directly with gold pins which were softly approached and fixed on the sample by a holding upper cap close to the corners of the sample, as depicted by yellow circles and a notation letter (i.e., A, B, E, D) in **Figure 9.16a**. The proper functionality of all contacts was carefully checked via I-V measurement. From the linear ohmic characteristic of all the contacts in **Figure 9.16b**, a lateral homogeneity in the VdP is inferred. From the standard approach of Hall effect measurement in the VdP setup, that is already significant since it verifies the crystallinity of the grown sample over mm-scale; otherwise, the sample would have deemed improper for the VdP measurements. A slight difference in diagonal I-V measurements for AE and DB is observed, which could be due to a bit of inequality of the local contact resistances, however, its influence is negligible in the VdP measurement.

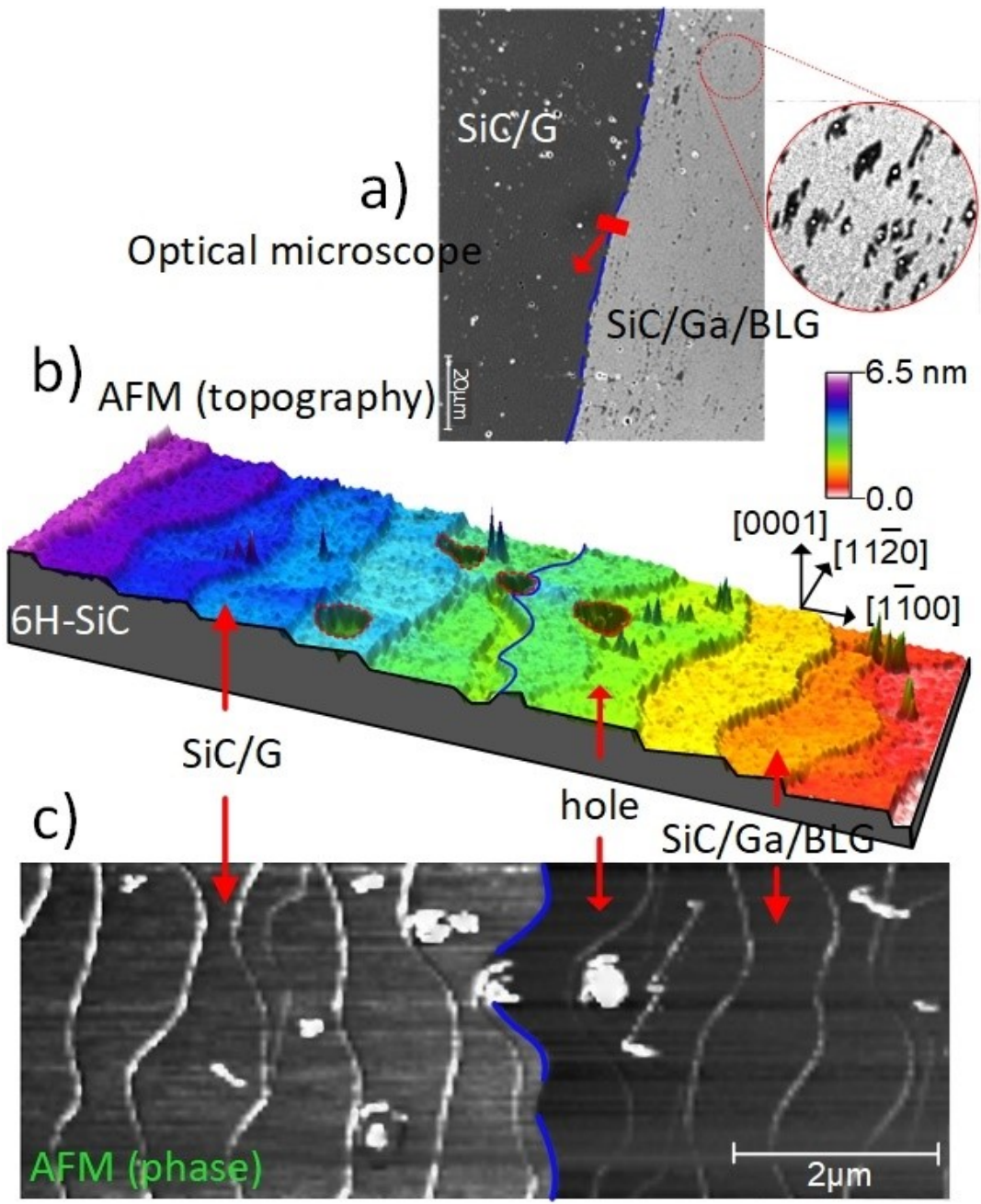

**Figure 9.15. Graphene-gallenene heterostack.**
(a) Optical microscopy of SiC/Ga/BLG during the Ga propagation. AFM topography (b) and phase (c) images of SiC/Ga/BLG heterostack. Phase contrast is observed as a result of different materials, i.e., the SiC/G (brighter on the left side) and SiC/Ga/BLG (darker on the right side) in the AFM phase image. The holes, as terminals where the formation of the gallenene initialized, are seen in topography and phase images.

For the Hall measurement, while the magnetic field switched on, the current was applied across one diagonal (i.e., DB or EA), and the voltage was measured along the other diagonal (i.e., EA or DB). For each case, the measurement was repeated for the inverted polarity of the magnetic field. The measurements were carried out in wide temperature and current ranges. **Figure 9.16c, d** demonstrate the VdP measurement performed at 1.7 K and 4.2 K both at a current of $I = 10$ μA. While the material system demonstrates a normal conductor state for $T > 4$ K (**Figure 9.16d**), interestingly, it exhibits a superconducting transition phase and Meissner state for $T < 4$ K.

This is the general feature of the sample in the presented measurements for different applied currents and temperatures up to 100 μA, and ~4 K, respectively. The SiC/Ga/BLG is settled in Meissner state for $B < {\sim}10$ mT (highlighted by the green circle in **Figure 9.16c**), and correspondingly, the Hall potential and resistance are zero. For $B > 10$ mT, the sample is in transitional phase (mixed state) before dropping into the normal phase where it behaves like a normal metal with the electron charge carrier ($n$-type) that can be inferred considering the ramp of the Hall resistance and magnetic field direction. In the mixed-phase, the sample indeed behaves like a type-II superconductor [376], wherein the resistance increases abruptly up to a second critical magnetic field. It begins then to pass to the normal phase. The slope of the mixed-phase is directly proportional to the applied current and carrier concentration in the system. Thereby, raising the applied current leads to a faster transition to the normal state. However, to what extent the temperature drifts is yet not known. This situation can be deduced from **Figure 9.16e**, which shows the VdP measurement (across EA) at 1.7 K for $I = 100$ μA (violet circles). Comparing the results in **Figure 9.16c and e**, it is seen when the current increases, the transition from superconducting to mixed-state occurs at lower B. Also, the resistance change in the mixed-state is slower. However, the slope of the Hall resistance in the normal phase remains almost unchanged for the whole range of current and correspondingly the charge carrier concentration ($n$ or $p$), mobility ($\mu$), and sheet resistance ($R_s$) too.

Considering the characteristic feature of the mixed-phase, it is important to take into account the main three components involved in the transverse Hall voltage: (i) geometrical misalignment of the contacts (ii) Hall voltage component, and (iii) contribution from the guided motion of vortex and antivortex. [377] The case (i) can be excluded, otherwise, its contribution could be detected as a non-zero Hall voltage at normal state measurements in zero magnetic fields. [378] However, the role of the two other contributors and possible interplays, e.g., a proximity effect, sign reversal effect [379], pinning effect [380], or weak coupling [381] for the observed behavior in the mixed-phase in this complex stacking system is not yet clear and requires further investigation.

The VdP method facilitates measuring the carrier density and mobility as a function of temperature in the sample.[210] Thereby, an $n \approx 1.3 \pm 0.1 \times 10^{12}$ cm$^{-2}$, and $\mu \approx 1400 \pm 50$ cm²/Vs ($T \approx 4.2$ K) are calculated under the conditions shown in **Figure 9.16d**. The same measurement at 50 K showed an $n \approx 1.8 \pm 0.1 \times 10^{12}$ cm⁻², and $\mu \approx 1000 \pm 50$ cm²/Vs, indicating that the mobility is temperature-dependent. The $n$-doping is a superposition of the carriers induced by the SiC as well as the metallic intercalation agent (Ga), which turns the free-standing top graphene bilayer to $n$-type but not $p$-type as it is the case, e.g., for hydrogen intercalation. [30,38] At room temperature, the mobility diminishes with a factor of approximately 5 times down to ~300 $\pm 50$ cm²/Vs with an $n \approx 4.5 \pm 0.1 \times 10^{12}$ cm⁻², following the saying "poor conductors at room temperature tends to be a better superconductor." This is in contrast to QFMLG, which its mobility is almost constant both at room and low temperatures. The electron-phonon coupling which mediates the superconducting behavior is also the main reason for the lower mobility of the gallenene sample measured at room temperature.

Not meeting gallenene, two other systems in the family of epitaxial graphene were investigated: one epitaxial SiC/G, which after applying H-intercalation, turned to QFBLG. **Figure 9.16g** shows the VdP measurements, performed on 5×5 mm² samples at both room- and cold temperatures.

From the slopes in **Figure 9.16g**, an $n$-type and $p$-type doping is deduced for the SiC/G and QFBLG samples, respectively. The so-called polarization doping originating from the bulk hexagonal SiC leads to negative pseudo-charges at the surface, which is compensated by holes in the quasi-freestanding graphene layers for the sake of charge neutrality, leading into $p$-doping. The polarization doping effect, however, is overcompensated by donor-like states from the buffer layer and interface states, which results in the Dirac point be located below the Fermi energy in epigraphene and thus its $n$-type conductivity (see section **2.4.4** for more information). [30,31]

Moreover, the SiC/Ga/BLG sample, in comparison with the QFBLG sample that was fabricated employing hydrogen intercalation in the same VdP measurement scales [38], exhibits much lower mobility; both at room and low temperatures, indicating higher scattering in SiC/Ga/BLG system. Additionally, the QFBLG graphene sample does not show any superconducting effect in the absence of 2D-Ga, which excludes the participation of low-angle twisted BLG effect in the samples. [382] Also, the significantly lower carrier density in our system compared to lithium [45] and calcium [283] doped SiC/G samples, rules out the argument of strong electron-phonon coupling as the origin of the superconductivity in the SiC/Ga/BLG. Therefore, it is concluded that the 2D-Ga is responsible for the superconductivity, which is in excellent agreement with the recent valuable study by Briggs et al. [41]

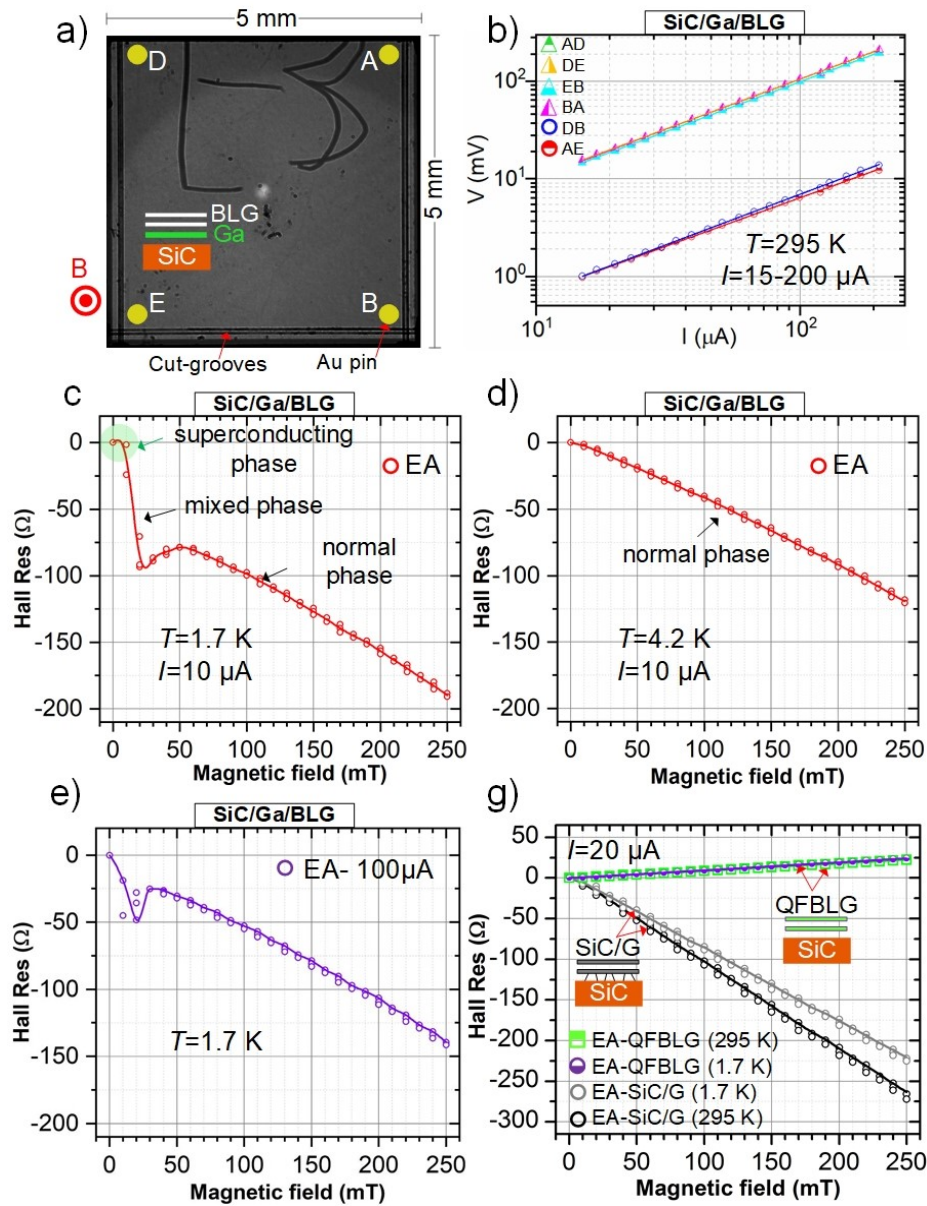

**Figure 9.16. The VdP measurement on bilayer graphene-gallenene heterostack.**

(a) Optical microscopy of SiC/Ga/BLG sample diced into 5 mm × 5mm and isolated from the graphene on the side and back of the sample by cut-grooves. Yellow circles represent the gold contacts softly approached to those corners of the sample and fixed by an upper cap. b) The linear ohmic feature of all I-V measurements (in total 8 configurations) indicates the proper functionality of the gold contacts. c) The Hall effect measurement (diagonal EA) shows a superconducting phase for B < 10 mT (marked with a green circle), a mixed state, and finally, a normal state. (d) The superconducting phase is broken for $T$ > 4 K. (e) The dependency of the superconducting phase and mixed-phase to the driven current. (See text for more details). (g) VdP measurement on SiC/G and after H-intercalation on SiC/QFBLG samples for both RT and cold. The measurements demonstrate the typical conductor phase in the absence of 2D-Ga. In all curves, the circles are actual measured values, and lines are interpolation results.

In summary, a dc Hall measurement in Van der Pauw configuration of a large-area 2D-gallium sheet was presented. The 2D-Ga sheet was grown on epigraphene, forming a stacking system of SiC/Ga/BLG. The measurements show the superconducting nature of 2D-Ga at $T < 4\,\text{K}$, in agreement with another recent study performed on a μm-scale. [41] The Hall effect measurements give evidence of type-II like superconducting behavior. [376] It is shown that the phase transition between the normal and superconductive states is not the first order in the presence of a magnetic field but up to four-order-of-magnitude. The complex behavior of the stacking system in the intermediate state is beyond the scope of the presented work and motivates further experimental and theoretical investigations. The high charge mobility value highlights the high-quality of the sample over mm-scales. Also, from the linear ohmic I-V characteristic of contacts, the high homogeneity of the sample is deduced, given further proof of the large-area growth of the 2D-Ga sheet. This study supports the potential metallic contacts in 2D-device electronic applications. Moreover, the atop graphene layers can prospectively be implemented for proceeding the creation of other exotic layers enabling exploring physical and electronic properties in such combined hybrid heterostack systems.

# List of acronyms

| | |
|---|---|
| **2D** | Two-dimensional |
| **2DEG** | Two-dimensional electron gas |
| **3D** | Three-dimensional |
| **AC** | Alternating current |
| **AFM** | Atomic force microscopy |
| **ALD** | Atomic layer deposition |
| **AlGaAs** | Aluminum gallium arsenide |
| **AM-KPFM** | Amplitude-modulated kelvin probe force microscopy |
| **ARPES** | Angle-resolved photoemission spectroscopy |
| **BF-LEEM** | Bright-field low-energy electron microscopy |
| **BLG** | Bilayer graphene |
| **BLG/Ga/SiC** | Bilayer graphene/gallium/silicon carbide heterostack |
| **BSE** | Backscattered electrons |
| **CCC** | Cryogenic current comparator |
| **CMP** | Chemical mechanical polishing |
| **CVD** | Chemical vapor deposition |
| **DC** | Direct current |
| **DF-LEEM** | Dark-field low-energy electron microscopy |
| **DFT** | Density functional theory |
| **DUV** | Deep ultraviolet |
| **ECP** | Electrochemical potential |
| **EG** | Epitaxial graphene |
| **epigraphene** | Epitaxial graphene grown on SiC (0001) |
| **F4-TCNQ** | Tetrafluoro-tetracyanoquinodimethane |
| **FET** | Field-effect transistor |
| **FLG** | Few-layer graphene |
| **FoV** | Field of view |
| **FWHM** | Full width half maximum |
| **GaAs** | Gallium arsenide |
| **H-Int.** | Hydrogen intercalated |
| **HOPG** | Highly Ordered Pyrolytic Graphite |
| **HRTEM** | High-resolution transmission electron microscopy |
| **IFL/BFL** | Interfacial layer (or buffer layer) |
| **IQHE** | Integer quantum Hall effect |
| **JCH** | Joint cubic-hexagonal |
| **KPFM** | Kelvin probe force microscopy |
| **LDOS** | Local density of states |
| **LEED** | Low energy electron diffraction |
| **LEEM** | Low-energy electron microscopy |
| **LLs** | Landau levels |
| **LPD** | Liquid phase deposition |
| **μ-LEED** | Micro-low energy electron diffraction |
| **MLG** | Monolayer graphene |
| **MPD** | Micro-pipe density |
| **N4PP** | Nano-four-point probe |
| **O-Int.** | Oxygen intercalated |
| **PASG** | Polymer assisted sublimation growth |
| **PMMA** | Poly-methyl-methacrylate |
| **PTB** | Physikalisch-Technische Bundesanstalt |
| **QFBLG** | Quasi-freestanding bilayer graphene |
| **QFMLG** | Quasi-freestanding monolayer graphene |
| **QFSG** | Quasi-freestanding graphene |
| **QHE** | Quantum Hall effect |
| **QHR** | Quantum Hall resistance |
| **RMS** | Root mean square |
| **RT** | Room temperature ($\approx$ 20 °C) |
| **SdH** | Shubnikov de Haas oscillation |
| **SEM** | Scanning electron microscopy |
| **SG** | Sublimation growth |
| **SiC** | Silicon carbide |
| **SiC/G** | Monolayer graphene on silicon carbide |
| **SPA-LEED** | Spot profile analysis low energy electron diffraction |
| **STM** | Scanning tunneling microscopy |
| **STP** | Scanning tunneling potentiometry |
| **STS** | Scanning tunneling spectroscopy |
| **TMDs** | Transition metal dichalcogenide |
| **UHV** | Ultra-high vacuum (pressure < $1 \times 10^{-9}$ mbar) |
| **vDOS** | Vibrational density of states |
| **VdP** | Van der Pauw |
| **XPEEM** | X-ray photoemission electron spectroscopy |
| **XPS** | X-ray photoelectron spectroscopy |

# List of symbols

| | |
|---|---|
| $\Psi$ | Wave function |
| $a$ | Lattice constant |
| $\phi_0$ | Flux quantum |
| $v_F$ | Fermi velocity |
| $\omega_C$ | Cyclotron frequency |
| $E$ | Energy |
| $E_F$ | Fermi level |
| $E_D$ | Energetic location of Dirac point |
| $E_g$ | Bandgap |
| $e$ | Elementary charge |
| $A_{Hall}$ | Hall coefficient |
| $\phi$ | Work function |
| $h$ | Planck constant |
| $\hbar$ | $h/2\pi$ |
| $I$ | Current |
| $I_t$ | Tunneling current |
| $I_D$ | Intensity of Raman D peak |
| $I_G$ | Intensity of Raman G peak |
| $v$ | Frequency |
| $\nu$ | Filling factor |
| $\mu$ | Charge carrier mobility |
| $n$ | Charge carrier density, electron density |
| $n_S$ | Two-dimensional sheet carrier density |
| $n_{LL}$ | Number of localized carriers |
| $p$ | Hole density |
| **q** | Elementary charge |
| *R* | Resistance |
| *R*$_{SH}$ | Sheet resistance |
| $\rho$ | Resistivity |
| $\rho_{xx}$ | Longitudinal resistivity |
| $\rho_{xy}$ | Transverse (Hall) resistivity |
| *T* | Temperature |
| *T*g | Glass transition temperature |
| **T** | Tesla |
| **t** | Thickness |
| *t* | Time |
| *P* | Pressure |
| *B* | Magnetic field |
| **m*** | Effective mass |
| *E*$_0$ | Energy of a Si–C pair without interaction between neighboring layers |
| *F*$_{es}$ | Electrostatic force |
| *k*$_B$ | Boltzmann constant |
| *n*$_{def}$ | Defect density |

# Curriculum vitae
## Davood Momeni

## Education

**Leibniz University Hannover**, Hannover, Germany
– Ph.D. Physics, 10/2016 – 09/2020

**Bremen University of Applied Sciences**, Bremen, Germany
– M.Sc. Electronics Engineering (Microsystems), 11/2013 – 12/2015

**Qazvin Azad University**, Qazvin, Iran
– B.Sc. Electrical Engineering (Electronics), 10/2004 – 07/2009

## Professional experience

**Physikalisch-Technische Bundesanstalt (PTB),** Brunswick, Germany
– Scientific researcher, since 10/2015

**Physikalisch-Technische Bundesanstalt (PTB),** Brunswick, Germany
– Student assistant, 03/2015 – 12/2015

## Awards & scholarships during graduate studies

– Best poster prize in the category "Synthesis & Growth"at Graphene Week 2018, European Union Graphene Flagship

– Scholarship holder – School for contacts in nanosystems (NTH-nano), 10/2015 – 10/2018

– Top-student award – QAU, 2009

# List of publications

### 1. Silicon carbide stacking-order-induced doping variation in epitaxial graphene

<u>**Davood Momeni Pakdehi**</u>, Philip Schädlich, Nguyen, T. T. Nhung Nguyen, Alexei. A. Zakharov, Stefan Wundrack, Emad Najafidehaghani, Florian Speck, Klaus Pierz, Thomas Seyller, Christoph Tegenkamp, Hans W. Schumacher
**Advanced Functional Materials**, 2004695 (2020)
DOI: 10.1002/adfm.202004695

**2. Minimum resistance anisotropy of epitaxial graphene on SiC**

<u>**Davood Momeni Pakdehi**</u>, Johannes Aprojanz, Anna Sinterhauf, Klaus Pierz, Mattias Kruskopf, Phillip Willke, Jens Baringhaus, Phillip Stöckmann, Georg A. Traeger, Frank Hohls, Christoph Tegenkamp, Martin Wenderoth, Franz J. Ahlers, Hans W. Schumacher
**ACS Applied Materials and Interfaces**, 10, 6, 6039-6045 (2018)
DOI: 10.1021/acsami.7b18641

**3. Homogeneous large-area quasi-freestanding monolayer and bilayer graphene on SiC**

<u>**Davood Momeni Pakdehi**</u>, Klaus Pierz, Stefan Wundrack, Johannes Aprojanz, Nguyen, Thi Thuy Nhung, Thorsten Dziomba, Frank Hohls, Andrey Bakin, Rainer Stosch, Christoph Tegenkamp, Franz J. Ahlers, Hans W. Schumacher
**ACS Applied Nano Materials**, 2, 2, 844-852 (2019)
DOI: 10.1021/acsanm.8b02093

**4. Substrate induced nanoscale resistance variation in epitaxial graphene**

Anna Sinterhauf, Georg A. Traeger, <u>**Davood Momeni Pakdehi**</u>, Philip Willke, Klaus Pierz, Frank Hohls, Franz J. Ahlers, Hans W. Schumacher, Thomas Seyller, Cristoph Tegenkamp, Martin Wenderoth
**Nature Communications** 11, 555 (2020)
DOI: 10.1038/s41467-019-14192-0

**5. Comeback of epitaxial graphene for electronics: large-area growth of bilayer-free graphene on SiC**

Mattias Kruskopf, <u>**Davood Momeni Pakdehi**</u>, Klaus Pierz, Stefan Wundrack, Rainer Stosch, Thorsten Dziomba, Martin Götz, Jens Baringhaus, Johannes Aprojanz, Christoph Tegenkamp, Frank Hohls, Franz J. Ahlers, Hans W. Schumacher
**IOP 2D Materials**, 3, 4, 041002 (2016)
DOI: 10.1088/2053-1583/3/4/041002

**6. Probing the structural transition from buffer layer to quasi-freestanding monolayer graphene by Raman spectroscopy**

Stefan Wundrack, <u>**Davood Momeni Pakdehi**</u>, Philip Schädlich, Florian Speck, Klaus Pierz, Thomas Seyller, Hans W. Schumacher, Andrey Bakin, Rainer Stosch
**Physical Review B**, 99, 4, 045443 (2019)
DOI: 10.1103/PhysRevB.99.045443

**7. Traceably calibrated scanning Hall probe microscopy at room temperature**

Manuela Gerken, Aurélie Solignac, <u>**Davood Momeni Pakdehi**</u>, Thomas Weimann, Klaus Pierz, Sibylle Sievers, and Hans Werner Schumacher
arXiv: 1910.12676 (2019)

**8. A morphology study on the epitaxial growth of graphene and its buffer layer**

Mattias Kruskopf, Klaus Pierz, **<u>Davood Momeni Pakdehi</u>**, Stefan Wundrack, Rainer Stosch, Andrey Bakin, Hans W. Schumacher
**Thin Solid Films**, 659, 7-15 (2018)
DOI: 10.1016/j.tsf.2018.05.025

**9. Liquid metal intercalation of epitaxial graphene: large-area gallenene layer fabrication through gallium self-propagation at ambient conditions**

Stefan Wundrack, **<u>Davood Momeni Pakdehi</u>**, Nils Schmidt, Klaus Pierz, Lena Michaliszyn, Hendrik Spende, Angelika Schmidt, Hans Schumacher, Rainer Stosch, Andrey Bakin
arXiv: 1905.12438 (2019)

**10. The Van Der Pauw investigation on large-area SiC/Ga/BLG superconducting heterostack**

**<u>Davood Momeni Pakdehi</u>**, Stefan Wundrack, Klaus Pierz, Rainer Stosch, Andrey Bakin, Hans W. Schumacher
(in preparation)

**Conference and workshop presentation during graduate studies**
- 12 contributed oral presentations as a first author
- 14 contributed poster presentations as a first author

# Acknowledgments

Since I have ever achieved anything worthwhile without help, I would like to thank all the people who made this thesis possible and supported my studies.

I would like to extend my deepest gratitude to Prof. Dr. Hans W. Schumacher and Prof. Dr. Rolf J. Haug for giving me excellent opportunities and guiding me throughout these years of academic research at the PTB and the Leibniz University Hannover. It was an incredibly beautiful time, and I learned a lot. I also would like to sincerely thank Prof. Dr. Christoph Tegenkamp for his support, fruitful discussions, and the many excellent pieces of advice that have contributed to the success of our joint publications. I am also very thankful to Prof. Dr. Siegner for the support of the projects and my position.

I would like to particularly thank Dr. Klaus Pierz for his continuous support, valuable discussions, and most importantly, his patience and friendship during my studies at the PTB. I have also learned and improved a lot my German during our discussions, Klaus, vielen Dank!

I also had the pleasure of collaborating with different excellent groups that I would like to address my thanks. Special thanks to Prof. Dr. Thomas Seyller and his group, Philip Schädlich, and Dr. Florian Speck from the University of Chemnitz for precious discussions, LEEM, and XPS measurements, and our successful joint publications. Many thanks to Prof. Dr. Christoph Tegenkamp and his group, Dr. Johannes Aprojanz, T. T. N. Nguyen, Jan P. Stöckmann, and Dr. Jens Baringhaus from University of Chemnitz/ Leibniz University Hannover for SEM, STM, and nano-four-point-probe measurements. I would like to thank PD. Dr. Martin Wenderoth and his group, Anna Sinterhauf, Georg A. Traeger, Dr. Philip Willke from the University of Göttingen for productive discussions, and STM/STP measurements. Thanks extend to Prof. Dr. Winfried Daum and his group, Wanja Dziony, Dr. Gerhard Lilienkamp, from the Technical University of Clausthal for XPS measurements. Many thanks to Prof. Dr. Andrey Bakin from TU Braunschweig for the valuable discussion and his willingness to collaborate. I am also very grateful to Dr. Alexei A. Zakharov for the indispensable discussions, LEEM, and XPEEM measurements.

I extend my gratitude to my colleagues at the PTB for their camaraderie, kindness, and support. I would like to express my thanks to Thorsten Dziomba, Stefan Wundrack, Dr. Mattias Kruskopf, and Dr. André Müller. It was a great pleasure for me to work with you. I could always rely on you, and that means a lot to me.

And of course, my thanks go to the other entire panel of experts at the PTB for their assistance in different parts of my thesis: Dr. Frank Hohls, Dr. Franz Josef

Ahlers, Dr. Thomas Weimann, Dr. Rainer Stosch, Dr. Martin Götz, Dr. Hansjorg Scherer, Eckart Pesel, Emad Najafidehaghani, Peter Hinze, Kathrin Störr, Bert Egeling, Holger Marx, Cristiano Cognolato, Judith Felgner, Christine Becker, and Katrin Volkmer.

I greatly acknowledge the scholarship from the school for contacts in nanosystems (NTH). This work was supported in part by the Joint Research Project "GIQS" (18SIB07). This project also received funding from the European Metrology Program for Innovation and Research (EMPIR) co-financed by the Participating States and from the European Unions' Horizon 2020 research and innovation program. The work was also co-funded by the Deutsche Forschungsgemeinschaft under Germany's Excellence Strategy – EXC-2123 Quantum Frontiers (390837967).

Last but not least, I would like to thank my parents, Zahra and Mahmood, lovely sister Taraneh, my dear brothers Mabood and Farzad, for their encouragement, support, and unconditional love. I would like to express my thanks to my beautiful wife, Swantje; thank you for your support and inspiration. Along with you, I thank our son, Karl Momeni, a great source of love and motivation. I love you!

## Selbstständigkeitserklärung

Hiermit versichere ich, die vorliegende Arbeit selbstständig verfasst und keine anderen als die angegebenen Quellen und Hilfsmittel benutzt zu haben.

Hannover, 17. Juli 2020 Davood Momeni Pakdehi